\documentclass[a4paper,12pt]{report}
\usepackage[utf8x]{inputenc}
\usepackage{a4wide}
\usepackage{anysize}
\usepackage{latexsym}
\usepackage{amssymb}
\usepackage{epsfig}
\usepackage{tabls}
\usepackage[dvips]{color}
\usepackage{url}
\usepackage{moreverb}
\usepackage{colortbl}
\usepackage{amsmath}
\usepackage{color}
\usepackage{colordvi}
\usepackage{lscape}
\usepackage{psfrag}
\usepackage{geometry}
\usepackage{float}
\usepackage{amsmath,amssymb}
\usepackage{appendix}
\usepackage{subfig}
\begin{document}

\def\title{Maneuvering, Multi-Target Tracking using Particle Filters}
\def \degree{Master of Technology}
\def\branch{Communication and Signal Processing}
\def\who{T M Feroz Ali}
\def\guide{Prof. V Rajbabu}

\newgeometry{top=30mm, bottom=22mm}
\begin{titlepage}
\large
\begin{center}
{\LARGE \bf \title}\\[0.5cm]
\textit{Submitted in partial fulfillment of the requirements}\\
   \textit{for the degree of}\\[0.3cm]
{\bf  \degree}\\
in \\
{\bf  \branch}\\[0.3cm]
by \\[0.2cm]
{\bf \who}\\
under the guidance of\\[0.4cm]
{\bf \guide}\\[0.5cm]
\includegraphics[width=2.5in]{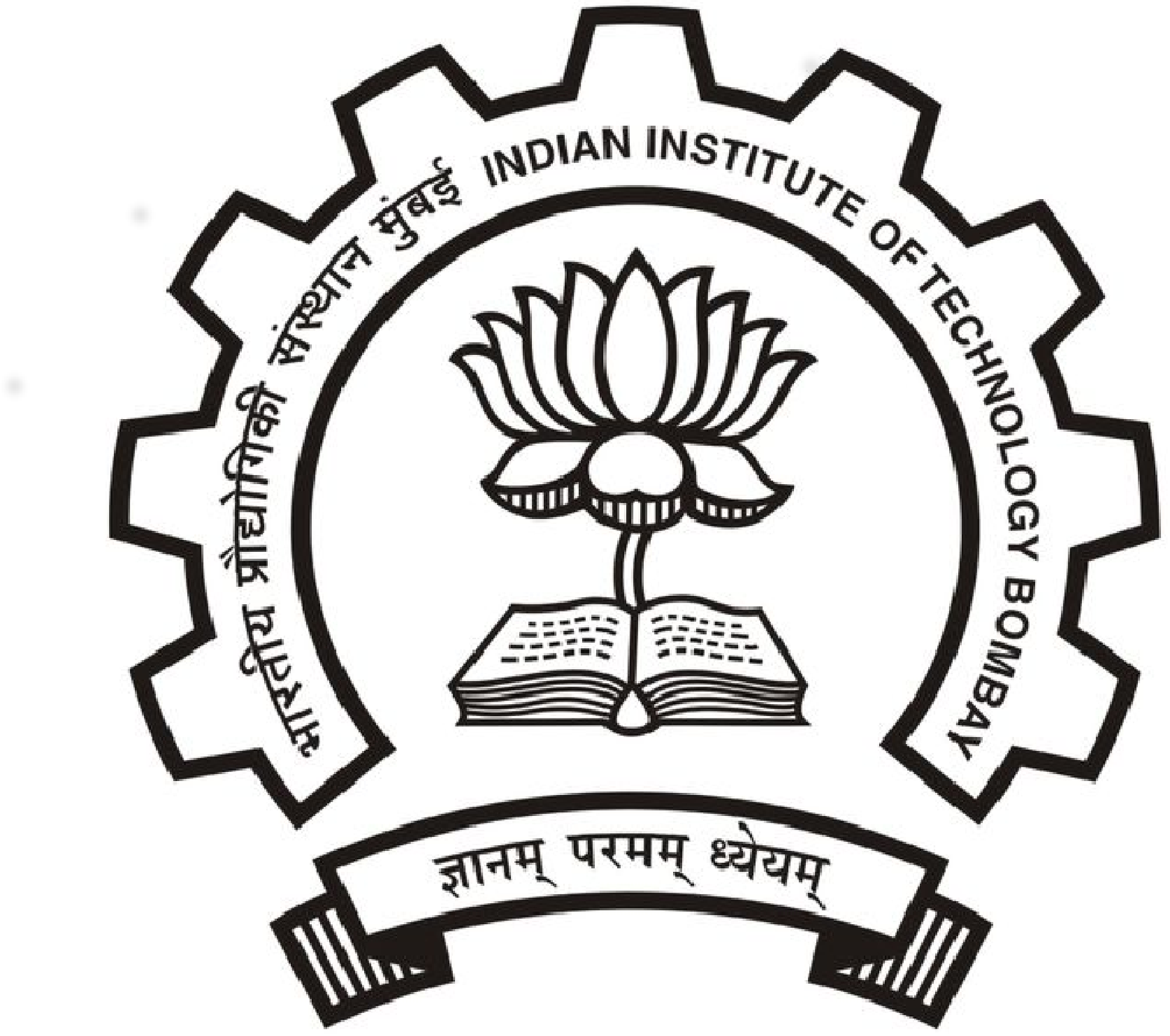}\\[0.5cm]
Department of Electrical Engineering \\ 
Indian Institute of Technology, Bombay \\ 
June, 2012
\end{center}
\end{titlepage}

\newgeometry{top=30mm, bottom=22mm, left=30mm, right=20mm, textheight=245mm, textwidth=160mm, headheight=3mm, headsep=12mm, footskip=10mm}
\pagenumbering{roman}


\chapter*{Acknowledgment}
I would like to thank my guide Prof. V Rajbabu for his valuable guidance, encouragement
and patience during the work.
\par
I would like to express my sincere thanks to Naval Research Board (NRB) for providing the necessary funding for the project. I am thankful to Bharti Center for Communication for providing me with the resources for my project activities. I
have used the resources of Bharti center to the fullest and I would like to thank them for providing
me with all the facilities.
\par
Finally I would like to thank everybody who have helped me in every way in my project
enabling me to complete this work.

\begin{flushright}\hspace{248pt}\textbf{T M Feroz Ali}\end{flushright}

\newpage
\tableofcontents
\newpage
\newpage
\listoffigures
\newpage

\bibliographystyle{IEEEtran}
\newpage
\chapter*{Abstract}
The aim of the project is to develop tracking and estimation techniques relevant to underwater targets. The received measurements of the targets have to be processed using the models of the target dynamics to obtain better estimates of the target states like position, velocity etc. This work includes exploration of particle filtering techniques for target tracking. Particle filter is a numerical approximation method for implementing a recursive Bayesian estimation procedure. It does not require the assumptions of linearity and Guassianity like the traditional Kalman filter (KF) based techniques. Hence it is capable of handling non-Gaussian noise distributions and non-linearities in the target's measurements as well as target dynamics. The performance of particle filters is verified using simulations and compared with EKF. 

Particle filters can track maneuvering targets by increasing the number of particles. Particle filter have higher computational load which increases in the case of multi-targets and highly maneuvering targets. The efficient use of particle filters for multi-target tracking using Independent Partition Particle Filter (IPPF) and tracking highly maneuvering targets using Multiple Model Particle Filter(MMPF) are also explored in this work. These techniques require only smaller number of particles and help in reducing the computational cost. The performance of these techniques are also simulated and verified.

Data association problem exists in multi-target tracking due to lack of information at the observer about the proper association between the targets and the received measurements. The problem becomes more involved when the targets move much closer and there are clutter and missed target detections at the observer. Monte Carlo Joint Probabilistic Data Association Filter (MCJPDAF) efficiently solves data association during the mentioned situation. MC-JPDAF also incorporates multiple observers. Its performance is simulated and verified. Due to the inability of the standard MCJPDAF to track highly maneuvering targets, Monte Carlo Multiple Model Joint Probabilistic Data Association Filter (MC-MMJPDAF) which combines the technique of Multiple Model Particle Filter(MMPF) in the framework of MC-JPDAF has been proposed. The simulation results shows the efficiency of the proposed method. 

The results from the silmulation of particle filter based methods show that it handles maneuvering, multiple target tracking and has been verified with some field data. 

\newpage
\pagenumbering{arabic}

\chapter{Introduction}
\section{Need for Estimation}
The need for estimation arises since the measurements of a system may be noisy, incomplete, or delayed and the exact modeling of the system is not always possible. Depending only on the measurements is not feasible in cases when measurements are highly noisy, the delay between the occurrence of the process and the time of arrival of measurement is large, some of the states of the system are not observable, clutters exist along with the targets, targets-measurements association ambiguity exists etc. Estimation can help in getting filtered inferences with lesser variance than from the noisy measurements, predict about the system behavior in the future etc. Hence estimation techniques are needed to infer better about the system with the given system models and measurements.

\section{Objective}
The objective of target tracking is to continuously estimate and track the states of the target like position, velocity, acceleration, etc, using the available measurements of the target. The target's motion may be one or two dimensional and can have constant velocity or maneuvering motions also.  The initial state of the target may be unknown. The possible motion models of the target is assumed to be known. There may be multiple targets which may be closer or far apart. The measurrement sensor is assumed to be stationary. The measurements of the target may be available as range, bearing and/or Doppler frequency measurements. The accuracy and noise distribution of the measurement sensors are also assumed to be known.

\section{Mathematical Formulation}
Given a discrete stochastic model of a dynamic system (moving target) using a state space representation
\begin{eqnarray}
\mathbf{x}_k & = & \mathbf{f}_{k-1}(\mathbf{x}_{k-1},\mathbf{w}_{k-1}) \label{eqn:state_model} \\
\mathbf{z}_k & = & \mathbf{h}_{k}(\mathbf{x}_{k},\mathbf{v}_{k}) \label{eqn:measurement_model} 
\end{eqnarray}
where $k$ is the time index, $\mathbf{x}_k$ is the state vector, $\mathbf{w}_k$ is the process noise, $\mathbf{z}_k$ is the measurement of the target, $\mathbf{v}_k$ is the measurement noise, $\mathbf{f}_k(\cdotp)$ is the time varying system, $\mathbf{h}_k(\cdotp)$ is the measurement equation and $T$ is the sampling interval of the discrete system, the task is to recursively estimate the state $\mathbf{x}_k$ of the system from its available measurements $\mathbf{Z}_k = \mathbf{z}_{1:k}\equiv \{\mathbf{z}_i;i=1,...,k\}$.
The state vector $\mathbf{x}_k$ contains all the information required to describe the target dynamics. The noise sequences $\mathbf{w}_k$ and $\mathbf{v}_k$ are assumed to be zero mean, white noise and mutually independent with known probability distribution function. The initial target state distribution $p(\mathbf{x}_0)$ is assumed to be known and to be independent of the noise sequences $\mathbf{w}_k$ and $\mathbf{v}_k$ \cite{4,7,9,11}.
Two fundamental assumptions about the system are that the dynamic variable $\mathbf{x}_k$ is Markov of order one.
\begin{equation}
p(\mathbf{x}_k|\mathbf{x}_{1:k-1},\mathbf{z}_{1:k-1})=p(\mathbf{x}_k|\mathbf{x}_{k-1})
\label{eqn:markov1} 
\end{equation}
and $\mathbf{z}_k$ is conditionally independent of past states and measurements.
\begin{equation}
p(\mathbf{z}_k|\mathbf{x}_{1:k},\mathbf{z}_{1:k-1})=p(\mathbf{z}_k|\mathbf{x}_{k})
\label{eqn:markov2} 
\end{equation}\\
where $\mathbf{x}_{1:k}\equiv \{\mathbf{x}_i;i=1,...,k\}$. 

\section{Measurements and System Models}
The states of the targets considered in this report are the positions and velocities $x, y, v_x, v_y$ in the cartesian co-ordinate system.
\begin{equation}
\mathbf{x}=
\begin{bmatrix}
x & v_{x} & y  & v_{y} 
 \end{bmatrix}^T
\end{equation}
 The motion models of the target considered are constant velocity model and constant turn rate model. The constant velocity model is described by 
\begin{align}
\mathbf{x}_{k} & = f(\mathbf{x}_{k-1})+\mathbf{w}_{k-1}\\
& = F_1 \mathbf{x}_{k-1}+\mathbf{w}_{k-1}
\end{align}
where $F_1$ is a matrix given by
\begin{equation}
 F_1=\begin{bmatrix}
1&T&0&0\\
0&1&0&0\\
0&0&1&T\\
0&0&0&1\\
\end{bmatrix}
\end{equation}
where $T$ is the sampling period of the target dynamics. The constant turn rate model with turn rate $\Omega$ $rad/s$ is given by 
\begin{align}
\mathbf{x}_{k} & = f(\mathbf{x}_{k-1})+\mathbf{w}_{k-1}\\
& = F_2 \mathbf{x}_{k-1}+\mathbf{w}_{k-1}
\end{align}
where $F_2$ is a matrix given by
\begin{equation}
F_2=
\begin{bmatrix}
1    &\dfrac{\sin (\Omega T)}{\Omega}       &0    &-\dfrac{1 - \cos(\Omega  T)}{\Omega}\\
        0    &\cos(\Omega  T)          &0     &-\sin(\Omega  T)\\
        0    &\dfrac{1-\cos(\Omega  T)}{\Omega}   &1      &\dfrac{\sin(\Omega  T)}{\Omega}\\
        0    &\sin(\Omega  T)          &0      &\cos(\Omega  T) \\
\end{bmatrix} 
\end{equation} 
where $T$ is the sampling period of the target dynamics. The available measurements of the target considered in this report are range and bearing.
They are related to the target states by the measurement model:
\begin{align}
z_{k} & = h(\mathbf{x}_{k})+\mathbf{v}_k\\
& = \begin{bmatrix}
\sqrt{x^2_k+y^2_k}\\
\tan^{-1}\left(\dfrac{y_k}{x_k}\right)
\end{bmatrix} + \mathbf{v}_k
\end{align}

\section{Organization of Report}
The chapter \ref{chap:Bayesian} describes about the Bayesian estimation and the conceptual solution for recursive Bayesian estimation. Particle filter which is a numerical Monte Carlo approximation method for the implementation of the recursive Bayesian solution is described in chapter \ref{chap:Particle_filtering}. An advanced particle filtering technique called Independent Partition Particle Filter (IPPF) for tracking multiple targets efficiently is described in chapter \ref{chap:IPPF}. In chapter \ref{chap:MMPF}, we have discussed Multiple Model Particle filter (MMPF) which is used for tracking highly maneuvering targets.
\chapter{Bayesian Estimation}
\label{chap:Bayesian}
The Bayesian approach to estimate the state $\mathbf{x}_k$ from the measurements $\mathbf{Z}_k$ is to calculate the posterior distribution of $\mathbf{x}_k$ conditioned on the measurements $\mathbf{Z}_k$. This conditional pdf is denoted as $p(\mathbf{x}_k|\mathbf{Z}_k)$. The estimation based on this posterior distribution is called Bayesian because it is constructed using Bayes rule.
\begin{equation}
p(\mathbf{x}_k|\mathbf{Z}_k)=\dfrac{p(\mathbf{Z}_k|\mathbf{x}_k)p(\mathbf{x}_k)}{p(\mathbf{Z}_k)}  
\end{equation}
where $p(\mathbf{x}_k)$ is the prior target distribution, $p(\mathbf{Z}_k|\mathbf{x}_k)$ is the measurement likelihood (measure of how likely the measurement is true, given the state), $p(\mathbf{Z}_k)$ is called the evidence which is a normalizing factor. Once $p(\mathbf{x}_k|\mathbf{Z}_k)$ is estimated, then we can estimate the statistical properties of the estimate of the target such as mean, median, covariance, etc.

\section{Recursive Bayesian Estimation}
The requirement is to recursively compute the posterior target density $p(\mathbf{x}_k|\mathbf{Z}_k)$ whose computation requires only the estimated target density at the previous time $p(\mathbf{x}_{k-1}|\mathbf{Z}_{k-1})$ and the current measurement $\mathbf{z}_k$. No history of observations or estimates is required. The first measurement is obtained at $k=1$. Hence the initial density of the state ${\mathbf{x}_0}$ can be written as 
\begin{equation}
p(\mathbf{x}_0)=p(\mathbf{x}_0|\mathbf{Z}_0)  
\end{equation}
where $\mathbf{Z}_0$ is the set of no measurements. The conditional pdf $p(\mathbf{x}_k|\mathbf{Z}_{k-1})$ can be written as
\begin{align}
p(\mathbf{x}_{k}|\mathbf{Z}_{k-1})& = \int p[(\mathbf{x}_k,\mathbf{x}_{k-1})|\mathbf{Z}_{k-1}]d\mathbf{x}_{k-1}  \\
& = \int p(\mathbf{x}_k|\mathbf{x}_{k-1},\mathbf{Z}_{k-1})p(\mathbf{x}_{k-1}|\mathbf{Z}_{k-1})d\mathbf{x}_{k-1}
\label{eqn:condpdf1} 
\end{align}
But according to \eqref{eqn:markov1}, under the Markovian assumption the state $\mathbf{x}_k$ is determined only by $\mathbf{x}_{k-1}$ and $w_{k-1}$. Hence  \eqref{eqn:condpdf1} can be written as 
\begin{equation}
p(\mathbf{x}_{k}|\mathbf{Z}_{k-1})=\int p(\mathbf{x}_k|\mathbf{x}_{k-1})p(\mathbf{x}_{k-1}|\mathbf{Z}_{k-1})d\mathbf{x}_{k-1}
\label{eqn:condpdf2} 
\end{equation}
The pdf $p(\mathbf{x}_k|\mathbf{x}_{k-1})$ is referred to as the transitional density and is available from the system equation $f_k(\cdotp)$ and the process noise $w_k$. The pdf $p(\mathbf{x}_{k-1}|\mathbf{Z}_{k-1})$ is available at the initial time as $p(\mathbf{x}_0|\mathbf{Z}_0)$. Then the posterior conditional pdf of $\mathbf{x}_k$,  $p(\mathbf{x}_k|\mathbf{Z}_k)$ can be written as
\begin{align}
p(\mathbf{x}_{k}|\mathbf{Z}_{k})& = p(\mathbf{x}_{k}|\mathbf{z}_k,\mathbf{Z}_{k-1}) \\
& = \dfrac{p(\mathbf{x}_{k},\mathbf{z}_k,\mathbf{Z}_{k-1})}{p(\mathbf{z}_{k},\mathbf{Z}_{k-1})}\label{eqn:condpdf3} \\
& = \dfrac{p(\mathbf{z}_{k}|\mathbf{x}_k,\mathbf{Z}_{k-1})p(\mathbf{x}_k|\mathbf{Z}_{k-1})p(\mathbf{Z}_{k-1})}{p(\mathbf{z}_{k}|\mathbf{Z}_{k-1})p(\mathbf{Z}_{k-1})}\label{eqn:condpdf4} \\
& = \dfrac{p(\mathbf{z}_{k}|\mathbf{x}_k,\mathbf{Z}_{k-1})p(\mathbf{x}_k|\mathbf{Z}_{k-1}))}{p(\mathbf{z}_{k}|\mathbf{Z}_{k-1})}\label{eqn:condpdf5} \\
& = \dfrac{p(\mathbf{z}_{k}|\mathbf{x}_k)p(\mathbf{x}_k|\mathbf{Z}_{k-1}))}{p(\mathbf{z}_{k}|\mathbf{Z}_{k-1})} \label{eqn:condpdf6} 
\end{align}
In \eqref{eqn:condpdf3} and \eqref{eqn:condpdf5}, Bayes rule is used and in \eqref{eqn:condpdf6}, \eqref{eqn:markov1} is used. The pdf $p(\mathbf{z}_k|\mathbf{x}_k)$ can be obtained using the measurement equation $h(\cdotp)$. The pdf $p(\mathbf{x}_k|\mathbf{Z}_{k-1})$ is available from \eqref{eqn:condpdf2}. The pdf $p(\mathbf{z}_k|\mathbf{Z}_{k-1})$ which is a normalizing constant, may be obtained as follows.
\begin{align}
p(\mathbf{z}_{k}|\mathbf{Z}_{k-1})& = \int p(\mathbf{z}_k,\mathbf{x}_k|\mathbf{Z}_{k-1})d\mathbf{x}_{k}  \\
& = \int p(\mathbf{z}_k|\mathbf{x}_k,\mathbf{Z}_{k-1})p(\mathbf{x}_k|\mathbf{Z}_{k-1})d\mathbf{x}_{k}  \\
& = \int p(\mathbf{z}_k|\mathbf{x}_k)p(\mathbf{x}_k|\mathbf{Z}_{k-1})d\mathbf{x}_{k}
\label{eqn:condpdf7} 
\end{align}
The pdf $p(\mathbf{z}_k|\mathbf{x}_k)$ and $p(\mathbf{x}_k|\mathbf{Z}_{k-1})$ in \eqref{eqn:condpdf7} are available as discussed previously. Hence all the pdfs of the right side of   
\eqref{eqn:condpdf6} are available. Hence formal solution to the recursive Bayesian estimation can be summarized as in Table 									\ref{tab:Recursive_Bayesian} \cite{7,9,11}.
\begin{table}[h] 
\caption{Recursive Bayesian Estimator \cite{4}} 
\centering          
\begin{tabular}{l}
  \hline
  \begin{minipage}{5in}
    \vskip 4pt
\begin{enumerate}
\item For $k=0$, initialize $p(\mathbf{x}_0|\mathbf{Z}_{0})=p(\mathbf{x}_0)$
\item For $k>0$
    \begin{itemize}
     \item Prediction step: Calculate the a priori pdf using \eqref{eqn:condpdf2}.
	  \begin{equation}
	  p(\mathbf{x}_{k}|\mathbf{Z}_{k-1})=\int p(\mathbf{x}_k|\mathbf{x}_{k-1})p(\mathbf{x}_{k-1}|\mathbf{Z}_{k-1})d\mathbf{x}_{k-1}   
	  \end{equation}
     \item Update step: Calculate the posterior pdf using \eqref{eqn:condpdf6} .
	  \begin{equation}
	  p(\mathbf{x}_{k}|\mathbf{Z}_{k})=\dfrac{p(\mathbf{z}_{k}|\mathbf{x}_k)p(\mathbf{x}_k|\mathbf{Z}_{k-1}))}{p(\mathbf{z}_{k}|\mathbf{Z}_{k-1})}	
	  \end{equation}
    \end{itemize}
\end{enumerate}
  \vskip 4pt
 \end{minipage}
 \\
  \hline
 \end{tabular}
\label{tab:Recursive_Bayesian} 
\end{table}
The measurement $\mathbf{z}_k$ is used to update the prior density $p(\mathbf{x}_k|\mathbf{Z}_{k-1})$ to obtain the posterior density. Thus, in principle the posterior pdf $p(\mathbf{x}_k|\mathbf{Z}_{k})$ can be obtained recursively by the two stages: prediction and update.

In general the implementation of this conceptual solution is not practically possible since it requires the storage of the entire pdf which is an infinite dimensional vector.  Analytical solution to these recursive equations cannot be determined in general because of complex and high dimensional integrals and are known only for few cases. For example in the system described by \eqref{eqn:state_model} and  \eqref{eqn:measurement_model}, if $f(\cdotp)$ and $h(\cdotp)$ are linear and initial density $p(\mathbf{x}_0)$ is Gaussian, noise sequences $w_k$ and $v_k$ are zero mean mutually independent, and  $p(\mathbf{x}_0)$, $w_k$ and $v_k$ are additive Gaussian, the optimal Bayesian solution is the Kalman filter. The exact implementation of the Kalman filter is feasible since its posterior density $p(\mathbf{x}_k|\mathbf{Z}_k)$ also turns out to be Gaussian and can be completely represented by its mean and covariance which are finite dimensional. Hence the storage of the posterior density $p(\mathbf{x}_k|\mathbf{Z}_{k})$ becomes convenient and the recursive Bayesian solution reduces to the recursive estimation of the mean and covariance of the posterior density $p(\mathbf{x}_k|\mathbf{Z}_{k})$. Thus Kalman filter is the optimal filter for the type of system mentioned above, and no other filter does better than it.

In practice $f(\cdot)$ and $h(\cdotp)$ may be nonlinear, and $p(\mathbf{x}_0)$, $w_k$ and $v_k$ may be non Gaussian. In such cases the posterior densities may be multi modal and/or non Gaussian. For such cases approximations or suboptimal Bayesian solutions are required for a practical realization. Analytical and numerical approximation methods for the implementation of the recursive Bayesian solution include extended Kalman filter, unscented Kalman filter, particle filter etc. The particle filter is explored in the subsequent chapters.

\section{Summary}
The Bayesian estimation problem can be conceptually solved recursively by two steps: prediction and update. Kalman filter is the optimal filter when the target state dynamics and measurement equation are linear and all the random elements in the model are additive Gaussian, and process and measurement noise are zero mean. In general, implementation of recursive Bayesian solution is not possible and hence analytical and numerical approximation techniques are required. A numerical approximation technique called particle filter for target tracking is explored in the subsequent chapters.

\chapter{Particle Filtering}
\label{chap:Particle_filtering}
Particle filter is a class of sequential Monte Carlo method to solve recursive Bayesian filtering problems. Monte Carlo methods are computational algorithms that are based on repeated random sampling to compute their results. Initially they define a domain of possible inputs, generate random input samples from a posterior distribution over this domain, perform the computation over this input samples to get the output samples and infer about the output probability distribution based on these output samples \cite{11}. Particle filters was initially developed for target tracking by  N.J. ~Gordon et.al \cite{7}. There have been significant modifications on the particle filter by A. ~Doucet et.al \cite{8,10,13}, B. ~Ristic et.al \cite{4} and are explored in this chapter.  Particle filter doesn't require the assumptions of linearity and Guassianity like the traditional Kalman filter (KF), Extended Kalman filter (EKF), etc. Hence it is capable of handling non-Gaussian noise distributions and non-linearities in the target's measurements as well as target dynamics.

The posterior distribution of the state of the system at every instant $k$ is represented by a set of $N$ random samples $\mathbf{x}_k^{(i)}$ called particles with associated weights $w_k^{(i)}$. The weights are normalized such that $\sum_{i=1}^N w_k^{(i)}=1$. This particle set  $\{\mathbf{x}_k^{(i)},w_k^{(i)}\}_{i=1}^N$ can then be regarded representing a probability distribution 
\begin{equation}
 p_N(\mathbf{x}_k)=\sum_{i=1}^N w_k^{(i)}\delta(\mathbf{x}-\mathbf{x}_k^{(i)}) 
\end{equation}
where $\delta(\cdotp)$ is the Dirac $\delta$-function. This particle set represents the probability distribution $p(\mathbf{x})$ if $p_N\rightarrow p$ as $N\rightarrow \infty$. Thus we have a discrete weighted approximation of a probability distribution function. The properties of the distribution $p(\mathbf{x})$ can be approximately calculated using these samples.

\section{Monte Carlo Approach}
Suppose $\pi(\mathbf{x})$ is a probability density function with $\mathbf{x}\in \mathbb{R}^{n_x}$ satisfying
\begin{eqnarray}
\pi(\mathbf{x}) & \geq & 0  \\
\int\pi(\mathbf{x})d\mathbf{x} & = & 1 
\end{eqnarray}
where $n_x$ is the dimension of the state vector and $\mathbb{R}$ is a set of real numbers.
If $N\gg1$ independent random samples $\{\mathbf{x}^{(i)};i=1,....,N\}$ are available from the distribution $\pi(\mathbf{x})$, then its discrete approximation is given by
\begin{equation}
 p_N(\mathbf{x})=\dfrac{1}{N}\sum_{i=1}^N \delta(\mathbf{x}-\mathbf{x}^{(i)}) 
\end{equation}
Then any integral function on the probability density function $\pi(\mathbf{x})$ can be approximated using an equivalent summation function on the samples from $p_N(\mathbf{x})$ and it converges to the true value as $N\rightarrow \infty$. Suppose it is required to evaluate a multidimensional integral 
\begin{equation}
 I=\int g(\mathbf{x})d\mathbf{x} 
\end{equation}
then the Monte Carlo approach will be to factorize $g(\mathbf{x})=f(\mathbf{x})p(\mathbf{x})$ such that $p(\mathbf{x})\geq0$ and $\int p(\mathbf{x})d\mathbf{x}=1$, where $p(\mathbf{x})$ is interpreted as a probability distribution from which samples can be drawn easily and $f(\mathbf{x})$ is a function on $\mathbf{x}$. Then the integral can be written as
\begin{eqnarray}
I & = & \int f(\mathbf{x})p(\mathbf{x})d\mathbf{x} \label{eqn:condpdf8} \\
 & = & E_{p(\mathbf{x})}[f(\mathbf{x})] 
\end{eqnarray}
where $E_{p(\mathbf{x})}[\cdotp]$ is the expectation w.r.t distribution $p(\mathbf{x})$. Hence the integral $I$ is the expectation of $f(\mathbf{x})$ with respect to the distribution $p(\mathbf{x})$.
 Then Monte Carlo estimate of $I$ can be obtained by generating $N$ samples $\{\mathbf{x}^{(i)}\}_{i=1}^N$ from distribution $p(\mathbf{x})$ and calculating the summation
\begin{eqnarray}
 I_N & = & \dfrac{1}{N}\sum_{i=1}^N f(\mathbf{x}^{(i)})\delta(\mathbf{x}-\mathbf{x}^{(i)}) \\
 & = & \dfrac{1}{N}\sum_{i=1}^N f(\mathbf{x}^{(i)}) \label{eqn:condpdf9}\\
& \approx & I 
\end{eqnarray}
This estimate is unbiased and converges to the true value $I$ as $N\rightarrow \infty$. 

If the distribution $p(\mathbf{x})$ is standard and has closed analytical form, then generation of random samples from it is possible. But since in target tracking the posterior distribution may be multivariate and non standard, it is not possible to sample efficiently from this distribution.
There are two problems in the basic Monte Carlo method as mentioned in \cite{10}.

\textit{Problem 1 :} Sampling from the distribution $p(\mathbf{x})$ is not possible if it is complex high dimensional probability distribution.

\textit{Problem 2 :} The computational complexity of sampling from target distribution $p(\mathbf{X_k})$ where $\mathbf{X}_k=\{\mathbf{x}_j;j=0,..,k$ increases at least linearly with the number of variables $k$.

\section{Importance Sampling}
Importance sampling helps in addressing the \textit{Problem 1} discussed above. Suppose we are interested in generating samples from $p(\mathbf{x})$ which is difficult to sample, importance sampling is a technique which helps to indirectly generate samples from a suitable distribution $q(\mathbf{x})$ that is easy to sample, and modify this samples by appropriate weighting so that it represents the samples from the distribution $p(\mathbf{x})$. Thus importance sampling makes the calculation of $E_{p(\mathbf{x})}[f(\mathbf{x})]$ feasible. The pdf $q(\mathbf{x})$ is referred to as proposal or importance density.
The integral in \eqref{eqn:condpdf8} can be modified as 
\begin{align}
\label{eq:PF_integral}
I & = \int f(\mathbf{x})p(\mathbf{x})d\mathbf{x} \\
 & = \int f(\mathbf{x})\dfrac{p(\mathbf{x})}{q(\mathbf{x})}q(\mathbf{x})d\mathbf{x} \\
  & = E_{q(\mathbf{x})}[f(\mathbf{x})\dfrac{p(\mathbf{x})}{q(\mathbf{x})}]
\end{align}
provided $p(\mathbf{x})>0 \Rightarrow q(\mathbf{x})>0$ for all $\mathbf{x}\in\mathbb{R}^{n_x}$ and $p(\mathbf{x})/q(\mathbf{x})$ has an upper bound. Then according to \eqref{eqn:condpdf9} Monte Carlo estimate of $I$ can be obtained by generating samples $\mathbf{x}^{(i)};i=1,....,N; N\gg1$ from the distribution $q(\mathbf{x})$ and evaluating
\begin{align}
 I_N & = \dfrac{1}{N}\sum_{i=1}^N f(\mathbf{x}^{(i)})\dfrac{p(\mathbf{x}^{(i)})}{q(\mathbf{x}^{(i)})} \\
 & = \dfrac{1}{N}\sum_{i=1}^N f(\mathbf{x}^{(i)})\tilde{w}(\mathbf{x}^{(i)}) \\
\tilde{w}(\mathbf{x}^{(i)}) & = \dfrac{p(\mathbf{x}^{(i)})}{q(\mathbf{x}^{(i)})}; \;\qquad i=1,..,N \label{eqn:condpdf10}
\end{align}
where $\tilde{w}(\mathbf{x}^{(i)})$ are called the importance weights. The weights are then normalized to qualify it to be a probability distribution.
\begin{eqnarray}
 w^{(i)}=\dfrac{{\tilde{w}}^{(i)}}{\sum_{i=1}^N {\tilde{w}}^{(i)}} \label{eqn:condpdf11}
\end{eqnarray}
Thus the random samples from distribution $p(\mathbf{x})$ are equivalent to the the random samples from distribution $q(\mathbf{x})$, with associated weights $w^{(i)}$ given in \eqref{eqn:condpdf11}. Thus the samples $\{\mathbf{x}^{(i)}\}_{i=1}^N$ from $q(\mathbf{x})$ with weights $\{w^{(i)}\}_{i=1}^N$ represent the probability distribution of $p(\mathbf{x})$ as $N\rightarrow \infty$ and can be used to compute estimate the integral $I$.

\section{Sequential Importance Sampling}
Sequential importance sampling helps in addressing the \textit{Problem 2} described above. Sequential Importance Sampling unlike importance sampling requires only a fixed computational complexity at every time step. It is also known as bootstrap filtering, particle filtering or condensation algorithm. It is the sequential version of the Bayesian filter using importance sampling.

Consider a joint posterior distribution $p(\mathbf{X}_k|\mathbf{Z}_k)$, where $\mathbf{X}_k=\{\mathbf{x}_j;j=0,...,k\}$ is the sequence of all target states upto time $k$ and $\mathbf{Z}_k=\{\mathbf{z}_j;j=0,...,k\}$ is the sequence of all target measurements upto time $k$. Let $\{\mathbf{X}_k^{(i)},w_k^{(i)}\}_{i=1}^N$ be the particles such that
\begin{equation}
 p(\mathbf{X}_k|\mathbf{Z}_k) \approx \sum_{i=1}^N w_k^{(i)} \delta (\mathbf{X}_k-\mathbf{X}_k^{(i)})                                                                                                                                                                                                                     
\end{equation}
If the importance density $q(\mathbf{X}_k|\mathbf{Z}_k)$ is used to generate particles $\{\mathbf{X}_k^{(i)}\}_{i=1}^N$, then its corresponding weights according to \eqref{eqn:condpdf10} can be written as
\begin{equation}
w^{(i)}\varpropto \dfrac{p(\mathbf{x}^{(i)})}{q(\mathbf{x}^{(i)})};  \;\qquad i=1,..,N \label{eqn:condpdf12}
\end{equation}
We can express the importance function using Bayes rule as
\begin{eqnarray}
 q(\mathbf{X}_k|\mathbf{Z}_k)& = & q(\mathbf{x}_k,\mathbf{X}_{k-1}|\mathbf{Z}_k)\\
& = & q(\mathbf{x}_k|\mathbf{X}_{k-1},\mathbf{Z}_k)q(\mathbf{X}_{k-1}|\mathbf{Z}_k) \label{eqn:condpdf13}\\
& = & q(\mathbf{x}_k|\mathbf{X}_{k-1},\mathbf{Z}_k)q(\mathbf{x}_{k-1}|\mathbf{X}_{k-2},\mathbf{Z}_k)......q(\mathbf{x}_{1}|\mathbf{X}_{0},\mathbf{Z}_k)q(\mathbf{X}{0}|\mathbf{Z}_k) \label{eqn:condpdf14}\\
& = & q(\mathbf{x}_{0}|\mathbf{Z}_k)\Pi_{n=1}^k q(\mathbf{x}_n|\mathbf{X}_{n-1},\mathbf{Z}_k)
\end{eqnarray}
In order to make the importance sampling recursive at every instant $k$ without modifying the previous simulated trajectories $\{\mathbf{X}_{k-1}^{(i)}\}_{i=1}^N$, the new set of samples at time $k$, $\mathbf{X}_{k}^{(i)}\sim q(\mathbf{X}_k|\mathbf{Z}_k)$ must be obtained using the previous set of samples $\mathbf{X}_{k-1}^{(i)}\sim q(\mathbf{X}_{k-1}|\mathbf{Z}_{k-1})$ and the importance density must be chosen such that $q(\mathbf{X}_{k-1}|\mathbf{Z}_{k})=q(\mathbf{X}_{k-1}|\mathbf{Z}_{k-1})$. Then \eqref{eqn:condpdf13} can be written as 
\begin{eqnarray}
 q(\mathbf{X}_k|\mathbf{Z}_k) & = & q(\mathbf{x}_k|\mathbf{X}_{k-1},\mathbf{Z}_k)q(\mathbf{X}_{k-1}|\mathbf{Z}_{k-1}) \label{eqn:condpdf15}\\
& = & q(\mathbf{x}_{0}|\mathbf{Z}_0)\Pi_{n=1}^k q(\mathbf{x}_n|\mathbf{X}_{n-1},\mathbf{Z}_n)
\end{eqnarray}
Thus the importance density at $k$ can be expressed in terms of importance density at $k-1$ so that new samples $\mathbf{X}_{k}^{(i)} \sim q(\mathbf{X}_k|\mathbf{Z}_k)$ can be obtained by augmenting each previous samples $\mathbf{X}_{k-1}^{(i)} \sim q(\mathbf{X}_{k-1}|\mathbf{Z}_{k-1})$ with the new state $\mathbf{x}_k^{(i)} \sim q(\mathbf{x}_k|\mathbf{X}_{k-1},\mathbf{Z}_{k})$. These particles along with their new importance weights can approximate the posterior distribution $p(\mathbf{X}_k|\mathbf{Z}_k)$ as $N\rightarrow \infty$.

In order to calculate the new importance weights for the above samples recursively, the pdf $p(\mathbf{X}_k|\mathbf{Z}_k)$ can be written using \eqref{eqn:condpdf5} as  \cite{4}.

\begin{eqnarray}
p(\mathbf{X}_{k}|\mathbf{Z}_{k})
& = & \dfrac{p(\mathbf{z}_{k}|\mathbf{X}_k,\mathbf{Z}_{k-1})p(\mathbf{X}_k|\mathbf{Z}_{k-1})}{p(\mathbf{z}_{k}|\mathbf{Z}_{k-1})}  \\
& = & \dfrac{p(\mathbf{z}_{k}|\mathbf{X}_k,\mathbf{Z}_{k-1})p(\mathbf{x}_k|\mathbf{X}_{k-1},\mathbf{Z}_{k-1})p(\mathbf{X}_{k-1}|\mathbf{Z}_{k-1})}{p(\mathbf{z}_{k}|\mathbf{Z}_{k-1})} \label{eqn:condpdf16}
\end{eqnarray}
Using the assumption in \eqref{eqn:markov1}, \eqref{eqn:condpdf16} can be written as
\begin{eqnarray} 
p(\mathbf{X}_{k}|\mathbf{Z}_{k})& = & \dfrac{p(\mathbf{z}_{k}|\mathbf{x}_k)p(\mathbf{x}_k|\mathbf{x}_{k-1})}{p(\mathbf{z}_{k}|\mathbf{Z}_{k-1})}p(\mathbf{X}_{k-1}|\mathbf{Z}_{k-1})\\
& \varpropto & p(\mathbf{z}_k|\mathbf{x}_k)p(\mathbf{x}_k|\mathbf{x}_{k-1})p(\mathbf{X}_{k-1}|\mathbf{Z}_{k-1}) \label{eqn:condpdf17} 
\end{eqnarray}
 The proportionality follows because $p(\mathbf{z}_{k}|\mathbf{Z}_{k-1})$ is a normalizing constant. Using \eqref{eqn:condpdf17} and \eqref{eqn:condpdf15}, \eqref{eqn:condpdf12} can be rewritten as
\begin{eqnarray} 
w_k^{(i)} & \varpropto & \dfrac{p(\mathbf{z}_k|\mathbf{x}_k^{(i)})p(\mathbf{x}_k^{(i)}|\mathbf{x}_{k-1}^{(i)})}{q(\mathbf{x}_k^{(i)}|\mathbf{X}_{k-1}^{(i)},\mathbf{Z}_{k})}\dfrac{p(\mathbf{X}_{k-1}^{(i)}|\mathbf{Z}_{k-1})}{q(\mathbf{X}_{k-1}^{(i)}|\mathbf{Z}_{k-1})} \\
& = & w_{k-1}^{(i)}\dfrac{p(\mathbf{z}_k|\mathbf{x}_k^{(i)})p(\mathbf{x}_k^{(i)}|\mathbf{x}_{k-1}^{(i)})}{q(\mathbf{x}_k^{(i)}|\mathbf{X}_{k-1}^{(i)},\mathbf{Z}_{k})}
\end{eqnarray} 
If the importance density also satisfies $q(\mathbf{x}_k|\mathbf{X}_{k-1},\mathbf{Z}_k)=q(\mathbf{x}_k|\mathbf{x}_{k-1},\mathbf{z}_k)$, then the importance weight can be calculated recursively as 
\begin{eqnarray}
\label{eq:imp_weights} 
w_k^{(i)}\varpropto w_{k-1}^{(i)}\dfrac{p(\mathbf{z}_k|\mathbf{x}_k^{(i)})p(\mathbf{x}_k^{(i)}|\mathbf{x}_{k-1}^{(i)})}{q(\mathbf{x}_k^{(i)}|\mathbf{x}_{k-1}^{(i)},\mathbf{z}_{k})} \label{eqn:condpdf18} 
\end{eqnarray} 
Thus sequential importance sampling filter consists of recursive propagation of particles $\mathbf{x}_k^{(i)}$ according to \eqref{eqn:condpdf15} and update of importance weights  $w_k^{(i)}$ according to \eqref{eqn:condpdf18}. Hence in order to obtain the particles at instant $k$, only the past particles $\{\mathbf{x}_{k-1}^{(i)},\mathbf{w}_{k-1}^{(i)}\}_{i=1}^N$ and measurement $\mathbf{z}_k$ are required and can discard the past trajectories $\mathbf{X}_{k-2}^{(i)}$ and measurements $\mathbf{Z}_{k-1}$, and requires only fixed computational complexity. Thus it addresses the \textit{Problem 2} discussed previously. Hence the posterior filtered density $p(\mathbf{x}_k|\mathbf{Z}_k)$ can be calculated recursively. The pseudo-code for the sequential importance sampling (SIS) filter is repeated in Table.\ref{tab:SIS} from \cite{8}. 
\begin{table}[t] 
\caption{Sequential Importance Sampling (SIS) \cite{8}} 
\centering          
\begin{tabular}{l}
  \hline
  \begin{minipage}{5in}
    \vskip 4pt
\begin{enumerate}
\item For $k=0$,
    \begin{itemize}
	  \item For $i=1,....,N$: Initialize
	  \begin{itemize}
		\item Sample $\mathbf{x}_0^{(i)}\sim q(\mathbf{x}_0|\mathbf{z}_0)$
		\item Evaluate the unnormalized importance weights
		    \begin{eqnarray}
			{\tilde{w}}_0^{(i)}& = & \dfrac{p(\mathbf{z}_0|\mathbf{x}_0^{(i)})p(\mathbf{x}_0^{(i)})}{q(\mathbf{x}_0^{(i)}|\mathbf{z}_{0})} 
		    \end{eqnarray}
	  \end{itemize}
	  \item For $i=1,....,N$:
	  \begin{itemize}
	        \item Normalize the importance weights
		\begin{eqnarray}
		    w^{(i)}& = & \dfrac{{\tilde{w}}_0^{(i)}}{\sum_{i=1}^N {\tilde{w}}_0^{(i)}}		    
		\end{eqnarray}
	  \end{itemize}
    \end{itemize}
\item For $k>0$
    \begin{itemize}
	\item For $i=1,....,N$:
	\begin{itemize} 
		\item Sample $\mathbf{x}_k^{(i)} \sim q(\mathbf{x}_k|\mathbf{X}_{k-1},\mathbf{Z}_{k})$
		\item Evaluate the unnormalized importance weights
		    \begin{eqnarray}
			{\tilde{w}}_k^{(i)} & \varpropto & w_{k-1}^{(i)}\dfrac{p(\mathbf{z}_k|\mathbf{x}_k^{(i)})p(\mathbf{x}_k^{(i)}|\mathbf{x}_{k-1}^{(i)})}{q(\mathbf{x}_k^{(i)}|\mathbf{x}_{k-1}^{(i)},\mathbf{z}_{k})} 	    
		    \end{eqnarray}
	\end{itemize}
	\item For $i=1,....,N$:
			    \begin{itemize}
				\item Normalize the importance weights
				\begin{eqnarray}
				    w^{(i)}& = & \dfrac{{\tilde{w}}^{(i)}}{\sum_{i=1}^N {\tilde{w}}^{(i)}}		    
				\end{eqnarray}
			    \end{itemize}	
    \end{itemize}
\end{enumerate}
  \vskip 4pt
 \end{minipage}
 \\
  \hline
 \end{tabular}
\label{tab:SIS} 
\end{table} 
The weight update and proposal for each particle in the sequential importance sampling filter can be calculated in parallel. Hence availability of parallel computational techniques like graphics processing unit (GPU) and FPGA facilitates the implementation of SIS filter without loosing time efficiency.

\section{Implementation Issues}
\subsection{Degeneracy}
\begin{figure}[t!]
 \centering 
   \includegraphics[scale=0.5]{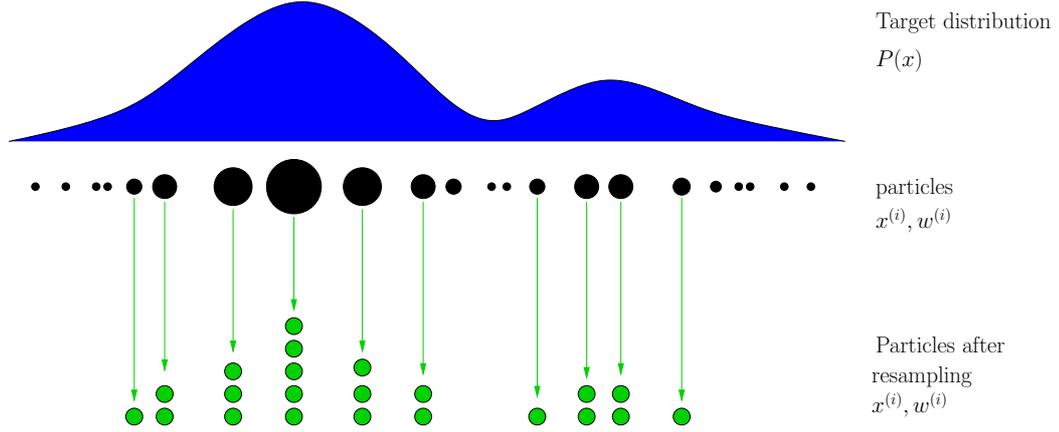}
\caption{Resampling of a set of particles representing a distribution $P(\mathbf{x})$ is illustrated. The size of the particles represents their weight.}
\label{fig:resampling}
\end{figure}

According to \cite{8}, the variance of importance weights increases over time if the importance density is of the form \eqref{eqn:condpdf15}. Hence after a certain number of recursive steps, the weights degrade or get degenerated such that most  particles have negligible weight. A large computational effort has to be wasted on updating these particles even though their contribution to the posterior estimate is negligible. Hence only a few high weight particles contribute to the posterior distribution $p(\mathbf{x}_k|\mathbf{Z}_k)$ effectively. One level of degeneracy can be estimated based on effective sample size($N_{eff})$
\begin{eqnarray}
N_{eff}& = & \dfrac{1}{\sum_{i=1}^N (w_k^{(i)})^2}	    
\end{eqnarray}
The two extreme cases are 
\begin{enumerate}
 \item If the weights are uniform, $w_k^{(i)}=\frac{1}{N}$, for $i=1,...,N$, then $N_{eff}=N$.
 \item If weights are such that $w_k^{(j)}=1$ and $w_k^{(i)}=0$ for $i\neq j$, then $N_{eff}=1$.
\end{enumerate}
For all other intermediate cases $1<N_{eff}<N$. Thus higher degeneracy implies lesser $N_{eff}$ and vice versa.\\\\
$Solution: Resampling$\\
Resampling is a technique to reduce degeneracy. If degeneracy is observed, i.e., $N_{eff}$ falls below some threshold $N_{thr}$, then resampling is done. It keeps as many samples with non-zero significant weights and neglects the negligible weights. It replaces the old set of particles and their weights with new set of particles and weights by removing the low weight particles and replicating the high weight particles and associating them with uniform weights such that the resultant particles represent the posterior pdf in a better form for later iterations. Thus it does a transformation of the set $\{\mathbf{x}_k^{(i)}, w_k^{(i)}\}_{i=1}^N$ to $\{\mathbf{x}_k^{(i)}, N^{-1}\}_{i=1}^N$ such that the final set represents the same distribution as of the first. The concept of resampling is illustrated in Fig \ref{fig:resampling}.\\

\begin{figure}[t!]
 \centering 
   \includegraphics[scale=0.4]{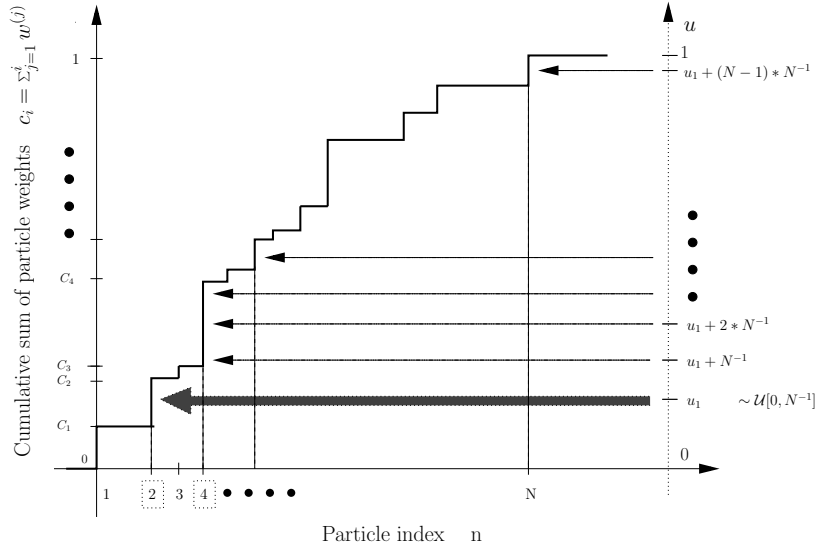}
\caption{Multinomial Resampling: The high weight particles such as particles with indices $2$, $4$, etc, are selected more number of times.}
\label{fig:Mresampling}
 \end{figure}

One way of implementation of resampling is multinomial resampling \cite{4,11} which involves generating uniformly distributed random samples in range $(0,1)$ and using them to obtain samples from the required target posterior density by inverse transformation. It has three main steps. First it generates independent uniform random samples $u_j\sim \mathcal{U}[0,1]$ for $j=1,...N$. Secondly it accumulates the weights $w_k^{(i)}$ into a sum until it is just greater than $u_j$.
\begin{equation}
 \sum_{i=1}^{m-1} w_k^{(i)} <u_j\leq \sum_{i=1}^{m} w^{(i)} 
\end{equation}
Hence it projects  $u_j$ to the cumulative sum of the weights $w_k^{(i)}$ as shown in Fig \ref{fig:Mresampling}. The new particle ${\tilde{\mathbf{x}}}_k^{(j)}$ is set equal to the old particles $\mathbf{x}_k^{(m)}$ with weight $1/N$ and is repeated until $N$ samples are obtained. The large weight particles have higher chance of being selected and multiplied. Its pseudo code is given in Table \ref{tab:Mresampling}.

\begin{table}[p!] 
\caption{Multinomial Resampling \cite{4,11}} 
\centering          
\begin{tabular}{l}
  \hline
  \begin{minipage}{4in}
    \vskip 4pt
$[\{\tilde{\mathbf{x}}_k^{(n)},\tilde{w}_k^{(n)}\}_{n=1}^{N}]=$RESAMPLE$[\{\mathbf{x}_k^{(i)}, w_k^{(i)}\}_{i=1}^N]$
\begin{itemize}
	\item $c(0)=0$ 
	\item FOR $i=1:N$,
	\begin{itemize}
		\item $c(i)=c(i-1)+ w_k^{(i)}$ 			
	\end{itemize}
	\item END FOR
	\item FOR  $n=1:N$,
	\begin{itemize}
	      \item Draw $u_n\sim \mathcal{U} [0,1]$  		
	      \item m=1
	      \item WHILE $(c(m)<u_n)$
	      \begin{itemize}
		    \item$ m=m+1$
	      \end{itemize}
	      \item END WHILE
	      \item Set $\tilde{\mathbf{x}}_k^{(n)}=\mathbf{x}_k^{(m)}$
	      \item Set $\tilde{w}_k^{(n)}=N^{-1}$
	      \end{itemize}
	\item END FOR
\end{itemize}
   \vskip 4pt
 \end{minipage}
 \\
  \hline
 \end{tabular}
\label{tab:Mresampling} 
\end{table}

\begin{figure}[t!]
 \centering 
   \includegraphics[scale=0.4]{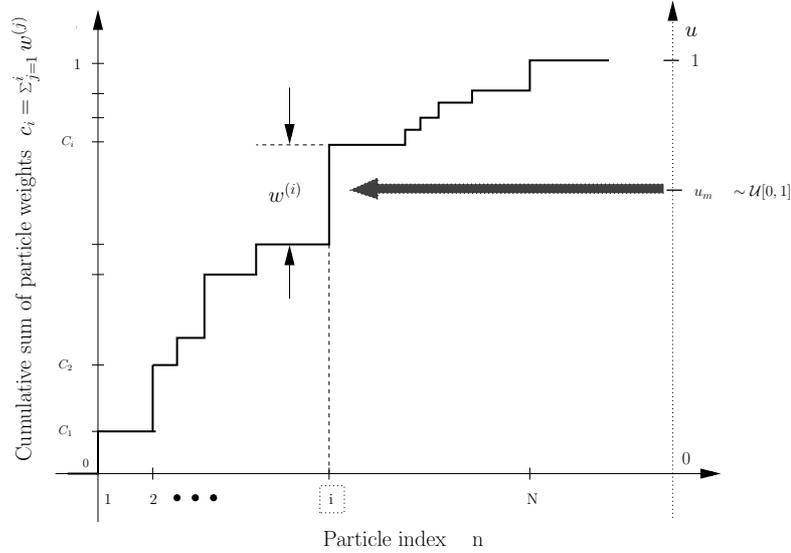}
\caption{Systematic Resampling}
\label{fig:Sresampling}
 \end{figure}

Another slightly different method of resampling is the systematic resampling\cite{4, 11}. It has the same procedure as multinomial sampling except that pseudo uniform random variables are generated instead of independent uniform random variables. Here a uniform random number $u_1\sim \mathcal{U}[0,N^{-1}]$ is generated once and the rest are generated by increasing this random number $u_1$ by $1/N$ cumulatively and then performing the inverse transformation as shown in Fig \ref{fig:Sresampling} similar to the multinomial sampling to get the required target posterior distribution. Its pseudo code is repeated in Table \ref{tab:Sresampling} from \cite{4}.
where $T$ is the sampling period of the target dynamics.

\begin{table}[p!] 
\caption{Systematic Resampling} 
\centering          
\begin{tabular}{l}
  \hline
  \begin{minipage}{4in}
    \vskip 4pt
$[\{\tilde{\mathbf{x}}_k^{(n)},\tilde{w}_k^{(n)}\}_{n=1}^{N}]=$RESAMPLE$[\{\mathbf{x}_k^{(i)}, w_k^{(i)}\}_{i=1}^N]$
\begin{itemize}
	\item $c(0)=0$ 
	\item FOR $i=1:N$,
	\begin{itemize}
		\item $c(i)=c(i-1)+ w_k^{(i)}$ 		
	\end{itemize}
	\item END FOR
	\item Draw the starting point $u_1\sim \mathcal{U} [0, \frac{1}{N}]$  		
	\item m=1
	\item FOR  $n=1:N$,
	\begin{itemize}	      
	      \item $u_n=u_1+R^{-1}(n-1)$
	      \item WHILE $(c(m)<u_n)$
	      \begin{itemize}
		    \item$ m=m+1$
	      \end{itemize}
	      \item END WHILE
	      \item Set $\tilde{\mathbf{x}}_k^{(n)}=\mathbf{x}_k^{(m)}$
	      \item Set $\tilde{w}_k^{(n)}=N^{-1}$
	      \end{itemize}
	\item END FOR
\end{itemize}
   \vskip 4pt
 \end{minipage}
 \\
  \hline
 \end{tabular}
\label{tab:Sresampling} 
\end{table}

 Thus resampling involves $N$ draws from the initial particles using their own probability distribution as the selection probabilities and assigning each particle a weight of $w^{(i)}=\frac{1}{N}$ for $i=1,...,N$. This strategy of resampling along with importance sampling is termed as sampling importance resampling(SIR).

Even though resampling helps to remove degeneracy, it introduces another issue known as sample impoverishment which is described next. The accuracy of any estimate of a function of the distribution decreases with resampling. It also limits the opportunity to parallelize the propagation and update of the particles since they have to be combined to find the cumulative density required for resampling. Hence in order to minimize the frequency of resampling, a proper proposal function has to be used so that there is significant overlap between the prior particles and the likelihood. Strategies of selecting good proposal function are explained in section \ref{sec:imp_fn_selection}.

\subsection{Sample Impoverishment}
When there is very less overlap between the prior and the likelihood, only few particles will have higher weight. A subsequent resampling causes loss of diversity among particles as particles with large weight are sampled many times with the result that resultant sample will contain many repeated points or less distinct points. This is called sample impoverishment. After some iterations it leads to a situation when all particles collapse to a single particle. \\\\
$Solution: Roughening$\\
One method to solve sample impoverishment is to increase the number of particle $N$. But it increases the computational demand. Roughening is an efficient method proposed in \cite{7} to solve sample impoverishment. Here random noise $\Delta \mathbf{x}$ is added to each component of the particle after the resampling process such that:
\begin{eqnarray}
 \mathbf{x}_k^{(i)}(m)& = & \mathbf{x}_k^{(i)}(m)+\Delta \mathbf{x}(m)  \\
\Delta \mathbf{x} & \sim & \mathcal{N}(0,KMN^{-1/d}) \\
M(m)& = & \max_{i,j}|\mathbf{x}_{k}^{(i)}(m)-\mathbf{x}_{k}^{(j)}(m)|; \;\;\; m=1,...,d \label{eqn:roughening} 
\end{eqnarray}
where $K$ is a scalar tuning parameter, $N$ is the number of particles, $d$ is the dimension of state space, $M$ is the vector containing maximum difference between each particle elements before roughening. Higher value of $K$ will blur the distribution and low value of $K$ will create group of points around the original samples. Hence $K$ is a compromise and has to be tuned. A value of $K=0.2$ has been used in \cite{7}. The pseudo code for roughening is shown in Table.\ref{tab:Roughening} 

\begin{table}[ht!] 
\caption{Roughening} 
\centering          
\begin{tabular}{l}
  \hline
  \begin{minipage}{4in}
    \vskip 4pt
$[\{\mathbf{x}_k^{(n)},w_k^{(n)}\}_{n=1}^{N}]=$ROUGHEN$[\{\mathbf{x}_k^{(i)},w_k^{(i)}\}_{i=1}^{N}]$
\begin{enumerate}
	      \item For $m=1,...,d$
		    \begin{equation}
			   M(m)=\max_{i,j}|\mathbf{x}_{k}^{(i)}(m)-\mathbf{x}_{k}^{(j)}(m)|
		    \end{equation}	  
	      \item For $i=1,...,N$
		  \begin{itemize}
			\item Calculate random noise vector
			\begin{eqnarray} 
			   \Delta \mathbf{x} & \sim & \mathcal{N}(0,KMN^{-1/d}) 
			\end{eqnarray}
		      \item For $m=1,...,d$			  
			      \begin{eqnarray}
				  \mathbf{x}_k^{(i)}(m)& = & \mathbf{x}_k^{(i)}(m)+\Delta \mathbf{x}(m)  				  
			      \end{eqnarray}			
		  \end{itemize}	  
\end{enumerate}
   \vskip 4pt
 \end{minipage}
 \\
  \hline
 \end{tabular}
\label{tab:Roughening} 
\end{table}

Other solutions for sample impoverishment include prior editing, Markov Chain Monte Carlo resampling, regularized particle filter, auxiliary particle filter etc.

\section{Selection of Importance function}
\label{sec:imp_fn_selection}
A good selection of importance density minimizes the frequency of resampling. Since increase in the variance of the weights of the particles causes degeneracy, the better method will be to select the importance density which minimizes the variance of the importance weights based on the available information $\mathbf{X}_{k-1}$ and $\mathbf{Z}_k$. 

\subsection{Optimal Importance function}
The best way of selecting an importance density is to choose the one which minimizes the variance of the weights. According to \cite{8}, the optimal importance density that minimizes the variance of the importance weights conditional upon the simulated trajectories $\mathbf{X}_{k-1}^{(i)}$ and observations $\mathbf{Z}_k$ is given by
\begin{eqnarray}
 q(\mathbf{x}_k|\mathbf{X}_{k-1}^{(i)},\mathbf{Z}_k)& = & p(\mathbf{x}_k|\mathbf{x}_{k-1}^{(i)},\mathbf{z}_k)\\
& = & \dfrac{p(\mathbf{x}_k,\mathbf{x}_{k-1}^{(i)},\mathbf{z}_k)}{p(\mathbf{x}_{k-1}^{(i)},\mathbf{z}_k)} \\
& = & \dfrac{p(\mathbf{z}_k|\mathbf{x}_k,\mathbf{x}_{k-1}^{(i)})p(\mathbf{x}_k|\mathbf{x}_{k-1}^{(i)})p(\mathbf{x}_{k-1}^{(i)})}{p(\mathbf{z}_k|\mathbf{x}_{k-1}^{(i)})p(\mathbf{x}_{k-1}^{(i)})} \\
& = & \dfrac{p(\mathbf{z}_k|\mathbf{x}_k,\mathbf{x}_{k-1}^{(i)})p(\mathbf{x}_k|\mathbf{x}_{k-1}^{(i)})}{p(\mathbf{z}_k|\mathbf{x}_{k-1}^{(i)})} \label{eqn:condpdf19} 
\end{eqnarray}
Then the weight update equation for particles drawn from this optimal importance density can be obtained using \eqref{eqn:condpdf18} and \eqref{eqn:condpdf19} as
\begin{eqnarray}
 w_k^{(i)} & \varpropto & w_{k-1}^{(i)}p(\mathbf{z}_k|\mathbf{x}_{k-1}^{(i)}) 
\end{eqnarray}
Another advantage of using the optimal importance function is that the importance weight at instant $k$ doesn't depend on $\mathbf{x}_k$ and hence evaluation of of weight $w_k^{(i)}$ and proposal of $\mathbf{x}_k^{(i)}$ can be parallelized for better practical results.

In order to use this optimal importance function, we should be able to sample particles from $p(\mathbf{x}_k|\mathbf{x}_{k-1}^{(i)},\mathbf{z}_k)$ and to evaluate 
\begin{eqnarray}
 p(\mathbf{z}_k|\mathbf{x}_{k-1}^{(i)}) & = & \int p(\mathbf{z}_k|\mathbf{x}_k)p(\mathbf{x}_k|\mathbf{x}_{k-1})d\mathbf{x}_k 
\end{eqnarray}
at least upto a normalizing constant. But these exact calculations are possible only for some special cases like systems of form
\begin{eqnarray}
\mathbf{x}_k & = & \mathbf{f}_{k-1}(\mathbf{x}_{k-1})+\mathbf{w}_{k-1} \label{eqn:condpdf20a}\\
\mathbf{z}_k & = & \mathbf{H}_{k}\mathbf{x}_{k}+\mathbf{v}_{k}\label{eqn:condpdf20b} 
\end{eqnarray}
where $\mathbf{f}_{k-1}(\cdotp)$ can be a non linear function, $H_k$ is a matrix, $v_k$ and $w_k$ are mutually independent zero mean white Gaussian noise with known covariances $Q_k$ and $R_k$

\subsection{Suboptimal Importance Functions}
\subsubsection{Importance Function Obtained by Local Linearization}
For systems of form \eqref{eqn:condpdf21a} and \eqref{eqn:condpdf21b}, where both the system and measurement equation are non linear, local linearization of function $\mathbf{h}_{k}(\cdotp)$ is done similar to Extended Kalman Filter to get the linearized matrix $H_k$ so that the problem becomes similar to the system defined in \eqref{eqn:condpdf20a} and \eqref{eqn:condpdf20b}.
\begin{eqnarray}
\mathbf{x}_k & = & \mathbf{f}_{k-1}(\mathbf{x}_{k-1})+\mathbf{w}_{k-1} \label{eqn:condpdf21a}\\
\mathbf{z}_k & = & \mathbf{h}_{k}(\mathbf{x}_{k})+\mathbf{v}_{k} \label{eqn:condpdf21b}\\
H_k & = & {\dfrac{\delta \mathbf{h}_{k}(\mathbf{x}_{k})}{\delta \mathbf{x}_{k}}}|_{\mathbf{x}_k=f(\mathbf{x}_{k-1})} 
\end{eqnarray}

\subsubsection{Prior Importance Function}
One popular choice of importance density is the transitional prior itself. 
\begin{eqnarray}
  q(\mathbf{x}_k|\mathbf{x}_{k-1}^{(i)},\mathbf{z}_k) & = & p(\mathbf{x}_k|\mathbf{x}_{k-1}^{(i)})\label{eqn:condpdf22}
\end{eqnarray}
For a system with state space representation of \eqref{eqn:condpdf20a} and \eqref{eqn:condpdf20b}, the prior becomes
\begin{equation}
 p(\mathbf{x}_k|\mathbf{x}_{k-1}^{(i)})=\mathcal{N}(\mathbf{x}_k;f_{k-1}(\mathbf{x}_{k-1}^{(i)},Q_{k-1}))
\end{equation}
Using \eqref{eqn:condpdf18} and \eqref{eqn:condpdf22}, the weight update equation simplifies to 
\begin{equation}
 w_k^{(i)}\varpropto w_{k-1}^{(i)}p(\mathbf{z}_k|\mathbf{x}_{k}^{(i)}) \label{eqn:condpdf23}
\end{equation}
This method has the advantage that importance weights are easily calculated and the importance density can be easily sampled. But this method is less efficient since the particles are proposed without the knowledge of the observation and hence the overlap between the prior and the likelihood might be less.

\section{Generic Particle Filters}
\begin{figure}[t!]
 \centering 
   \includegraphics[scale=0.5]{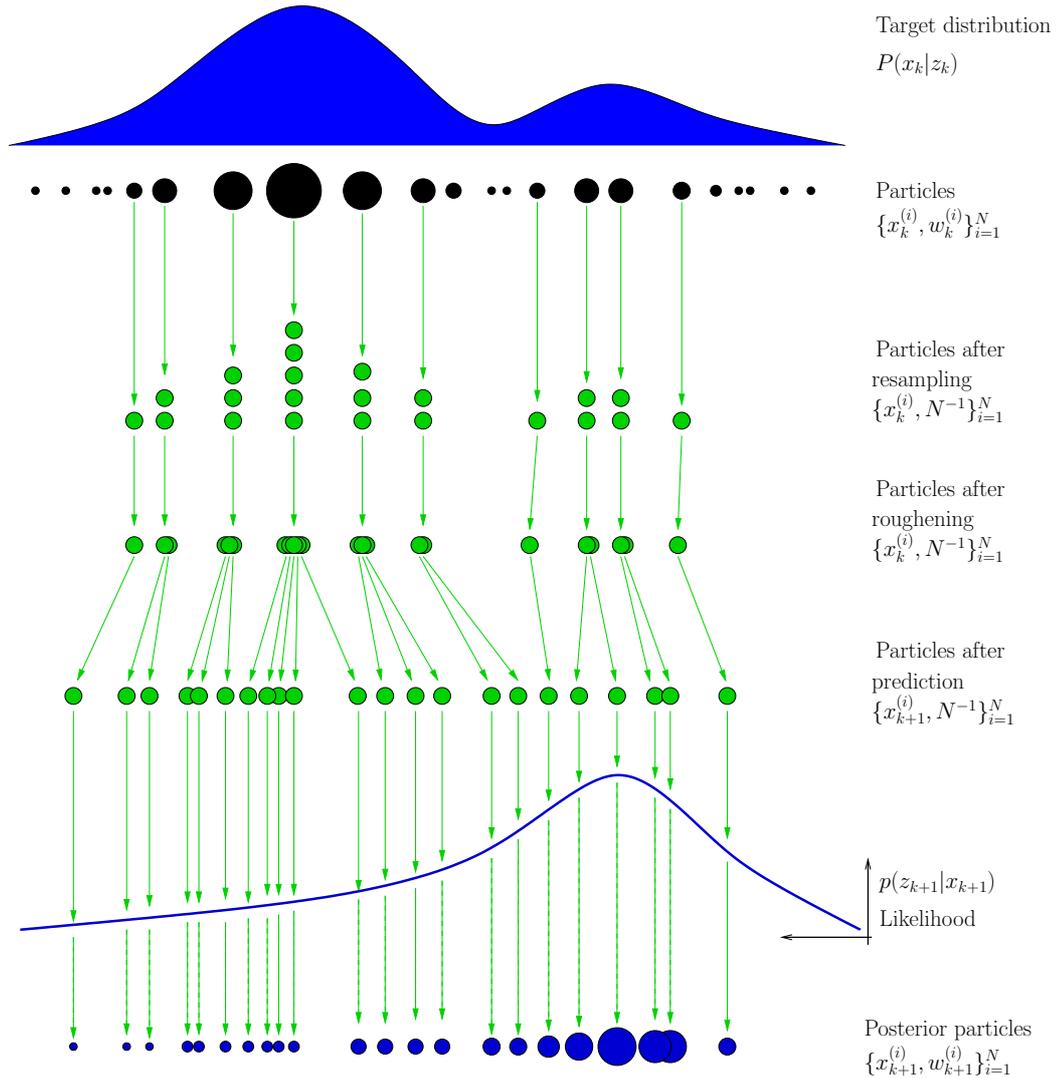}
\caption{A single cycle of a particle filter with $N=22$ and transitional prior as the importance density}
\label{fig:BootstrapPF}
 \end{figure}

The pseudo code for a generic particle filter which incorporates resampling and roughening is shown in Table.\ref{tab:GPF} \cite{4,8,11}.  A graphical representation of a PF with $N=22$ samples and using the transitional prior as the importance density is shown in Fig \ref{fig:BootstrapPF}. At the top we have the target distribution $p(\mathbf{x}_k|\mathbf{z}_k)$ which is approximated using the particles $\{\mathbf{x}_k^{(i)},w_k^{(i)}\}_{i=1}^{N}$. If $N_{eff}<N_{thr}$ resampling is executed on these particles to obtain uniform weight particles $\{\mathbf{x}_k^{(i)},N^{-1}\}_{i=1}^{N}$, which still approximates  the target distribution $p(\mathbf{x}_k|\mathbf{z}_k)$. Resampling is followed by roughening to modify duplicate particles. The resultant particles are used for prediction using the transitional prior to get particles $\{\mathbf{x}_{k+1}^{(i)},N^{-1}\}_{i=1}^{N}$ that approximate the density $p(\mathbf{x}_{k+1}|\mathbf{z}_k)$. Next the weight update is carried out using the likelihood $p(\mathbf{z}_{k+1}|\mathbf{x}_{k+1})$ to obtain particles $\{\mathbf{x}_{k+1}^{(i)},w_{k+1}^{(i)}\}_{i=1}^{N}$ that approximate the density $p(\mathbf{x}_{k+1}|\mathbf{z}_{k+1})$.

\begin{table}[ht!] 
\caption{Generic Particle Filter \cite{4,11}} 
\centering          
\resizebox{!}{4.5in} {
\begin{tabular}{l}
  \hline
  \begin{minipage}{6in}
    \vskip 4pt
\begin{enumerate}
\item For $k=0$,
    \begin{itemize}
	  \item For $i=1,....,N$: Initialize
	  \begin{itemize}
		\item Sample $\mathbf{x}_0^{(i)}\sim q(\mathbf{x}_0|\mathbf{z}_0)$
		\item Evaluate the unnormalized importance weights
		    \begin{eqnarray}
			{\tilde{w}}_0^{(i)}& = & \dfrac{p(\mathbf{z}_0|\mathbf{x}_0^{(i)})p(\mathbf{x}_0^{(i)})}{q(\mathbf{x}_0^{(i)}|\mathbf{z}_{0})} 
		    \end{eqnarray}
	  \end{itemize}
	  \item For $i=1,....,N$:
	  \begin{itemize}
	        \item Normalize the importance weights
		\begin{eqnarray}
		    w^{(i)}& = & \dfrac{{\tilde{w}}^{(i)}}{\sum_{i=1}^N {\tilde{w}}^{(i)}}	    
		\end{eqnarray}
	  \end{itemize}
    \end{itemize}
\item For $k>0$
    \begin{itemize}
	\item For $i=1,....,N$:
	\begin{itemize} 
		\item Sample $\mathbf{x}_k^{(i)} \sim q(\mathbf{x}_k|\mathbf{X}_{k-1},\mathbf{Z}_{k-1})$
		\item Evaluate the unnormalized importance weights
		    \begin{eqnarray}
			{\tilde{w}}_k^{(i)} & \varpropto & w_{k-1}^{(i)}\dfrac{p(\mathbf{z}_k|\mathbf{x}_k^{(i)})p(\mathbf{x}_k^{(i)}|\mathbf{x}_{k-1}^{(i)})}{q(\mathbf{x}_k^{(i)}|\mathbf{x}_{k-1}^{(i)},\mathbf{z}_{k})} 	    
		    \end{eqnarray}
		\item For $i=1,....,N$:
			    \begin{itemize}
				\item Normalize the importance weights
				\begin{eqnarray}
				    w^{(i)}& = & \dfrac{{\tilde{w}}^{(i)}}{\sum_{i=1}^N {\tilde{w}}^{(i)}}		    
				\end{eqnarray}
			    \end{itemize}
	\end{itemize}
    \end{itemize}
\item Calculate $N_{eff}$ 
    \begin{eqnarray}
	N_{eff}& = & \dfrac{1}{\sum_{i=1}^N (w_k^{(i)})^2}	    
    \end{eqnarray}

\item If $N_{eff}<N_{thr}$
    \begin{itemize}
      \item Resample the particles using algorithm in Table \ref{tab:Sresampling} or Table \ref{tab:Mresampling}
	\begin{eqnarray}
	    [\{\mathbf{x}_k^{(n)},w_k^{(n)}\}_{n=1}^{N}] & = & RESAMPLE[\{\mathbf{x}_k^{(i)},w_k^{(i)}\}_{i=1}^{N}]
	\end{eqnarray}

      \item Roughen the particles using algorithm in Table \ref{tab:Roughening}
	  \begin{eqnarray}
	  [\{\mathbf{x}_k^{(n)},w_k^{(n)}\}_{n=1}^{N}]] & = & ROUGHEN[\{\mathbf{x}_k^{(i)},w_k^{(i)}\}_{i=1}^{N}]
	  \end{eqnarray}
    \end{itemize}  
\end{enumerate}
  \vskip 4pt
 \end{minipage}
 \\
  \hline
 \end{tabular}
}
\label{tab:GPF} 
\end{table}

\section{Bootstrap Filter}
Bootstrap filter proposed in \cite{7} is also known as sequential importance resampling (SIR) filter. It is a modification of the above generic particle filter. It uses transitional prior as the importance density and performs resampling at every step. For this choice of importance density the weight update equation is given by \eqref{eqn:condpdf23}. Since the resampling is done at every step, the resampled particles at the previous instant have weights $w_{k-1}^{(i)}=N^{-1}$ for a $i=1,...,N$. Hence the weight update equation reduces to 
\begin{equation}
 w_k^{(i)}\varpropto p(\mathbf{z}_k|\mathbf{x}_{k}^{(i)}) 
\end{equation}
The bootstrap filter has the advantage that the importance weights can be easily calculated and the importance density can be easily sampled. The pseudocode for bootstrap filter is shown in Table \ref{tab:BootstrapPF}.
\begin{table}[ht!] 
\caption{Bootstrap Particle Filter \cite{4,11}} 
\centering          
\begin{tabular}{l}
  \hline
  \begin{minipage}{6in}
    \vskip 4pt
\begin{enumerate}
\item For $k=0$,
    \begin{itemize}
	  \item For $i=1,....,N$: Initialize
	  \begin{itemize}
		\item Sample $\mathbf{x}_0^{(i)}\sim p(\mathbf{x}_0)$
		\item Assign particle weights
		    \begin{eqnarray}
			w_0^{(i)}& = & N^{-1} 
		    \end{eqnarray}
	  \end{itemize}	  
    \end{itemize}
\item For $k>0$
    \begin{itemize}
	\item For $i=1,....,N$:
	\begin{itemize} 
		\item Sample $\mathbf{x}_k^{(i)} \sim p(\mathbf{x}_k|\mathbf{X}_{k-1})$
		\item Evaluate the unnormalized importance weights
		    \begin{eqnarray}
			{\tilde{w}}_k^{(i)} & \varpropto p(\mathbf{z}_k|\mathbf{x}_k^{(i)})		    
		    \end{eqnarray}
	\end{itemize}
	\item For $i=1,....,N$:
			    \begin{itemize}
				\item Normalize the importance weights
				\begin{eqnarray}
				    w^{(i)}& = & \dfrac{{\tilde{w}}^{(i)}}{\sum_{i=1}^N {\tilde{w}}^{(i)}}		    
				\end{eqnarray}
			    \end{itemize}
	
	\item Resample the particles using algorithm in Table \ref{tab:Sresampling} or Table \ref{tab:Mresampling}
		  \begin{eqnarray}
		      [\{\mathbf{x}_k^{(n)},w_k^{(n)}\}_{n=1}^{N}] & = & RESAMPLE[\{\mathbf{x}_k^{(i)},w_k^{(i)}\}_{i=1}^{N}]
		  \end{eqnarray}

	\item Roughen the particles using algorithm in Table \ref{tab:Roughening}
		    \begin{eqnarray}
		    [\{\mathbf{x}_k^{(n)},w_k^{(n)}\}_{n=1}^{N}]] & = & ROUGHEN[\{\mathbf{x}_k^{(i)},w_k^{(i)}\}_{i=1}^{N}]
		    \end{eqnarray}
    \end{itemize}
\end{enumerate}
  \vskip 4pt
 \end{minipage}
 \\
  \hline
 \end{tabular}
\label{tab:BootstrapPF} 
\end{table}

\section{Other Particle Filters}
The variations in the selection of importance density and/or modification of the resampling step has resulted in various versions of particle filters like
\begin{enumerate}
 \item Auxiliary SIR filter
 \item Regularized particle filter
 \item MCMC particle filter
 \item Multiple Model particle filter(MMPF)
 \item Independent partition particle filter(IPPF) etc.
\end{enumerate}
Of these particle filters, the IPPF and MMPF will be considered in later chapters.
\section{Simulation Results}
A target motion scenario and its measurements are simulated according to the given models and the estimates using the generic particle filter algorithm is compared with the true trajectories. For comparison, estimation is done using the extended Kalman filter also on the same target tracking problem and the results are compared.
We have a target which has constant velocity and constant turn motions. The state vector consists of position and velocities of the target,
\begin{eqnarray}
\mathbf{\mathbf{x}}=
\begin{bmatrix}
x & v_{x} & y  & v_{y}
 \end{bmatrix}^T
\end{eqnarray}
 The initial true state of the target was 
$\mathbf{x}_{0}=\begin{bmatrix}
100&20&100&20
 \end{bmatrix}^T$. From time $k=0s$ to $k=20s$, $k=61s$ to $k=70s$, $k=91s$ to $k=100s$, the target has constant velocity motion.  From $k=21s$ to $k=60s$, $k=11s$ to $k=90s$, it moves in clockwise constant turn rate motion of $6 rad/s$. The measurement sensor is located at the origin. The target's range $r$ and bearing $\theta$ at time $k$ are available as the measurement $\mathbf{z}_k$.
\begin{eqnarray}
\mathbf{z}_k=h(\mathbf{x}_{k})+\mathbf{v}_k \\
\mathbf{v}_k\sim \mathcal{N}(0,Q_v) 
\end{eqnarray}
 where $\mathbf{v}_k$ is the measurement error, $h(\cdotp)$ is the measurement model . The measurement error $\mathbf{v}_k$ is uncorrelated and has zero mean Gaussian distribution with covariance matrix $Q_v$. 

\begin{eqnarray}
\mathbf{z}_k=
\begin{bmatrix}
r\\
\theta\\
\end{bmatrix}
\end{eqnarray}
\begin{eqnarray}
\mathbf{Q_v}=
\begin{bmatrix}
\sigma_{r}^{2} &0 \\
0 &\sigma_{\theta}^{2} \\
\end{bmatrix}=
\begin{bmatrix}
10 &0 \\
0 &1\\
\end{bmatrix} 
\end{eqnarray}

The measurement model $h(\cdotp)$ for the target is given by:
\begin{eqnarray}
h(\mathbf{x}_{k})=
\begin{bmatrix}
\sqrt{x^2_k+y^2_k}\\
\tan^{-1}\left(\dfrac{y_k}{x_k}\right)
\end{bmatrix} 
\end{eqnarray}

 The initial state estimate is assumed to be a Gaussian vector with mean $\mathbf{x}_{0}$ and error covariance $P_{0}$, such that
\begin{equation}
 \mathbf{x}_{0}=\begin{bmatrix}
100&20&100&20
 \end{bmatrix}^T 
\end{equation}
\begin{equation}
 P_{0}=\begin{bmatrix}
100 &10 &100 &10 
 \end{bmatrix}^T 
\end{equation}
Hence initial particles $\{\mathbf{x}_{0}^{(i)}\}_{i=1}^N$ were generated based on the distribution 
\begin{equation}
 \mathbf{x}_{0} \sim \mathcal{N}(\mathbf{x}_{0},P_{0})
\end{equation}
In this implementation of the particle filter, the transitional prior which is a suboptimal choice of the importance density is  used to propose particles. The state transition model $f(\cdotp)$ for estimation of state at time $k$ is such that: 
\begin{eqnarray}
\mathbf{x}_{k} & = & f(\mathbf{x}_{k-1})+\mathbf{w}_{k-1} 
\end{eqnarray}
where $\mathbf{w}_{k-1}$ is the process noise with zero mean. 
The state transition model $f(\cdotp)$ used in this implementation of the generic particle filter is constant velocity model.
 Hence $f(\cdotp)$ is a matrix $F$ given by:
\begin{equation}
 F=\begin{bmatrix}
1&T&0&0\\
0&1&0&0\\
0&0&1&T\\
0&0&0&1\\
\end{bmatrix} 
\end{equation}\\
where $T$ is the sampling period of the target dynamics. 
The process noise assumed has a diagonal covariance matrix $Q_{w}$ as:
\begin{equation} 
Q_w = diag \left(
\begin{array}{cccccccc}
5 ,&1, &5 ,& 1
\end{array} \right) 
\end{equation}

The number of particles used was $N=500$. The detailed implementation algorithm for the target tracking problem is given in Table.\ref{tab:GPF_impltn}. Since the resampling can only reduce the accuracy of the estimates of the distribution, the estimates such as conditional mean, covariance of samples, mean square error(MSE) are calculated before resampling. Results shown are calculated for 100 Monte Carlo runs.

\begin{table}[ht!] 
\caption{Implementation of GPF} 
\centering          
\resizebox{!}{4in} {
\begin{tabular}{l}
  \hline
  \begin{minipage}{6in}
    \vskip 4pt
\begin{enumerate}
      \item For $k=0$, initialize all particles:
      \begin{itemize}
	    \item For $i=1,...,100$, generate samples $\mathbf{x}_{0}^{(i)} \sim \mathcal{N}(\mathbf{x}_{0},P_{0})$
	    \item For $i=1,...,100$, assign weights $w_{0}^{(i)}=\dfrac{1}{100}$
      \end{itemize}
      \item For  $k>0$,
      \begin{itemize}
		  \item For $i=1, 2,..., 100$  
		  \begin{itemize}
		      \item Draw sample $\mathbf{x}_{k}^{(i)}$ using the transitional prior.
		      \begin{equation}
			      a_k^{(i)}= F\mathbf{x}_{k-1}^{(i)}
		      \end{equation}
			\begin{equation}
			      \mathbf{x}_{k}^{(i)} \sim p(\mathbf{x}_{k}\mid \mathbf{x}_{k-1}^{(i)})=\mathcal{N}(a_k^{(i)},Q_{w}) 
			\end{equation}	
		    
		      \item Evaluate the unnormalized importance weights
		    \begin{eqnarray}
			{\tilde{w}}_k^{(i)} & \varpropto & w_{k-1}^{(i)}\dfrac{p(\mathbf{z}_k|\mathbf{x}_k^{(i)})p(\mathbf{x}_k^{(i)}|\mathbf{x}_{k-1}^{(i)})}{q(\mathbf{x}_k^{(i)}|\mathbf{x}_{k-1}^{(i)},\mathbf{z}_{k})} 	    
		    \end{eqnarray}				    
		  \end{itemize}

		 \item For $i=1,....,100$:
			    \begin{itemize}
				\item Normalize the importance weights
				\begin{eqnarray}
				    w^{(i)}& = & \dfrac{{\tilde{w}}^{(i)}}{\sum_{i=1}^N {\tilde{w}}^{(i)}}		    
				\end{eqnarray}
			    \end{itemize}

		\item Calculate the target estimates such as conditional mean, covariances, mean square error MSE etc.

		\item Calculate $N_{eff}$ 
		    \begin{eqnarray}
			N_{eff}& = & \dfrac{1}{\sum_{i=1}^N (w_k^{(i)})^2}		    
		    \end{eqnarray}

		\item If $N_{eff}<N_{thr}$
		    \begin{itemize}
		      \item Resample the particles using algorithm in Table \ref{tab:Sresampling} or Table \ref{tab:Mresampling}
			\begin{eqnarray}
			    [\{\mathbf{x}_k^{(n)},w_k^{(n)}\}_{n=1}^{N}] & = & RESAMPLE[\{\mathbf{x}_k^{(i)},w_k^{(i)}\}_{i=1}^{N}]
			\end{eqnarray}

		      \item Roughen the particles using algorithm in Table \ref{tab:Roughening}
			  \begin{eqnarray}
			  [\{\mathbf{x}_k^{(n)},w_k^{(n)}\}_{n=1}^{N}]] & = & ROUGHEN[\{\mathbf{x}_k^{(i)},w_k^{(i)}\}_{i=1}^{N}]
			  \end{eqnarray}
		    \end{itemize}  
	\end{itemize}
\end{enumerate}
  \vskip 4pt
 \end{minipage}
 \\
  \hline
 \end{tabular}
}
\label{tab:GPF_impltn} 
\end{table}

\begin{figure}[h]
\centering
\subfloat [$xy$ track]
{\label{GPF_traj}\includegraphics[scale=0.4]{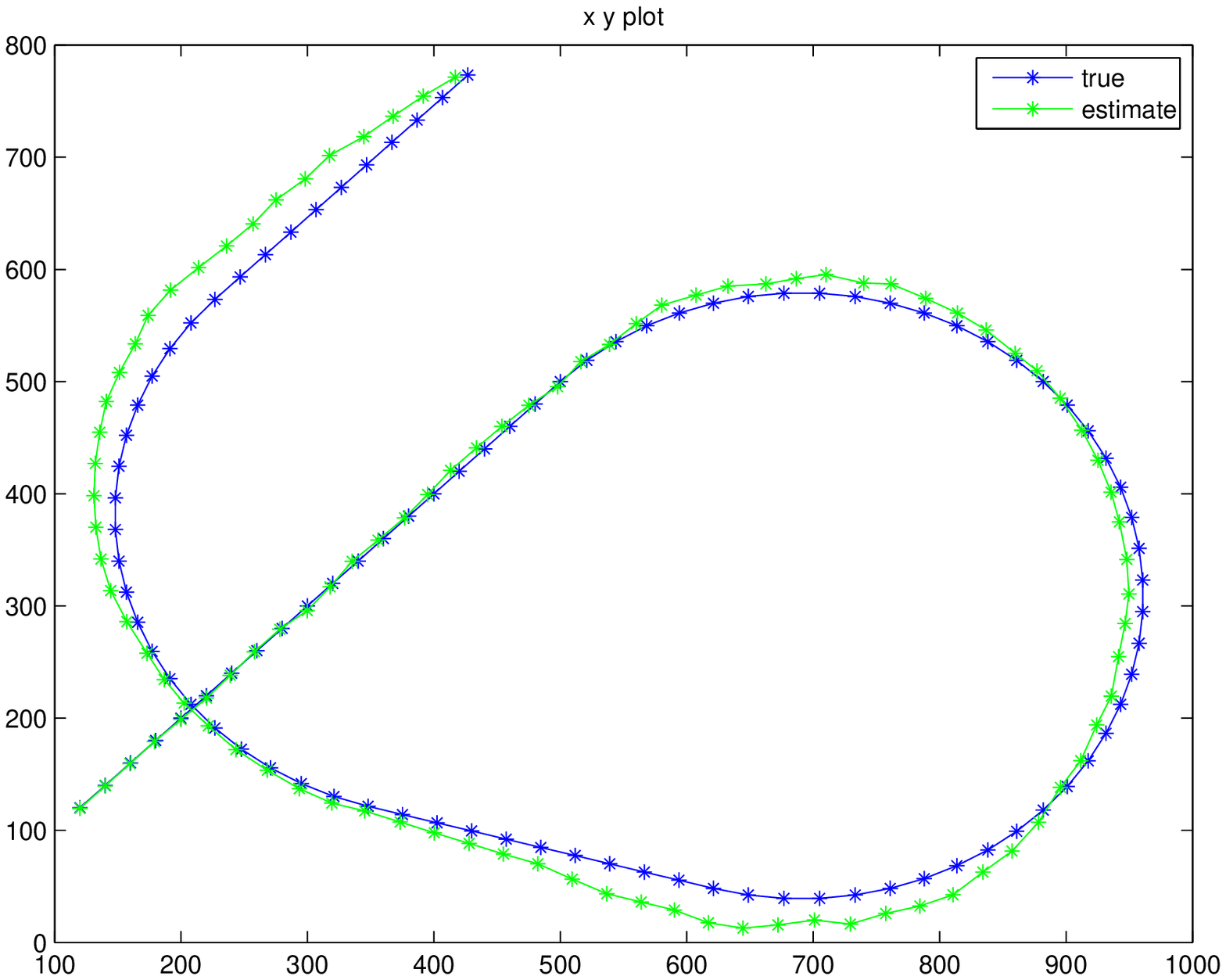} 
} 
\subfloat [MSE]
{\label{GPF_MSE}\includegraphics[scale=0.4]{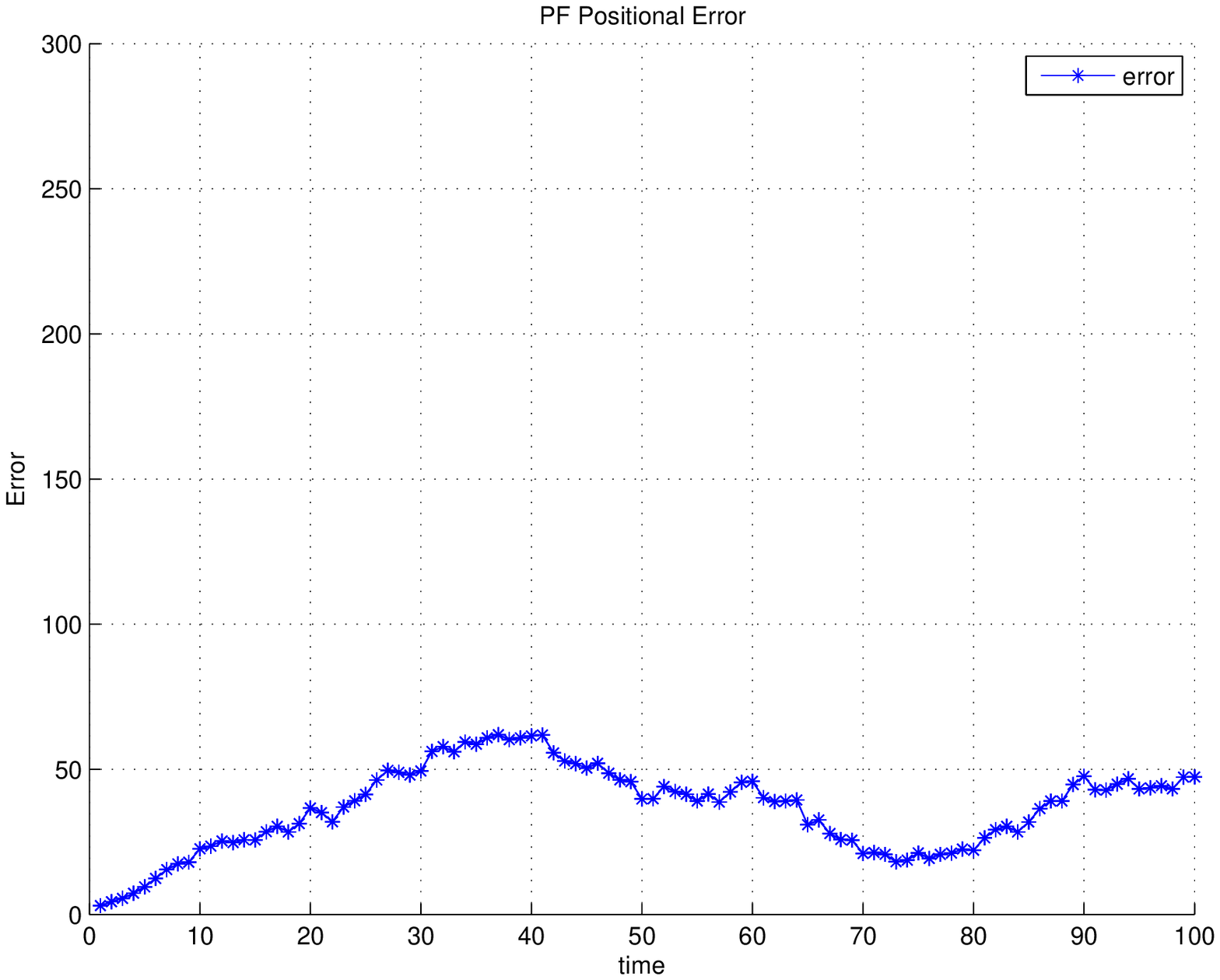}  
}\\   
\caption{Target's true $xy$ track with its estimate and MSE obtained using generic PF.}
\end{figure}

\begin{figure}[p]
\centering
\subfloat [position $x$ ]
{\includegraphics[scale=0.4]{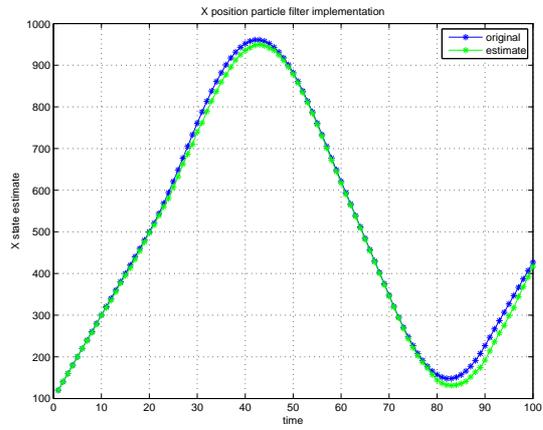}  
}\\
\subfloat [position $y$ ]
{\includegraphics[scale=0.4]{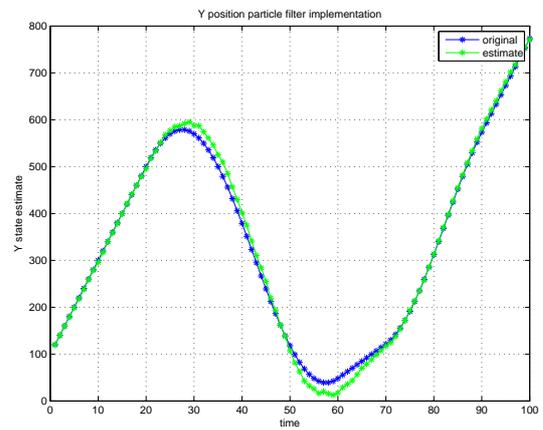}  
}\\
\subfloat [velocity $v_x$]
{\includegraphics[scale=0.4]{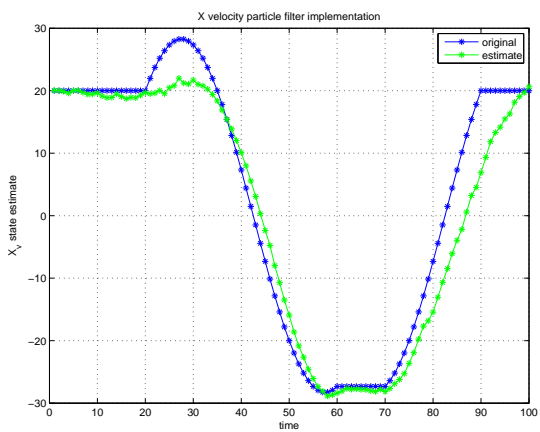}  
}%
\subfloat [velocity $v_y$]
{\includegraphics[scale=0.4]{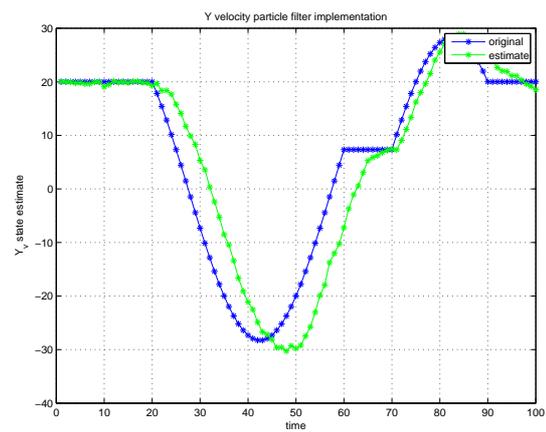}  
}\\
\caption{Target's true states and their estimates obtained using generic PF.}
\label{GPF_state}
 \end{figure}

The true trajectory of the target and its estimates are shown in Fig.\ref{GPF_traj}. The state estimates of the target are shown in Fig. \ref{GPF_state}. The mean square error MSE of the position estimates are shown in Fig.\ref{GPF_MSE}.
%
%
\begin{figure}[h]
\centering
\subfloat [$xy$ track]
{\label{EKF_traj}\includegraphics[scale=0.4]{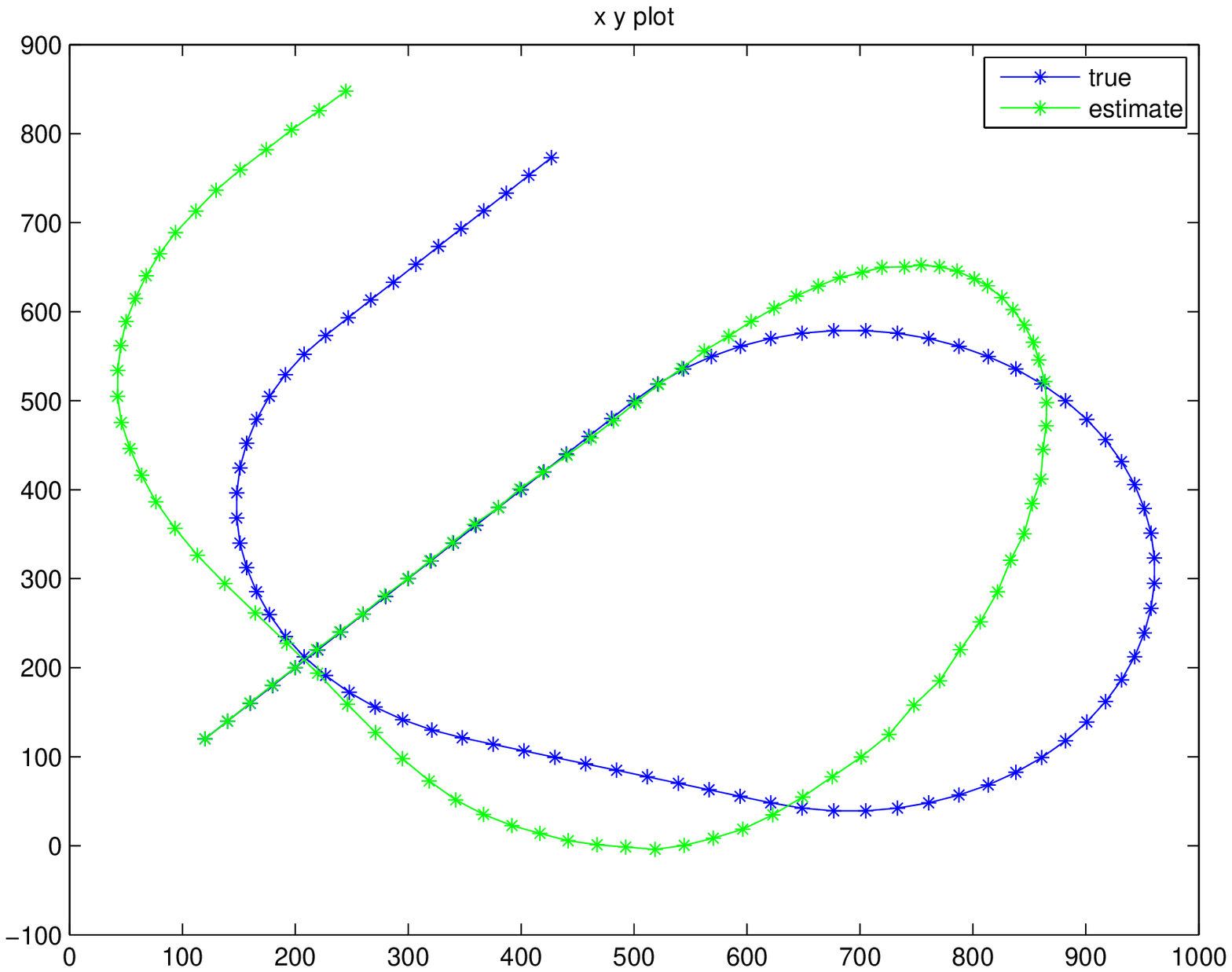}
} 
\subfloat [MSE]
{\label{EKF_MSE}\includegraphics[scale=0.4]{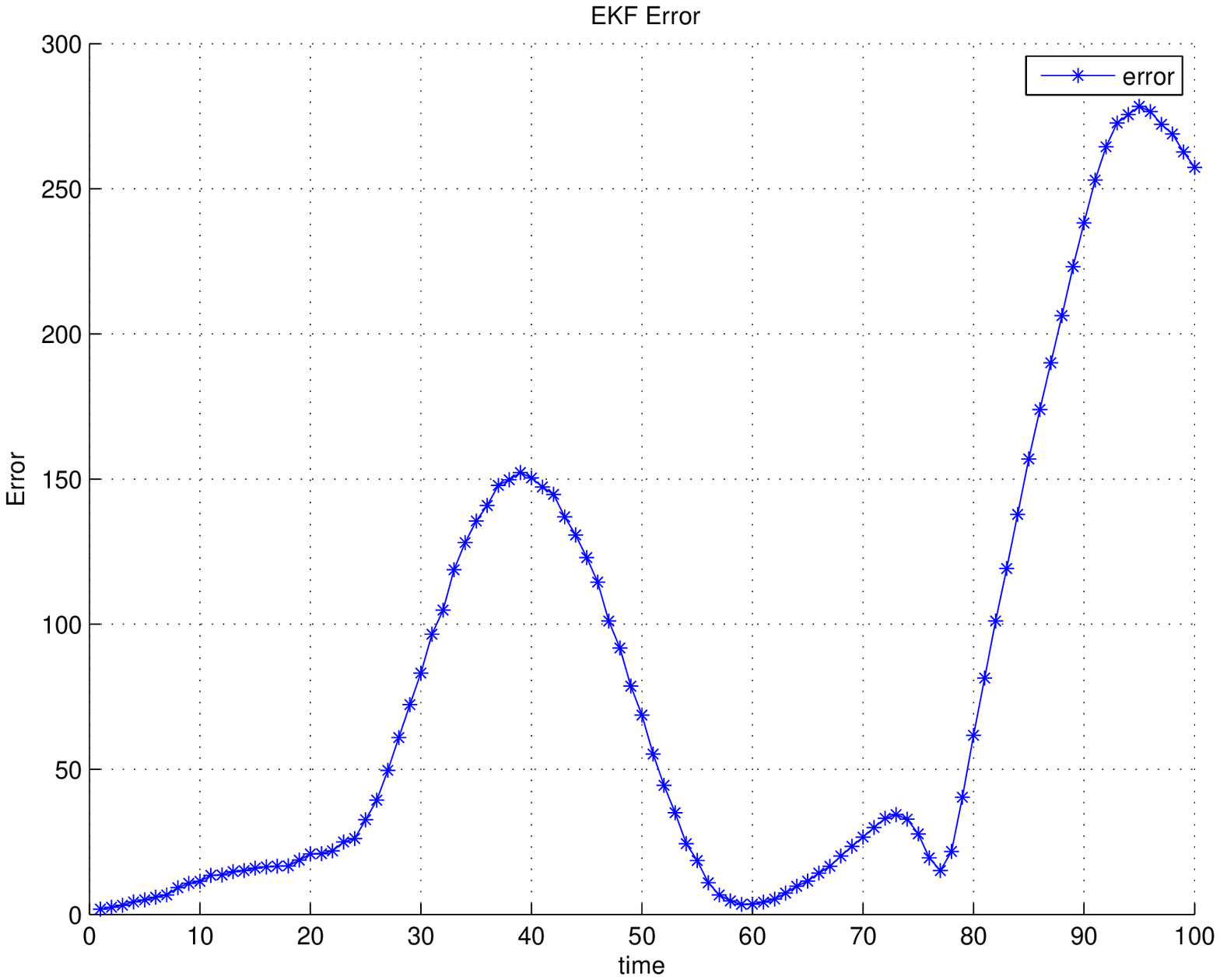}  
}\\   
\caption{Target's true $xy$ track with its estimate and MSE obtained using EKF.}
\end{figure}


\subsection{Comparison of Particle Filter with EKF}
\begin{figure}[p]
\centering
\subfloat [position $x$ ]
{\includegraphics[scale=0.4]{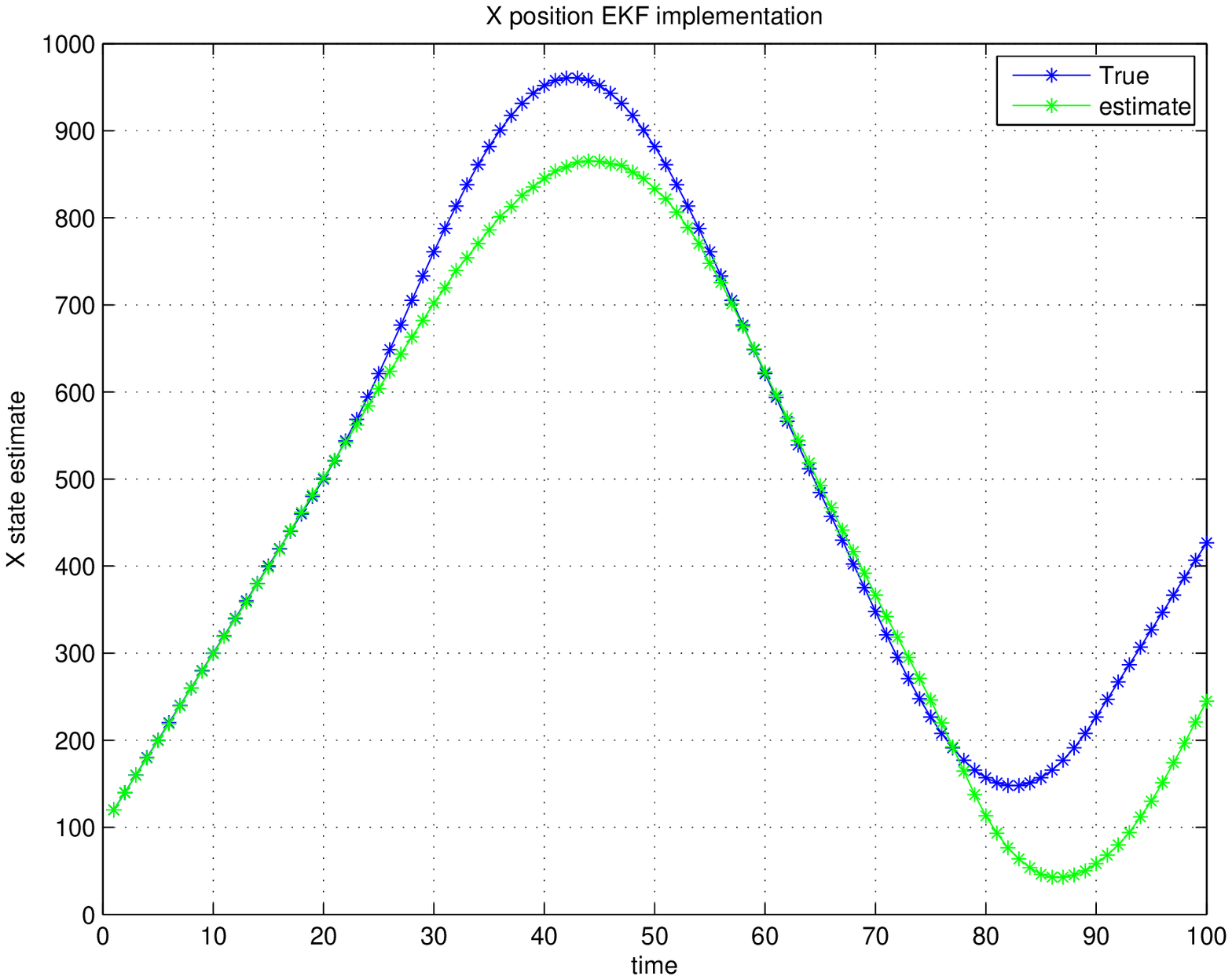}  
}\\
\subfloat [position $y$ ]
{\includegraphics[scale=0.4]{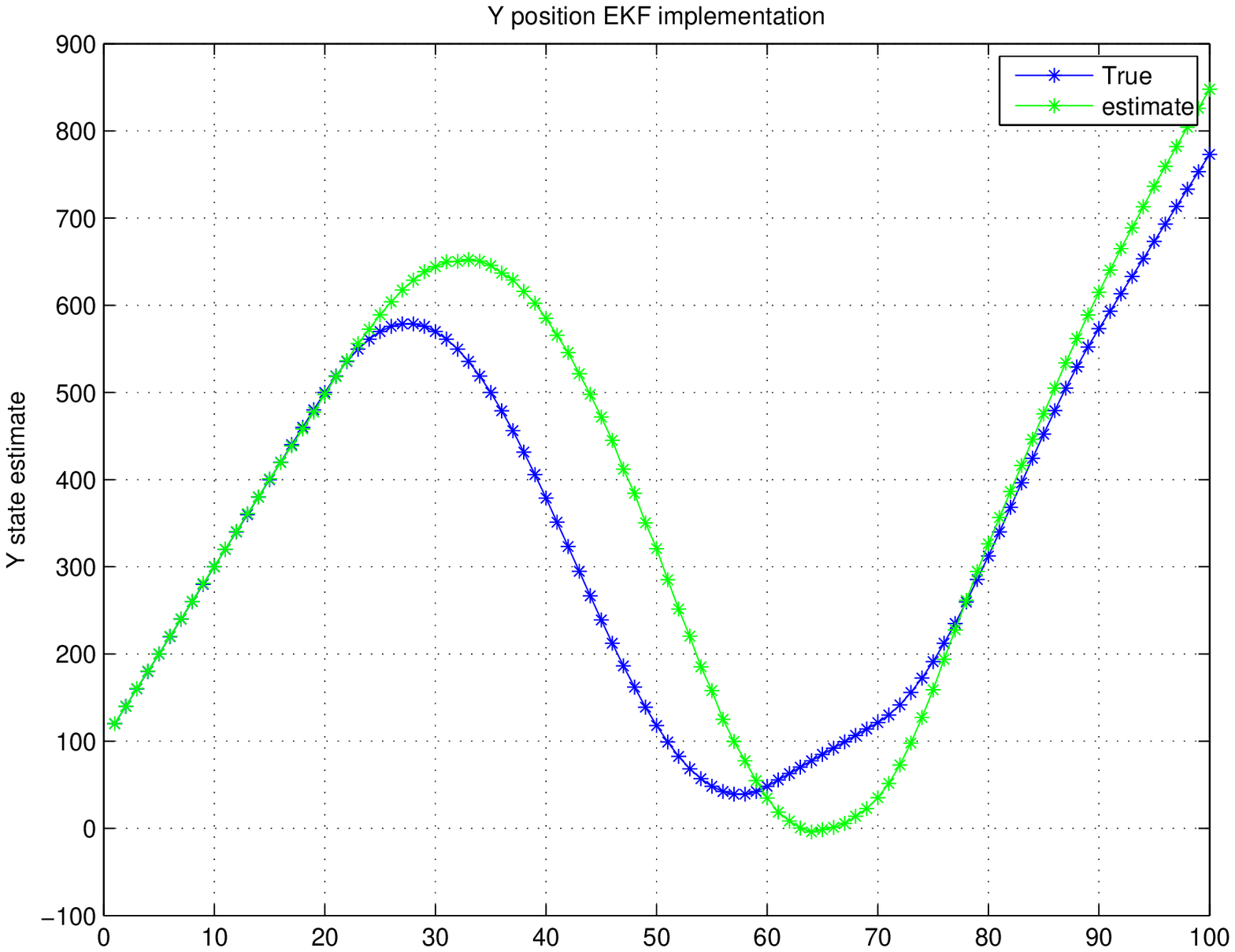}  
}\\
\subfloat [velocity $v_x$]
{\includegraphics[scale=0.4]{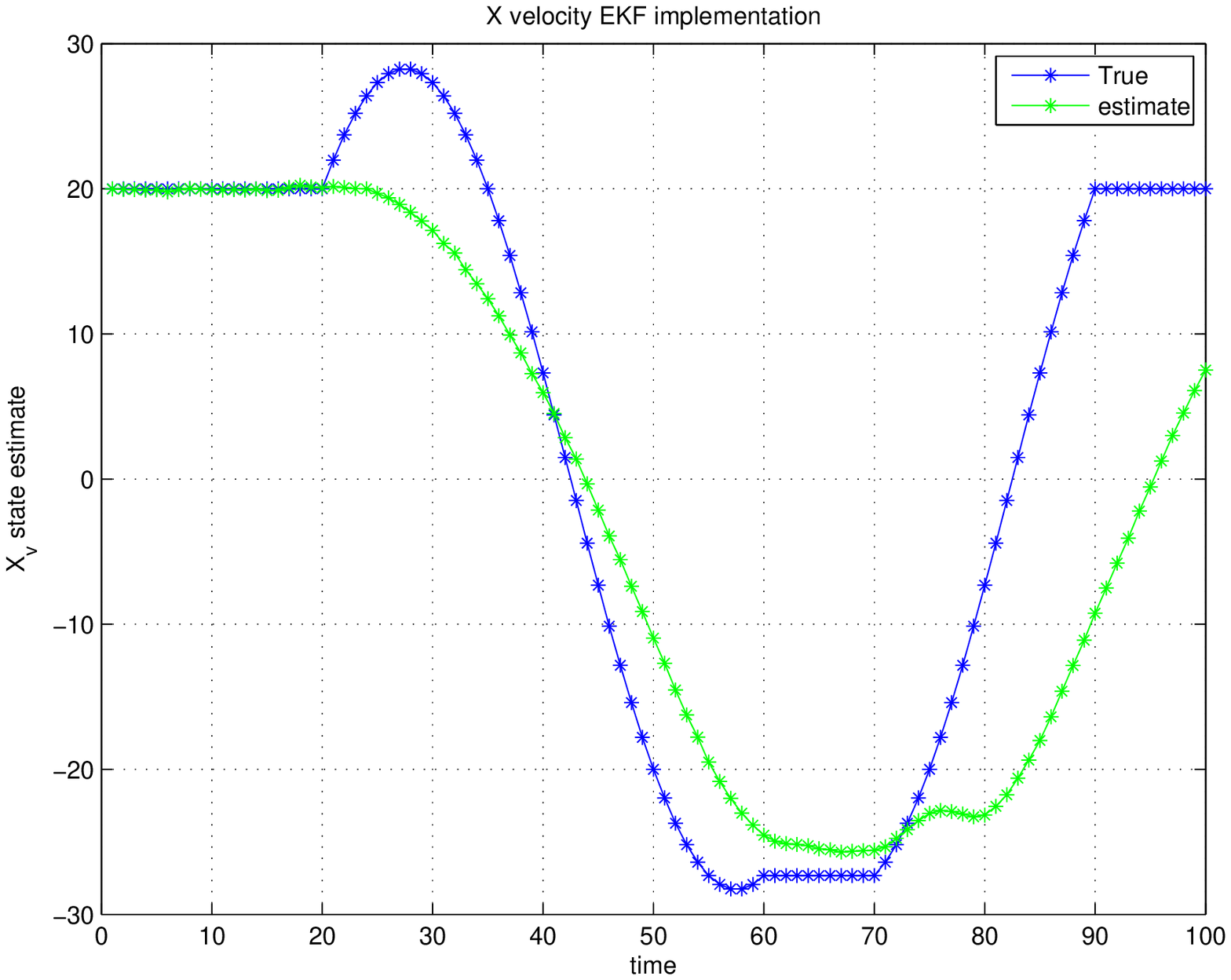}  
}%
\subfloat [velocity $v_y$]
{\includegraphics[scale=0.4]{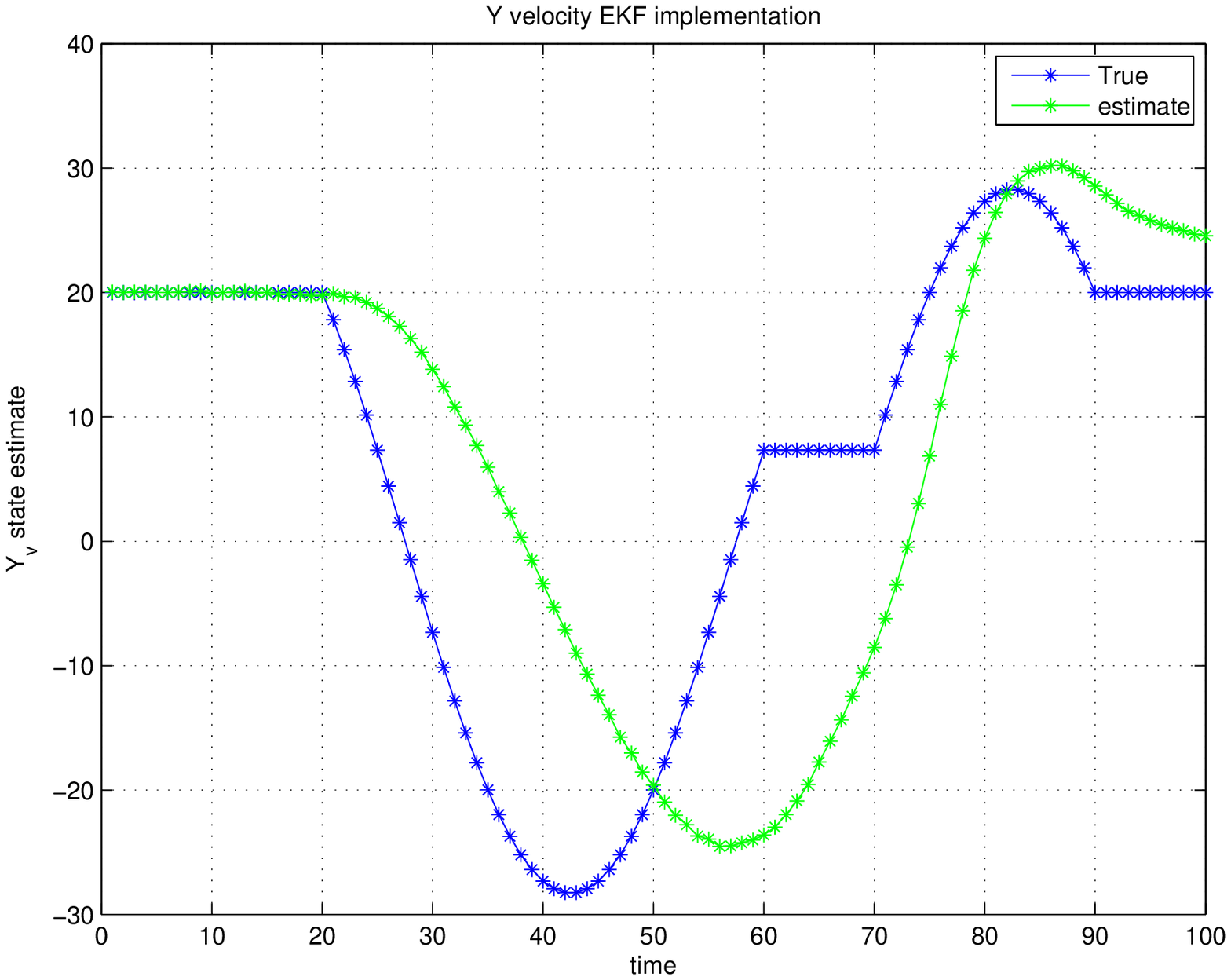}  
}\\
\caption{Target's true states and their estimates obtained using EKF.}
\label{EKF_state}
\end{figure}

For comparison the extended Kalman filter(EKF) is also implemented for the same target motion scenario. The true trajectory of the target and its estimate is shown in Fig.\ref{EKF_traj}.  The state estimates are calculated after 100 Monte Carlo runs and are shown in Fig.\ref{EKF_state}. The mean square error MSE of the position estimate is shown in Fig.\ref{EKF_MSE}. The results show that the estimates obtained using EKF diverge. Thus it shows that particle filter has better tracking accuracy under nonlinear target motions and it can handle moderate maneuvers of the target  by using only constant velocity models without the need of maneuvering models. 

\section{Summary}
Particle filter is a class of Monte Carlo method to solve recursive Bayesian estimation. It represents the probability distribution of a target using particles and associated weights. It doesn't require the assumptions of linearity and Guassianity and is capable of handling complex noise distributions and non-linearities in the target's measurements as well as target dynamics. Importance sampling provide the alternative to sample particles from a complex distribution using an another suitable easy to sample distribution called importance density. Sequential importance sampling helps to perform importance sampling recursively and reduce its computational complexity. Particle filter consists of proposing particles using importance function and weight update of these particles at every iteration and is capable of parellel implementation. Implementation issues like degeneracy and sample impoverishment are addressed by resampling and roughening respectively. Selection of good importance density also reduces the frequency of resampling. Simulations confirm that particle filter outperforms EKF in tracking maneuvering targets at the expense of increased computational cost. Independent partition particle filter (IPPF) for multi-target tracking and Multiple Model Particle filter (MMPF) for maneuvering target tracking are explored in the subsequent chapters.

\chapter{Multi-target Tracking using Independent Partition Particle Filter (IPPF)}
\label{chap:IPPF}
\thispagestyle{empty}

 Partitioned sampling is developed by J. ~Maccormick et.al \cite{3} for tracking more than one target. The independent partition Particle Filter (IPPF) is given by M.~Orton et.al \cite{1} is a convenient way to propose particles when part or all of the joint multi-target density factors. These techniques are explored in this chapter. In particle filters, the number of particles required to model a distribution increases with dimension
 $n_{x}$ of the state space. The upper bound on the variance of the estimation error has the form $cN^{-1}$,
 where $c$ is a constant and $N$ is the number of particles used by the particle filter.
The constant $c$ depends heavily on the state vector dimension $n_{x}$ of the system \cite{4}. The variance of the estimation error
 for particle filter becomes exponential in $n_{x}$ for poorly chosen importance density and is referred
 to as \textquotedblleft curse of dimensionality\textquotedblright. Hence the number of required particles $N$ should be higher for higher dimensional systems like multi-target tracking systems.

In the case of multi-targets, the proportion of state space that is filled by the region of the likelihood
 with reasonably high probability gets smaller. A particle with one very improbable state, and all the remaining states
 being probable may be rejected during resampling step of particle filter since overall this particle is 
improbable. It is the low probability of the bad estimates that determines the fate of the whole particle.
 Hence parts of the particle are penalized at the expense of other parts. A better approach is to ensure that 
either whole particle is probable or the whole particle is improbable. This can be done by redistributing the
 set of weighted particles so as to increase the density of particles in certain regions of interest, and 
account for redistribution by suitable weights such that it doesn't alter the underlying distribution 
described by the former particles. This is accomplished by Weighted Resampling technique described in \cite{1,2,3}.

\section{Weighted Resampling}
Weighted resampling with respect to a function  $g(\mathbf{x})$, is an operation on the particle set
which populates peaks of  $g(\mathbf{x})$ with particles without altering the distribution actually represented by the particle set. Given a weighted set of particles, the weighted resampling populates certain parts of the configuration space with particles in the desired manner so that representation is more efficient for future operations. It has the advantage that subsequent operations
on this particle set will produce more accurate representation of the desired probability distributions. Weighted 
resampling is carried out with respect to a strictly positive weighting function $g(\mathbf{x})$. It is analogous to the importance function used in standard importance sampling. Let the $i^{th}$ particles be $\mathbf{x}^{(i)}$ with weight $w^{(i)}$. 

\begin{equation}
 \mathbf{x}=\{\mathbf{x}^{(1)},\mathbf{x}^{(2)},\mathbf{x}^{(3)},....,\mathbf{x}^{(N)}\}
\end{equation}
\begin{equation}
\mathbf{w}=\{w^{(1)}, w^{(2)}, w^{(3)}, . . . . , w^{(N)}\}
\end{equation}

  Given a set of $N$ particles $\mathbf{x}$, with corresponding weights  $\mathbf{w}$,
 it produces a new particle set by resampling from $\mathbf{x}$,
 using secondary weights which are proportional to $g(\mathbf{x})$. This has the effect of selecting many particles
in regions where $g(\mathbf{x})$ is peaked. The weights of the resampled particles are calculated in such a way that overall distribution
represented by the new particle set is same as the old one.
Thus 
 asymptotically any strictly positive function is acceptable as the weighting
function $g(\mathbf{x})$, but it is better to select a function which has advantage in our application.
We would like the weighted resampling step to position as many particles as possible near peaks in the posterior. Hence a 
natural choice therefore is to take  $g(\mathbf{x})$ to be the likelihood function of the target itself. The algorithm for one dimensional
 weighted resampling with respect to importance function  $g(\mathbf{x})$ is repeated in Table.\ref{tab:WR} from \cite{3}. Here $\mathbf{x}_{k},w_{k}$ represents the particles at time $k$.

\begin{table}[h] 
\caption{Weighted Resampling \cite{3}} 
\centering          
\begin{tabular}{l}
  \hline
  \begin{minipage}{4.5in}
    \vskip 4pt
$[\{\mathbf{x}_{k}^{(j)},w_{k}^{(j)}\}_{j=1}^{N}]$ =
 Weighted Resampling $[\{\mathbf{x}_{k}^{(i)}, w_{k}^{(i)}\}_{i=1}^{N}, g(\cdotp)]$
\begin{itemize}
\item Define secondary weights $\rho^{(i)} =  \dfrac{g(\mathbf{x}^{(i)})}{\sum^{N}_{j=1}g(\mathbf{x}^{(j)})}$.
\item Sample indices $j(i)$ from the distribution formed by $\rho^{(i)}$ for $i=1,2,3,...,N$ as explained in Appendix A
\item Set $\mathbf{x}^{(i)} = \mathbf{x}^{(j(i))}$
\item Set $w^{(i)} = \dfrac{w^{(j(i))}}{\rho^{(j(i))}}$ 
\item Normalize $w^{(i)}$ such that $\sum w=1$
$$w^{(i)} =  \dfrac{w^{(j(i))}}{\sum^{N}_{i=1}w^{(j(i))}}$$
\end{itemize}
  \vskip 4pt
 \end{minipage}
 \\
  \hline
 \end{tabular}
\label{tab:WR} 
\end{table}

 \begin{figure}[h]
 \centering 
   \includegraphics[scale=0.5]{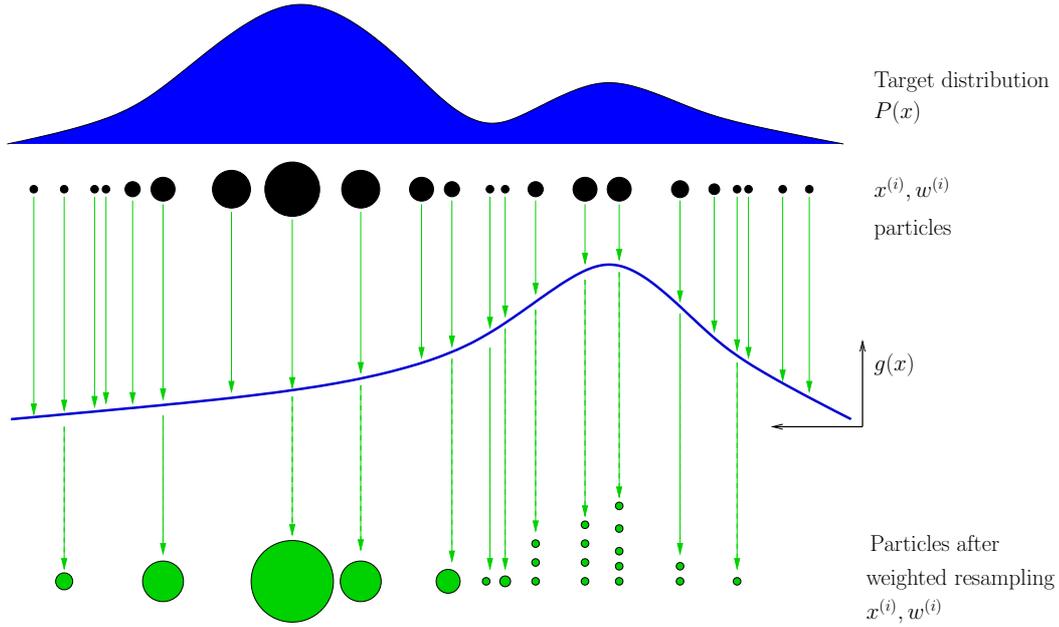}
\caption{Weighted resampling of a set of particles representing a distribution $P(x)$ with respect to function $g(x)$: After weighted resampling, more number of particles get populated near the peak of the function $g(x)$. The resultant particles may have non uniform weights and still have the same initial distribution $P(x)$.}
\label{fig:weighted_resampling}
 \end{figure}

The fourth step in the Weighted Resampling algorithm has the effect of counteracting the extent
 to which the particles were biased by the secondary weights. An intuitive proof that weighted
 resampling doesn't alter underlying distribution is given in \cite{3}. Thus Weighted Resampling has the similar objective and effect as the importance resampling. The difference of weighted resampling and resampling is illustrated in Fig.\ref{fig:weighted_resampling} and Fig.\ref{fig:resampling}.

\section{Independent Partitioned Sampling}

Partitioned sampling is a general term for the method which consists of dividing the state vector into two or more partitions 
and sequentially applying dynamics for each partition and then followed by an appropriate resampling operation. Objective of the partitioned 
sampling is to use one's intuition about the problem to choose a decomposition of the dynamics which simplify the problem, and a weighting function  $g(\mathbf{x})$ to have better rearrangement of the particles.
If the weighting function for the intermediate resampling is chosen to be highly peaked close to the peak in the likelihood for
 that partition, then the weighted resampling step will increase the number of particles close to the peak in the likelihood for
 that partition. After applying this method to each partition, the result is that more particles are likely to contain mostly
good states so that fewer are rejected at the final resampling step.

For independent targets, \cite{3} introduces the Independent Partition Particle Filter. Here the state  $\mathbf{x}_{t}$  is assumed to be separable into
independent partitions, each partition containing the state for one target .Thus  $\mathbf{x}_{t}$ is the union of several partitions,
$\mathbf{x}_{t} \equiv \{\mathbf{x}_{t}(1),\mathbf{x}_{t}(2),\mathbf{x}_{t}(3),.....,\mathbf{x}_{t}(K_{t})\}$ where we have $K_{t}$ partitions, which is the same as the number of targets. 
If the prior is assumed to be independent, and if the
 likelihood and the importance function are also independent
with respect to the same partitioning, then the posterior will have the same independence. 
In this scenario, weighted resampling allows the particles to interact and swap target states. Thus it is used to do the crossover of the targets among the particles implicitly.
Suppose there are two targets $A$ and $B$, represented using five particles $\{A_{1},B_{1}\},\{A_{2},B_{2}\},\{A_{3},B_{3}\},\{A_{4},B_{4}\},\{A_{5},B_{5}\}$. 
Suppose $A_{3},A_{4},B_{2}, B_{5}$ are less probable states and $A_{1}, A_{2},A_{5}, B_{1}, B_{3},B_{4} $ are highly probable, then weighted
resampling applied to each partition does the crossover among the particles and can generate five new particles $\{A_{2},B_{1}\},\{A_{1},B_{3}\},\{A_{2},B_{4}\},\{A_{5},B_{5}\},\{A_{3},B_{1}\}$
such that these particles have more probable states. Hence the new particles get more concentrated at the peak of the posterior.
The algorithm for Independent partition particle filter from \cite{4} is repeated in Table.\ref{tab:IPPF}. 
\begin{table}[ht] 
\caption{Independent Partition Particle Filter (IPPF) \cite{4}} 
\centering          
\resizebox{!}{4in} {
\begin{tabular}{l}
  \hline
  \begin{minipage}{7in}
    \vskip 4pt
\begin{enumerate}
	\item For $t=0$, initialize all particles:
	\begin{itemize}
	    \item For $i=1,...,N$ sample $\mathbf{x}_{0}^{(i)} \sim p(\mathbf{x}_{0})$ where $p(\mathbf{x}_{0})$ is the prior distribution of the target.
	    \item For $i=1,...,N$ calculte weights $w_{0}^{(i)}$ according to $p(\mathbf{x}_{0})$.
	\end{itemize}
	\item For  $t>0$,
	\begin{itemize}
	      \item For $k=1, 2,..., K_{t}$
		\begin{itemize}
			\item For $i=1, 2,..., N$ 
			\begin{itemize}
			      \item Draw sample from the importance density. $\mathbf{x}(k) \sim q_{k}(\mathbf{x}_{t}(k)\mid \mathbf{x}_{t-1}^{(i)}(k),z_{t})$
			      \item Compute secondary weights $g(\mathbf{x}_{t}^{(i)}(k))$
			\end{itemize}
		\end{itemize}
		\item For $k=1, 2,..., K_{t}$ 
		\begin{itemize}
			\item For $i=1, 2,..., N$
			\begin{itemize}
				\item Normalize the secondary weights $$\rho^{(i)}(k)= \dfrac{g(\mathbf{x}_{t}^{(i)}(k)}{\sum^{N}_{j=1}g(\mathbf{x}^{(i)}(k)}$$
			\end{itemize}
		\end{itemize}
		\item For $k=1, 2,..., K_{t}$ 
		\begin{itemize}
			\item For $i=1, 2,..., N$
			\begin{itemize}
				\item Sample indices $j_{k}(i)$ from the distribution formed by $\rho^{(n)}(k)$ for $n=1,2,3,...,N$ by any of the method given in Appendix.\ref{appendix:a}.
			\end{itemize}
		\end{itemize}
		\item For $i=1, 2,..., N$ 
		\begin{itemize}
		    \item Set the new particles $\mathbf{x}_{t}^{(i)} \equiv \{\mathbf{x}_{t}^{(j_{1}(i))}(1),\mathbf{x}_{t}^{(j_{2}(i))}(2),x_{t}^{(j_{3}(i))}(3),.....,\mathbf{x}_{t}^{(j_{K_{t}}(i))}(K_{t})\}$
			    and compute their corresponding particle weights.
		\end{itemize}
		\item For $i=1, 2,..., N$ evaluate the importance weights $$ w_{t}^{(i)}=\dfrac{w_{t-1}^{(i)} p(z_{t}\rvert  \mathbf{x}_{t}^{(i)}) p(\mathbf{x}_{t}^{(i)}\rvert \mathbf{x}_{t-1}^{(i)})}{q(\mathbf{x}_{t}^{(i)}\mid \mathbf{x}_{t-1}^{(i)},z_{t}) \prod_{k=1}^{K_{t}}\rho^{(j_{k}(i))}(k)} $$
		\item For $i=1, 2,..., N$, normalize weights:
		  $$w^{(i)} =  \dfrac{w^{(i)}}{\sum^{N}_{j=1}w^{(j)}}$$
	\end{itemize}
	\item If required resample the particles and do roughening.
\end{enumerate}
 \vskip 4pt
 \end{minipage}
 \\ 
  \hline
 \end{tabular}
}
\label{tab:IPPF} 
\end{table}

\section{Simulation Results}
To verify the effectiveness of the algorithm, targets' motion scenario and their measurements are simulated according to the given models and the estimates obtained using the algorithm is compared with the true trajectories. For comparison, estimation is done using the standard bootstrap particle filter also on the same target tracking problem and the results are compared.

\subsection{Multi-target tracking using IPPF}
\begin{figure}[h]
\centering
\subfloat [position $x$ ]
{\includegraphics[scale=0.4]{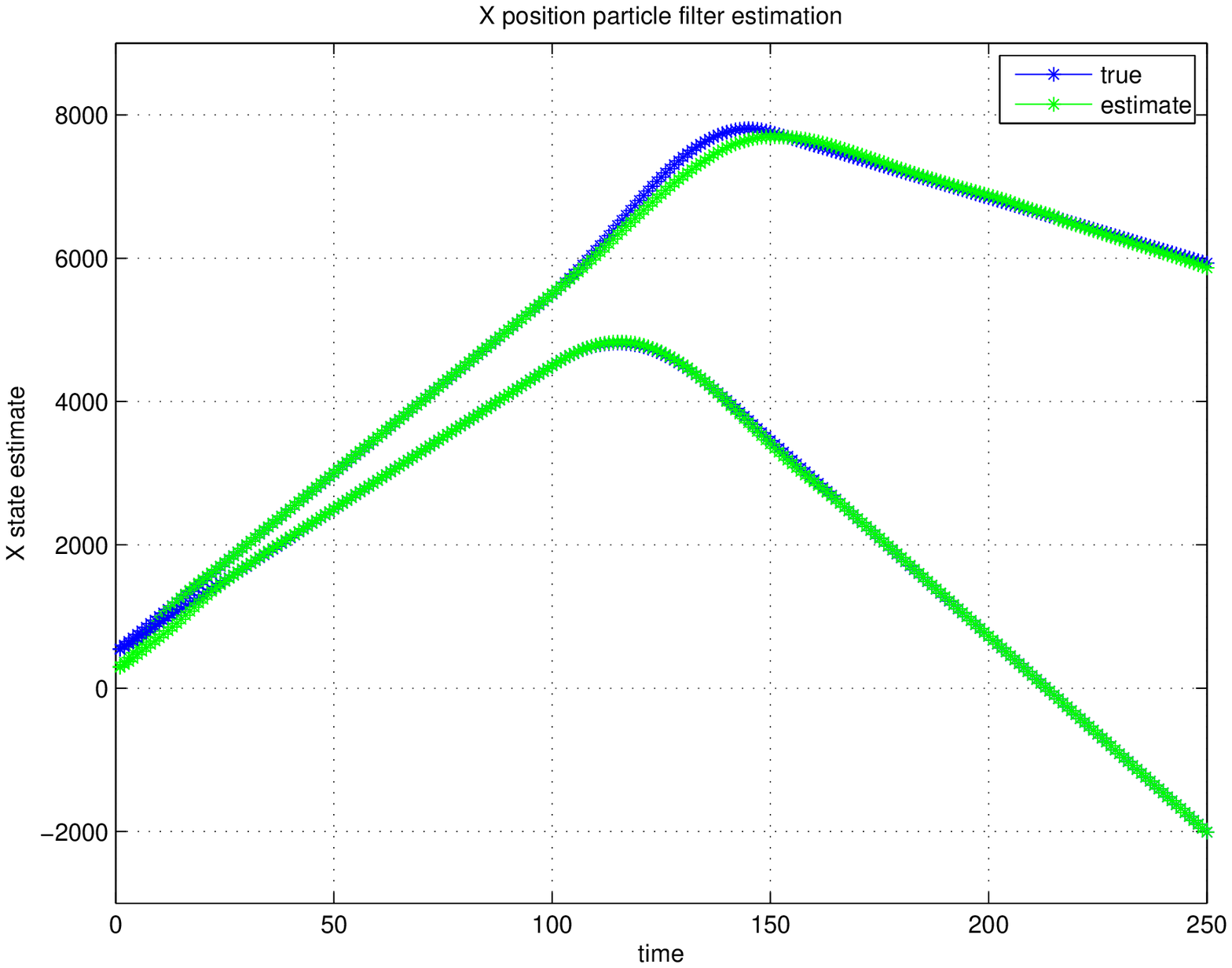}  
}\\
\subfloat [position $y$ ]
{\includegraphics[scale=0.4]{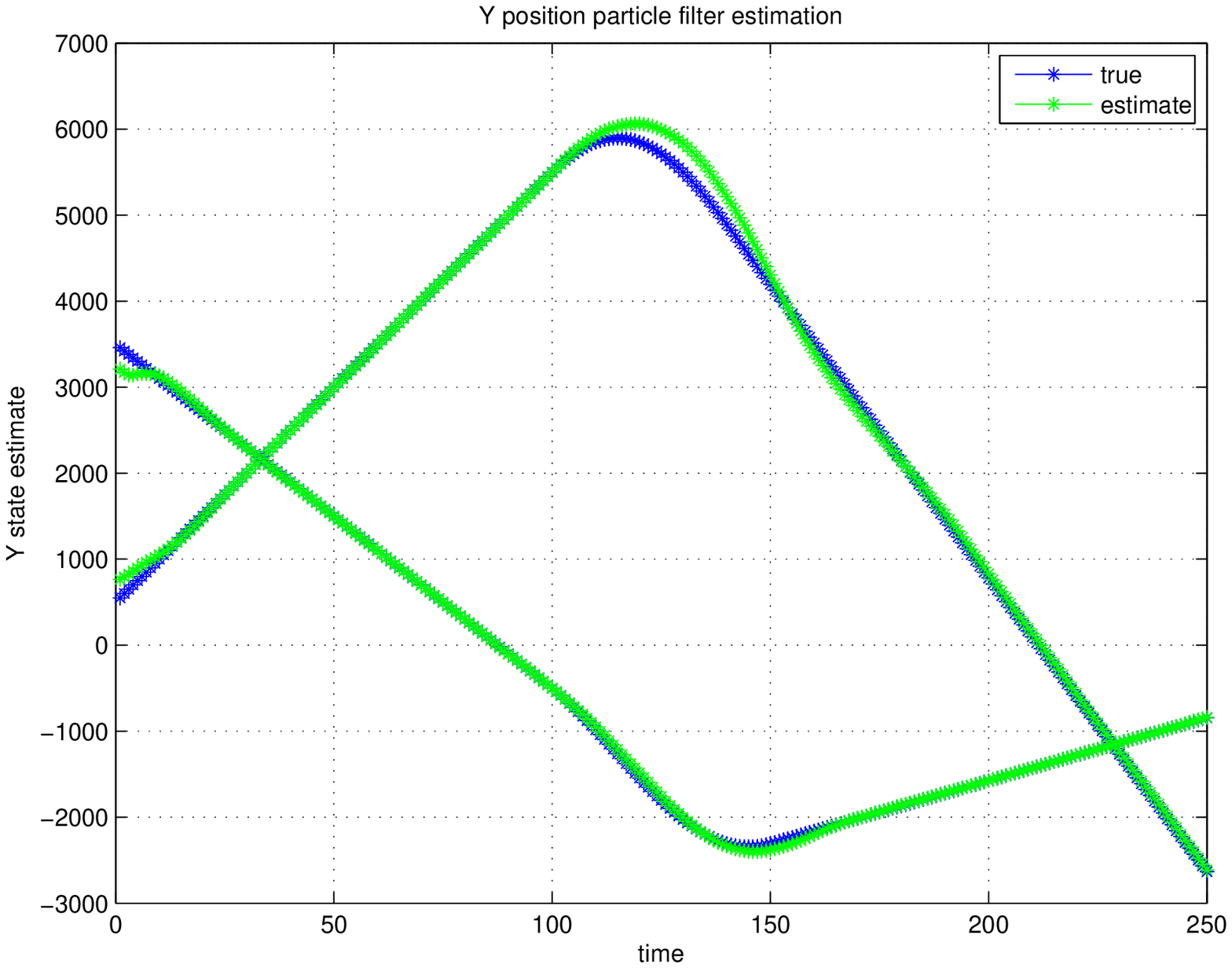}  
}\\
\subfloat [velocity $v_x$]
{\includegraphics[scale=0.4]{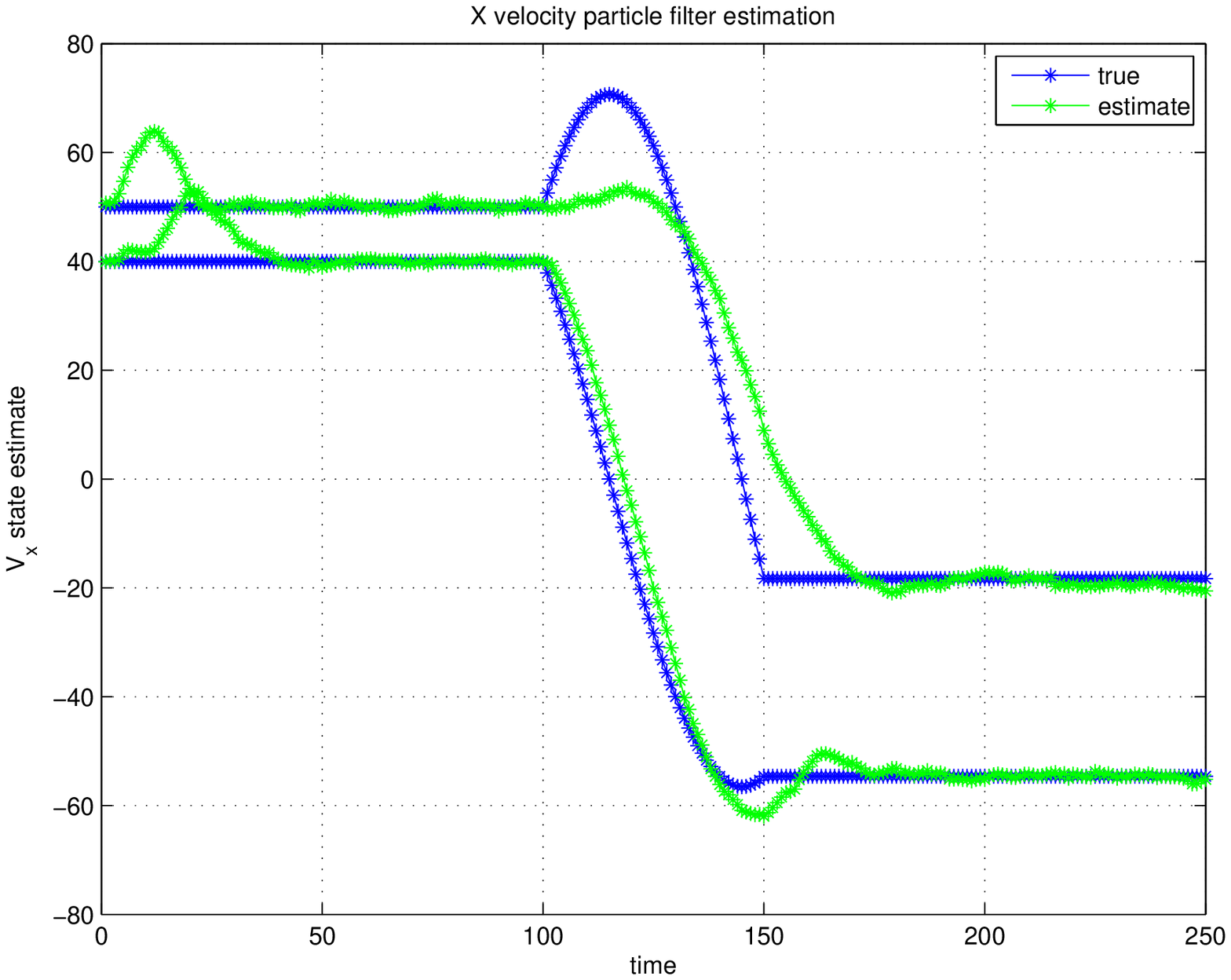}  
}%
\subfloat [velocity $v_y$]
{\includegraphics[scale=0.4]{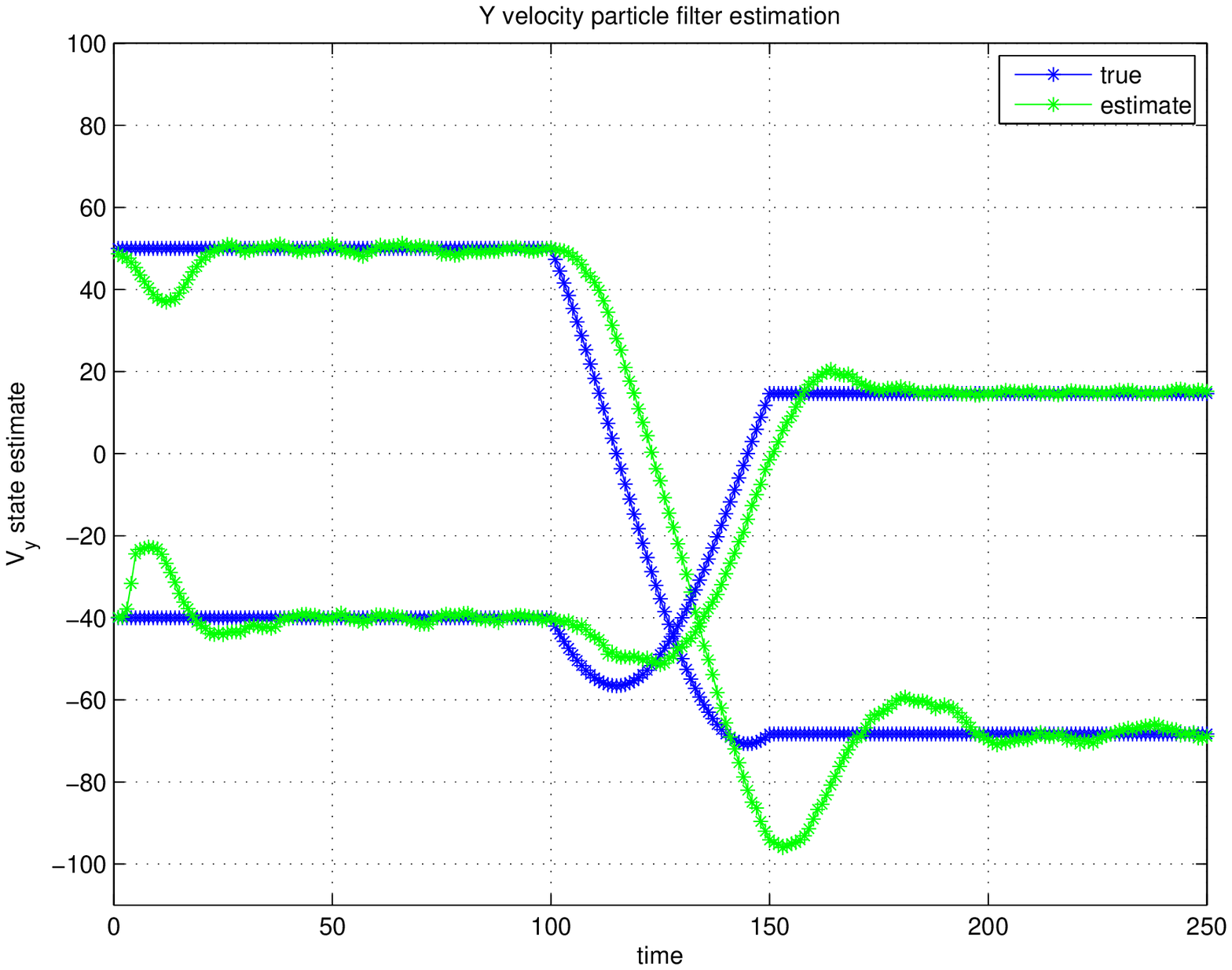}  
}\\
\caption{Targets' true states and their estimates obtained using IPPF.}
\label{IPPF_state}
\end{figure}

We have two independent targets $A$ and $B$ which have constant velocity and constant turn motions. The state vector consists of position and velocities of two targets,
\begin{eqnarray}
\mathbf{x}=
\begin{bmatrix}
 \mathbf{x}(1) &\mathbf{x}(2)
\end{bmatrix}^T=
\begin{bmatrix}
x(1) & v_{x}(1) & y(1)  & v_{y}(1) &\vdots& x(2) &  v_{x}(2) &y(2) & v_{y}(2)
 \end{bmatrix}^T
\end{eqnarray}
 The initial true state of the targets are  
$\mathbf{x}_{0}=\begin{bmatrix}
500&50&500&50&450&40&350&-40
 \end{bmatrix}^T$.
 From time $t=0s$ to $t=100s$, both targets have constant velocity motion.  From $t=101s$ to $t=150s$, they move in clockwise constant turn rate motion of $3 rad/s$. From $t=151s$ to $t=250s$, both targets have again constant velocity motion.
The measurement sensor is located at the origin. The target's ranges $r_{1},r_{2}$ and bearings $\theta_{1},\theta_{2}$ at time $t$ are available as the measurement $\mathbf{z}_t$
\begin{eqnarray}
\mathbf{z}_t=h(\mathbf{x}_{t})+\mathbf{v}_t
\end{eqnarray}
\begin{eqnarray}
\mathbf{v}_t\sim \mathcal{N}(0,Q_v)
\end{eqnarray}
 where $\mathbf{v}_t$ is the measurement error, $h(\cdotp)$ is the measurement model . We assume that the data association of the targets are already done and we know exactly which measurements belong to which targets. The measurement error $\mathbf{v}_t$ is uncorrelated and has zero mean Gaussian distribution with covariance matrix $Q_v$. $z_{t}(k)$ represents measurement of the target $k$.

\begin{eqnarray}
\mathbf{z}_t=
\begin{bmatrix}
z_{t}(1)&z_{t}(2)
\end{bmatrix}=
\begin{bmatrix}
r_{1}\\
\theta_{1}\\
r_{2}\\
\theta_{2}
\end{bmatrix}
\end{eqnarray}
\begin{eqnarray}\\
\mathbf{Q_v}=
\begin{bmatrix}
 Q_{v:1} &\mathbf{0}\\
 \mathbf{0} &Q_{v:2} \\
\end{bmatrix}=
\begin{bmatrix}
\sigma_{r_{1}}^{2} &0 &0 &0\\
0 &\sigma_{\theta_{1}}^{2} &0 &0\\
0 &0 &\sigma_{r_{2}}^{2} &0\\
0 &0 &0 &\sigma_{\theta_{2}}^{2}
\end{bmatrix}=
\begin{bmatrix}
10 &0 &0 &0\\
0 &0.5 &0 &0\\
0 &0 &10 &0\\
0 &0 &0 &0.5\\
\end{bmatrix}
\end{eqnarray}
The measurement model $h(\cdotp)$ for the targets is given by:
\begin{eqnarray}
h(\mathbf{x}_{t})=
\begin{bmatrix}
h_1(\mathbf{x}_{t}(1))\\
h_2(\mathbf{x}_{t}(2))\\
\end{bmatrix}
\end{eqnarray}
The measurement model $h_k(\cdotp)$ for target $k=1,2$ is given by:
\begin{eqnarray}
z_{t}(k)=
h_k(\mathbf{x}_{t}(k))=
\begin{bmatrix}
\sqrt{x^2_t(k)+y^2_t(k)}\\
\tan^{-1}\left(\dfrac{y_t(k)}{x_t(k)}\right)
\end{bmatrix}
\end{eqnarray}
 The initial state estimate is assumed to be a Gaussian vector with mean $\mathbf{x}_{0}$ and error covariance $P_{0}$, such that
\begin{equation}
 \mathbf{x}_{0}=\begin{bmatrix}
250&50&750&50&250&40&250&-40
 \end{bmatrix}^T
\end{equation}
\begin{equation}
 P_{0}=diag \left(100,10,100,10,100,10,100,10\right)
\end{equation}
Hence initial particles $\{\mathbf{x}_{0}^{(i)}\}_{i=1}^N$ were generated based on the distribution 
\begin{equation}
 \mathbf{x}_{0} \sim \mathcal{N}(\mathbf{x}_{0},P_{0})
\end{equation}
In this implementation of the particle filter, the transitional prior which is a suboptimal choice of importance density is  used to propose particles. The state transition model $f(\cdotp)$ for estimation of state at time $t$ is such that: 
\begin{eqnarray}
\mathbf{x}_{t} & = & f(\mathbf{x}_{t-1})+\mathbf{w}_{t-1}\\
f(\mathbf{x}_{t}) & = & 
\begin{bmatrix}
f_1(\mathbf{x}_{t}(1))\\
f_2(\mathbf{x}_{t}(2))\\
\vdots\\
f_{K_t}(\mathbf{x}_{t}(K_t))\\
\end{bmatrix}
\end{eqnarray}
where $\mathbf{w}_{t-1}$ is the process noise with zero mean.
For the target $k$, the state transition model $f_{k}(\cdotp)$ for estimation of state at time $t$ is such that:
\begin{equation}
 \mathbf{x}_{t}(k)=f_{k}(\mathbf{x}_{t-1}(k))
\end{equation}
The state transition model $f_{k}(\cdotp)$ used in this implementation of the IPPF is constant velocity model.
 Hence $f_{k}(\cdotp)$ is a matrix $F$ given by:
\begin{equation}
 F=\begin{bmatrix}
1&T&0&0\\
0&1&0&0\\
0&0&1&T\\
0&0&0&1\\
\end{bmatrix}
\end{equation}\\
where $T$ is the sampling period of the target dynamics. Since both the targets are estimated based on the same type of state transition model, the importance density used is the same for both the targets, i.e. $f_{1}(\cdotp)=f_{2}(\cdotp)$.
The process noise assumed has a diagonal covariance matrix $Q_{w}$ as:
\begin{equation} 
Q_w = diag \left(10 ,2.5, 35 , 2.5, 10 , 2, 10, 2 \right)
\end{equation}
A total of $N=100$ particles were used. The detailed implementation algorithm for the two target tracking problem is given in Table.\ref{tab:IPPF_impltn}.
%
%
\begin{table}[ht] 
\caption{Implementation of IPPF} 
\centering          
\resizebox{!}{4.5in} {
\begin{tabular}{l}
  \hline
  \begin{minipage}{8in}
    \vskip 4pt
\begin{enumerate}
      \item For $t=0$, initialize all particles:
      \begin{itemize}
	    \item For $i=1,...,100$, generate samples $\mathbf{x}_{0}^{(i)} \sim \mathcal{N}(\mathbf{x}_{0},P_{0})$
	    \item For $i=1,...,100$, assign weights $w_{0}^{(i)}=\dfrac{1}{100}$
      \end{itemize}
      \item For  $t>0$,
      \begin{itemize}
	      \item For $k=1, 2$
	      \begin{itemize}
		  \item For $i=1, 2,..., 100$  
		  \begin{itemize}
		      \item Draw sample $\mathbf{x}_{t}^{(i)}(k)$ using the transitional prior.
		      \begin{equation}
			      a_t^{(i)}(k)= F\mathbf{x}_{t-1}^{(i)}(k)
		      \end{equation}
			\begin{equation}
			      \mathbf{x}_{t}^{(i)}(k) \sim p(\mathbf{x}_{t}(k)\mid \mathbf{x}_{t-1}^{(i)}(k))=\mathcal{N}(a_t^{(i)}(k),Q_{w})
			\end{equation}
			\item Compute secondary weights using the likelihood $p(z_{t}(k)\mid 	\mathbf{x}_{t}^{i}(k))$ and the observation model $h_k$ for the target $k$.
			\begin{equation}
				b_t(k)=h_k(\mathbf{x}_{t}^{(i)}(k))
			\end{equation}
			\begin{equation}
			      g(\mathbf{x}_{t}^{(i)}(k))=\mathcal{N}(z_t(k);b_t(k),Q_{v:k})
			\end{equation}				    
		  \end{itemize}
	      \end{itemize}

	      \item For $k=1, 2$
	      \begin{itemize}
		  \item For $i=1, 2,..., 100$, Normalize the secondary weights
			\begin{equation}
			      \rho^{(i)}(k)= \frac{g(\mathbf{x}_{t}^{(i)}(k)}{\sum^{100}_{j=1}g(\mathbf{x}_{t}^{(j)}(k)}
			\end{equation}
	      \end{itemize}

	      \item For $k=1, 2$
	      \begin{itemize}
		  \item For $i=1, 2,..., 100$  
		  \begin{itemize}
			  \item Sample indices $j_{k}(i)$ from the distribution formed by $\rho^{(n)}(k)$ for $n=1,2,3,...,100$ by any of the method given in Appendix.\ref{appendix:a}.
		  \end{itemize}	
	      \end{itemize}
	      \item For $i=1, 2,..., 100$
	      \begin{itemize} 
		    \item Set the new particles $\mathbf{x}_{t}^{(i)} \equiv \{\mathbf{x}_{t}^{(j_{1}(i))}(1),\mathbf{x}_{t}^{(j_{2}(i))}(2)\}$
		    and compute their corresponding particle weights, 
		    $\mathbf{w}_{t}^{(i)} \equiv w_{t}^{(j_{1}(i))} \times w_{t}^{(j_{2}(i))}$
	      \end{itemize}

	      \item For $i=1, 2,..., 100$
		      \begin{itemize}
			    \item Evaluate likelihood of the particles
				  $$p(\mathbf{z}_{t}\rvert \mathbf{x}_{t}^{(i)})=\mathcal{N}(\mathbf{z}_{t};h(\mathbf{x}_{t}^{(i)}),\mathbf{Q_v})$$
			    \item Evaluate the importance weights
			      $$ w_{t}^{(i)}=\frac{w_{t-1}^{(i)} p(z_{t}\rvert \mathbf{x}_{t}^{(i)})}{\prod_{k=1}^{2}\rho^{(j_{k}(i))}(k)} $$

		      \end{itemize}

	      \item For $i=1, 2,..., 100$, normalize weights:
				$$w^{(i)} =  \frac{w^{(i)}}{\sum^{100}_{j=1}w^{(j)}}$$
        \end{itemize}
       \item If required resample the particles and do roughening.
\end{enumerate}
  \vskip 4pt
 \end{minipage}
 \\
  \hline
 \end{tabular}
}
\label{tab:IPPF_impltn} 
\end{table}
The true trajectories of the targets and their estimates are shown in Fig.\ref{IPPF_xy_state}. The state estimates of the targets are shown in Fig.\ref{IPPF_state}. The mean square error MSE of the position estimates for 100 Monte Carlo runs are shown in Fig.\ref{IPPF_T1_MSE} and Fig.\ref{IPPF_T2_MSE}.


\subsection{Comparison of IPPF with Standard Bootstrap PF}
\begin{figure}[h]
\centering
\subfloat [IPPF estimate]
{\label{IPPF_xy_state}\includegraphics[scale=0.4]{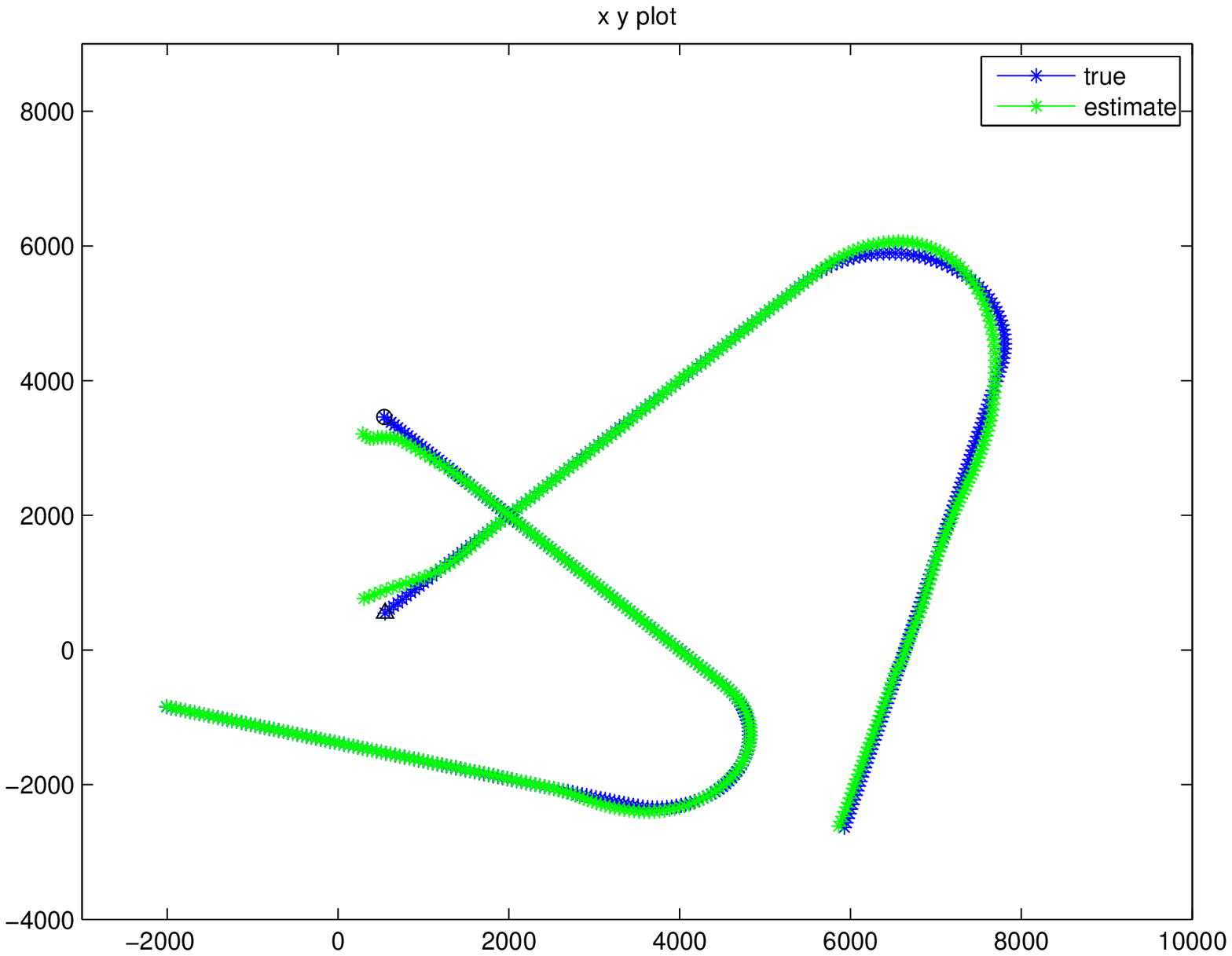}  
} 
\subfloat [PF estimate]
{\label{PF_xy_state}\includegraphics[scale=0.4]{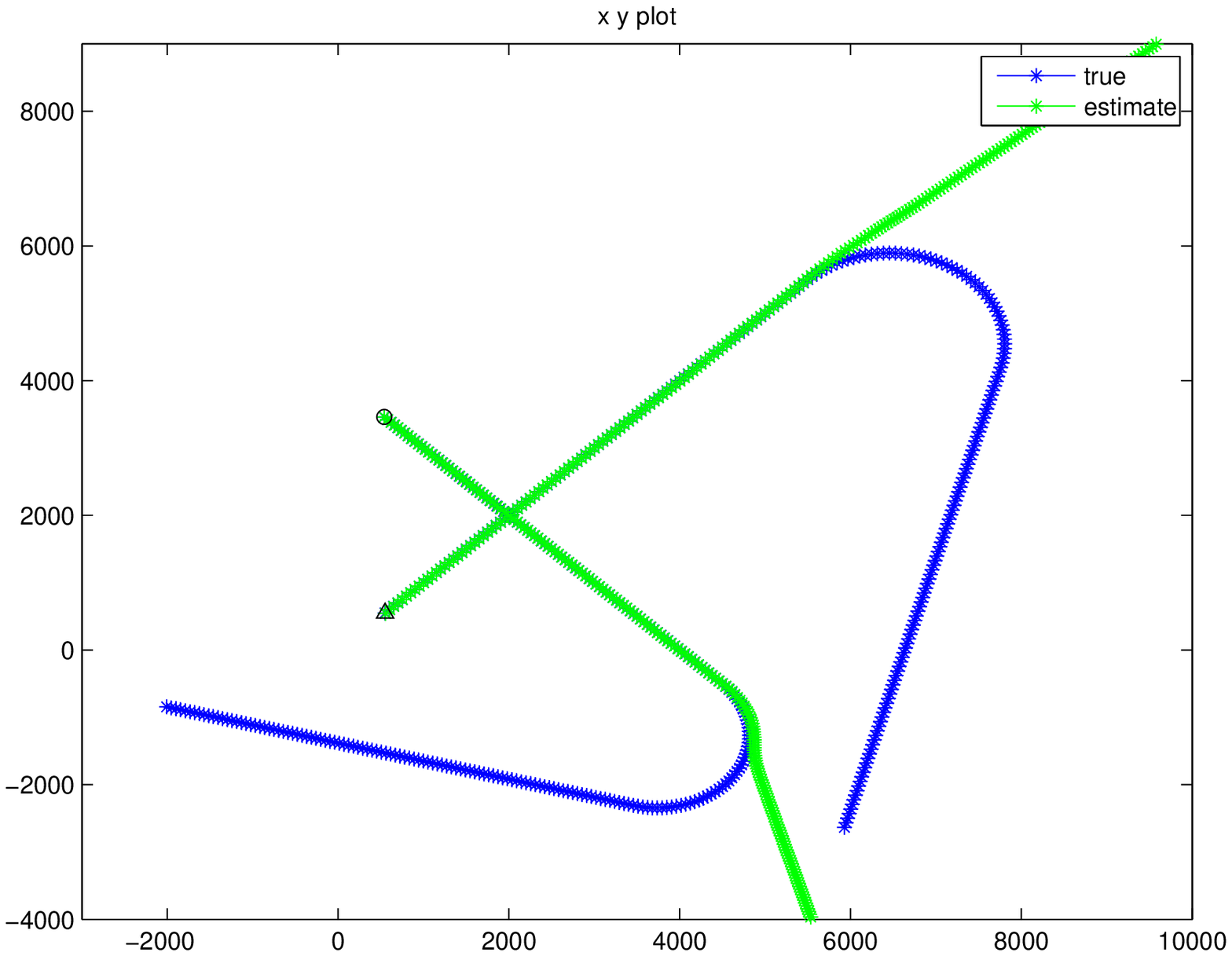}  
}\\   
\caption{Targets' true $xy$ track and their estimated track obtained using IPPF and PF: The PF diverges during multi target tracking, but IPPF has good performance even with the same number of particles.}
\label{IPPF_traj}
\end{figure}

\begin{figure}[h]
\centering
\subfloat [IPPF Target 1 estimate]
{\label{IPPF_T1_MSE}\includegraphics[scale=0.4]{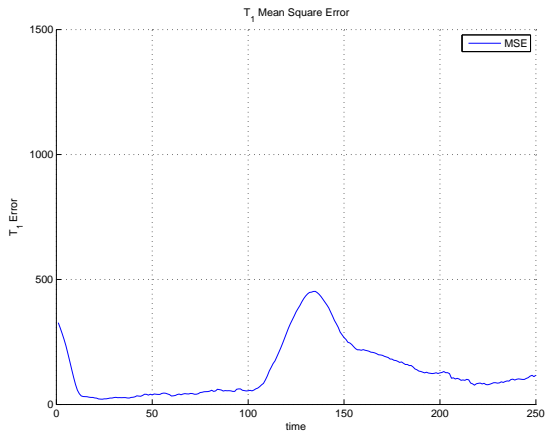}  
}
\subfloat [IPPF Target 2 estimate]
{\label{IPPF_T2_MSE}\includegraphics[scale=0.4]{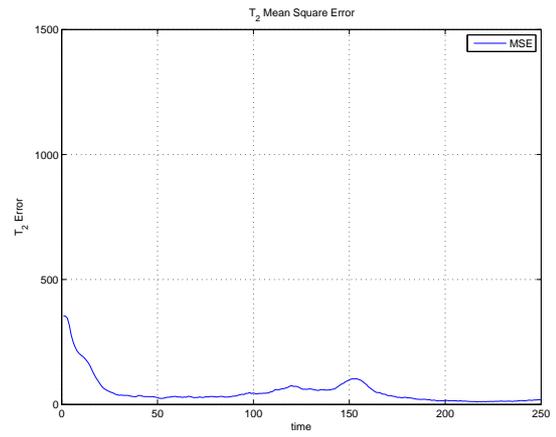}  
}\\
\subfloat [PF Target 1 estimate]
{\label{PF_T1_MSE}\includegraphics[scale=0.4]{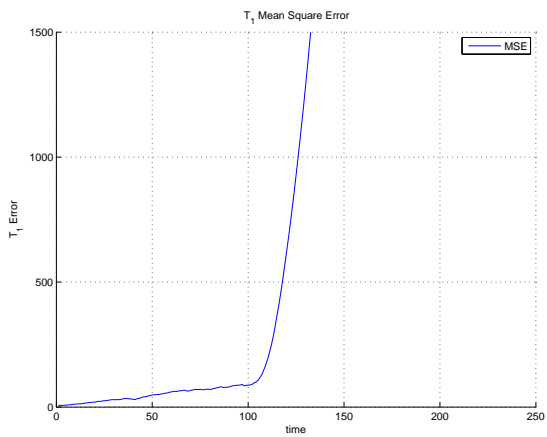}  
}
\subfloat [PF Target 2 estimate]
{\label{PF_T2_MSE}\includegraphics[scale=0.4]{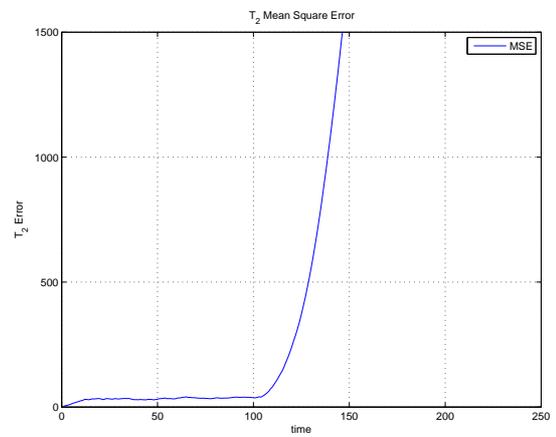}  
}\\
\caption{MSE of the position estimates obtained using IPPF and PF.}
\label{IPPF_MSE}
\end{figure}

\begin{figure}[h]
\centering
\subfloat [position $x$ ]
{\includegraphics[scale=0.4]{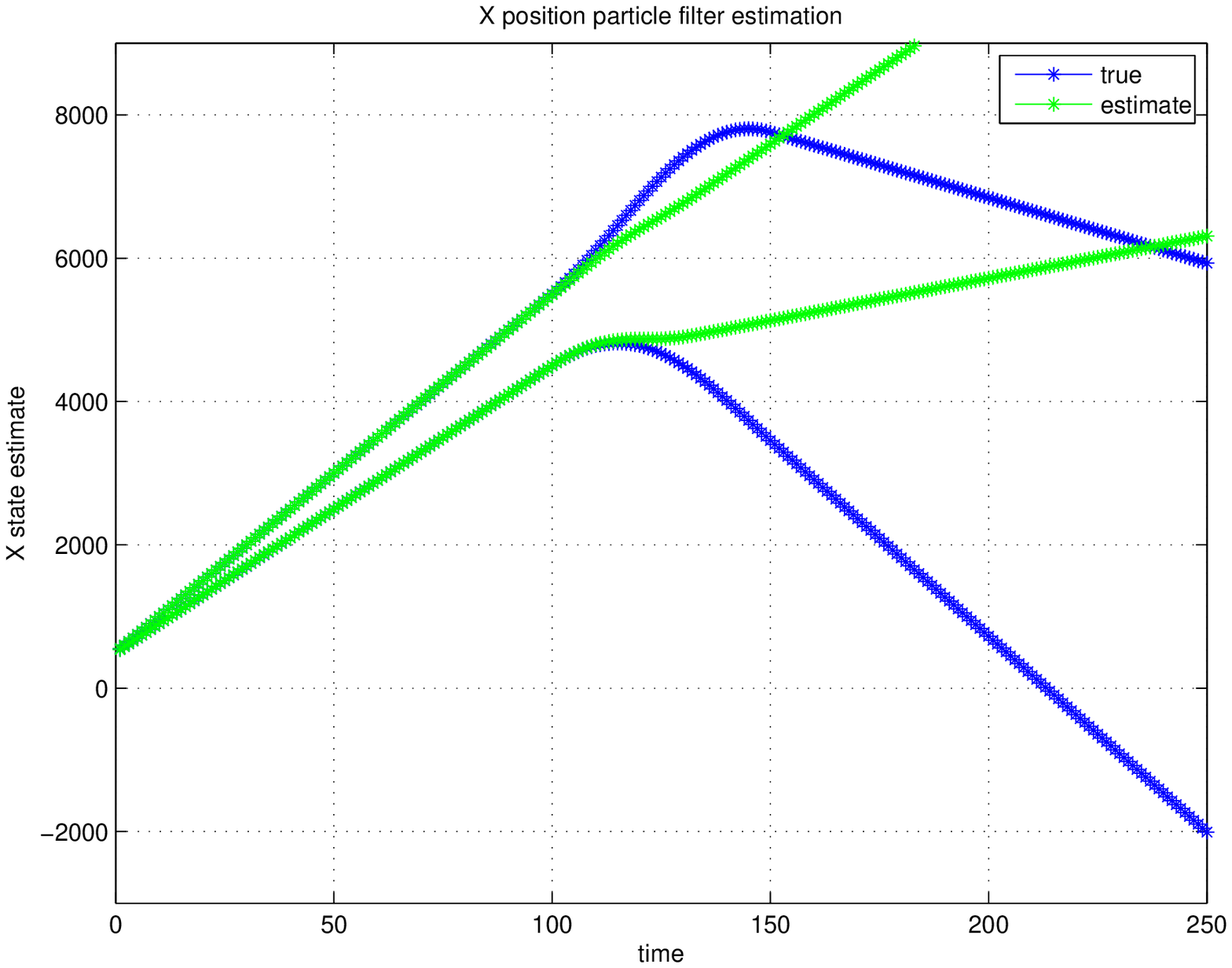}  
}\\
\subfloat [position $y$ ]
{\includegraphics[scale=0.4]{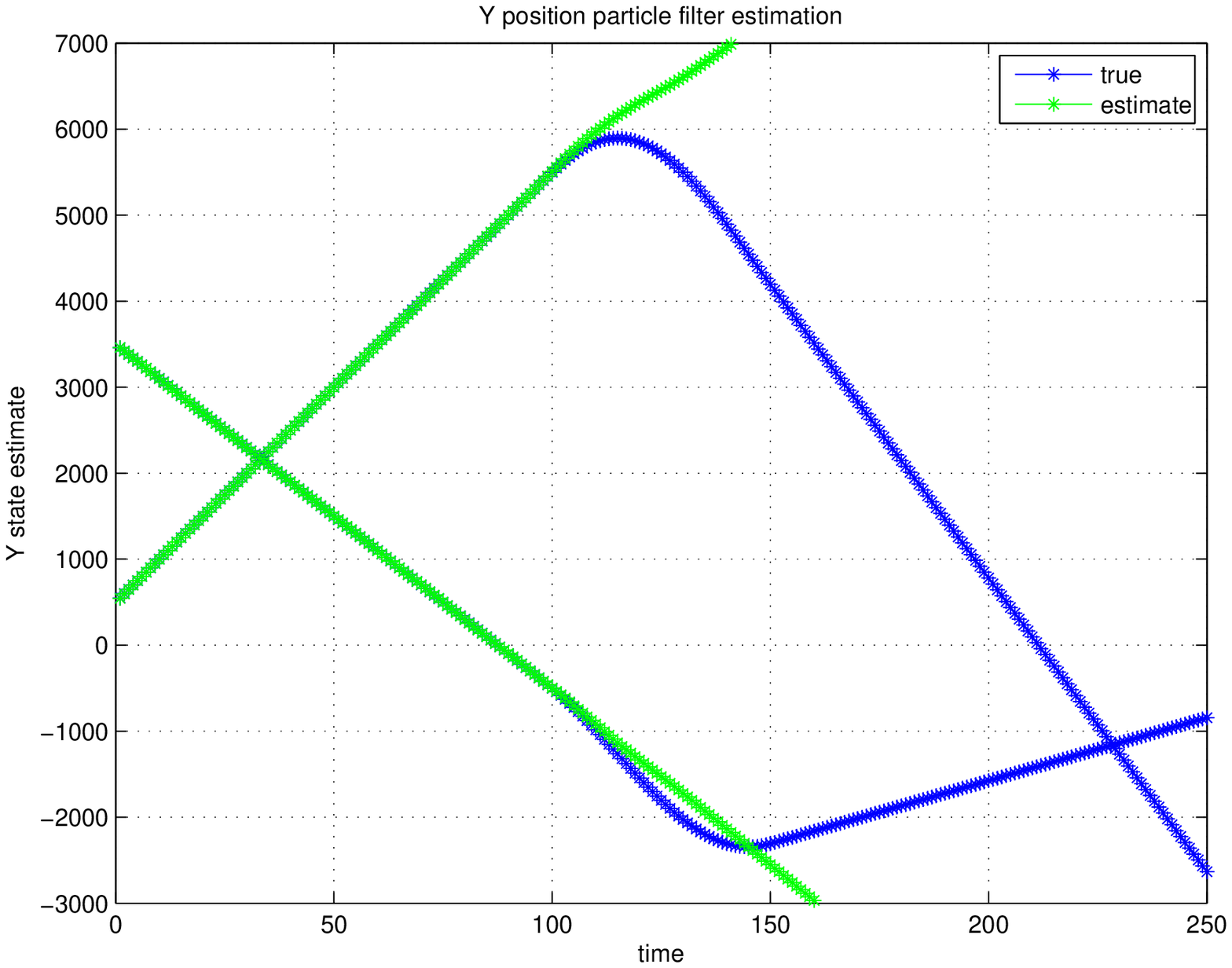}  
}\\
\subfloat [velocity $v_x$]
{\includegraphics[scale=0.4]{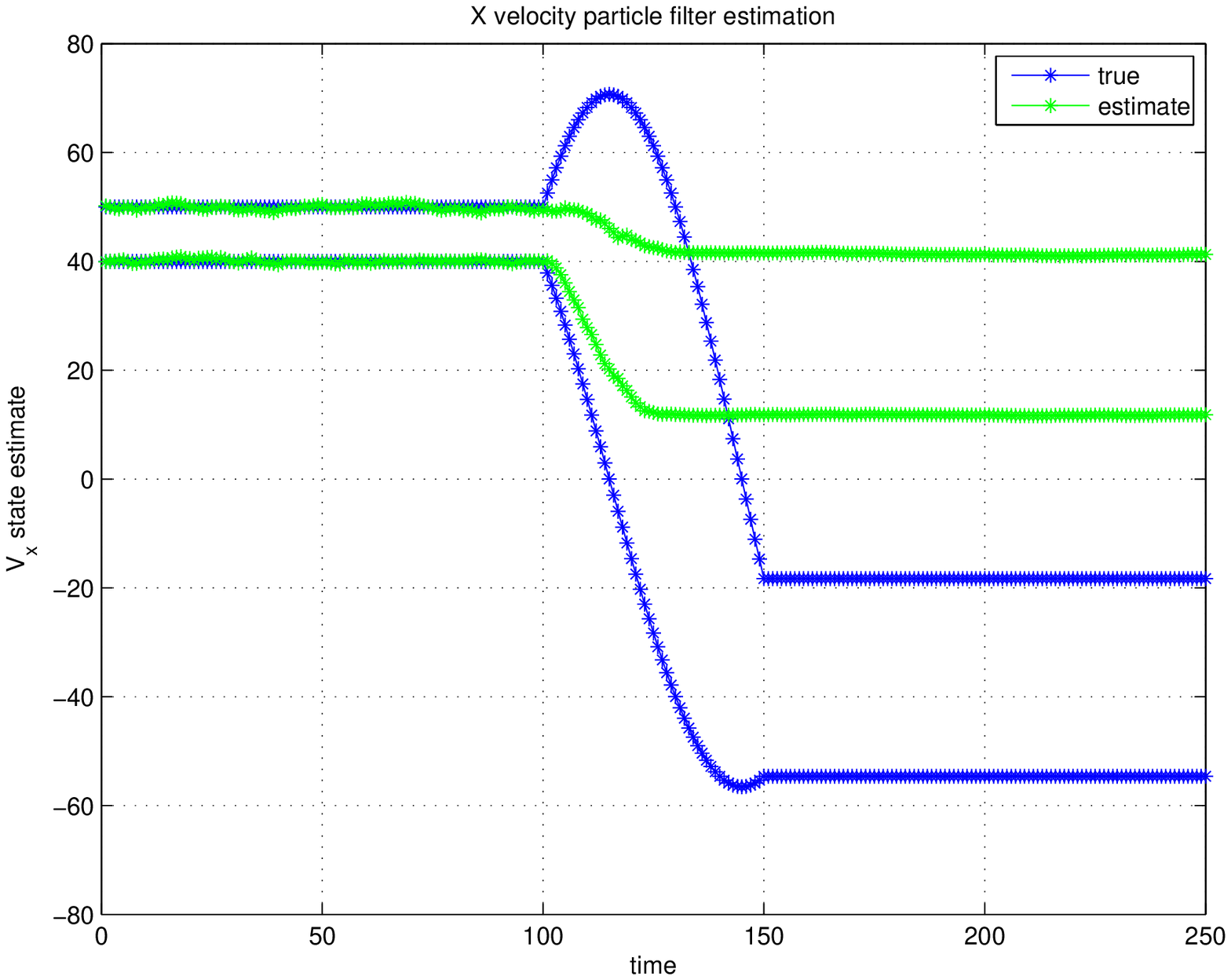}  
}%
\subfloat [velocity $v_y$]
{\includegraphics[scale=0.4]{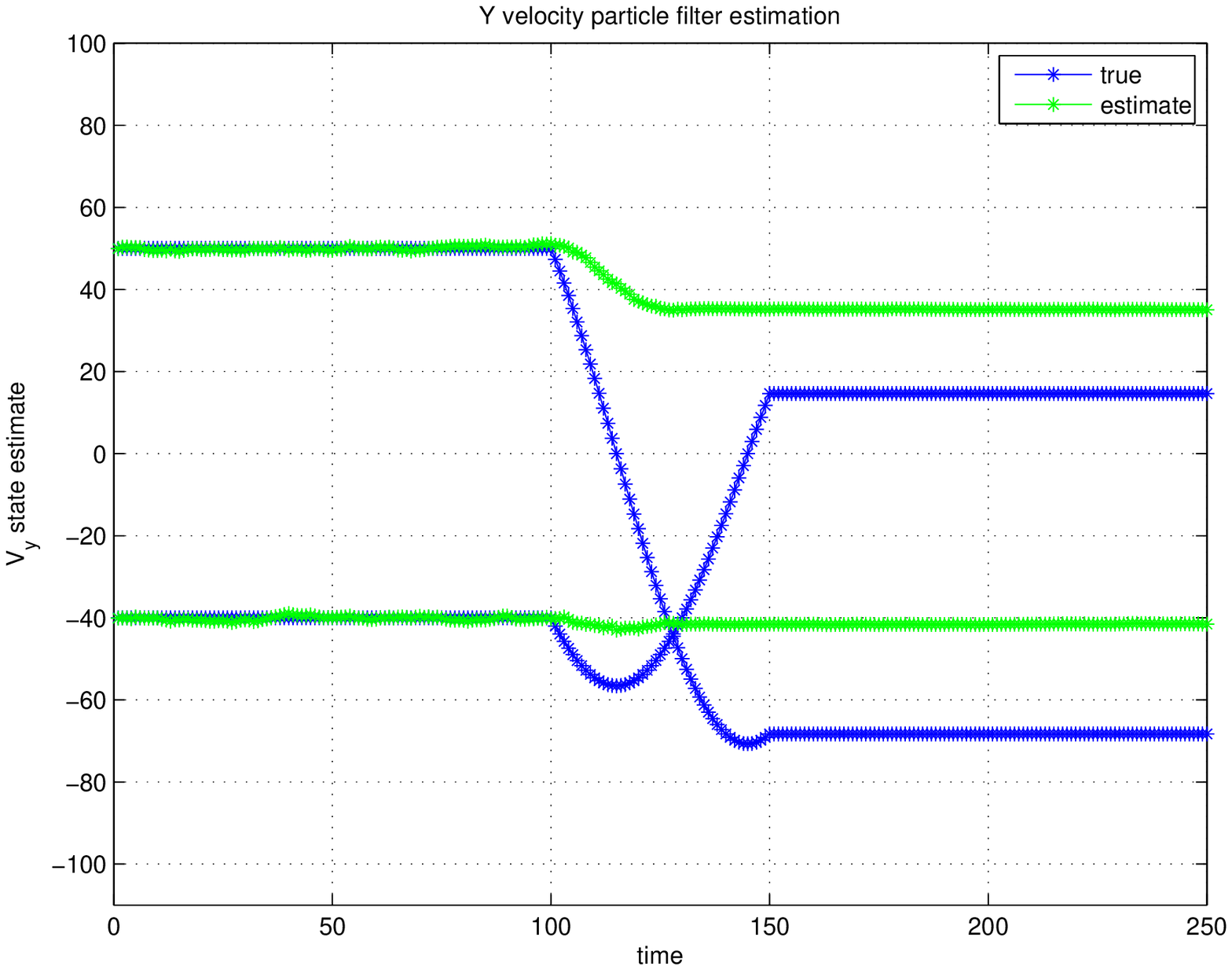}  
}\\
\caption{Targets' true states and their estimates obtained using standard bootstrap PF.}
\label{PF_state}
\end{figure}

For comparison the standard bootstrap particle filter is also implemented for the same target scenario with N=100 particles. The true trajectories of the targets and their estimates are shown in Fig.\ref{PF_xy_state}. The state estimates of the targets are shown in Fig.\ref{PF_state}. The mean square error (MSE) of the position estimates for 100 Monte Carlo runs are shown in Fig.\ref{PF_T1_MSE} and Fig.\ref{PF_T2_MSE}. The results show that estimates are highly diverged compared to the IPPF estimates.
Thus it shows that IPPF improves particle survival rate of the particles when there are multiple targets and hence we can use fewer particles while maintaining robustness.

\section{Summary}
 In high dimensional systems, the proportion of high likelihood particles are smaller. Hence higher number of particles are required for high dimensional systems like multi target tracking. Weighted resampling is used to efficiently modify particles using the measurement likelihood so that less number of particles are rejected during resampling. Weighted resampling doesn't alter the underlying probability distribution of the particles. Independent partitioned sampling facilitates the application of target dynamics and measurement update individually on each independent target and allows the use of weighted resampling on each target to have better rearrangement of the particles with the result that most particles are likely to contain mostly good states and fewer are rejected during resampling. Incorporation of independent partition sampling and weighted resamplimg helps the IPPF to track multiple targets with lesser number of particles.

\chapter{Multiple Model Particle Filter (MMPF)}
\label{chap:MMPF}
\thispagestyle{empty}

Multiple Model Bootstrap filter (MMPF) proposed by S. ~McGinnity et.al \cite{6}, is an extension of the standard particle filter to the multiple model target tracking problem. In maneuvering targets, apart from the straight line motion, the target can have different types of dynamics similar to circular motion, accelerated motions etc. Also they can have abrupt deviation from one type of motion to another. Such processes are difficult to represent using a single kinematic model of the target. Hence filters with multiple models representing different possible maneuvering states are run in parallel, operating simultaneously on the measurements. The validity of these models are evaluated and the final target state estimate is a probability weighted combination of the individual filters. 

In multiple model particle filter, each particle consists of a state vector augmented by an index vector representing the model. Thus particles have continuous valued vector $\mathbf{x}_t$ of target kinematics variables, like position, velocity, acceleration, etc, and a discrete valued regime variable $A_t$ that represents the index of the model which generated $\mathbf{x}_t$ during the time period $(t-1^+,t]$. The regime variable can be one of the fixed set of $s$ models i.e., $A_t\in S=\{1,2, . . . .,s\}$. The posterior density $p(y_t \mid z_t)$ is represented  using $N$ particles $\{y_t^n,w_t^n\}_{n=1}^N$, i.e., the augmented state vector and the weight. The posterior model probabilities $\{\pi_i(t)\}_{i=1}^s$ are approximately equal to the proportion of the samples from each model in the index set $\{A_t^n\}_{n=1}^N$.

It will be assumed that model switching is a Markovian process with known mode transition probabilities $\pi_{ij}$.
	
\begin{equation}
 \pi_{ij}=P[A_t=j \mid A_{t-1}=i]\;;\qquad  i,j \in S=\{1,2, . . . .,s\} \\
\end{equation}
\begin{equation}
  \pi_{ij}\geq0\\
\end{equation}
\begin{equation}
\sum_{j=1}^s \pi_{ij} = 1\\
\end{equation}

The mode transition probabilities will be assumed time invariant and independent of the base state and hence the system is assumed to have an $s$-state homogeneous Markov chain with mode transition probability matrix $\Pi=[ \pi_{ij}]_{s\times s}$, where $i,j \in S$. These mode transition probabilities are designed based on the estimator performance requirements. A lower value of $\pi_{ij}$ will contribute for less peak error during maneuver but higher RMS error during the quiescent period. Similarly a higher value of $\pi_{ij}$ will contribute for more peak error during maneuver but lower RMS error during the quiescent period \cite{5}.
\begin{table}[H] 
\caption{Multiple Model Particle Filter, MMPF \cite{4}} 
\centering          
\begin{tabular}{l}
  \hline
  \begin{minipage}{4in}
    \vskip 4pt
$[\{y_t^n, w_t^n\}_{n=1}^N]=$MMPF$[\{y_{t-1}^n, w_{t-1}^n\}_{n=1}^{N},z_t]$
\begin{itemize}
	\item Regime transition (Table.\ref{tab:RT}):\\
$[\{A_t^n\}_{n=1}^N]=$RT$[\{A_{t-1}^n\}_{n=1}^{N},\Pi]$	  
	\item Regime Conditioned SIS (Table.\ref{tab:RTSIS}):\\
$[\{\mathbf{x}_t^n, w_t^n\}_{n=1}^N]=$RC-SIS$[\{\mathbf{x}_{t-1}^n, A_t^n, w_{t-1}^n\}_{n=1}^{N},z_t]$
        \item If required resample the particles and do roughening.
\end{itemize}
   \vskip 4pt
 \end{minipage}
 \\
  \hline
 \end{tabular}
\label{tab:MMPF} 
\end{table}
The algorithm for multiple model particle filter is repeated in Table.\ref{tab:MMPF} from \cite{4,6}. The first step is to generate the index set 
$\{A_t^n\}_{n=1}^N$ based on the transition probability matrix $\Pi$. Thus it gives the appropriate model and importance density to be used by each particle at time $k-1$ for generating the particle at time $k$. This is called regime transition.  Its pseudo code is repeated in Table.\ref{tab:RT} from \cite{4}. 
\begin{table}[H] 
\caption{Regime Transition \cite{4}} 
\centering          
\begin{tabular}{l}
  \hline
  \begin{minipage}{4in}
    \vskip 4pt
$[\{A_t^n\}_{n=1}^N]=$RT$[\{A_{t-1}^n\}_{n=1}^{N},\Pi]$
\begin{itemize}
	\item FOR $i=1:s$,
	\begin{itemize}
	    \item $c_i(0)=0$ 
	    \item FOR $j=1:s$,
	    \begin{itemize}
		\item $c_i(j)=c_i(j-1)+ \pi_{ij}$ 		
	    \end{itemize}
	    \item END FOR
	\end{itemize}
	\item END FOR
	\item FOR  $n=1:N$,
	\begin{itemize}
	      \item Draw $u_n\sim \mathcal{U} [0,1]$  		
	      \item Set $i=A_{t-1}^n$ 
	      \item m=1
	      \item WHILE $(c_i(m)<u_n)$
	      \begin{itemize}
		    \item$ m=m+1$
	      \end{itemize}
	      \item END WHILE
	      \item Set $A_{t}^n=m$
	\end{itemize}
	\item END FOR
\end{itemize}
   \vskip 4pt
 \end{minipage}
 \\
  \hline
 \end{tabular}
\label{tab:RT} 
\end{table}
It implements the rule that if $A_{t-1}^n=i$, then $A_t^n$ should be set to $j$ with probability $\pi_{ij}$. It finds the cumulative distribution function of random variable $A_t$ conditioned on $A_{t-1}=i$, i.e. $\sum_{j=1}^{m} \pi_{ij}$ for $1\leq m \leq s$. It generates a uniform random variable  $u_n\sim \mathcal{U} [0,1]$ and set $A_t^n$ to $m\in S=\{1,2, . . . .,s\}$ such that 
\begin{equation}
\sum_{j=1}^{m-1} \pi_{ij} < u_n \leq \sum_{j=1}^{m} \pi_{ij}
\end{equation}
 The regime conditioned SIS filtering is done next. Its pseudo code is repeated in Table.\ref{tab:RTSIS} from \cite{4}. The optimal regime conditioned importance density is 
\begin{equation}
q(\mathbf{x}_{t}\mid \mathbf{x}_{t-1}^{(n)},A_t^n,z_{t})_{opt}=p(\mathbf{x}_{t}\mid \mathbf{x}_{t-1}^{(n)},A_t^n,z_{t})
\end{equation}
A suboptimal choice of the regime conditioned importance density is the transitional prior.
\begin{equation}
q(\mathbf{x}_{t}\mid \mathbf{x}_{t-1}^{(n)},A_t^n,z_{t})_{sub-opt}=p(\mathbf{x}_{t}\mid \mathbf{x}_{t-1}^{(n)},A_t^n)
\end{equation}
 The posterior prediction density is formed by transforming each particle using the model indexed by its corresponding augmented regime variable. After regime conditioned SIS filtering, posterior densities will automatically be weighted towards high likelihood as well as towards more appropriate models. If necessary resampling is done on the posterior density to reduce the effect of degeneracy.

\begin{table}[H] 
\caption{Regime Conditioned SIS \cite{4}} 
\centering          
\begin{tabular}{l}
  \hline
  \begin{minipage}{4in}
    \vskip 4pt
$[\{\mathbf{x}_t^n, w_t^n\}_{n=1}^N]=$RC-SIS$[\{\mathbf{x}_{t-1}^n, A_t^n, w_{t-1}^n\}_{n=1}^{N},z_t]$
\begin{itemize}
	\item FOR $n=1:N$,
	\begin{itemize}
	    \item Draw $\mathbf{x}_t^n\sim q(\mathbf{x}_{t}\mid \mathbf{x}_{t-1}^{(n)},A_t^n,z_{t})$ 
	    \item Evaluate the importance weights upto a normalizing constant 
		  \begin{equation}w_{t}^{(n)}=\frac{w_{t-1}^{(n)} p(z_{t}\rvert  \mathbf{\mathbf{x}}_{t}^{(n)},A_{t}^{(n)}) p(\mathbf{x}_{t}^{(n)}\rvert  \mathbf{x}_{t-1}^{(n)},A_{t}^{(n)})}{q(\mathbf{x}_{t}\mid \mathbf{x}_{t-1}^{(n)},A_t^n,z_{t})}		   
		  \end{equation}
	\end{itemize}
	\item END FOR
        \item FOR $n=1, 2,..., N$, normalize weights:
		  \begin{equation}
		      w^{(n)}_{t} =  \dfrac{w^{(n)}_t}{\sum^{N}_{j=1}w^{(j)}}		   
		  \end{equation}	
	\item END FOR
\end{itemize}
   \vskip 4pt
 \end{minipage}
 \\
  \hline
 \end{tabular}
\label{tab:RTSIS} 
\end{table}


%
%
\section{Simulation Results}
To verify the effectiveness of the algorithm, targets' motion scenario and their measurements are simulated according to the given models and the estimates using the algorithm is compared with the true trajectories. For comparison, estimation is done using the standard bootstrap particle filter and interacting multiple model-extended Kalman filter (IMM-EKF) for the same target tracking scenario.

\subsection{Target Tracking using MMPF}
We have one target which has constant velocity and constant turn motions. The augmented state vector consists of position $x,y$, velocities $v_x, v_y$ of the target, and the regime variable $A$, 
\begin{equation}
\mathbf{x}=
\begin{bmatrix}
x & v_{x} & y  & v_{y} &A
 \end{bmatrix}^T 
\end{equation}
The initial unaugmented true state of the target is 
$\mathbf{x}_{0}=\begin{bmatrix}
500&100&500&0
 \end{bmatrix}^T$.
 From $t=0s$ to $t=20s$, $t=49s$ to $t=60s$, $t=81s$ to $t=100s$  the target follows constant velocity motion.  From $t=21s$ to $t=48s$, $t=61s$ to $t=80s$, it moves in clockwise constant turn rate motion of $60 rad/s$. 
The measurements are target's range $r$ and bearing $\theta$ available as $\mathbf{z}$.
\begin{eqnarray}
\mathbf{z}_t=h(\mathbf{x}_{t})+\mathbf{v}_t
\end{eqnarray}
\begin{eqnarray}
\mathbf{v}_t\sim \mathcal{N}(0,Q_v)
\end{eqnarray}
 where $\mathbf{v}_t$ is the measurement error, $h(\cdotp)$ is the measurement model. The measurement error $\mathbf{v}_t$ is uncorrelated and has zero mean Gaussian distribution with covariance matrix $Q_v$
\begin{eqnarray}
\mathbf{z}=
\begin{bmatrix}
r\\
\theta\\
\end{bmatrix}
\end{eqnarray}
\begin{eqnarray}\\
\mathbf{Q_v}=
\begin{bmatrix}
\sigma_{r}^{2} &0\\
0  &\sigma_{\theta}^{2}
\end{bmatrix}=
\begin{bmatrix}
10 &0 \\
0 &0.1\\
\end{bmatrix}
\end{eqnarray}
The sensor is located at the origin. The initial state estimate is assumed to be a Gaussian vector with mean $\mathbf{x}_{0}$ and error covariance $P_{0}$, such that
\begin{equation}
 \mathbf{x}_{0}=\begin{bmatrix}
500&100&500&0
 \end{bmatrix}^T
\end{equation}
\begin{equation}
 P_{0}=\begin{bmatrix}
100 &10 &100 &10 
 \end{bmatrix}^T
\end{equation}
Hence initial unaugmented particles $\{\mathbf{x}_{0}^{(i)}\}_{i=1}^{N}$ were generated based on the distribution
\begin{equation}
\mathbf{x} \sim \mathcal{N}(\mathbf{x}_{0},P_{0})
\end{equation}
 
The process noise assumed has the diagonal covariance matrix $Q_{w}$ as:
\begin{equation} 
Q_w = diag(20,\;10,\;35,\;10)
\end{equation}

\begin{figure}[p]
 \centering 
  
\subfloat [MMPF estimate ]
{ \label{fig:MMPF_mode_prob}\includegraphics[scale=0.4]{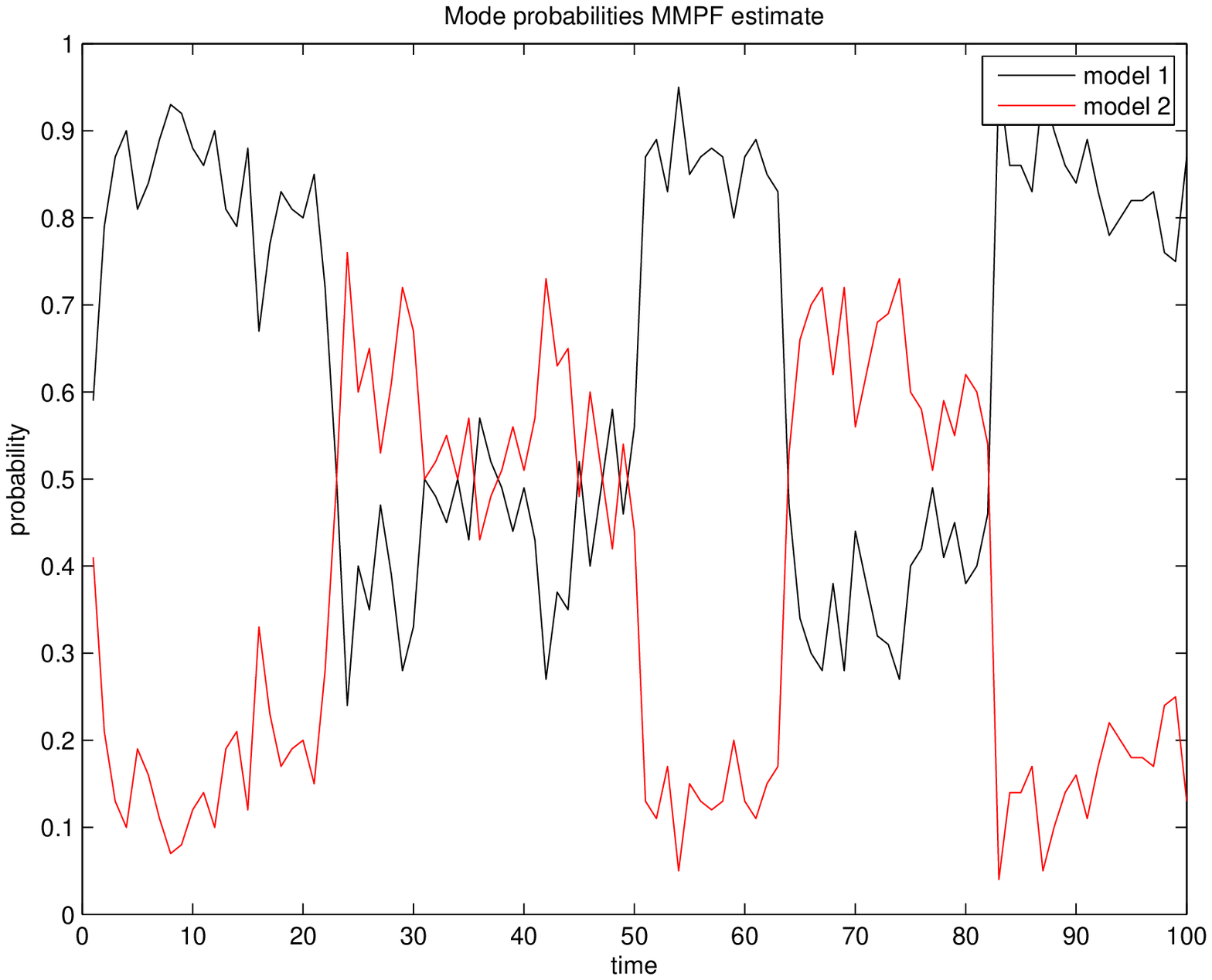}
}
\subfloat [IMM-EKF estimate ]
{ \label{fig:IMMEKF_mode_prob} \includegraphics[scale=0.4]{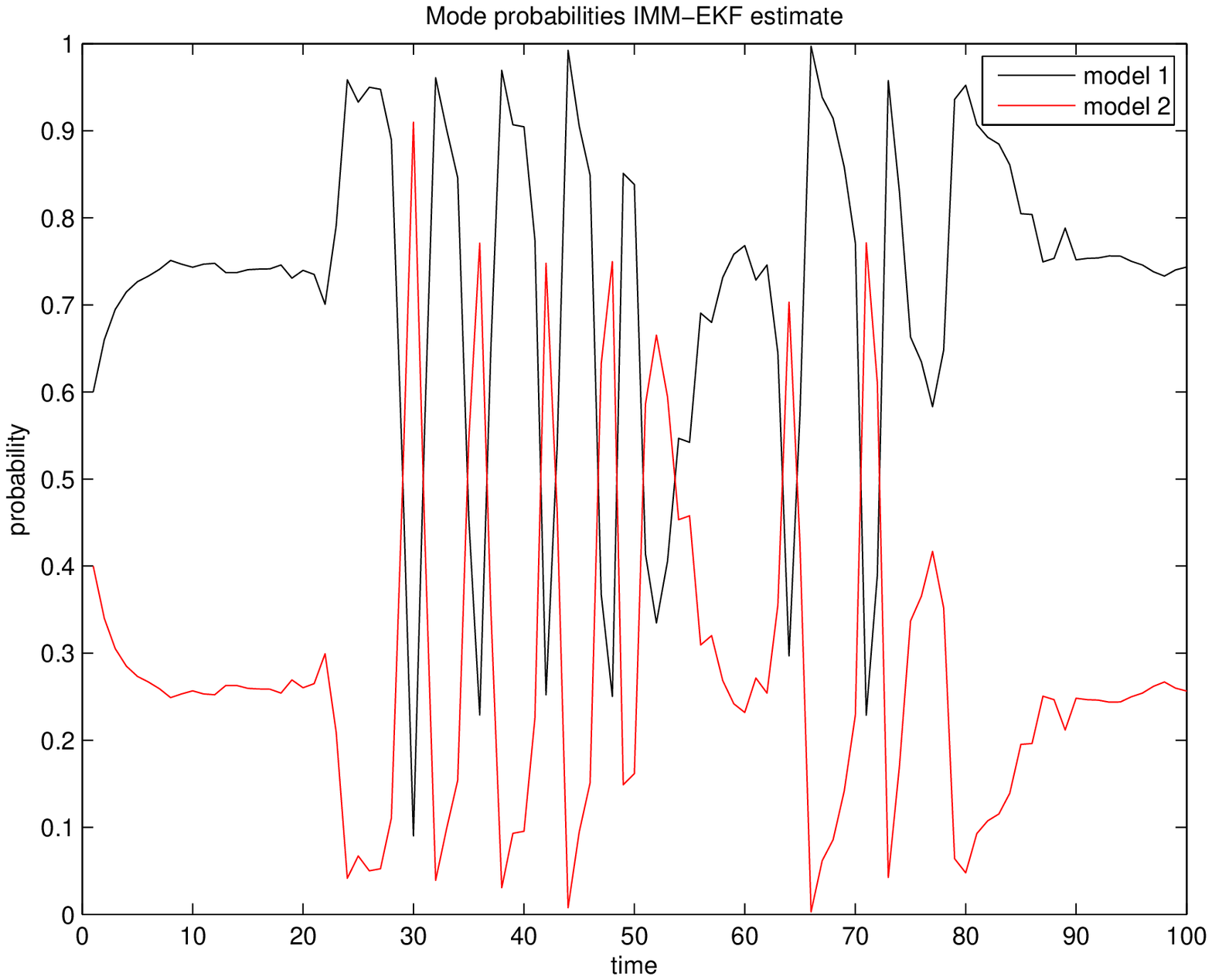}
}\\
\caption{Mode probabilities estimated using MMPF and IMM-EKF: The error is less in MMPF compared to IMM-EKF: The mode probabilities calculated by the MMPF have more match with the true mode probabilities. Thus the MMPF mode probabilities can give the information about the current target motion model.}
\label{fig:Ch_MMPF_mode_prob}
 \end{figure}

The two target motion models used by this imlementation of MMPF are constant velocity model and constant turn rate model with turn rate of $60 rad/s$. Hence the regime variable can take any of the two values, $A=1$ for constant velocity model and $A=2$ for constant turn rate model. The state transition model $f_{k}(\cdotp)$ for estimation of target state at time $t$ using $k^{th}$ model is such that:
\begin{equation}
\mathbf{x}_{t}=f_{k}(\mathbf{x}_{t-1})+\mathbf{w}_{t-1}
\end{equation}
where $\mathbf{w}_{t-1}$ is the process noise with zero mean and covariance $Q_w$.
The $f_{1}(\cdotp)$ is the constant velocity model and $f_{2}(\cdotp)$ is the constant turn rate model with turn rate of $60 rad/s$.
Hence $f_{1}(\cdotp)$ and $f_{2}(\cdotp)$ are matrices $F_1$ and $F_2$ repectively given by:
\begin{equation}
 F_1=\begin{bmatrix}
1&T&0&0\\
0&1&0&0\\
0&0&1&T\\
0&0&0&1\\
\end{bmatrix}
\end{equation}
\begin{equation}
F_2=
\begin{bmatrix}
1    &\dfrac{\sin (\Omega T)}{\Omega}       &0    &-\dfrac{1 - \cos(\Omega  T)}{\Omega}\\
        0    &\cos(\Omega  T)          &0     &-\sin(\Omega  T)\\
        0    &\dfrac{1-\cos(\Omega  T)}{\Omega}   &1      &\dfrac{\sin(\Omega  T)}{\Omega}\\
        0    &\sin(\Omega  T)          &0      &\cos(\Omega  T) \\
\end{bmatrix}
\end{equation}
where $T$ is the sampling period of the target dynamics and $\Omega$ is the turn rate.
In this implementation of the particle filter, the transitional prior which is a suboptimal choice of importance density $q(\mathbf{x}_{t} \mid \mathbf{x}_{t-1}^{(i)},z_{t})$ is  used to propose the particles. Thus the importance density used is:\\
\begin{eqnarray*}
q(\mathbf{x}_{t}\mid \mathbf{x}_{t-1}^{(n)},A_t^n,z_{t})&=&p(\mathbf{x}_{t}\mid \mathbf{x}_{t-1}^{(n)},A_t^n)\\
&=&\left\{
 \begin{array}{rl}
  \mathcal{N}(f_{1}(\mathbf{x}_{t-1}),Q_{w}) & \text{if } A_t^n = 1\\
  \mathcal{N}(f_{2}(\mathbf{x}_{t-1}),Q_{w}) & \text{if } A_t^n = 2\\
  \end{array} \right.
\end{eqnarray*}
The mode transition probability matrix assumed by the filter for the target was
\begin{equation}
\pi_{ij}=\begin{bmatrix}
.9 &.1\\
.3 &.7
\end{bmatrix}
\end{equation}
A total of $N=100$ particles were used. The initial mode probability is assumed to be uniform.
\begin{equation}
 \pi_i(0)=0.5  \;;\qquad i=1,\;2
\end{equation}
Hence particles were equally divided and associated with the considered target motion models, i.e. 50 particles' regime variable were associated with constant velocity model ($A=1$) and the rest were associated with constant turn rate model ($A=2$). The true trajectories of the target and its track estimate are shown in Fig. \ref{fig:MMPF_xy_plot}. The state estimates of the targets are shown in Fig. \ref{fig:MMPF_x_state}, Fig. \ref{fig:MMPF_y_state}, Fig. \ref{fig:MMPF_vx_state} and Fig. \ref{fig:MMPF_vy_state}. The mean square error (MSE) of the position estimates for 100 Monte Carlo runs are shown in Fig. \ref{fig:MMPF_MSE}. The simulation results show that MMPF can successfully track maneuvering targets if the information about the various maneuvering models are given. The ratio of regime variables corresponding to each model gives the mode probabilities and are plotted in Fig. \ref{fig:MMPF_mode_prob}. It clearly indicates that when the target is in a particular motion model, particles resembling this motion model are automatically selected more number of times by the MMPF and given more weightage. Thus the model probability gives the information about the current target motion model.
%


\subsection{Comparison of MMPF with Standard Bootstrap PF}
The standard bootstrap particle filter is implemented for the same target and measurement scenario with N=100 particles. The same states $\mathbf{x}=
\begin{bmatrix}
x & v_{x} & y  & v_{y} 
 \end{bmatrix}^T$, constant velocity model, the process noise $Q_w$ and the transitional prior as the importance density were used by the filter. The true trajectories of the targets and the track estimates are shown in Fig.\ref{fig:PF_xy_plot}. The state estimates of the targets are shown in Fig. \ref{fig:MMPF_x_state}, Fig. \ref{fig:MMPF_y_state}, Fig. \ref{fig:MMPF_vx_state} and Fig. \ref{fig:MMPF_vy_state}. The MSE of the position estimates for 100 Monte Carlo runs are shown in Fig.\ref{fig:PF_MSE}. The results show that the estimates diverge and the standard bootstrap particle filter is not able to track high maneuvering targets using single model. PF can have proper tracking only with higher number of particles,  but the same performance can be achieved using MMPF with lesser number of particles. Thus it shows that MMPF improves tracking of targets with high maneuvers when compared to the standard bootstrap particle filter. 

\subsection{Comparison of MMPF with IMM-EKF}
\begin{figure}[p]
 \centering 
\subfloat [MMPF estimate ]
{\label{fig:MMPF_xy_plot}\includegraphics[scale=0.4]{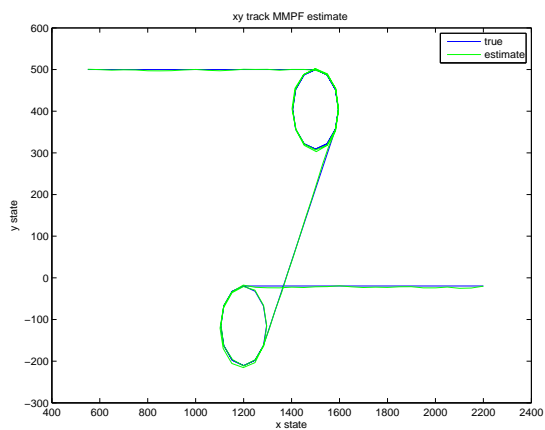}
}\\
\subfloat [PF estimate ]
{ \label{fig:PF_xy_plot}\includegraphics[scale=0.4]{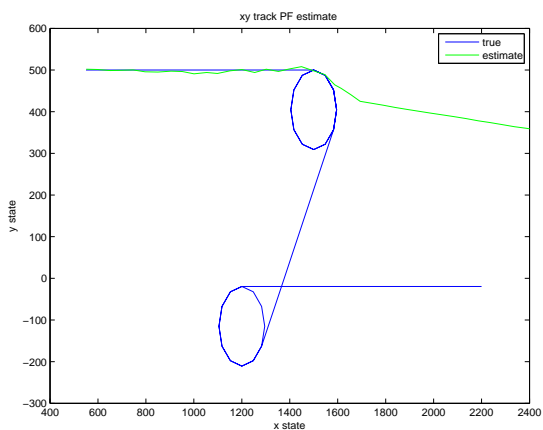}
}
\subfloat [IMM-EKF estimate ]
{ \label{fig:IMMEKF_xy_plot}\includegraphics[scale=0.4]{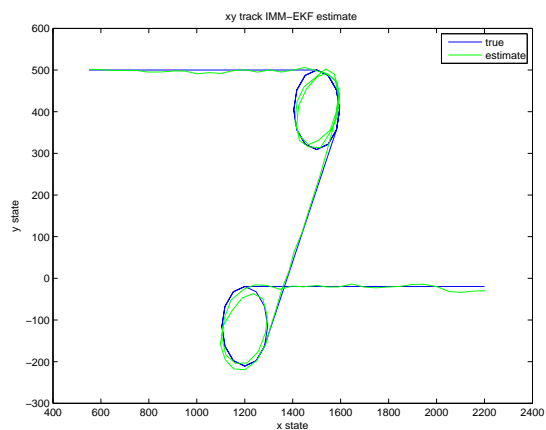}
}\\
\caption{Target's true $xy$ track and its estimated tracks obtained using MMPF, PF and IMM-EKF: The MMPF estimate is more accurate than PF and IMM-EKF estimate. The PF estimate diverges completely during the maneuver of the target. (PF can have proper tracking only with higher number of particles,  but the same performance can be achieved using MMPF with lesser number of particles.)}
\label{fig:Ch_MMPF_xy_plot}
 \end{figure}

\begin{figure}[p]
\centering
\subfloat [MMPF estimate]
{\label{fig:MMPF_MSE}\includegraphics[scale=0.4]{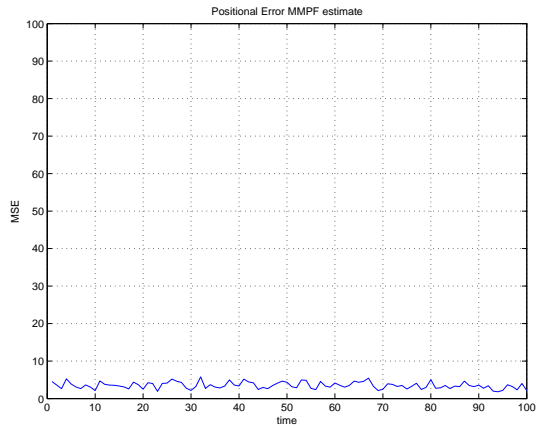}  
}\\
\subfloat [PF estimate]
{\label{fig:PF_MSE}\includegraphics[scale=0.4]{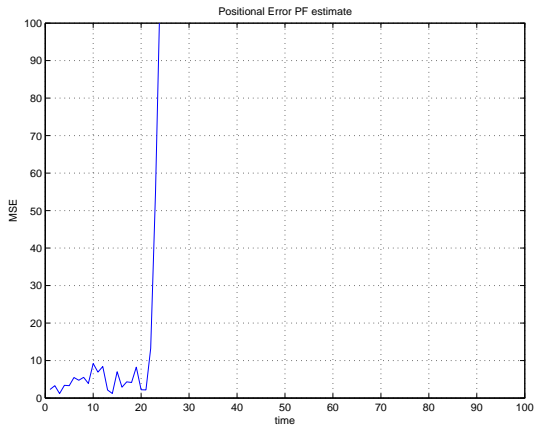}   
}
\subfloat [IMM-EKF estimate]
{\label{fig:IMMEKF_MSE}\includegraphics[scale=0.4]{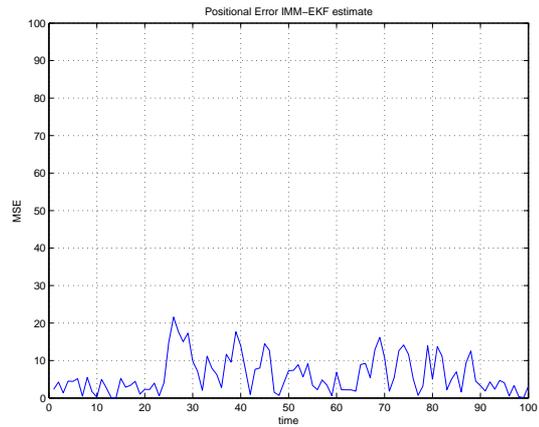}  
}\\
\caption{MSE of the position estimate for 100 Monte carlo runs obtained using MMPF, PF and IMM-EKF: MMPF and IMM-EKF have similar performance, but PF estimates are diverged. (PF can have proper tracking only with higher number of particles,  but the same performance can be achieved using MMPF with lesser number of particles.)}
\label{fig:Ch_MMPF_MSE}
 \end{figure}

\begin{figure}[p]
\centering
\subfloat [MMPF estimate ]
{\includegraphics[scale=0.4]{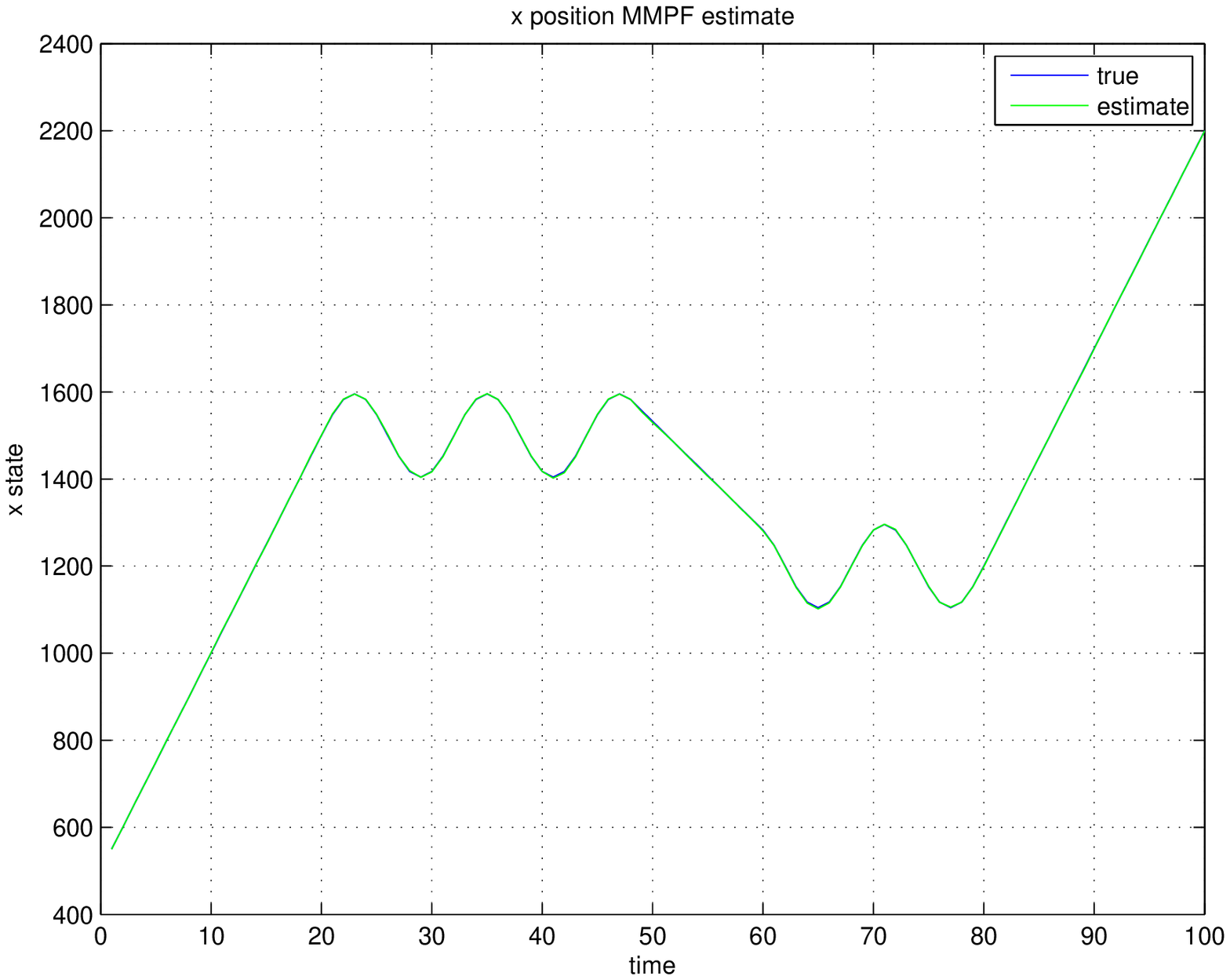}  
}\\
\subfloat [PF estimate ]
{\includegraphics[scale=0.4]{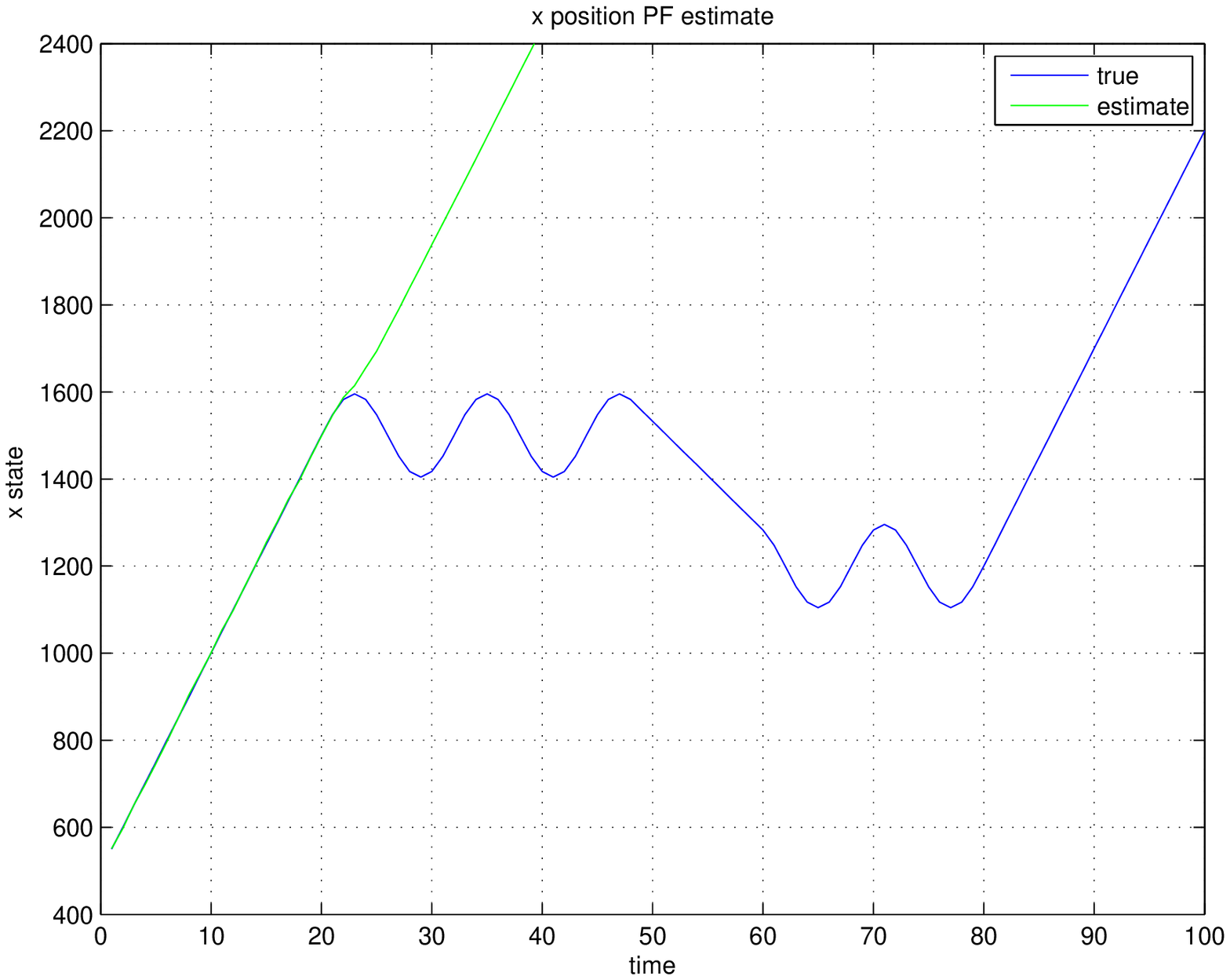} 
}
\subfloat [IMM-EKF estimate]
{\includegraphics[scale=0.4]{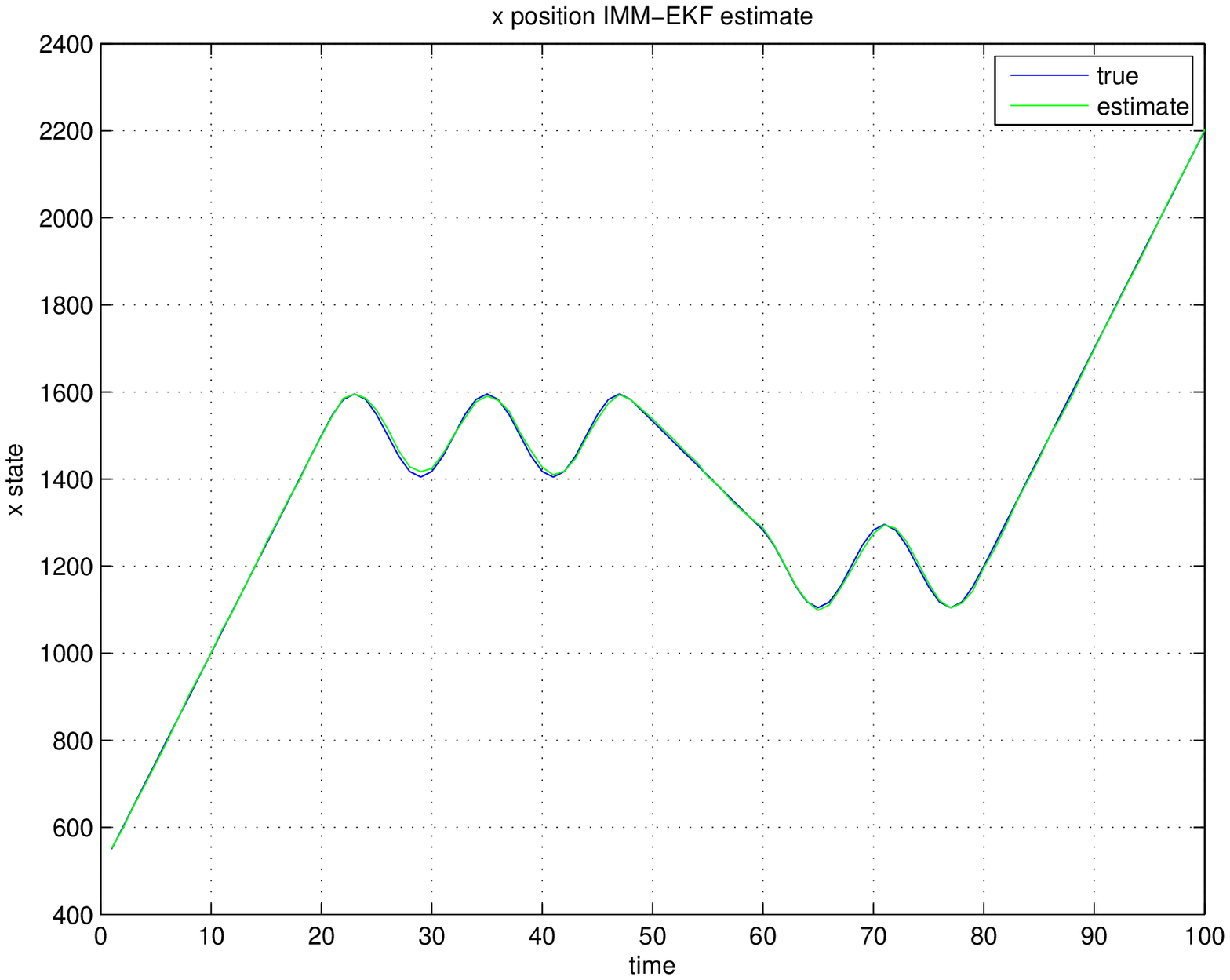} 
}\\
\caption{Target's true state $x$ and its estimates obtained using MMPF, PF and IMM-EKF: The performance of MMPF and IMM-EKF in estimating state $x$ are similar. PF estimates are diverged.}
\label{fig:MMPF_x_state}
 \end{figure}

\begin{figure}[p]
\centering
\subfloat [MMPF estimate ]
{\includegraphics[scale=0.4]{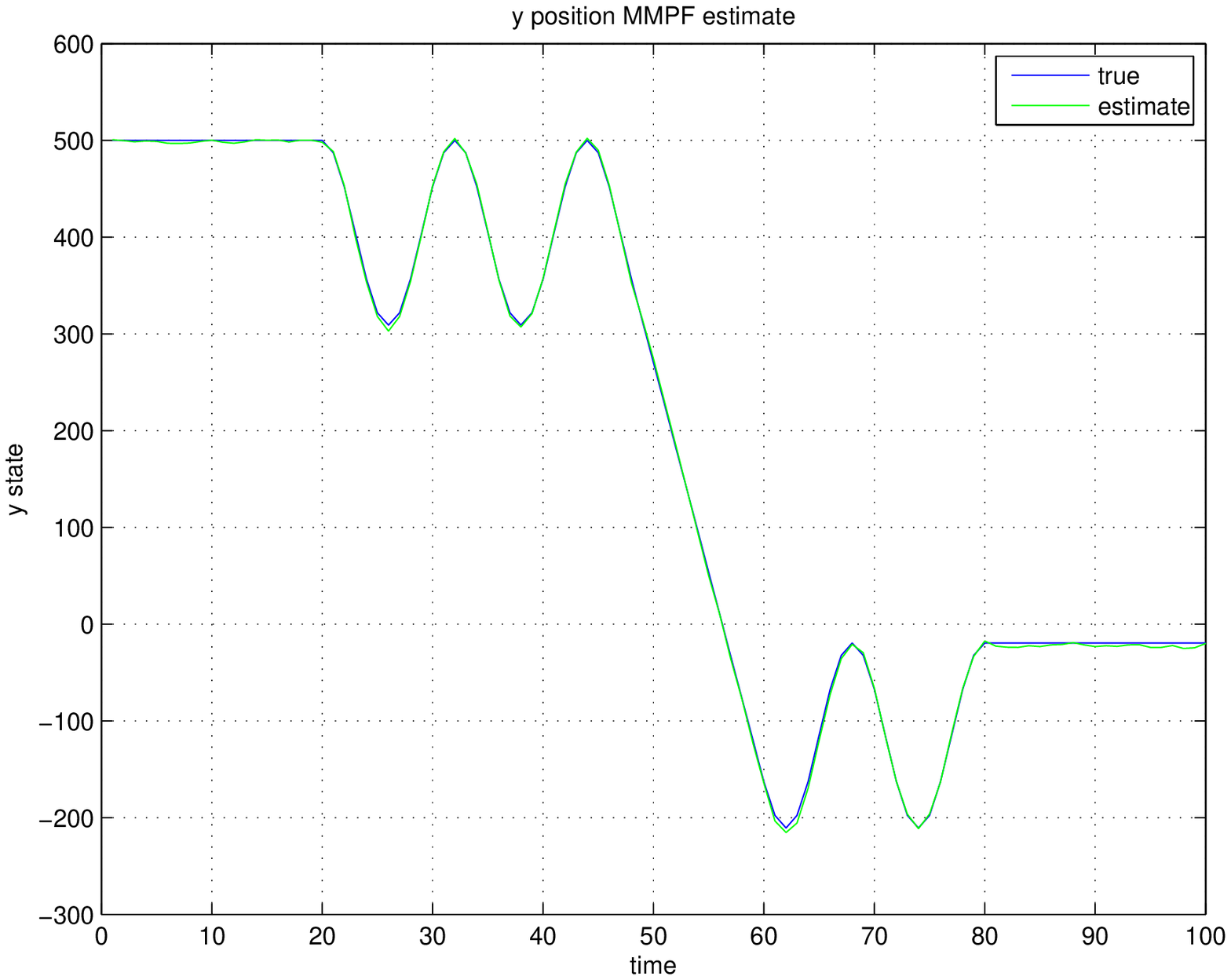}  
}\\
\subfloat [PF estimate ]
{\includegraphics[scale=0.4]{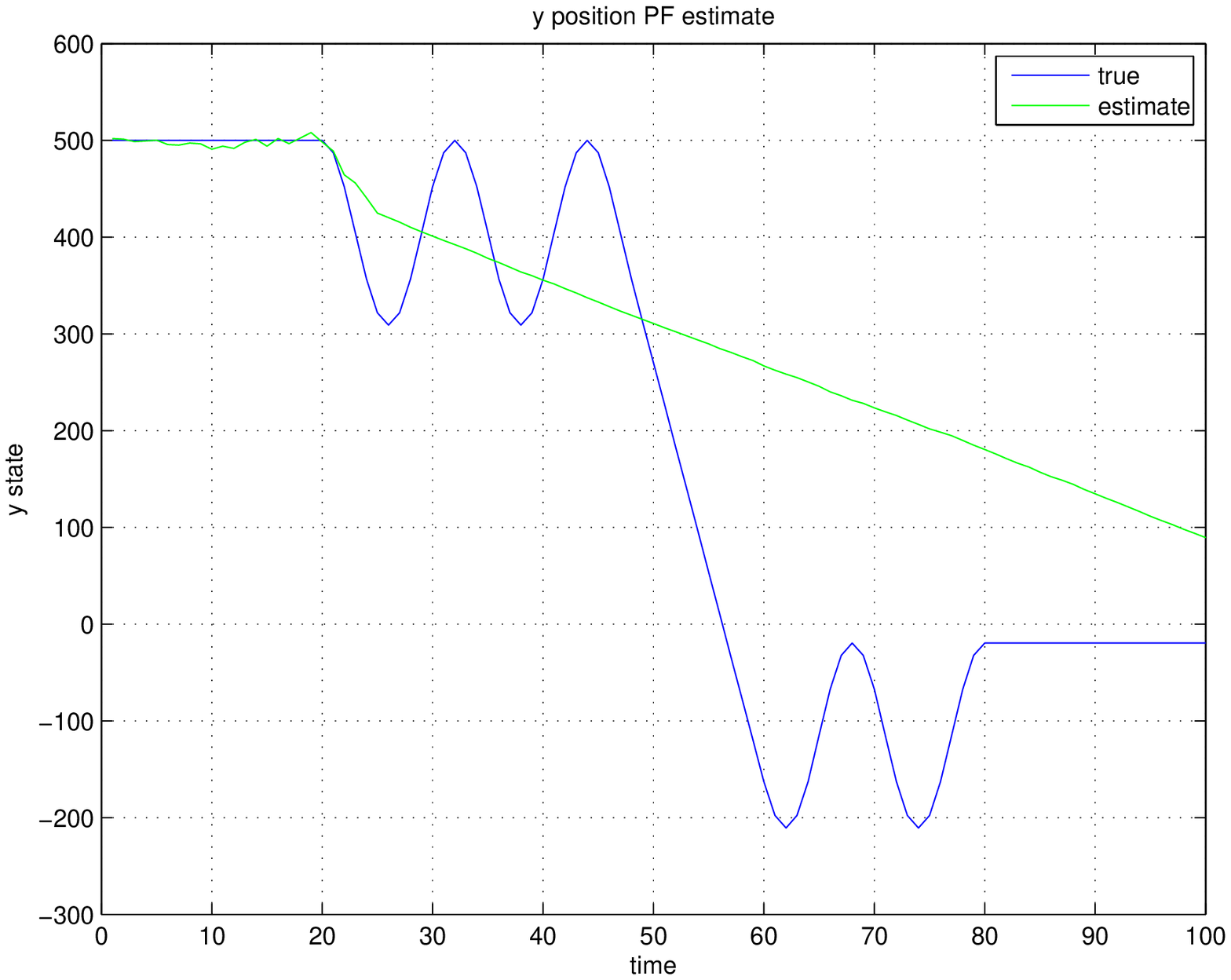} 
}
\subfloat [IMM-EKF estimate]
{\includegraphics[scale=0.4]{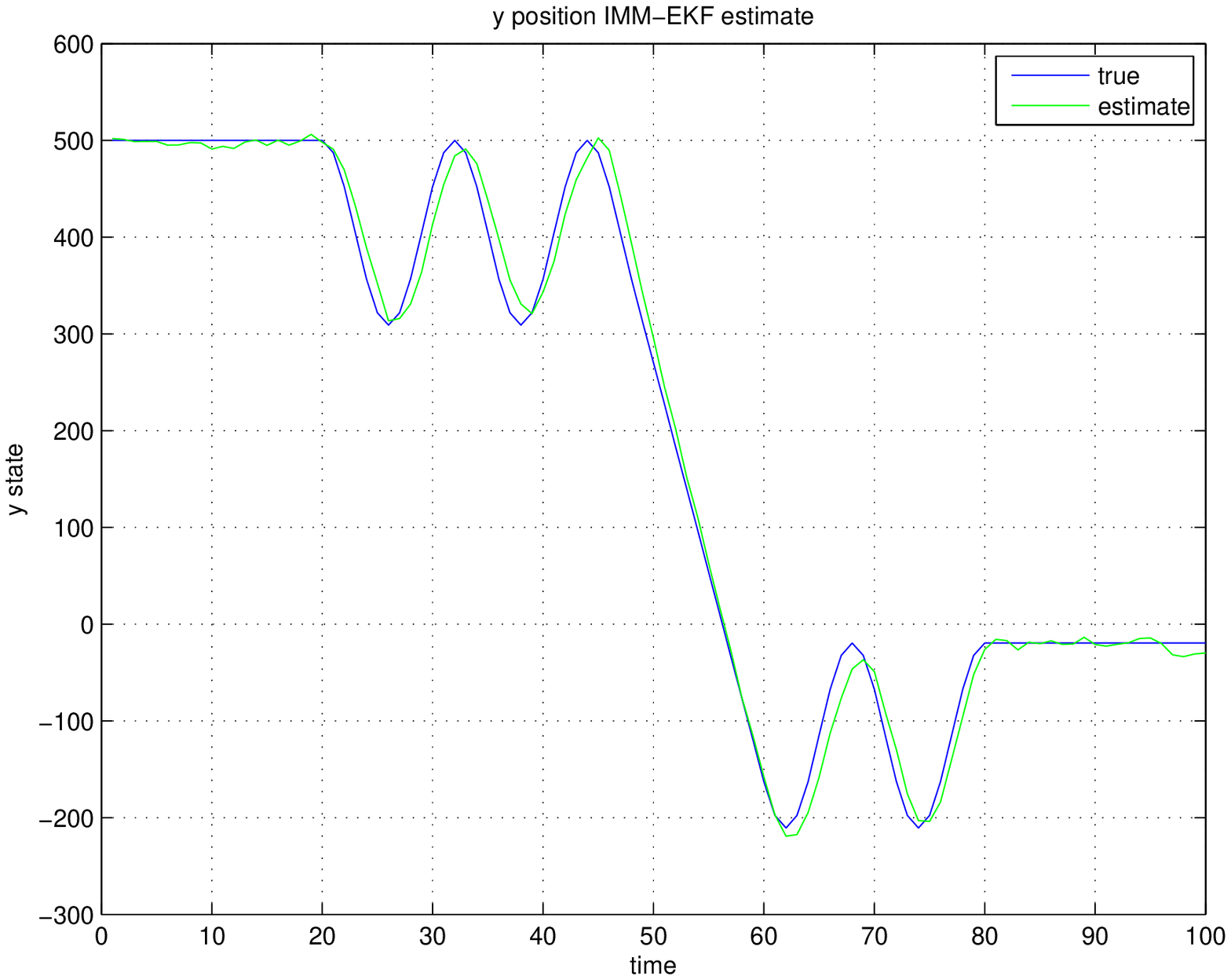} 
}\\
\caption{Target's true state $y$ and its estimates obtained using MMPF, PF and IMM-EKF: The performance of MMPF and IMM-EKF in estimating state $y$ are similar. PF estimates are diverged.}
\label{fig:MMPF_y_state}
 \end{figure}

\begin{figure}[p]
\centering
\subfloat [MMPF estimate ]
{\includegraphics[scale=0.4]{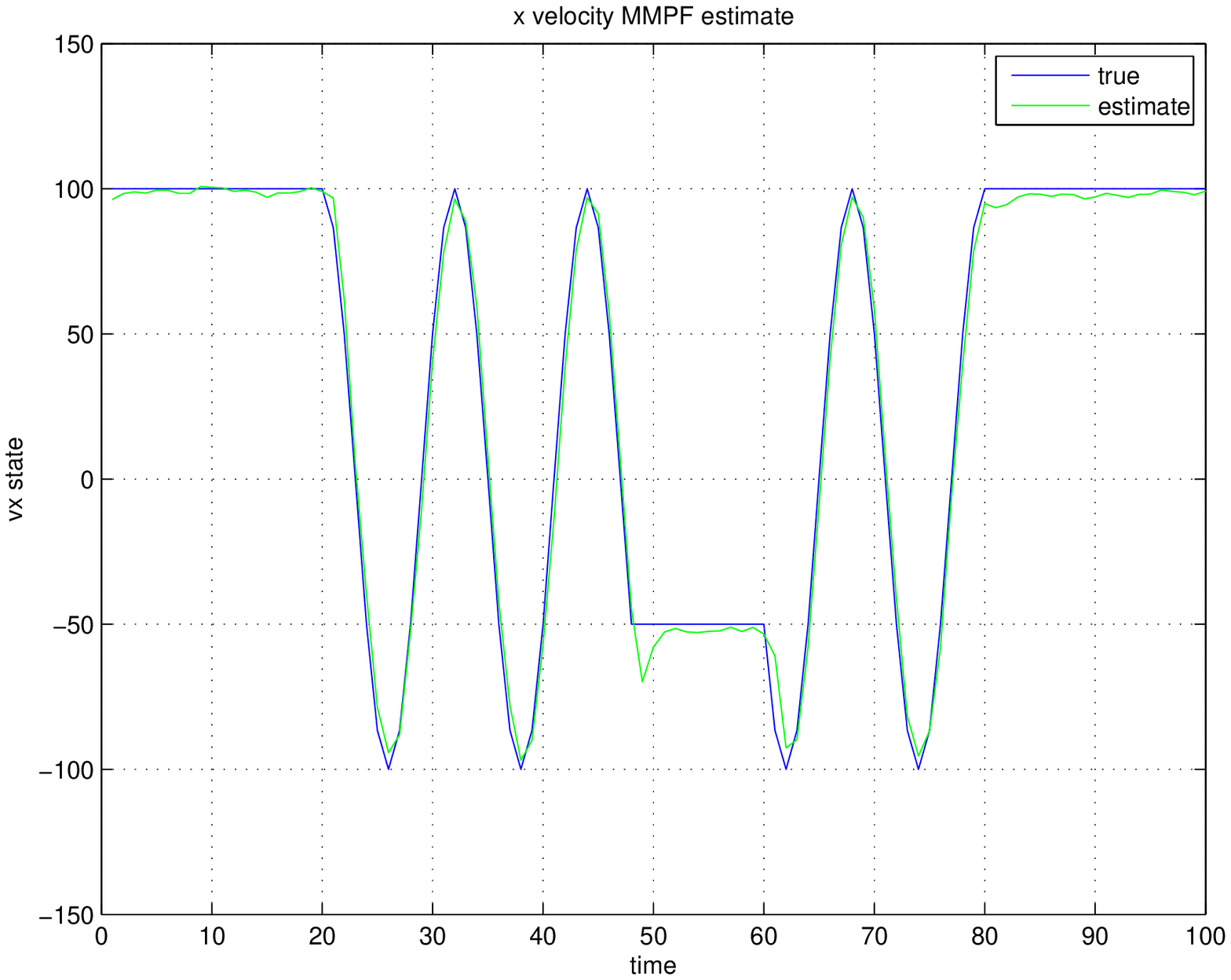}  
}\\
\subfloat [PF estimate ]
{\includegraphics[scale=0.4]{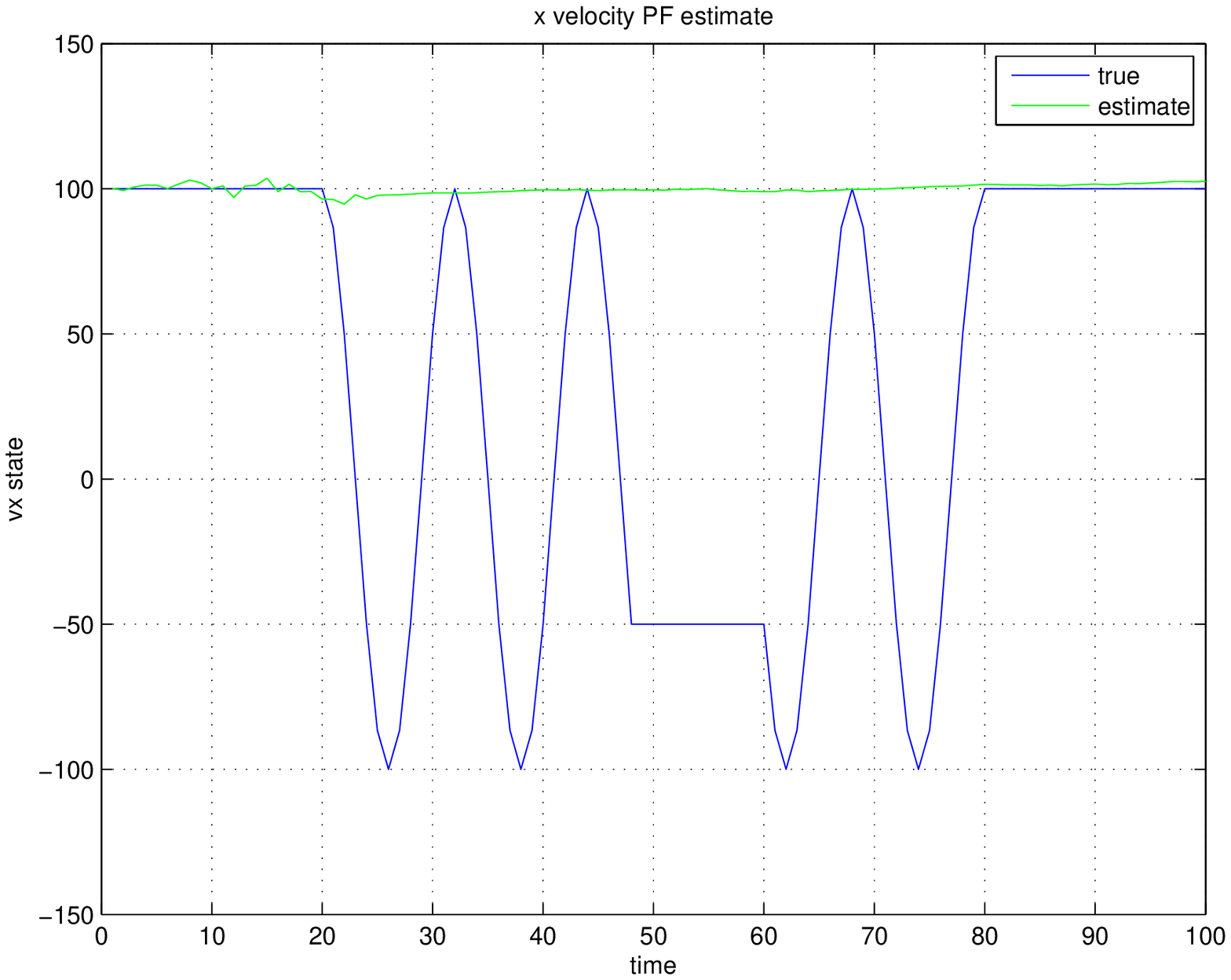} 
}
\subfloat [IMM-EKF estimate]
{\includegraphics[scale=0.4]{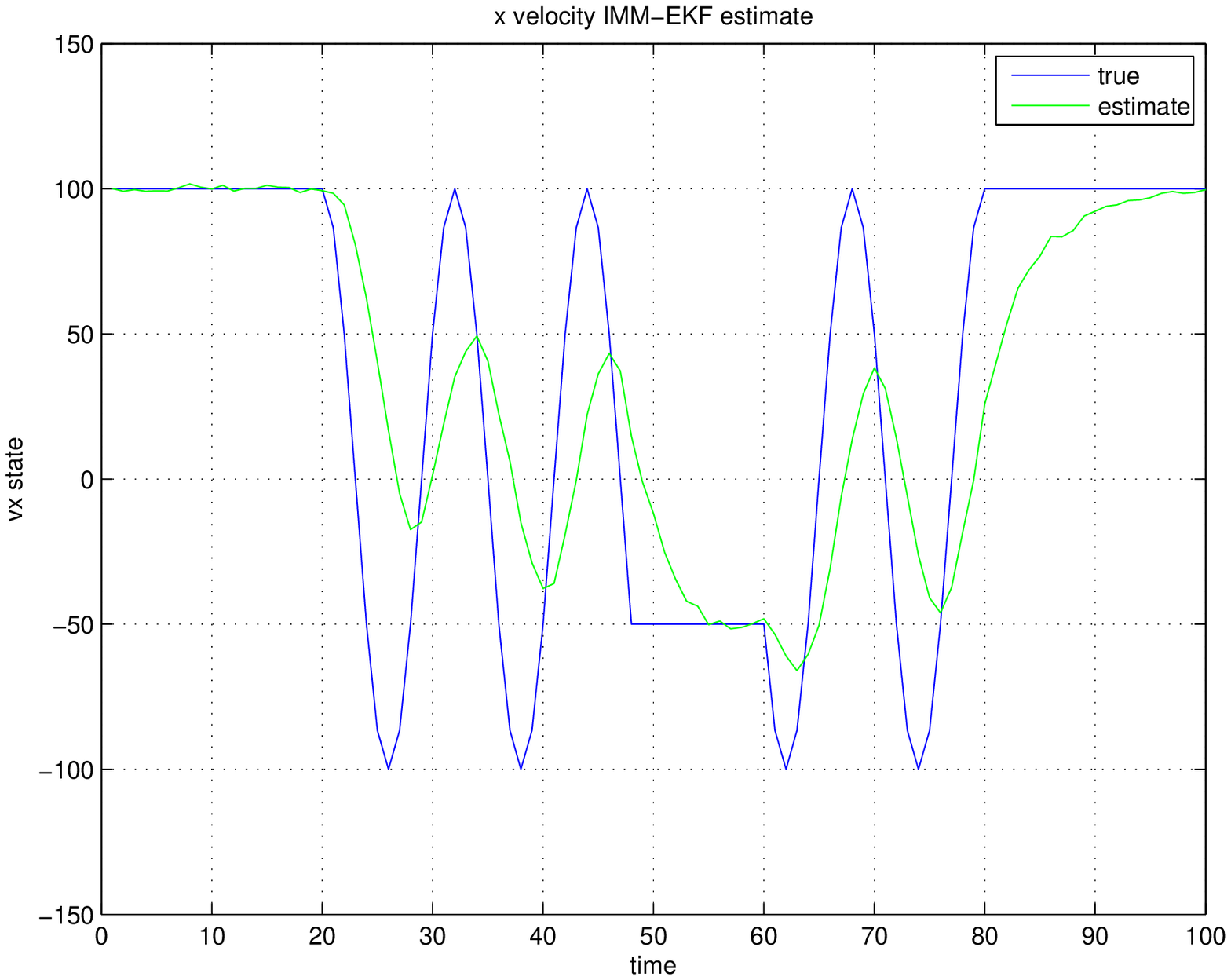} 
}\\
\caption{Target's true state $v_x$ and its estimates obtained using MMPF, PF and IMM-EKF: Velocity estimates of MMPF is better than EKF-IMM. PF $v_x$ velocity estimates are diverged.}
\label{fig:MMPF_vx_state}
 \end{figure}

\begin{figure}[p]
\centering
\subfloat [MMPF estimate ]
{\includegraphics[scale=0.4]{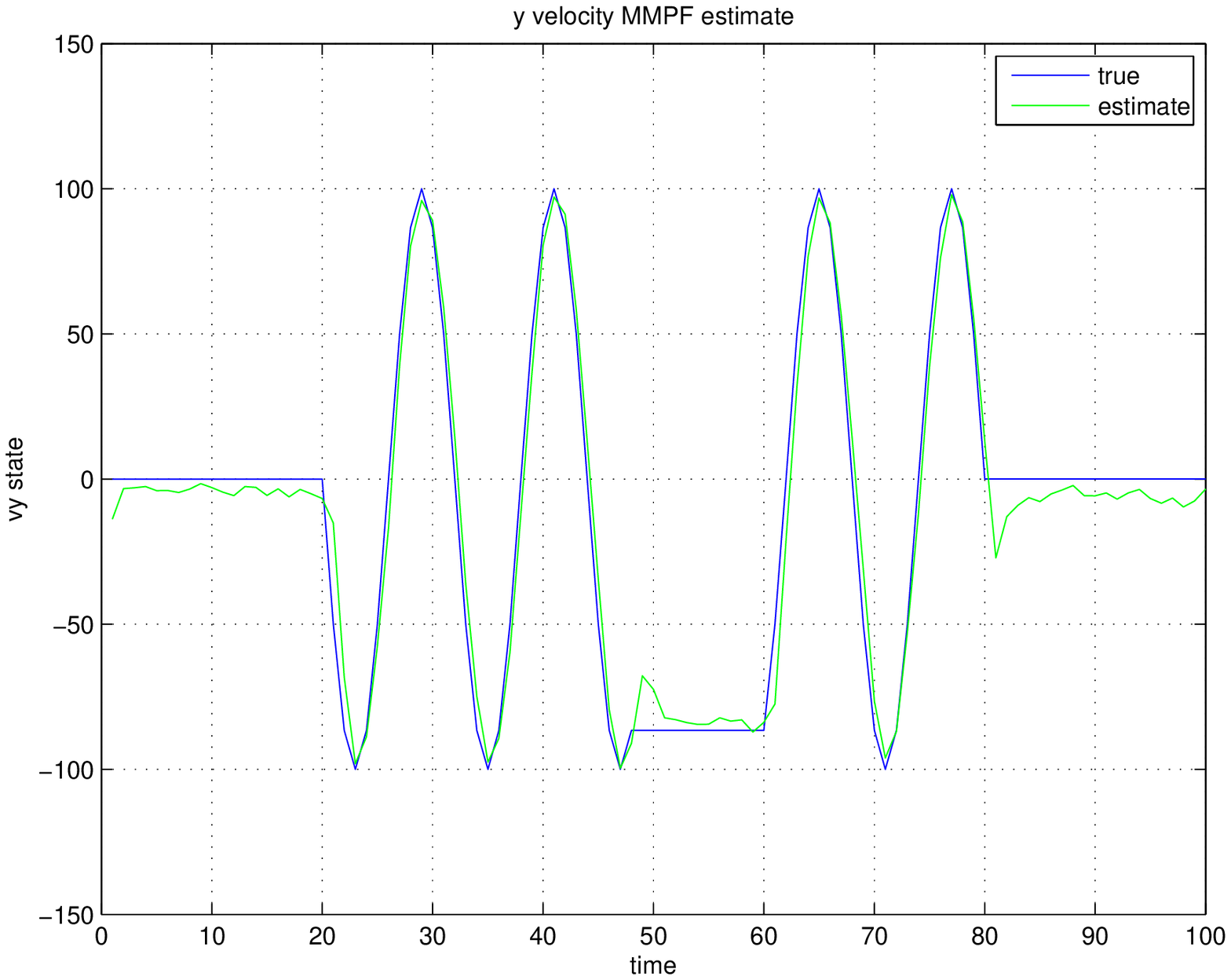}  
}\\
\subfloat [PF estimate ]
{\includegraphics[scale=0.4]{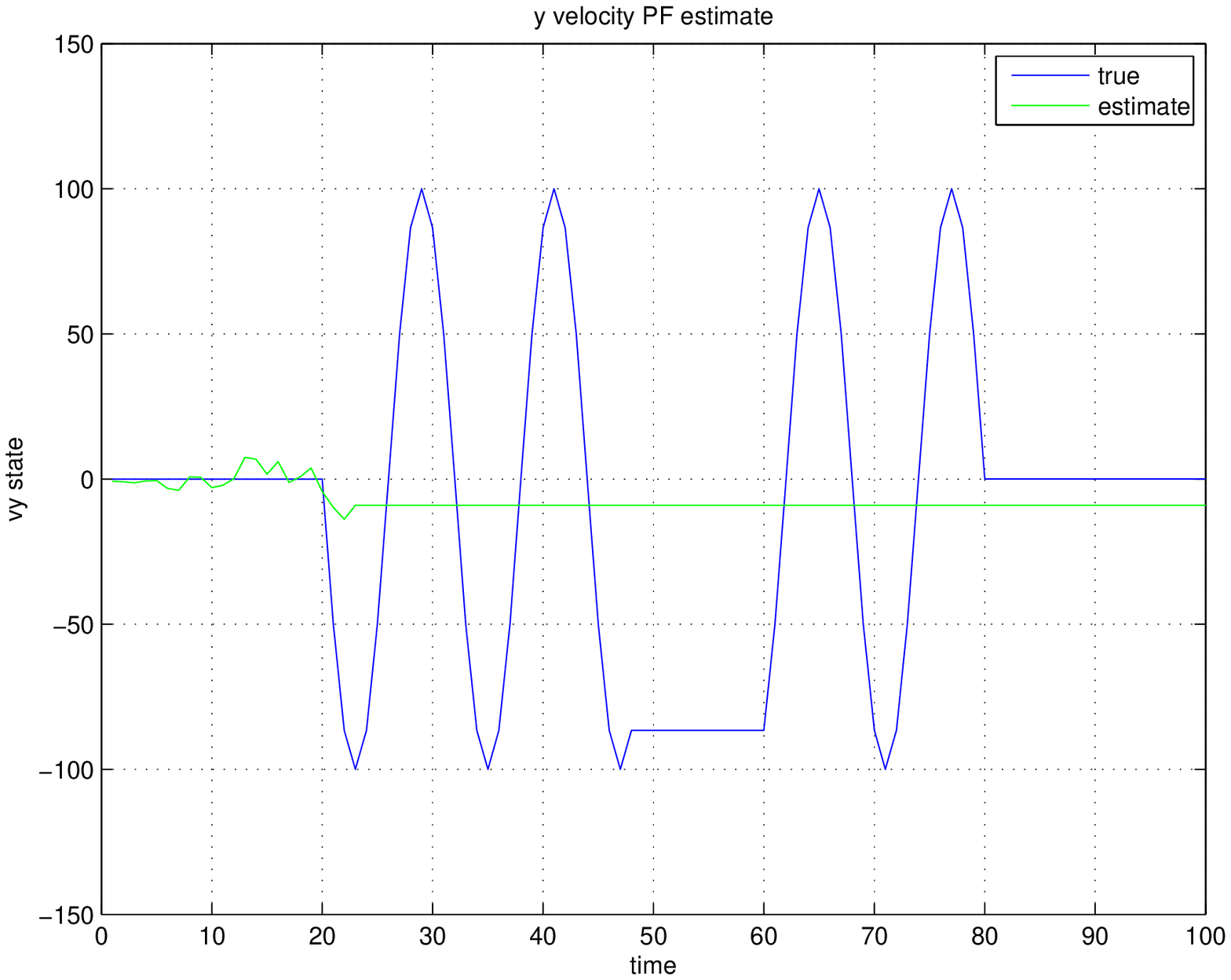} 
}
\subfloat [IMM-EKF estimate]
{\includegraphics[scale=0.4]{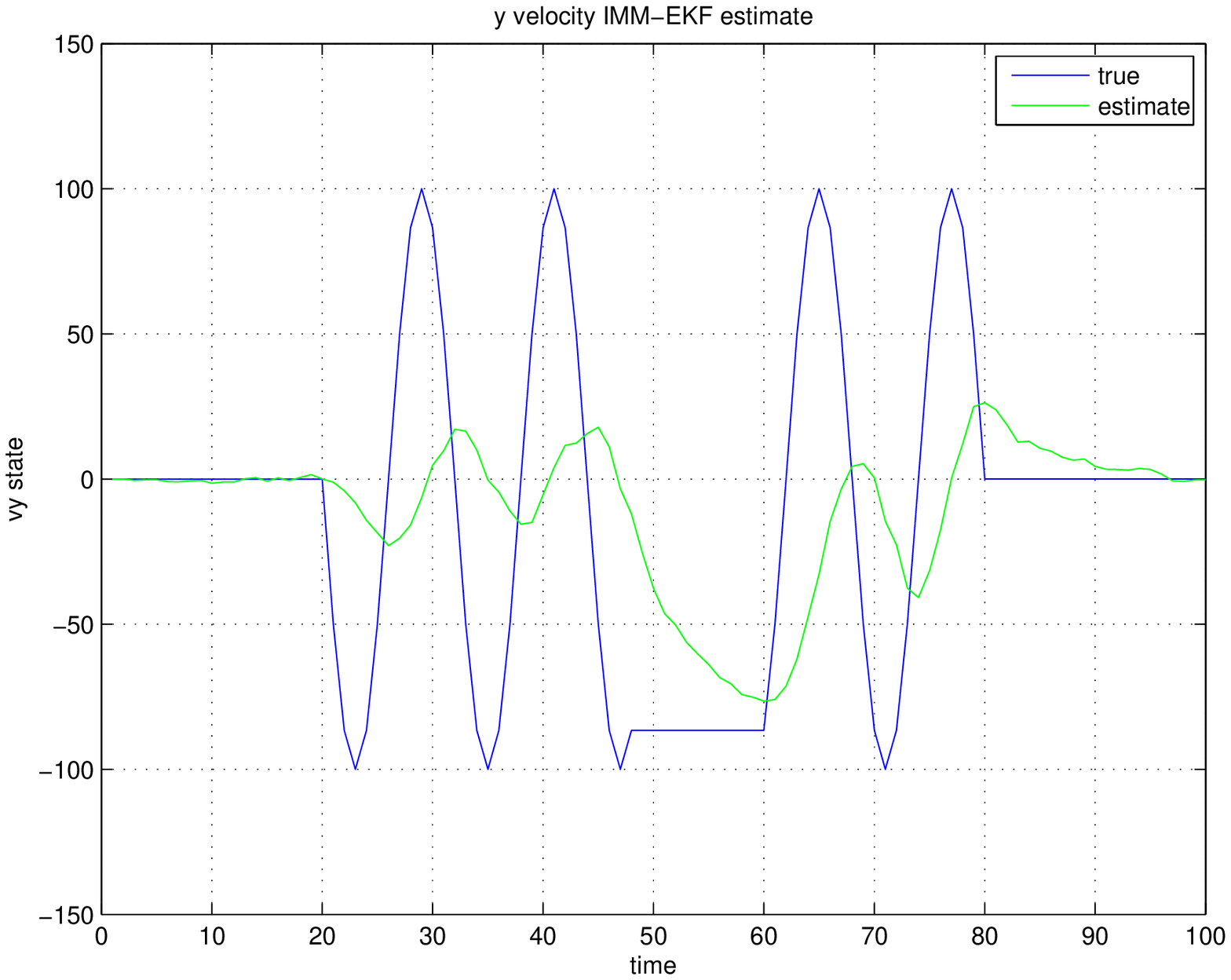} 
}\\
\caption{Target's true state $v_y$ and its estimates obtained using MMPF, PF and IMM-EKF: Velocity estimates of MMPF is better than EKF-IMM. PF $v_y$ velocity estimates are diverged.}
\label{fig:MMPF_vy_state}
 \end{figure}

The IMM-EKF filter is implemented for the same target and measurement scenario. The same states $\mathbf{x}=
\begin{bmatrix}
x & v_{x} & y  & v_{y} 
 \end{bmatrix}^T$, constant velocity and constant turn models, and the process noise $Q_w$ were used by the filter. The true trajectories of the targets and the track estimates are shown in Fig.\ref{fig:IMMEKF_xy_plot}. The state estimates of the targets are shown in Fig. \ref{fig:MMPF_x_state}, Fig. \ref{fig:MMPF_y_state}, Fig. \ref{fig:MMPF_vx_state} and Fig. \ref{fig:MMPF_vy_state}. The MSE of the position estimates for 100 Monte Carlo runs are shown in Fig.\ref{fig:IMMEKF_MSE}. The results show that state estimates have larger MSE compared to the MMPF estimates. The velocity estimates particularly have very large deviation from the true states. Also the mode probabilities calculated by the filter do not always match with the true mode probabilities.  Thus it is clear that the capability of IMM-EKF filter to track maneuvering targets using single model is less compared to multiple model particle filter (MMPF).

\section{Summary}
Targets can have abrupt deviation which are difficult to represent using single kinematic model. Particle filters can track maneuvering targets by using constant velocity model alone by increasing the number of particles. The number of particles for tracking highly maneuvering targets can be considerably reduced by incorporating multiple kinematic models. Thus multiple model particle filter (MMPF) proposes particles using multiple models. The modal that is used by a particular particle is determined by its regime/mode variable. These mode variables have transition between the models according to the transition probability matrix. The particles with correct mode have large likelihood and are selected more number of times during resampling. Thus the MMPF filters out and multiplies the particles which are closer to the true dynamics of the targets and use them efficiently. Hence lesser number of particles are enough to track highly maneuvering targets. The simulations show that MMPF have better tracking capability than standard PF and interacting multiple model Extended Kalman filter (IMM-EKF).
\chapter{Monte Carlo Joint Probabilistic Data Association Filter (MC-JPDAF)}
\label{chap:MC-JPDAF}
Bar Shalom at.al \cite{14}  developed the  Joint Probabilistic Data Association Filter (JPDAF) 
for solving the data association problem in multi-target tracking. It is the most widely applied method for multi-target tracking under data association uncertainty. Monte Carlo Joint Probabilistic Data Association Filter (MC-JPDAF) was developed by J.~Vermaak et.al \cite{15} for solving the data association problem in multi-target tracking using particle filter framework. It incorporates clutter and missing measurements and also measurements from multiple observers. Data association problem arises due to the lack of information at the observer about the proper association between the targets and the received measurements. The problem becomes more involved when the targets move much closer and there are clutter and missed target detections at the observer. 

In the literature, there are various other strategies to solve the data association problem like Multiple Hypothesis Tracking(MHT), Nearest Neighbour Standard Filter(NNSF), etc. MHT keep track of all possible association hypothesis over time. Its computational complexity increases with time since the number of hypothesis grows exponentially. NNSF associates each measurement with the nearest target and neglect many other feasible hypotheses. JPDAF considers all possible hypotheses at each time step. The infeasible hypotheses are neglected using a gating procedure to reduce computational complexity. It calculates the posterior hypotheses probability of the remaining hypotheses. The filtered estimate of each hypothesis is calculated and is combined by weighting each with their corresponding posterior hypothesis probability. For estimation using extended Kalman filter framework, JPDAF relies on linear Gaussian models for evaluation of target measurement hypotheses. Non-linear models can be accommodated by suitably linearizing using EKF. But its performance degrades as non-linearity becomes severe. MC-JPDAF combines JPDAF with particle filtering technique to accommodate non-linear and non-Gaussian models. The remaining part of the chapter explores the MC-JPDAF and is organized as follows. Section \ref{sec:Model_Description} describes the hypothesis models for the target and measurement association, models for association prior and the likelihood model. Section \ref{sec:MCJPDAF} describes the MC-JPDAF. The general JPDAF framework is described and the MC-JPDAF algorithm is explained later.


\section{Model Description}
\label{sec:Model_Description}
This section describes target and measurement model, two types of data association hypothesis model and the conversion between them.

\subsection{Target model}
The number of targets $K$ is assumed to be known and fixed. The state of the target $k$ at time $t$ is represented by $\mathbf{x}_{k,t},k=1,2,\ldots,K$. The combined state of all targets at time $t$ is represented by $\mathbf{x}_t=\{\mathbf{x}_{1,t},\mathbf{x}_{2,t},\ldots,\mathbf{x}_{K,t}\}$. Each target has independent Markov dynamics $p_k(\mathbf{x}_{k,t}|\mathbf{x}_{k,t-1})$. Hence the dynamics of the combined state factorizes over individual targets
\begin{equation}
p(\mathbf{x}_{k,t}|\mathbf{x}_{k,t-1})=\displaystyle\prod_{k=1}^{K} p_k(\mathbf{x}_{k,t}|\mathbf{x}_{k,t-1}) 
\end{equation}

\subsection{Measurement and data association model}
It is assumed that there are $N_o$ observers whose locations are given by $P_0^1,P_0^2,P_0^3,\ldots,P_0^{N_o}$. The observers are assumed to be static. The total number of measurements from an observer $i$ at a given time is denoted by $M^i$ which can vary with time due to missed target measurements and clutter measurements. Hence the measurement from a given observer $i$ is denoted by $\mathbf{y}^i=(\mathbf{y}_1^i,\mathbf{y}_2^i,\mathbf{y}_3^i,\ldots,\mathbf{y}_{M^i}^i)$. The combined set of measurements from all the $N_o$ observers are denoted as $\mathbf{y}=(\mathbf{y}^1,\mathbf{y}^2,\mathbf{y}^3,\ldots,\mathbf{y}^{N_o})$. The clutter measurements occur due to the multi path effects and observer errors etc. It is also assumed that every measurement at an observer can have only one source and more than one measurement cannot originate from a target. The targets can also be undetected. All the measurements can be clutter and there may be no measurements at a particular time.

The data association is represented using a set of association variables. There are two types of representation for data association hypothesis.
\begin{enumerate}
\item Measurement-to-Target association ($M{\rightarrow}T$)
\item Target-to-Measurement association ($T{\rightarrow}M$)
\end{enumerate}
Both carry same information and have one to one mapping between them. They can be converted from one type of representation to another.

\subsubsection{Measurement-to-Target association($M{\rightarrow}T$) hypothesis}
It is denoted by $\lambda=(\lambda^1,\lambda^2,\ldots,\lambda^{N_o})$, where $\lambda^i=(\mathbf{r}^i,M_C^i,M_T^i)$ is the hypothesis for the measurements from observer $i$. The hypothesis $\lambda^i$ indicates that the measurement has $M_C^i$ clutter measurements and $M_T^i$ target detected measurements. The sum of $M_C^i$ and $M_T^i$ gives the total number of measurements $M^i$, at the observer $i$
\begin{equation}
M^i=M_C^i+M_T^i.
\end{equation}
The measurements are indexed from $1$ to $M^i$ and targets are indexed from $1$ to $K$. The association vector $\mathbf{r}^i=(r_1^i,r_2^i,\ldots,r_{M^i}^i)$ gives the index of the targets which has caused the measurements $1$ to $M^i$. The association vector at observer $i$ is given by
\begin{equation}
r_j^i=
\begin{cases}
0 \hfill \text{  if measurement $j$ is due to clutter}\\
k \hfill \text{  if measurement $j$ is due to target $k$}
\end{cases} 
\end{equation}

\textit{Example }: $\mathbf{r}^i=(3,4,0,1,0,5)$

Here there are $M^i=6$ measurements, out of which the third and fifth measurements are due to clutter. The detected targets are $1,3,4$ and $5$. The first measurement correspond to target $3$. The second measurement correspond to target $4$. Fourth measurement correspond to target $1$ and sixth measurement correspond to target $5$.

\subsubsection{Target-to-Measurement association($T{\rightarrow}M$) hypothesis}
It is denoted by $\tilde{\lambda}=(\tilde{\lambda}^1,\tilde{\lambda}^2,\ldots,\tilde{\lambda}^{N_o})$, where $\tilde{\lambda}^i=(\mathbf{\tilde{r}}^i,M_C^i,M_T^i)$ is the target to measurement association hypothesis at observer $i$. It is similar to $M{\rightarrow}T$ association hypothesis except for the association vector $\tilde{\mathbf{r}}^i$. The association vector $\tilde{\mathbf{r}}^i=(\tilde{r}_1^i,\tilde{r}_2^i,\ldots,\tilde{r}_{K}^i)$ gives the measurements corresponding to the targets $1$ to $K$. Missed target detections are denoted as $0$. The association vector at observer $i$ is given by
\begin{equation}
\tilde{r}_j^i=
\begin{cases}
0 \hfill \text{  if target $k$ is undetected}\\
j\in(1,\ldots,M^i) \hfill \text{  if target $k$ generated measurement $j$}
\end{cases} 
\end{equation}

\textit{Example }: $\mathbf{\tilde{r}}^i=(2,4,0,1,5)$

The above association hypothesis denotes that there are $K=5$ targets out of which third target is undetected. First target correspond to second measurement. Second target correspond to fourth measurement. The fourth target correspond to first measurement and fifth target correspond to fifth measurement. 

\subsubsection{Conversion between $M{\rightarrow}T$ and $T{\rightarrow}M$ hypothesis}
Under the previously discussed assumptions, both representation are equivalent and carry same information. One can be uniquely converted to the other representation. The pseudo code for the conversion between $M{\rightarrow}T$ and $T{\rightarrow}M$ hypothesis are given in Table \ref{tab:T2M} and Table \ref{tab:M2T}.

\begin{table}[h] 
\caption{$M{\rightarrow}T$ to $T{\rightarrow}M$ conversion} 
\centering          
\begin{tabular}{l}
  \hline
  \begin{minipage}{4.5in}
    \vskip 4pt
$[(\mathbf{\tilde{r}}^i,M_C^i,M_T^i)]$ =
 $T{\rightarrow}M$ CONVERSION $[(\mathbf{r}^i,M_C^i,M_T^i),K,M^i]$
\begin{itemize}
\item $\mathbf{\tilde{r}}^i = zeros(1,K)$.
\item FOR $m=1:M^i$,
	\begin{itemize}
		\item IF($\mathbf{r}_m^i\neq0$)
		    \begin{itemize}
			\item $\mathbf{\tilde{r}}^i_{\mathbf{r}_m^i} = m$
		    \end{itemize}
		\item END IF
	\end{itemize}
\item END FOR
\end{itemize}
  \vskip 4pt
 \end{minipage}
 \\
  \hline
 \end{tabular}
\label{tab:T2M} 
\end{table}

\begin{table}[h] 
\caption{ $T{\rightarrow}M$ to $M{\rightarrow}T$ conversion} 
\centering          
\begin{tabular}{l}
  \hline
  \begin{minipage}{4.5in}
    \vskip 4pt
  $[(\mathbf{r}^i,M_C^i,M_T^i)]$ = 
 $M{\rightarrow}T $ CONVERSION $[(\mathbf{\tilde{r}}^i,M_C^i,M_T^i),K,M^i]$
\begin{itemize}
\item $\mathbf{r}^i = zeros(1,M^i)$.
\item FOR $k=1:K$,
	\begin{itemize}
		\item IF($\mathbf{\tilde{r}}_k^i\neq0$)
		    \begin{itemize}
			\item $\mathbf{r}^i_{\mathbf{\tilde{r}}_k^i} = k$
		    \end{itemize}
		\item END IF
	\end{itemize}
\item END FOR
\end{itemize}
  \vskip 4pt
 \end{minipage}
 \\
  \hline
 \end{tabular}
\label{tab:M2T} 
\end{table}

\textit{Example }: $[(\mathbf{r}^i=\{0,3,1,2\}),K=3]$=CONVERSION $[(\mathbf{\tilde{r}}^i=\{3,4,2\}),M^i=4]$

\subsection{Association prior}
The prior distribution of association hypothesis is assumed independent of state and past values of the association hypothesis. The prior distribution at observer $i$ can be written as
\begin{align}
p(\tilde{\lambda}^i) 
& = p(\tilde{\lambda}^i,M_C^i,M_T^i) \\
& = p(\tilde{\lambda}^i\mid M_T^i,M_C^i)p(M_T^i,M_C^i) \\
& = p(\tilde{\mathbf{r}}^i\mid M_T^i,M_C^i)p(M_T^i) p(M_C^i) 
\end{align}
The number of valid hypotheses conditional on the number of target and clutter measurements is given by
\begin{equation}
 N_{\tilde{\lambda}^i}(M_C^i,M_T^i)=  {^K\mathrm{C}_{M_T^i}}  {^{M^i}\mathrm{P}_{M_T^i}} 
\end{equation}
and follows from the number of ways of choosing $M_T^i$ targets from the $K$ targets, multiplied by the number of possible associations between $M^i$ measurements and $M_T^i$ target detections. The prior for the association vector is assumed to be uniform over all the valid hypotheses and is given by
\begin{equation}
p(\tilde{\mathbf{r}}^i\mid M_T^i,M_C^i)= {[N_{\tilde{\lambda}^i}(M_C^i,M_T^i)]}^{-1} 
\end{equation}
The clutter measurements are assumed to have Poisson distribution with mean ${\lambda}_C^i = \mu^i \tilde{V}^i$, where $\tilde{V}^i$ is the volume of space observed by the sensor and $\mu^i$ is the spatial density of clutter. The prior for the target measurements are assumed to follow binomial distribution.
\begin{eqnarray}
p(M_C^i)& = & {({\lambda}_C^i)}^{M_C^i}exp(-{\lambda}_C^i)/M_C^i!  \\
p(M_T^i)& = & \displaystyle \binom{K}{M_T^i}P_D^{M_T^i}(1-P_D)^{K-M_T^i}  
\end{eqnarray}
From an implementation point of view, a sequential factorized form of the association prior is used. It helps to calculate the association prior directly from a given target to measurement association hypothesis.
\begin{eqnarray}
 p(\tilde{\lambda}^i)=p(M_C^i)\displaystyle\prod_{k=1}^{K}p(\tilde{\mathbf{r}}_k^{i}\mid\tilde{\mathbf{r}}_{k-1}^{i})
\label{eq:association_prior1}
\end{eqnarray}
where
\begin{equation}
p(\tilde{\mathbf{r}}_k^{i}\mid\tilde{\mathbf{r}}_{k-1}^{i})\propto
\begin{cases}
1-P_D \;\qquad\text{if $j=0$,}\\
0 \;\qquad\text{if $j>0$ and $j \in \{\tilde{r}_1^i\cdot\cdot\cdot\tilde{r}_{k-1}^i\}$},\\
\frac{P_D}{M_k^i} \;\qquad \text{otherwise.}
\end{cases}
\label{eq:association_prior2}
\end{equation}

\section{Monte Carlo JPDAF}
\label{sec:MCJPDAF}
In JPDAF, the distribution of interest is the marginal filtering distribution for each of the targets rather than the joint distribution. It recursively updates the marginal filtering distribution for each of the targets using the recursive Bayesian estimator. The prediction step is done independently for each target. Due to the uncertainty in the data association, the update step can't be performed independently for individual target. Hence a soft assignment of the target to measurements is performed. 

JPDAF calculates all possible hypotheses at each time step. The infeasible hypotheses are neglected using a gating procedure to reduce computational complexity. For estimation using Kalman filter framework, JPDAF relies on linear Gaussian models for evaluation of target measurement hypotheses. Non-linear models can be accommodated by suitably linearizing using EKF. But its performance degrades as non-linearity becomes severe. MC-JPDAF implements the JPDAF using Monte Carlo technique to accommodate non-linear and non-Gaussian models. In this section, the general JPDAF framework and its Monte Carlo implementation MC-JPDAF are discussed.

\subsection{General JPDAF framework}
JPDAF is a sub-optimal method for data association in tracking multiple targets under target measurement uncertainty. It assumes independent targets. The recursive Bayesian estimation for multiple targets proceeds similar to the estimation of single target previously discussed in Table  \ref{tab:Recursive_Bayesian}. Estimation proceeds independently for individual target $k$ except the update step where the likelihood $p(\mathbf{y}_{t}|\mathbf{x}_k)$ can't be calculated independently for each target due to the target data association uncertainty.

At each time step $t$, JPDAF solves this data association problem by a soft assignment of targets to measurements according to the posterior marginal association probability $\beta_{jk}^i$,
\begin{equation}
 \beta_{jk}^i=p(\mathbf{\tilde{r}}_{k,t}^{i}=j\mid y_{1:t}) 
\end{equation}
where $\beta_{jk}$ is the posterior probability that the measurement $j$ is associated with target $k$ and $\beta_{0k}$ is the posterior probability of the target $k$ being undetected. 
JPDAF uses the posterior marginal association probability to define the likelihood of the target $k$ as 
\begin{equation}
 p_k(\mathbf{y}_{t}|\mathbf{x}_{k,t})=\displaystyle\prod_{i=1}^{N_o}\left[\beta_{0k}^i+\displaystyle\sum_{j=1}^{M^i}\beta_{jk}^ip_T^i(\mathbf{y}_{j,t,}^i\mid \mathbf{x}_{k,t})\right] 
\end{equation}
Here the likelihood of each target is assumed to be independent over the observers. The likelihood of the target with respect to a given observer is a mixture of the likelihood for the various target to measurement associations weighted by their posterior marginal association probability. The posterior marginal association probability $\beta_{jk}$ is computed by summing over all the posterior probabilities of the valid joint association hypotheses in which the same association event exists.
\begin{align}
\beta_{jk}^i & = p(\mathbf{\tilde{r}}_{k,t}^{i}=j\mid \mathbf{y}_{1:t})\\
& = \displaystyle\sum_{\{\lambda_t^i:\tilde{r}_{k,t}^i=j\}} p(\tilde{\lambda}_t^i\mid\mathbf{y}_{1:t}) 
\end{align}
The joint association probability $p(\tilde{\lambda}_t^i\mid\mathbf{y}_{1:t}^i)$ can be expressed as 
\begin{align}
 p(\tilde{\lambda}_t^i\mid\mathbf{y}_{1:t}) & = p(\tilde{\lambda}_t^i\mid\mathbf{y}_t
\mathbf{y}_{1:t-1})\\
 & = \frac{1}{c}p(\mathbf{y}_t\mid\tilde{\lambda}_t^i,\mathbf{y}_{1:t-1})p(\tilde{\lambda}_t^i\mid 
\mathbf{y}_{1:t-1})\\
& \propto p(\tilde{\lambda}_t^i) p(\mathbf{y}_t\mid\tilde{\lambda}_t^i,\mathbf{y}_{1:t-1})\\
& \propto p(\tilde{\lambda}_t^i)\displaystyle\prod_{j=1}^{M^i}p_{r_{j,t}^i}(\mathbf{y}_{j,t}^i\mid\mathbf{y}_{1:t-1})
\label{eq:JointAsscnProb1} 
\end{align} 
The clutter likelihood model for the observer is assumed to be uniform over the measurement space $V^i$, where $V^i=2\pi R_{max}^i$ and $R_{max}^i$ is the maximum range of the sensor $i$. Since there are $M_C^i$ clutter measurements, \eqref{eq:JointAsscnProb1} becomes 
\begin{equation}
  p(\tilde{\lambda}_t^i\mid\mathbf{y}_{1:t})\propto p(\tilde{\lambda}_t^i) (V^i)^{-M_C^i}\displaystyle\prod_{j\in \mathcal{I}^i}p_{r_{j,t}^i}(\mathbf{y}_{j,t}^i\mid \mathbf{y}_{1:t-1})
\label{eq:JointAsscnProb2} 
\end{equation}
where $\mathcal{I}^i=\{ j\in \{1,\ldots,M^i\}:r_j^i\neq0\}$. The number of clutter measurements in each hypothesis is calculated by converting $T{\rightarrow}M$ hypotheses to $M{\rightarrow}T$ hypotheses and finding the total number of zero entries in each. The association prior $p(\tilde{\lambda}_t^i)$ is calculated using \eqref{eq:association_prior1} and \eqref{eq:association_prior2}.
 $p_k(\mathbf{y}_{j,t}^i\mid \mathbf{y}_{1:t-1})$ is the predictive likelihood for the measurement $j$ associated with target $k$. The $M{\rightarrow}T$ hypothesis representation helps to obtain the target $r_{j,t}^i$ associated with the measurement $j$ in \eqref{eq:JointAsscnProb2}. The predictive likelihood can be calculated using the following integral.
\begin{equation}
 p_k(\mathbf{y}_{j,t}^i\mid \mathbf{y}_{1:t-1})\propto \int p_T^i(\mathbf{y}_{j,t}^i\mid \mathbf{x}_{k,t})p_k(\mathbf{x}_{k,t}\mid \mathbf{y}_{1:t-1})d\mathbf{x}_{k,t} 
\end{equation}
The recursive Bayesian estimation of the general JPDAF framework is repeated in Table \ref{tab:General JPDAF} from \cite{15}.

\begin{table}[ht] 
\caption{General JPDAF Algorithm \cite{15}} 
\centering          
\resizebox{!}{4in} {
\begin{tabular}{l}
  \hline
  \begin{minipage}{7in}
    \vskip 4pt
\begin{enumerate}
      \item Prediction step: FOR $k=1..K$, calculate the a priori pdf
	      \begin{equation}
	      p_k(\mathbf{x}_{k,t}|\mathbf{y}_{1:t-1})=\int p_k(\mathbf{x}_{k,t}|\mathbf{x}_{k,t-1})p_k(\mathbf{x}_{k,t-1}|\mathbf{y}_{1:t-1})d\mathbf{x}_{k,t-1}  
	      \end{equation}

      \item FOR $k=1..K$, calculate target likelihood by the below method:
	      \begin{itemize}
	      \item FOR $k=1..K$,$i=1..N_o$, $j=1..M^i$, calculate the predictive likelihood
		      \begin{equation}		      
		      p_k(\mathbf{y}_{j,t}^i\mid \mathbf{y}_{1:t-1}) \approx \int p_T^i(\mathbf{y}_{j,t,}^i\mid \mathbf{x}_{k,t})p_k(\mathbf{y}_{j,t}^i\mid \mathbf{y}_{1:t-1})d\mathbf{x}_{k,t}
		      \label{eq:pred_like}
		      \end{equation}

	      \item FOR observer $i=1..N_o$, enumerate all valid target to measurement association hypotheses $\tilde{\lambda}^i_t$. Convert $T{\rightarrow}M$ hypotheses to $M{\rightarrow}T$ hypotheses and calculate the number of clutter measurements $M_C^i$ in each hypothesis.

	      \item FOR observer $i=1..N_o$, calculate association prior of all hypotheses.
		      \begin{equation}
		      p(\tilde{\mathbf{r}}_k^{i}\mid\tilde{\mathbf{r}}_{k-1}^{i})\propto
		      \begin{cases}
		      1-P_D \;\qquad\text{if $j=0$}\\
		      0 \;\qquad\text{if $j>0$ and $j \in \{\tilde{r}_1^i\cdot\cdot\cdot\tilde{r}_{k-1}^i\}$}\\
		      \frac{P_D}{M_k^i} \;\qquad \text{otherwise}
		      \end{cases} 
		      \end{equation}
		      \begin{eqnarray}		      p(\tilde{\lambda}^i)=p(M_C^i)\displaystyle\prod_{k=1}^{K}p(\tilde{\mathbf{r}}_k^{i}\mid\tilde{\mathbf{r}}_{k-1}^{i}) 
		      \end{eqnarray}

	      \item FOR $i=1..N_o$, compute joint association posterior probability and normalize it at each observer $i$.
		      \begin{eqnarray}
			p(\tilde{\lambda}_t^i\mid\mathbf{y}_{1:t}) & \propto p(\tilde{\lambda}_t^i) (V^i)^{-M_C^i}\displaystyle\prod_{j\in \mathcal{I}^i}p_{r_{j,t}^i}(\mathbf{y}_{j,t}^i\mid \mathbf{y}_{1:t-1}) 
		      \end{eqnarray}
		      \begin{eqnarray}
			\displaystyle\sum p(\tilde{\lambda}_t^i\mid\mathbf{y}_{1:t}) & = 1 \qquad \text{at each observer $i$.}  
		      \end{eqnarray}

	      \item FOR $k=1..K$, $i=1..N_o$, $j=0..M^i$, calculate the marginal association posterior probability
	      \begin{equation}
	      \beta_{jk}^i = \displaystyle\sum_{\{\lambda_t^i:\tilde{r}_{k,t}^i=j\}} p(\tilde{\lambda}_t^i\mid\mathbf{y}_{1:t}) 
	      \end{equation}

	      \item FOR $k=1..K$, compute target likelihood.
		      \begin{equation}
		      p_k(\mathbf{y}_{t}|\mathbf{x}_{k,t})=\displaystyle\prod_{i=1}^{N_o}\left[\beta_{0k}^i+\displaystyle\sum_{j=1}^{M^i}\beta_{jk}^ip_T^i(\mathbf{y}_{j,t,}^i\mid \mathbf{x}_{k,t})\right] 
		      \end{equation}
	      \end{itemize}

     \item Update step: FOR $k=1..K$ calculate the posterior pdf.
	  \begin{equation}
	  p_k(\mathbf{x}_{k,t}|\mathbf{y}_{1:t})=\dfrac{p_k(\mathbf{y}_{t}|\mathbf{x}_{k,t})p_k(\mathbf{x}_{k,t}|\mathbf{y}_{1:t-1}))}{p_k(\mathbf{y}_{t}|\mathbf{y}_{1:t-1})}	
	  \end{equation}

\end{enumerate}
 \vskip 4pt
 \end{minipage}
 \\ 
  \hline
 \end{tabular}
}
\label{tab:General JPDAF} 
\end{table}

\subsection{Monte Carlo implementation of JPDAF}
Similar to JPDAF, the distributions of interest in MC-JPDAF are the marginal distribution for each of the targets. MC-JPDAF implements the general JPDAF in the particle filter approach. It approximates the marginal filtering distribution of each target using particles. The target $k$ is represented using $N$ samples, $\{\mathbf{x}_{k,t}^{(n)},w_{k,t}^{(n)}\}_{n=1}^N$. The recursive Bayesian estimation in JPDAF for each target $k$ is implemented using the sequential importance sampling used in the standard particle filter. The new samples at every time step is obtained using the proposal distribution,
\begin{equation}
 \mathbf{x}_{k,t}^{(n)}\sim q_k(\mathbf{x}_{k,t}\mid \mathbf{x}_{k,t-1}^{(n)},\mathbf{y}_t)
\end{equation}
The importance weights $w_{k,t}^{(n)}$ are obtained for each target $k$ recursively using the sequential importance sampling, similar to \eqref{eq:imp_weights}.
\begin{equation}
 w_{k,t}^{(n)}\varpropto w_{k,t-1}^{(n)}\dfrac{p_k(\mathbf{y}_t|\mathbf{x}_{k,t}^{(n)})p_k(\mathbf{x}_{k,t}^{(n)}|\mathbf{x}_{k,t-1}^{(n)})}{q(\mathbf{x}_{k,t}^{(n)}|\mathbf{x}_{k,t-1}^{(n)},\mathbf{y}_{t})}  ; \;\qquad\displaystyle \sum_{n=1}^N w_{k,t}^{(n)}=1 
\end{equation}
The target likelihood $p_k(\mathbf{y}_{t}\mid\mathbf{x}_{k,t}^{(n)})$ is calculated using the algorithm described in Table \ref{tab:General JPDAF}. The integral 
of equation \eqref{eq:pred_like} is also implemented using sequential importance sampling.
\begin{align}
p_k(\mathbf{y}_{j,t}^i\mid \mathbf{y}_{1:t-1}) & = \int p_T^i(\mathbf{y}_{j,t,}^i\mid \mathbf{x}_{k,t})p_k(\mathbf{x}_{k,t}\mid \mathbf{y}_{1:t-1})d\mathbf{x}_{k,t}\\
%
& = \int p_T^i(\mathbf{y}_{j,t}^i\mid \mathbf{x}_{k,t})\frac{p_k(\mathbf{x}_{k,t}\mid \mathbf{y}_{1:t-1})}{q_k(\mathbf{x}_{k,t}\mid\mathbf{x}_{k,t-1},\mathbf{y}_t)}q_k(\mathbf{x}_{k,t}\mid\mathbf{x}_{k,t-1},\mathbf{y}_t)d\mathbf{x}_{k,t}
\label{eq:pred_like_eval}
\end{align}
The above integral is similar to equation \eqref{eq:PF_integral}. The Monte Carlo estimate of it can be obtained by generating $N$ samples $\{\mathbf{x}_{k,t}^{(n)}\}_{n=1}^N$ from the the proposal distribution $q_k(\mathbf{x}_k\mid\mathbf{x}_{k,t-1},\mathbf{y}_t)$, calculating the summation and normalizing the weights.
\begin{align}
p_k(\mathbf{y}_{j,t}^i\mid \mathbf{y}_{1:t-1})
& \propto \displaystyle \sum_{i=1}^N \alpha_{k,t}^{(n)} p_T^i(\mathbf{y}_{j,t}^i\mid \mathbf{x}_{k,t}^{(n)}) 
\end{align}
where
\begin{equation}
\alpha_{k,t}^{(n)}\propto w_{k,t-1}^{(n)}\frac{p_k(\mathbf{x}_{k,t}^{(n)}\mid\mathbf{x}_{k,t-1}^{(n)})}{q_k(\mathbf{x}_{k,t}^{(n)}\mid\mathbf{x}_{k,t-1}^{(n)},\mathbf{y}_t)}  ; \;\qquad \displaystyle \sum_{n=1}^N \alpha_{k,t}^{(n)}=1 
\end{equation}

\subsection{Gating of hypotheses}
The number of hypotheses at a given observer $i$ is given by 
\begin{align}
 N_{\tilde{\lambda}}& = \displaystyle\sum_{M_T=0}^{min(K,M^i)} N_{\tilde{\lambda}}(M_C,M_T) \\
& = \displaystyle{\sum_{M_T=0}^{min(K,M^i)}} {^K\mathrm{C}_{M_T^i}} {^{M^i}\mathrm{P}_{M_T^i}}
\end{align}
The number of hypotheses increases exponentially with increasing number of targets $K$, and number of measurements $M^i$. This increases computational complexity and is almost infeasible for practical scenarios. Hence gating is used to reduce the number of hypotheses to a feasible level. A validation region is calculated for each target $k$ using the available information. All the measurement which fall inside the validation region are considered to be possible measurements and the measurements which fall outside the validation region are considered to be impossible measurements for the target $k$. The hypotheses containing impossible target measurements are ignored. Thus the number of valid hypotheses gets reduced. 

\begin{figure}[t!]
\centering
{\includegraphics[scale=0.5]{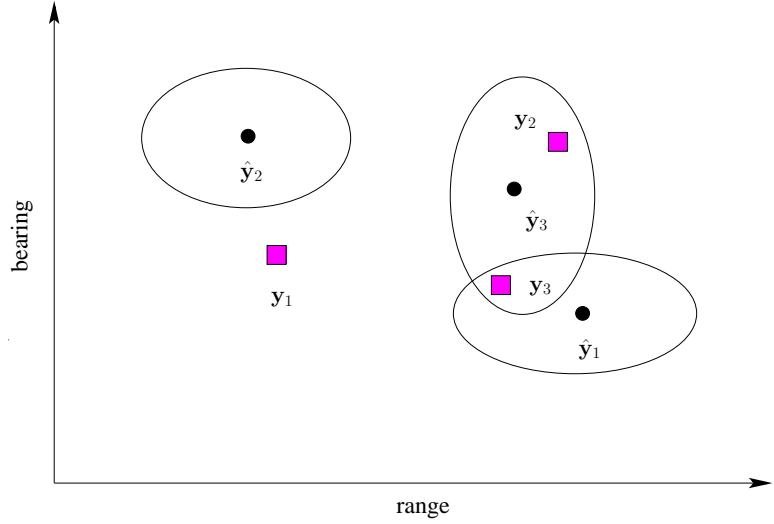}}    
\caption{Gating of measurement: Targets are shown in their measurement space using circles. The ellipses indicate their validation region. The measurements are shown in squares.}
\label{gating}
\end{figure}

Suppose $\hat{\mathbf{y}}_k=g(\mathbf{x}_k,\mathbf{p_0})$ is the measurement of the target $k$, then for Gaussian assumption of likelihood model, the Monte Carlo approximation of the predictive likelihood can be expressed as
\begin{align}
 p_k(\mathbf{y}\mid \mathbf{y}_{1:t-1}) & \approx \displaystyle{\sum_{n=1}^N}\mathcal{N}(\mathbf{y}\mid \hat{\mathbf{y}}_k^{(n)},\Sigma_{\mathbf{y}})\\
& \approx \mathcal{N}(\mu_{\hat{\mathbf{y}}_k},\Sigma_{\hat{\mathbf{y}}_k}) 
\label{eq:appr_pr_lh}
\end{align}
where 
\begin{eqnarray}
\mu_{\hat{\mathbf{y}}_k} & = & \displaystyle{\sum_{n=1}^N}\alpha_k^{(n)}\mathbf{g}(\mathbf{x}_k^{(n)},\mathbf{p_0}) \\
\Sigma_{\hat{\mathbf{y}}_k} & = & \Sigma_{\mathbf{y}}+\displaystyle{\sum_{n=1}^N}\alpha_k^{(n)}[\mathbf{g}(\mathbf{x}_k^{(n)},\mathbf{p_0})-\mu_{\hat{\mathbf{y}}_k}][\mathbf{g}(\mathbf{x}_k^{(n)},\mathbf{p_0})-\mu_{\hat{\mathbf{y}}_k}]^T 
\end{eqnarray}
Given a measurement $\mathbf{y}_j$, the squared distance of the measurement with respect to the predicted measurement of the target $k$ can be calculated as
\begin{equation}
 d_k^2(\mathbf{y}_j)=(\mathbf{y}_j-\mu_{\hat{\mathbf{y}}_k})^T \Sigma_{\hat{\mathbf{y}}_k}^{-1}(\mathbf{y}_j-\mu_{\hat{\mathbf{y}}_k}) 
\end{equation}
The set of validated measurements for target $k$ are those such that 
\begin{equation}
    \mathcal{Y}_k=\{\mathbf{y}_j:d_k^2(\mathbf{y}_j)\leq\varepsilon\}
\end{equation}
where $\varepsilon$ is the parameter which decides the volume of validation region. $d_k^2$ is chi square distributed approximately with degrees of freedom equal to the dimension of $\mathbf{y}_j$. Chi-square hypothesis testing is performed on the proposed target-measurement association hypotheses. A hypothesis is accepted if its chi-square statistics $d_k^2$ satisfies the relation $d_k^2<\chi^2_\alpha$ to obtain the set of gated hypotheses $\tilde{\Lambda}_t^i$ at each observer $i$. The gating reduces the number of hypotheses to a feasible level. For example, if we consider the situation in Fig.\ref{gating}, where there are three targets and three measurements, an exhaustive enumeration will result in 34 hypotheses as explained in Table.\ref{tab:all_hyp}. After gating, the number of hypotheses reduces to 5 as shown in Table \ref{tab:gated_hyp}. The summary of MC-JPDAF with gating is discussed in Table \ref{tab:MCJPDAF}.

\begin{table}[H] 
\caption{Enumeration of hypotheses for $K=3,$ $M^i=3$}  
\centering            
\begin{tabular}{c c c c c}      
\hline\hline                           

 Cases & \multicolumn{3}{c}{Hypotheses} & No. of hypotheses\\
   
& $\tilde{r}_1$& $\tilde{r}_2$& $\tilde{r}_3$& \\
\hline \hline                     
 
 $M_T=3$, $M_C=0$& $M_p$ & $M_q$& $M_r$&6  \\[1ex]

 $M_T=2$, $M_C=1$& $M_p$ & $M_q$& $0$&6  \\[-1ex] 
& $M_p$ & $0$& $M_r$&6  \\[-1ex] 
& $0$ & $M_q$& $M_r$&6  \\[1ex] 

 $M_T=1$, $M_C=2$& $M_p$ & $0$& $0$&3  \\[-1ex] 
& $0$ & $M_q$& $0$&3  \\[-1ex] 
& $0$ & $0$& $M_r$&3  \\[1ex]

 $M_T=0$, $M_C=3$& $0$ & $0$& $0$&1  \\[1ex] 

\hline
&&&&Total=34\\[1ex]
\hline\hline

\hline                          
\end{tabular} 
\label{tab:all_hyp} 
\end{table}

\begin{table}[H]
\caption{Hypotheses after gating}
\centering            
\begin{tabular}{ccc}
\hline
$\tilde{r}_1$ & $\tilde{r}_2$& $\tilde{r}_3$ \\
\hline
0&0&0\\
0&0&2\\
0&0&3\\
3&0&0\\
3&0&2\\
\hline
\end{tabular}
\label{tab:gated_hyp}
\end{table}

\begin{table}[ht] 
\caption{Monte Carlo JPDAF Algorithm \cite{15}} 
\centering          
\resizebox{!}{4in} {
\begin{tabular}{l}
  \hline
  \begin{minipage}{7in}
    \vskip 4pt
\begin{itemize}
      \item Prediction step: FOR $k=1..K$, $n=1:N$, draw samples 
	      \begin{equation}
	      \mathbf{x}_{k,t}^{(n)}\sim q_k(\mathbf{x}_{k,t}\mid \mathbf{x}_{k,t-1}^{(n)},\mathbf{y}_t)
	      \end{equation}

		    \item Evaluate the predictive weights upto a normalizing constant
			    \begin{equation}
			    \alpha_{k,t}^{(n)}\propto w_{k,t-1}^{(n)}\frac{p_k(\mathbf{y}_t\mid \mathbf{x}_{k,t}^{(n)})p_k(\mathbf{x}_{k,t}^{(n)}\mid\mathbf{x}_{k,t-1}^{(n)})}{q_k(\mathbf{x}_{k,t}^{(n)}\mid\mathbf{x}_{k,t-1}^{(n)},\mathbf{y}_t)} 
			    \end{equation}
		    \item Normalize the predictive weights
			    \begin{equation}
				\displaystyle \sum_{n=1}^N \alpha_{k,t}^{(n)}=1
			    \end{equation}

		    \item FOR $k=1..K$, $i=1..N_o$, $j=1..M^i$, calculate the predictive likelihood
			    \begin{equation}
				  p_k(\mathbf{y}_{j,t}^i\mid \mathbf{y}_{1:t-1}) \approx \displaystyle \sum_{i=1}^N \alpha_{k,t}^{(n)} p_T^i(\mathbf{y}_{j,t}^i\mid \mathbf{x}_{k,t}^{(n)}) 
			    \end{equation}

		    \item FOR observer $i=1..N_o$, enumerate all valid target to measurement association hypotheses $\tilde{\lambda}^i_t$.

		    \item Perform gating on the valid target to measurement hypotheses by the following procedure:
			  \begin{itemize}
				\item For $k=1..K$, calculate the approximation for the predictive likelihood of target $k$ using \eqref{eq:appr_pr_lh}
				      \begin{eqnarray}
				      p_k(\mathbf{y}\mid \mathbf{y}_{1:t-1}) & \approx & \mathcal{N}(\mu_{\hat{\mathbf{y}}_k},\Sigma_{\hat{\mathbf{y}}_k}) \\ 
				      \mu_{\hat{\mathbf{y}}_k} & = & \displaystyle{\sum_{n=1}^N}\alpha_k^{(n)}\mathbf{g}(\mathbf{x}_k^{(n)},\mathbf{p_0}) \\
				      \Sigma_{\hat{\mathbf{y}}_k} & = & \Sigma_{\mathbf{y}}+\displaystyle{\sum_{n=1}^N}\alpha_k^{(n)}[\mathbf{g}(\mathbf{x}_k^{(n)},\mathbf{p_0})-\mu_{\hat{\mathbf{y}}_k}][\mathbf{g}(\mathbf{x}_k^{(n)},\mathbf{p_0})-\mu_{\hat{\mathbf{y}}_k}]^T 
				      \end{eqnarray}
				\item For $k=1..K$, $i=1..N_o$, $j=1..M^i$, calculate the squared distance $d_k^i(\mathbf{y}_j)$ between the predicted and observed measurements using measurement innovations.
				      \begin{equation}
				      d_k^2(\mathbf{y}_j)=(\mathbf{y}_j-\mu_{\hat{\mathbf{y}}_k})^T \Sigma_{\hat{\mathbf{y}}_k}^{-1}(\mathbf{y}_j-\mu_{\hat{\mathbf{y}}_k})
				      \end{equation}
				\item Perform chi-square hypothesis testing on the proposed target-measurement association hypotheses. Accept a hypothesis if its chi-square statistics $d_k^2$ satisfies the relation $d_k^2<\chi^2_\alpha$	
				to obtain the set of gated hypotheses $\tilde{\Lambda}_t^i$ at each observer $i$.

			  \end{itemize}
\end{itemize}
  \vskip 4pt
 \end{minipage}
 \\
  \hline
 \end{tabular}
}
\label{tab:MCJPDAF} 
\end{table}

\begin{table}[ht] 
\caption{Monte Carlo JPDAF Algorithm (contd..)\cite{15}} 
\centering          
\resizebox{!}{4in} {
\begin{tabular}{l}
  \hline
  \begin{minipage}{7in}
    \vskip 4pt
\begin{itemize}

		  \item Convert $T{\rightarrow}M$ hypotheses to $M{\rightarrow}T$ hypotheses and calculate the number of clutter measurements $M_C^i$ in each hypothesis. 

		  \item FOR observer $i=1..N_o$, calculate association prior of all hypotheses.
			  \begin{equation}
			  p(\tilde{\mathbf{r}}_k^{i}\mid\tilde{\mathbf{r}}_{k-1}^{i})\propto
			  \begin{cases}
			  1-P_D \;\qquad\text{if $j=0$} \\
			  0 \;\qquad\text{if $j>0$ and $j \in \{\tilde{r}_1^i\cdot\cdot\cdot\tilde{r}_{k-1}^i\}$} \\
			  \frac{P_D}{M_k^i} \;\qquad \text{otherwise}
			  \end{cases} 
			  \end{equation}
			  \begin{eqnarray}		      p(\tilde{\lambda}^i)=p(M_C^i)\displaystyle\prod_{k=1}^{K}p(\tilde{\mathbf{r}}_k^{i}\mid\tilde{\mathbf{r}}_{k-1}^{i}) 
			  \end{eqnarray}
		  \item FOR $i=1..N_o$, compute joint association posterior probability and normalize it at each observer $i$.
			  \begin{eqnarray}
			    p(\tilde{\lambda}_t^i\mid\mathbf{y}_{1:t}) & \propto p(\tilde{\lambda}_t^i) (V^i)^{-M_C^i}\displaystyle\prod_{j\in \mathcal{I}^i}p_{r_{j,t}^i}(\mathbf{y}_{j,t}^i\mid \mathbf{y}_{1:t-1}) 
			  \end{eqnarray}
			  \begin{eqnarray}
			    \displaystyle\sum p(\tilde{\lambda}_t^i\mid\mathbf{y}_{1:t}) & = 1 \qquad \text{at each observer $i$.}
			  \end{eqnarray}
		  \item FOR $k=1..K$, $i=1..N_o$, $j=0..M^i$, calculate the marginal association posterior probability
		  \begin{equation}
		  \beta_{jk}^i = \displaystyle\sum_{\{\tilde{\lambda}_t^i\in\tilde{\Lambda}_t^i:\tilde{r}_{k,t}^i=j\}} p(\tilde{\lambda}_t^i\mid\mathbf{y}_{1:t}) 
		  \end{equation}
		  \item FOR $k=1..K$, compute target likelihood.
			  \begin{equation}		      p_k(\mathbf{y}_{t}\mid\mathbf{x}_{k,t}^{(n)})=\displaystyle\prod_{i=1}^{N_o}\left[\beta_{0k}^i+\displaystyle\sum_{j=1}^{M^i}\beta_{jk}^ip_T^i(\mathbf{y}_{j,t,}^i\mid \mathbf{x}_{k,t}^{(n)})\right]
			  \end{equation}

     \item Update step: FOR $k=1..K$, $n=1..N$, calculate and normalize particle weights.
	      \begin{equation}
	      w_{k,t}^{(n)}\varpropto w_{k,t-1}^{(n)}\dfrac{p_k(\mathbf{y}_t|\mathbf{x}_{k,t}^{(n)})p_k(\mathbf{x}_{k,t}^{(n)}|\mathbf{x}_{k,t-1}^{(n)})}{q(\mathbf{x}_{k,t}^{(n)}|\mathbf{x}_{k,t-1}^{(n)},\mathbf{y}_{t})}  ; \;\qquad\displaystyle \sum_{n=1}^N w_{k,t}^{(n)}=1
	      \end{equation}

     \item FOR $k=1..K$, if required, resample the particles $\{\mathbf{x}_{k,t}^{(n)}\}_{n=1}^N$ and do roughening.

\end{itemize}
  \vskip 4pt
 \end{minipage}
 \\
  \hline
 \end{tabular}
}
\label{tab:MCJPDAF_cntd} 
\end{table}

\section{Simulation Results}
To verify the effectiveness of the algorithm, targets' motion scenario and their measurements are simulated according to the given models and the estimates obtained using the algorithm is compared with the true trajectories. 

\subsection{Multi-target tracking using MC-JPDAF}
We have three independent targets which have nearly constant velocity motion. The state vector consists of position and velocities of the targets. The state of the $k$-th target at time $t$ is given by
\begin{eqnarray}
\mathbf{x}_{k,t}=
\begin{bmatrix}
x_{k,t} & \dot{x}_{k,t} & y_{k,t}  & \dot{y}_{k,t}
 \end{bmatrix}^T 
\end{eqnarray}
The initial true positions of the targets are $(-50,50)$, $(-50,0)$, $(-50,-50)$ in meters and their velocities are $(1,-1.5)$, $(1,0)$, $(1,0.75)$ in meters per second respectively. The targets move with near constant velocity model with $\sigma_x=\sigma_y=5\times10^{-4}$. All the targets have state transition model $F$ such that:
\begin{eqnarray}
\mathbf{x}_{k,t} & = & F(\mathbf{x}_{k,t-1})+\mathbf{w}_{k,t-1} 
\end{eqnarray}
where $\mathbf{w}_{k,t}$ is the process noise with zero mean and covariance $Q_{w,k}$. The matrices $F$ and $Q_{w,k}$ are given by,
\begin{equation}
 F=\begin{bmatrix}
1&T&0&0\\
0&1&0&0\\
0&0&1&T\\
0&0&0&1\\
\end{bmatrix}
\end{equation} 
\begin{equation} 
Q_{w,k} = \begin{bmatrix}
 \sigma_x^2(T^3)/3    &\sigma_x^2*(T^2)/2   &0 &0\\
        \sigma_x^2(T^2)/2   &\sigma_x^2T               &0 &0\\
        0 &0 &\sigma_y^2(T^3)/3    &\sigma_y^2(T^2)/2 \\
        0 &0 &\sigma_y^2(T^2)/2   &\sigma_y^2T         \\
\end{bmatrix} 
\end{equation}
where $T$ is the sampling period of the target dynamics. The measurement sensors are located at $(-45,-45)$, $(45,45)$ meters respectively. The $k$-th targets' range $r_{k}$ and bearing $\theta_{k}$ at time $t$ are available as the measurement $\mathbf{y}_{k,t}^i$ at time step of $T=1$ at each observer $i$. 
\begin{eqnarray}
\mathbf{y}_{k,t}^i=
\begin{bmatrix}
r_{k,t}\\
\theta_{k,t}\\
\end{bmatrix}
\end{eqnarray}
The errors in the range and bearing are such that $\sigma_R=5$ and $\sigma_\theta=0.05$. The maximum range detected by the sensor is $100 m$. The probability of detection of a target is $P_D=0.9$ and the clutter rate is $\lambda_C=5$. The exact association of the measurements to the targets is unknown at the observers.

\begin{figure}[h]
\centering
{\includegraphics[scale=0.5]{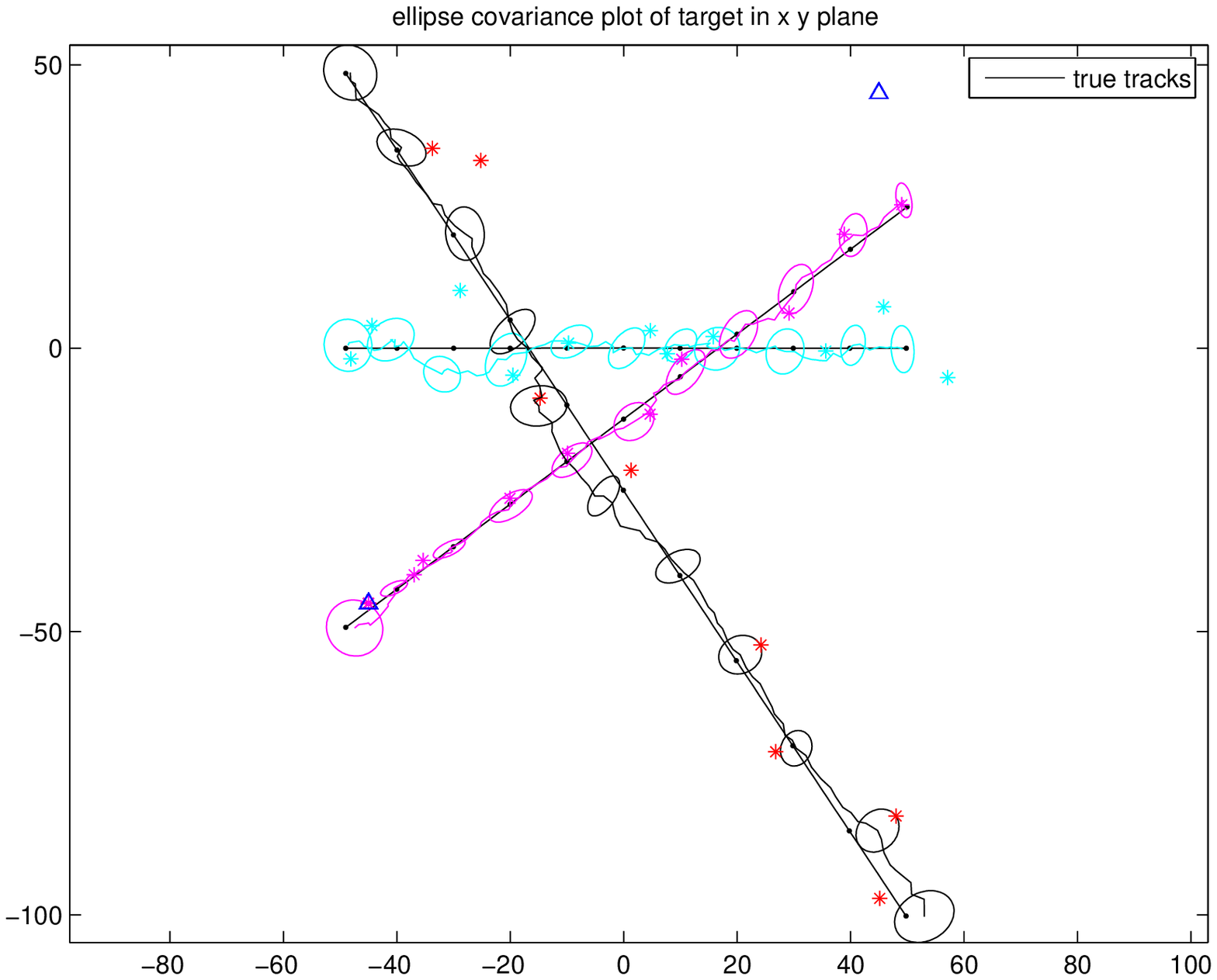}}    
\caption{Targets' true $xy$ track and their estimated track covariance obtained using MCJPDAF for a single run: The location of the sensors are shown in triangles.}
\label{MCJPDAF_cov_plot}
\end{figure}

\begin{figure}[h]
\centering
{\label{MCJPDAF_T1_MSE}\includegraphics[scale=0.5]{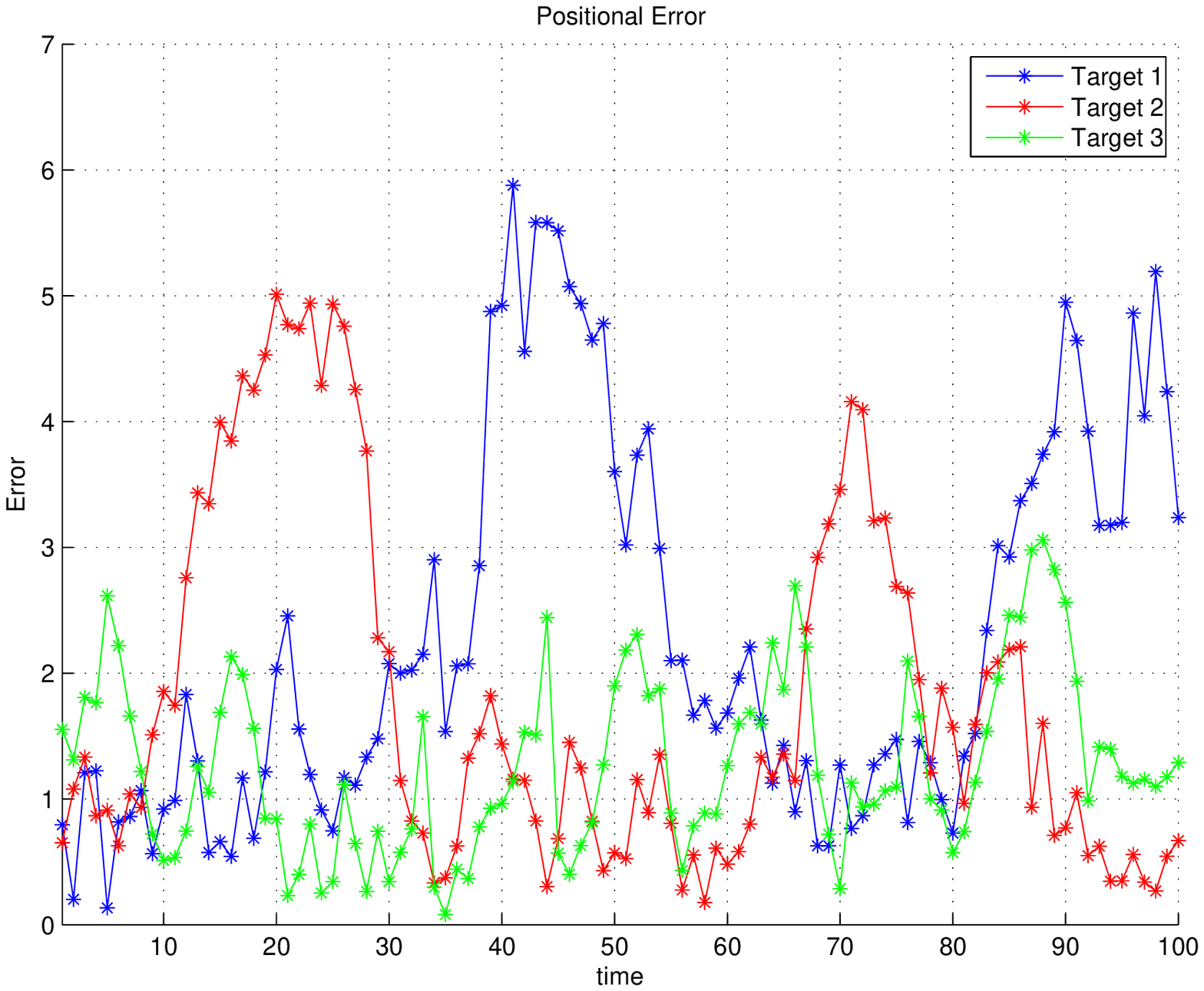}  
}
\caption{MSE of the position estimates from $20$ Monte Carlo runs, obtained using MCJPDAF.}
\label{MCJPDAF_MSE}
\end{figure}

\begin{figure}[h] 
\centering
\subfloat [position $x$ ]
{\includegraphics[scale=0.4]{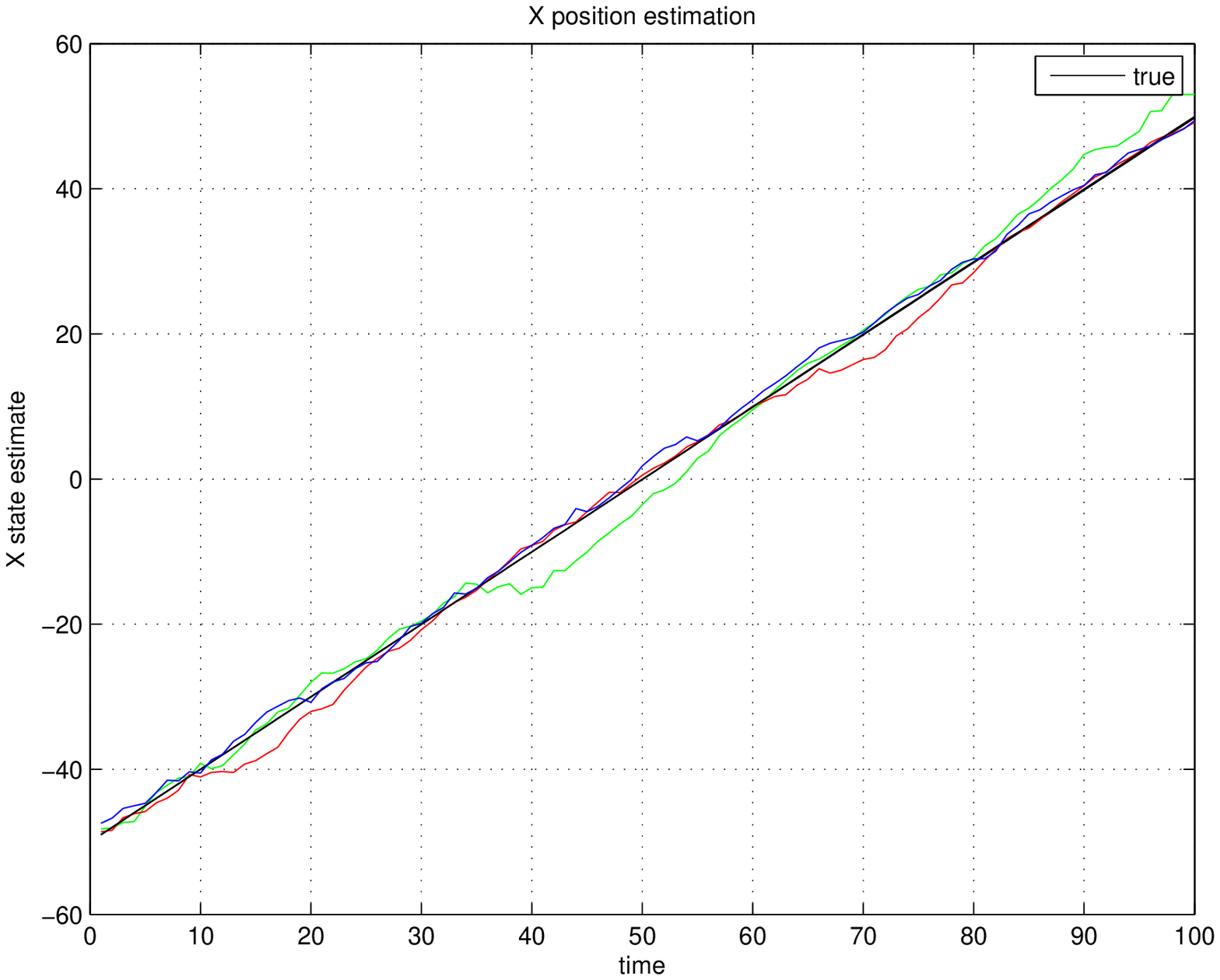}  
}\\
\subfloat [position $y$ ]
{\includegraphics[scale=0.4]{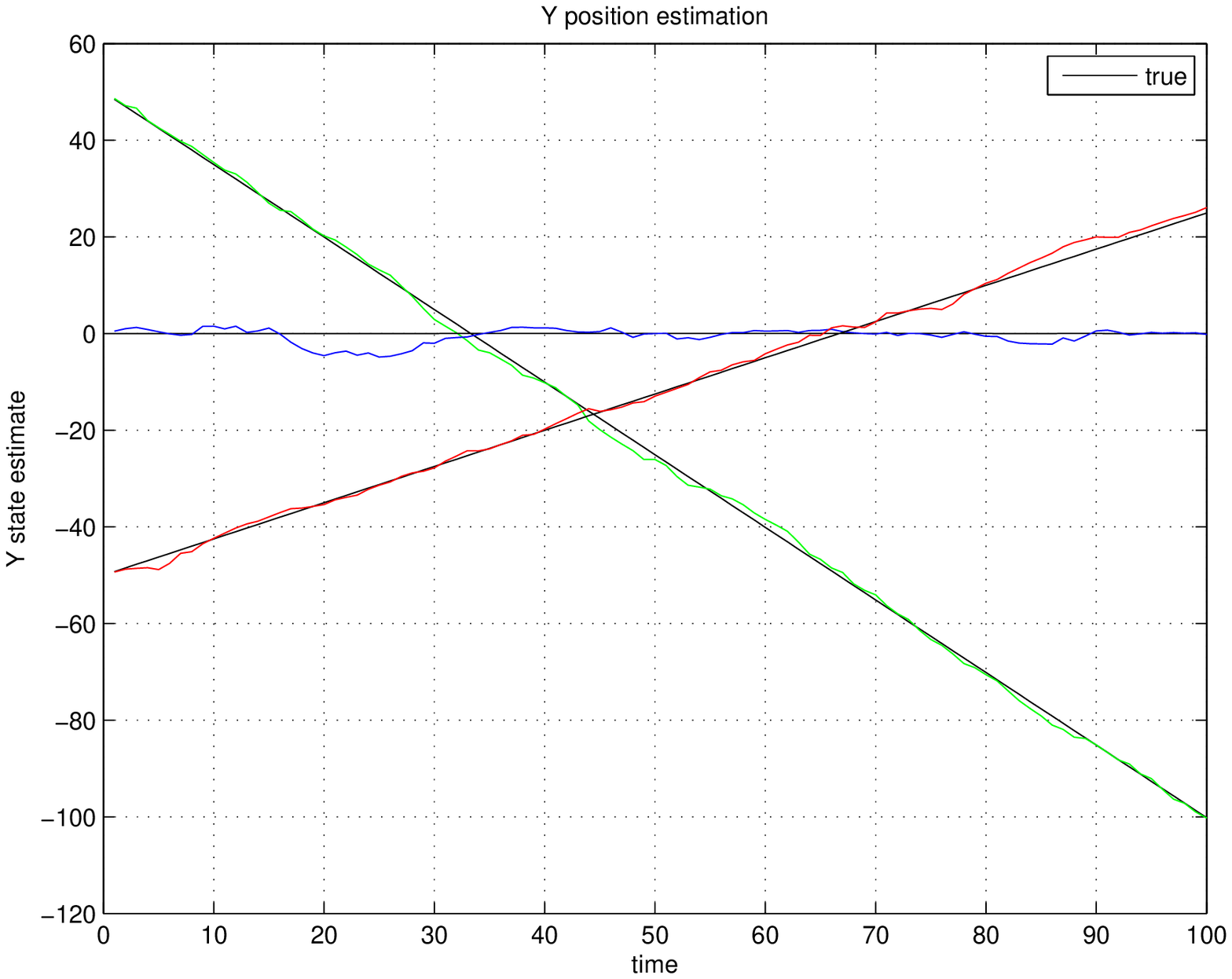}  
}\\
\subfloat [velocity $v_x$]
{\includegraphics[scale=0.4]{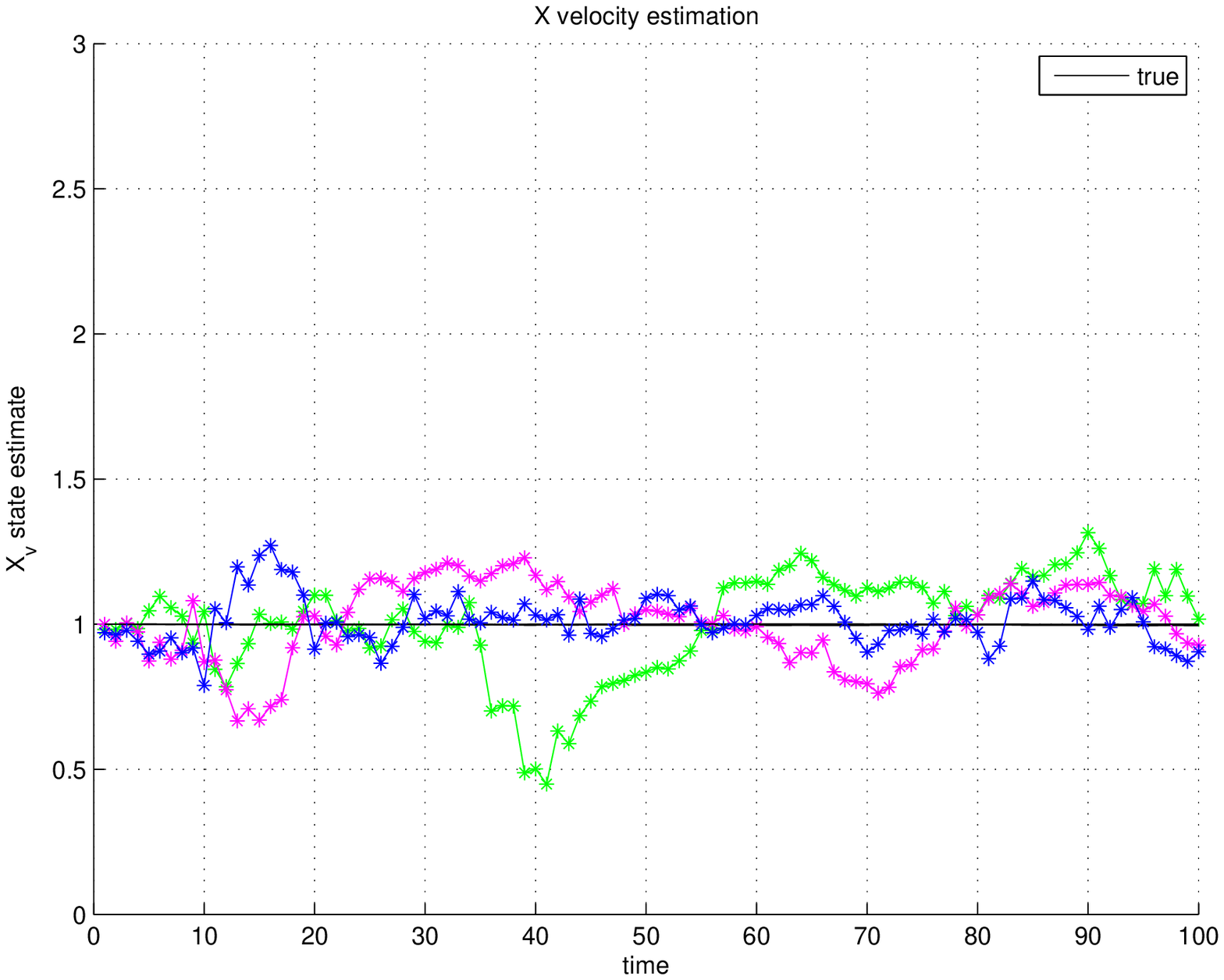}  
}%
\subfloat [velocity $v_y$]
{\includegraphics[scale=0.4]{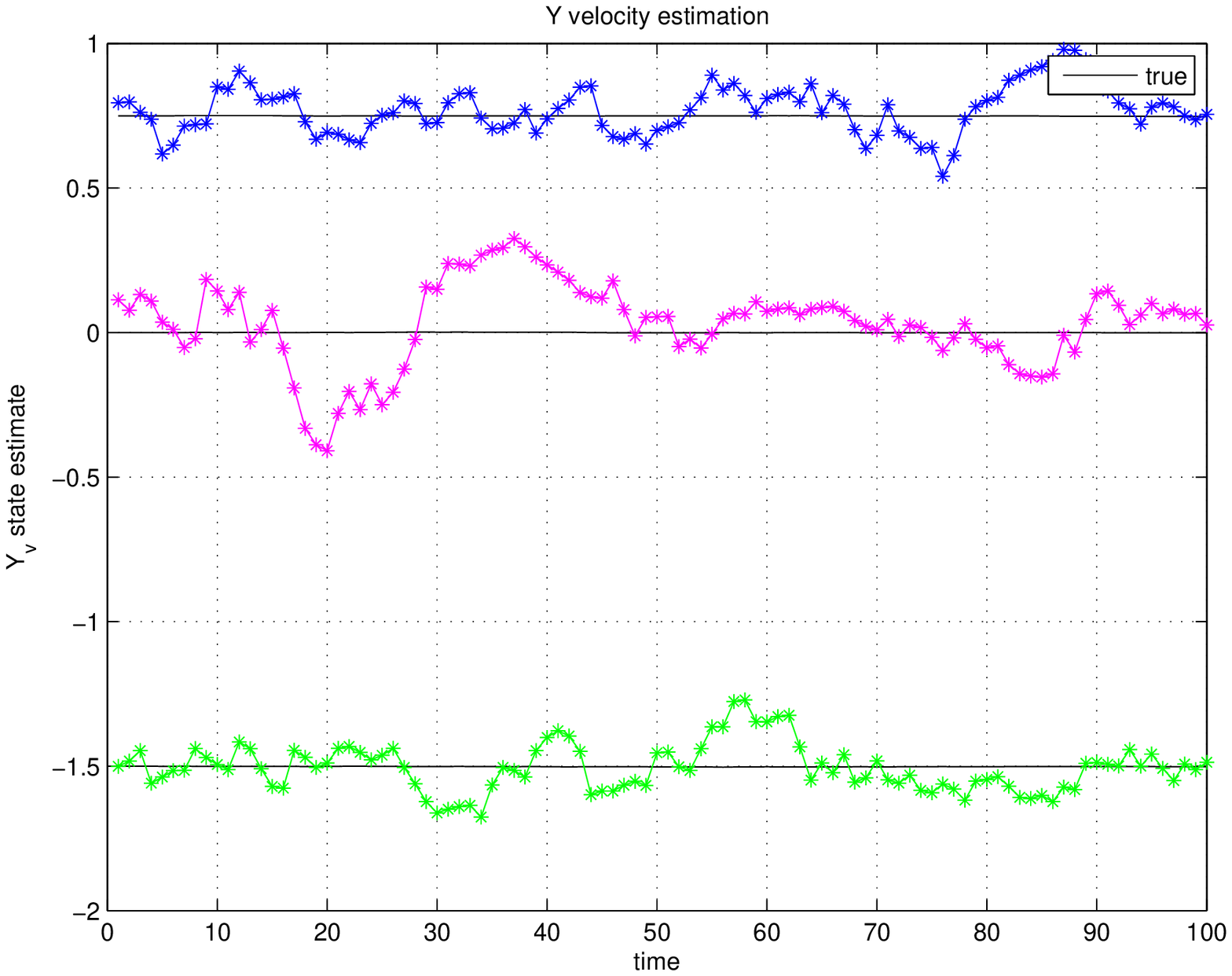}  
}\\
\caption{Targets' true states and their estimates from single run obtained using MCJPDAF.}
\label{MCJPDAF_state}
\end{figure}

The measurement model $h(\cdotp)$ for the target $k$ at the $i$-th observer is given by:
\begin{eqnarray}
\mathbf{y}_{k,t}^i=
h_k(\mathbf{x}_{k,t})^i+\mathbf{v}_{k,t}^i=
\begin{bmatrix}
\sqrt{(x_{k,t}-x_o^i)^2+(y_{k,t}-y_o^i)^2}\\
\tan^{-1}\left(\dfrac{y_{k,t}-y_o^i}{x_{k,t}-x_o^i}\right)
\end{bmatrix}
\end{eqnarray}
with $\mathbf{p}_0^i=(x_o^i,y_o^i)$. The maximum range of sensor is $R_{max}^i=100$ and the volume of measurement space is $V^i=2\pi R_{max}^i$.
The measurement error $\mathbf{v}_{k,t}^i$ is uncorrelated and has zero mean Gaussian distribution with covariance matrix $\Sigma_{\mathbf{y}_k}$. 
\begin{eqnarray}
\Sigma_{\mathbf{y}_k}=
\begin{bmatrix}
\sigma_R^{2} &0 \\
0 &\sigma_\theta^2 \\
\end{bmatrix} 
\end{eqnarray}
The measurement errors are assumed to be the same at all the observers. The initial state estimate is assumed to be a Gaussian vector with mean $ \hat{\mathbf{x}}_{k,0}=\mathbf{x}_{k,0}$ and error covariance $P_{k,0} = diag(5,0.1,5,0.1)$. Hence initial particles for each target $\{\mathbf{x}_{k,0}^{(n)}\}_{n=1}^N$ were generated based on the distribution 
\begin{equation}
 \mathbf{x}_{k,0} \sim \mathcal{N}(\mathbf{x}_{k,0},P_{k,0}) 
\end{equation}
In this implementation of the particle filter, the transitional prior which is a sub-optimal choice of importance density is  used to propose particles.
\begin{eqnarray}
 q(\mathbf{x}_{k,t}^{(n)}|\mathbf{x}_{k,t-1}^{(n)},\mathbf{y}_{t}) & = & p(\mathbf{x}_{k,t}|\mathbf{x}_{k,t-1}^{(n)})
\end{eqnarray}
The process noise used for estimation is such that $\sigma_x=\sigma_y=5\times10^{-2}$. The squared distance of the measurement with respect to the predicted measurement $d_k^2$, follows chi-square distribution with 2 degrees of freedom. The significance level used for the gating of hypotheses is $\alpha=0.01$. The chi-square critical value comes to be $\chi^2_\alpha=9.21$. A hypothesis is rejected if its chi-square statistics $d_k^2$ satisfies the relation $d_k^2>\chi^2_\alpha$. A total of $N=100$ particles were used. The simulation was carried for $20$ Monte Carlo runs and the estimates were obtained.
The true trajectories of the targets and their estimates of a single run are shown in Fig.\ref{MCJPDAF_cov_plot}. The ellipses indicate the 2-$\sigma$ region of the estimate covariances. The state estimates of the targets are shown in Fig.\ref{MCJPDAF_state}. The mean square error (MSE) of the position estimates are shown in Fig.\ref{MCJPDAF_MSE}. The results show that MCJPDAF handles data association uncertainty efficiently. It had good track of the target states in all the Monte Carlo runs and there were no diverged track estimates. The missing measurements and clutters didn't have any significant effect in the estimates. 


\begin{figure}[h]
\centering
\subfloat [Sensor 1]
{\includegraphics[scale=0.4]{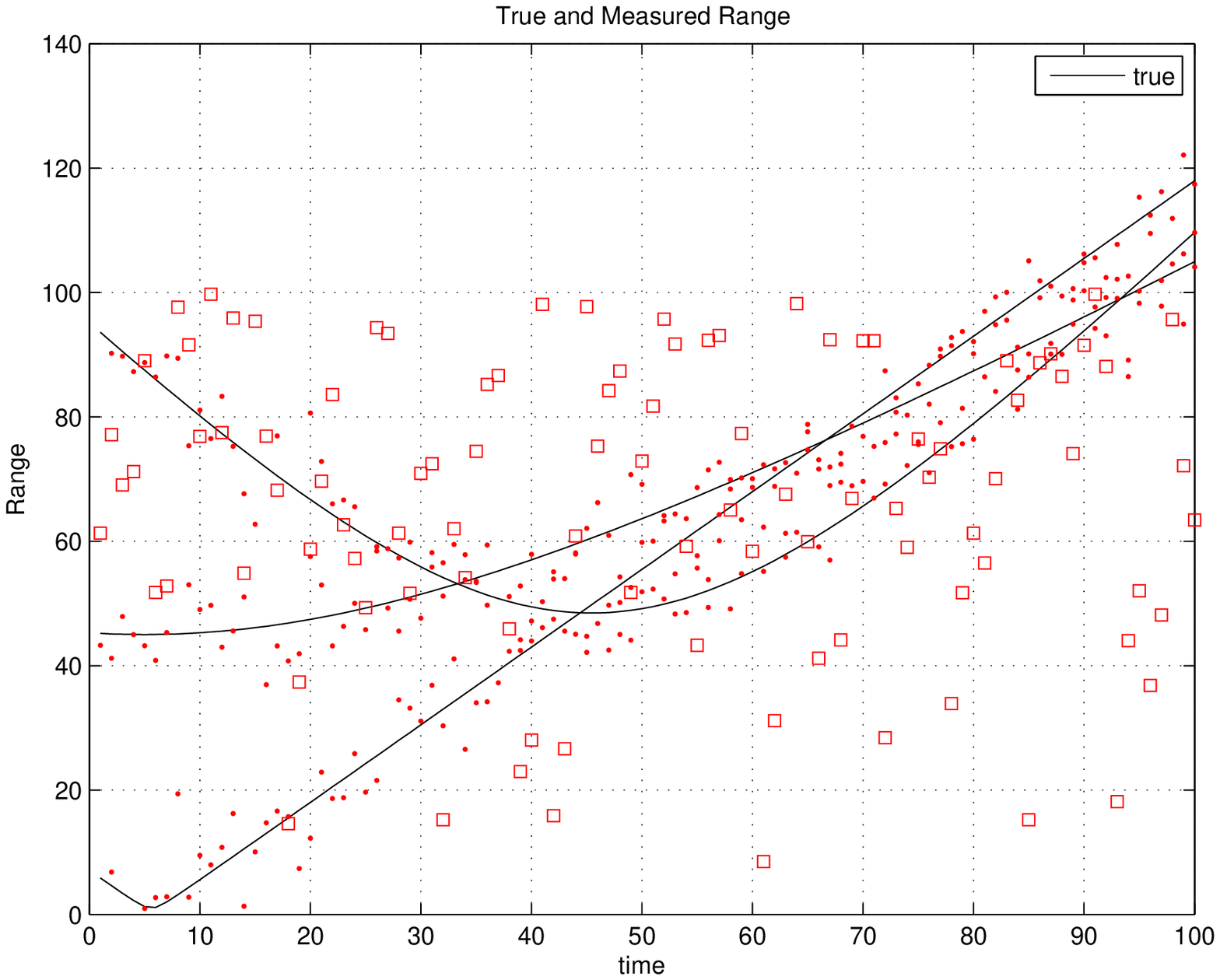}  
} 
\subfloat [Sensor 2]
{\includegraphics[scale=0.4]{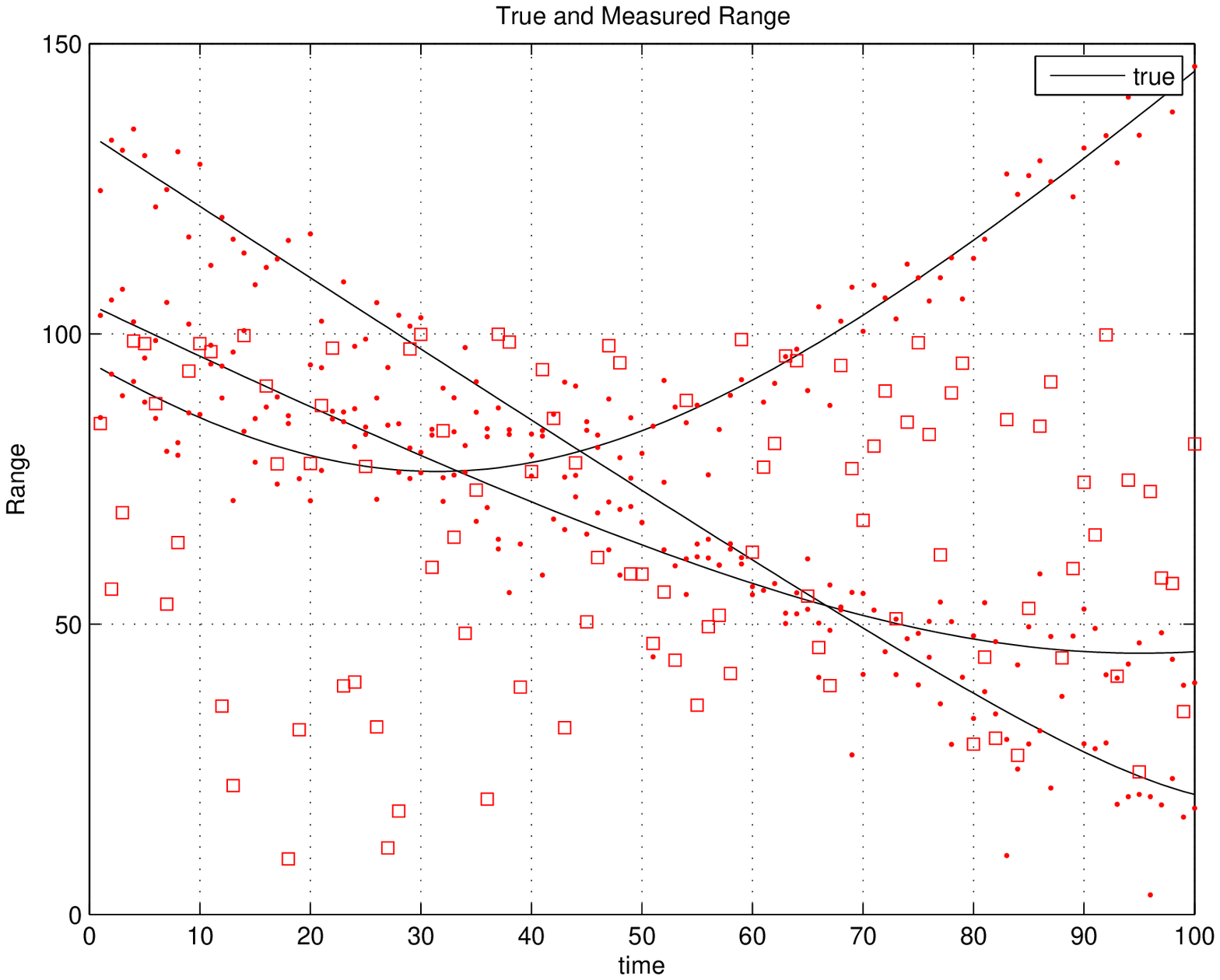}  
}\\   
\caption{Targets' true range and their measurements: The target measurements are shown in dots and the clutter measurements are shown in squares.}
\label{MCJPDAF_range_meas}
\end{figure}

\begin{figure}[h]
\centering
\subfloat [Sensor 1]
{\includegraphics[scale=0.4]{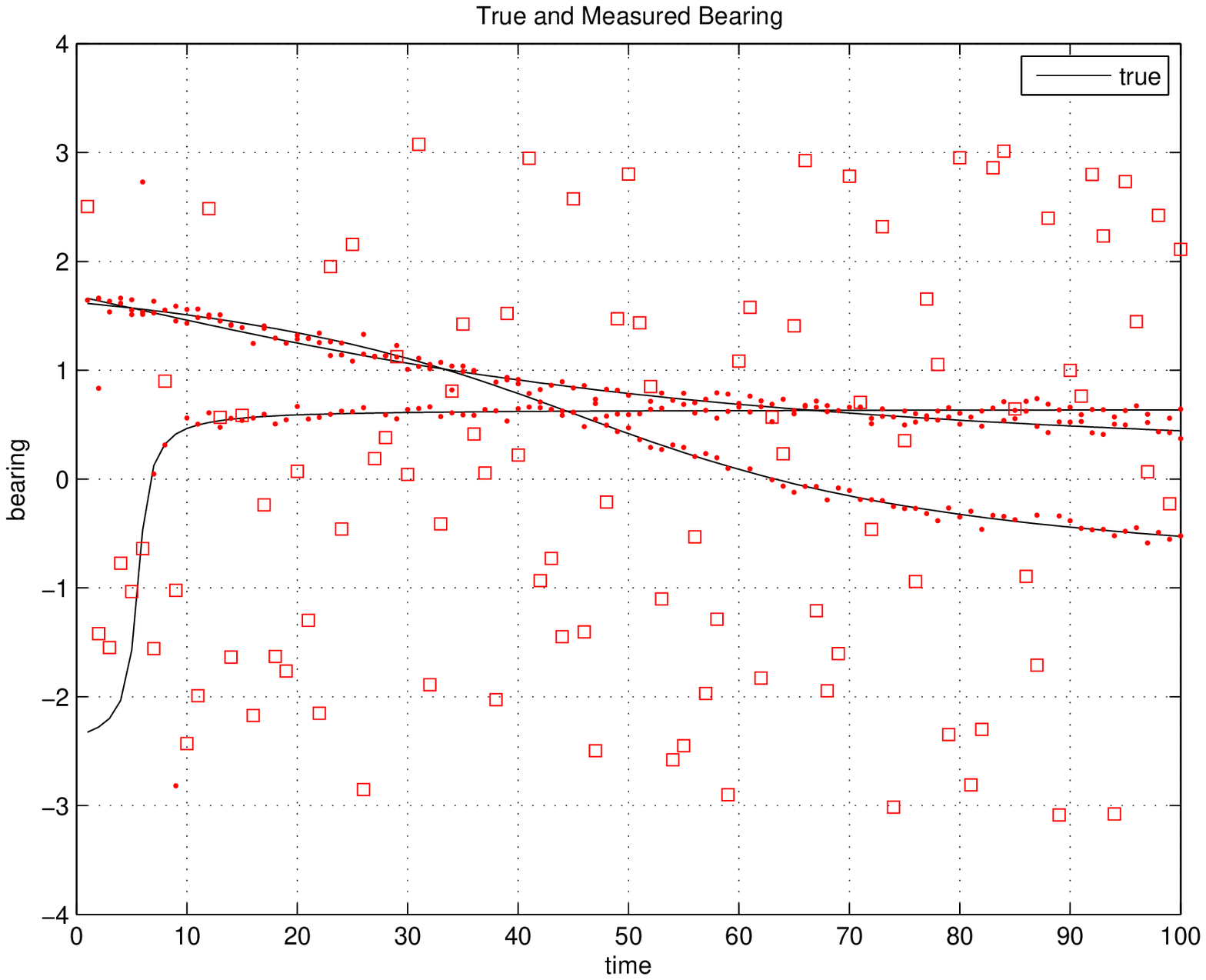}  
} 
\subfloat [Sensor 2]
{\includegraphics[scale=0.4]{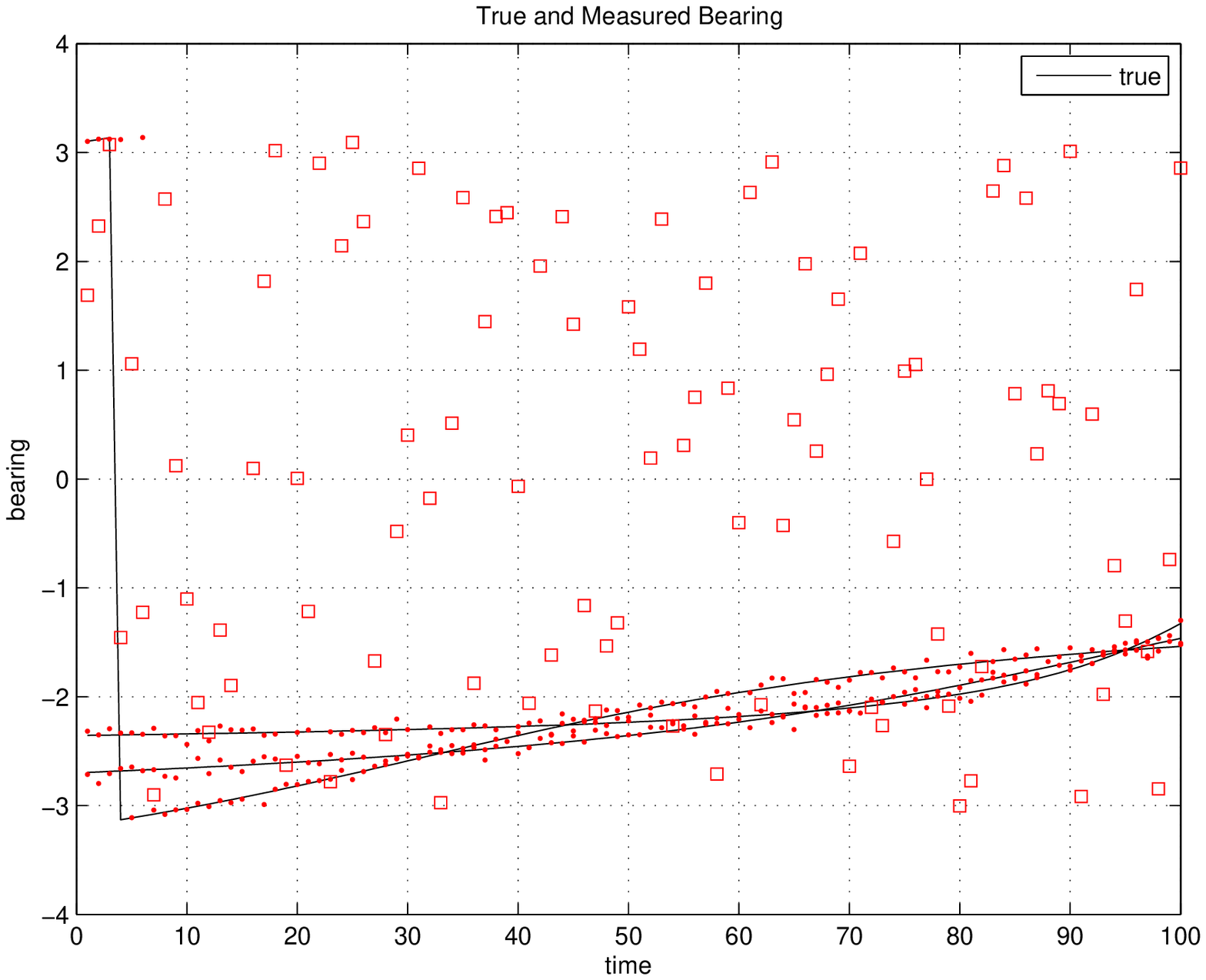}  
}\\   
\caption{Targets' true bearing and their measurements: The target measurements are shown in dots and the clutter measurements are shown in squares.}
\label{MCJPDAF_bearing_meas}
\end{figure}

 \begin{figure}[p]
 \centering 
   \includegraphics[scale=0.7]{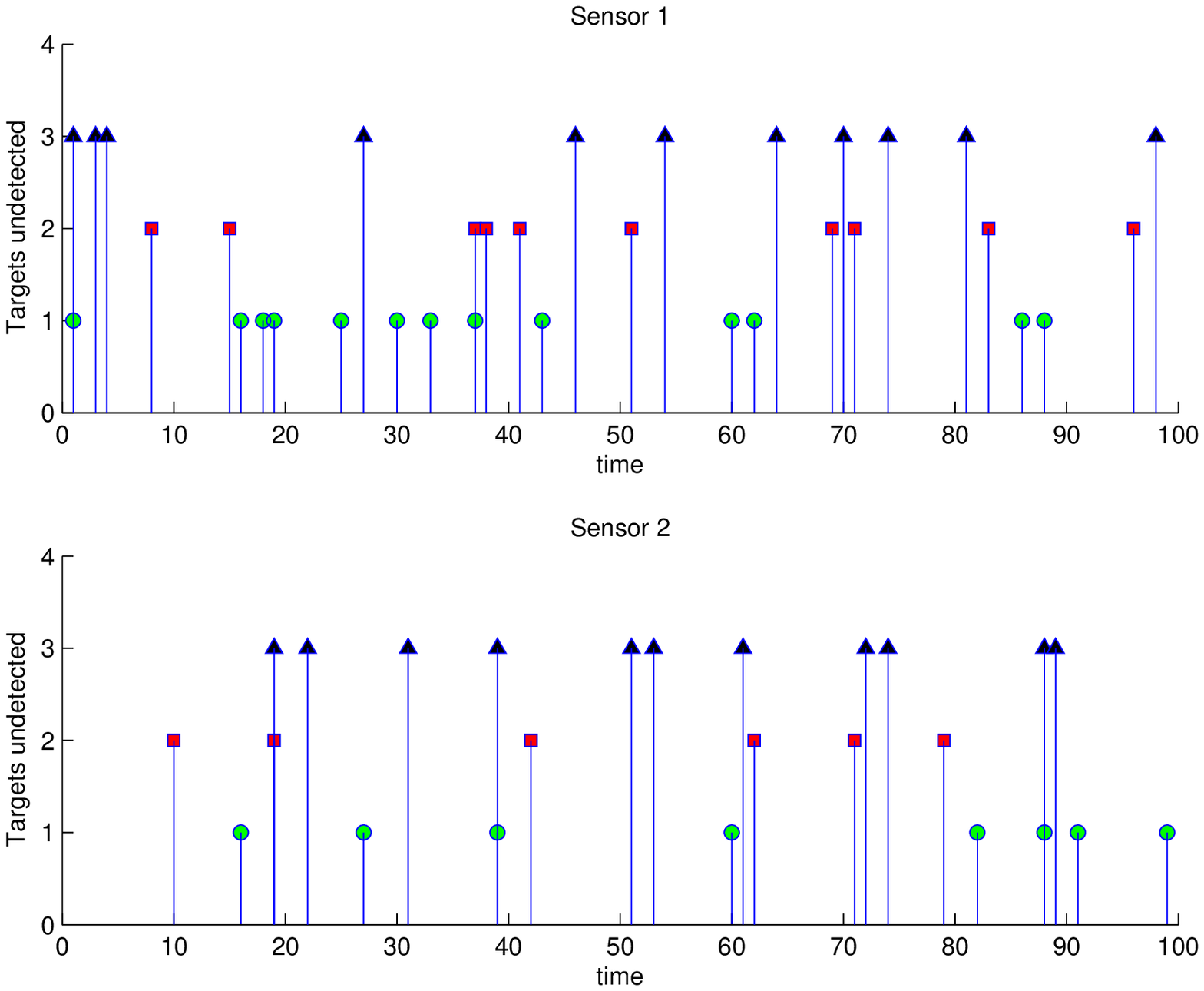} 
\caption{The index of the targets which were undetected at the sensors for a single run.}
\label{MCJPDAF_missed_targets}
 \end{figure}

\subsection{Performance of MCJPDAF with varying $\lambda_C$ and $P_D$}
To study the effect of missing target detections and clutter measurements on the estimates, simulations were carried out with varying clutter rate $\lambda_C$ and detection probability $P_D$. An easy problem $(\lambda_C=0.5,P_D=1.0)$, a medium problem $(\lambda_C=2.0,P_D=0.8)$ and a difficult problem $(\lambda_C=5,P_D=0.5)$ were simulated for 20 Monte Carlo runs and their results are compared.

\begin{figure}[h]
\centering
\subfloat [Easy problem]
{\includegraphics[scale=0.4]{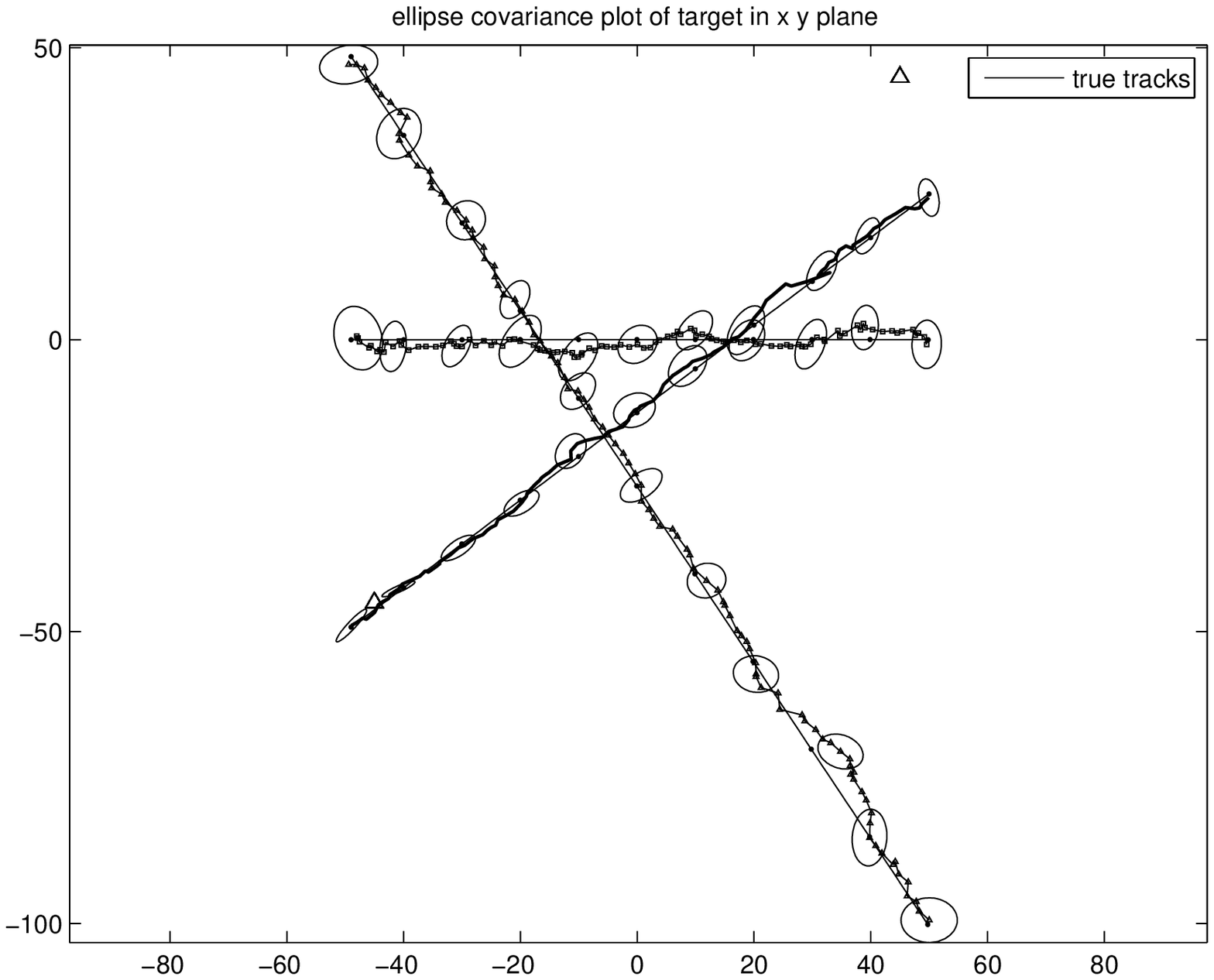}  
}\\
\subfloat [Medium problem]
{\includegraphics[scale=0.4]{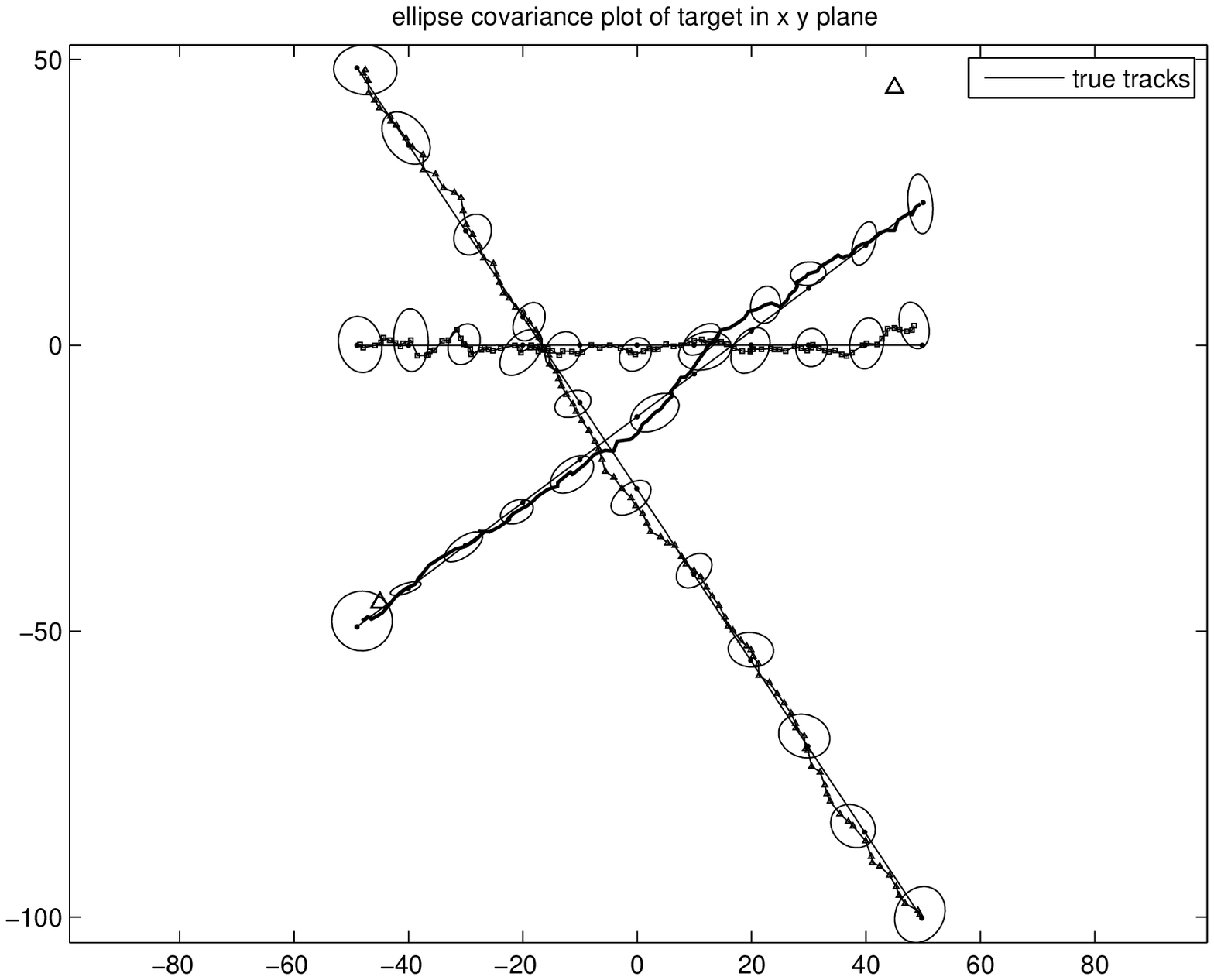}  
}\\
\subfloat [Difficult problem]
{\includegraphics[scale=0.4]{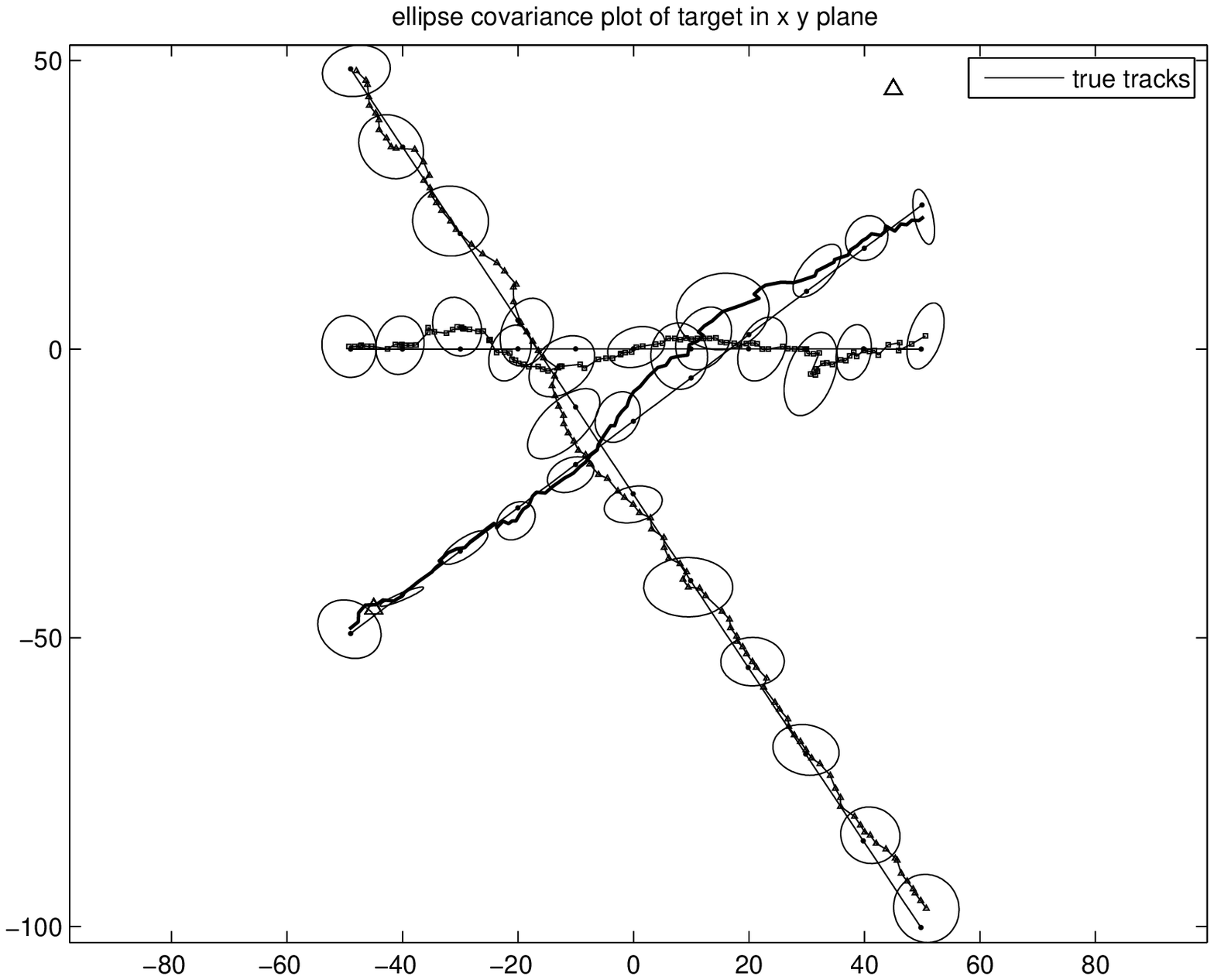}  
}%
\caption{Targets' true $xy$ track and their estimated track covariance obtained using MCJPDAF for a single run: The location of the sensors are shown in triangles.}
\label{MCJPDAF_cov_plot_case_all}
\end{figure}

\begin{figure}[h]
\centering
\subfloat [Easy problem]
{\includegraphics[scale=0.4]{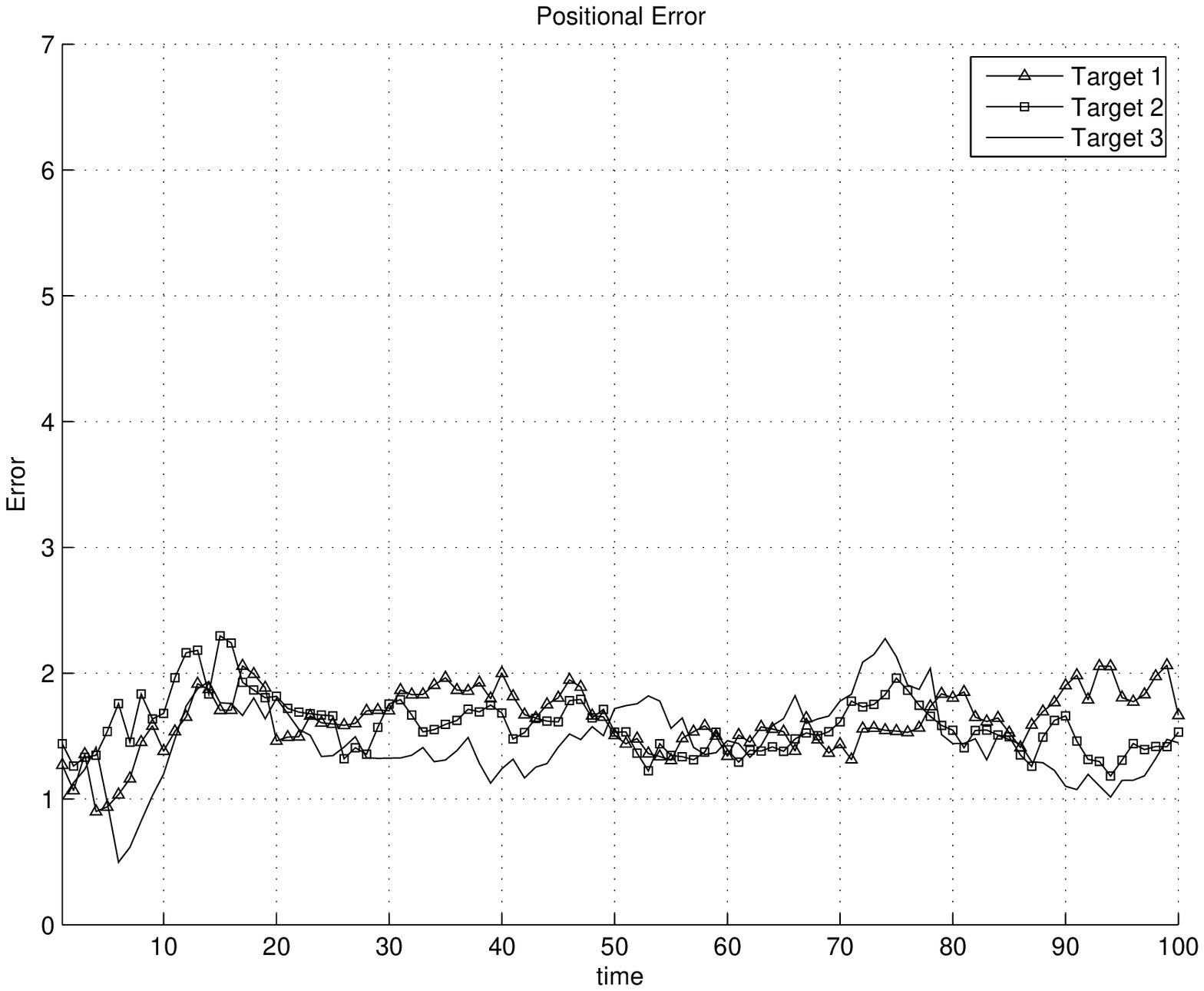}  
}\\
\subfloat [Medium problem]
{\includegraphics[scale=0.4]{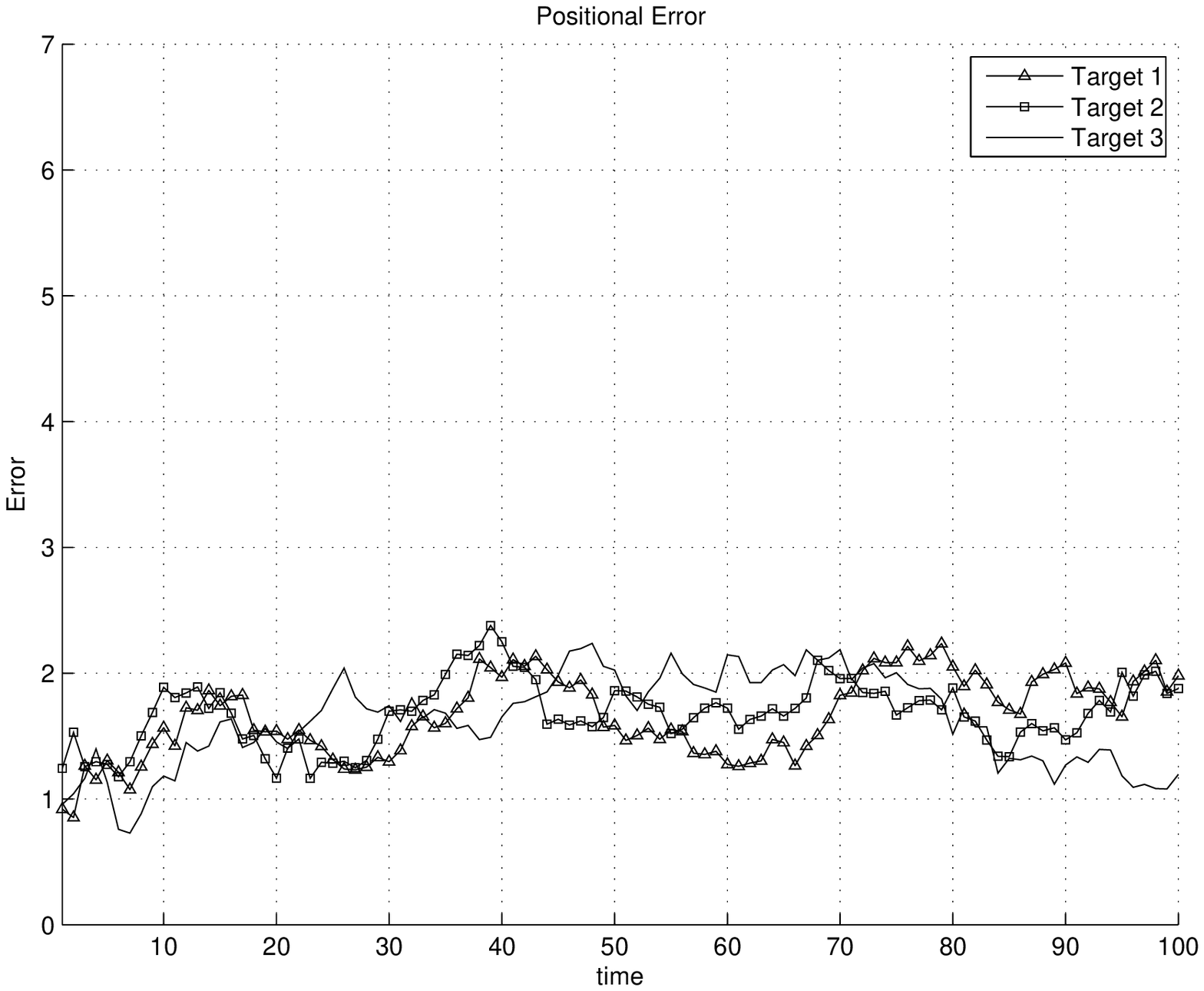}  
}\\
\subfloat [Difficult problem]
{\includegraphics[scale=0.4]{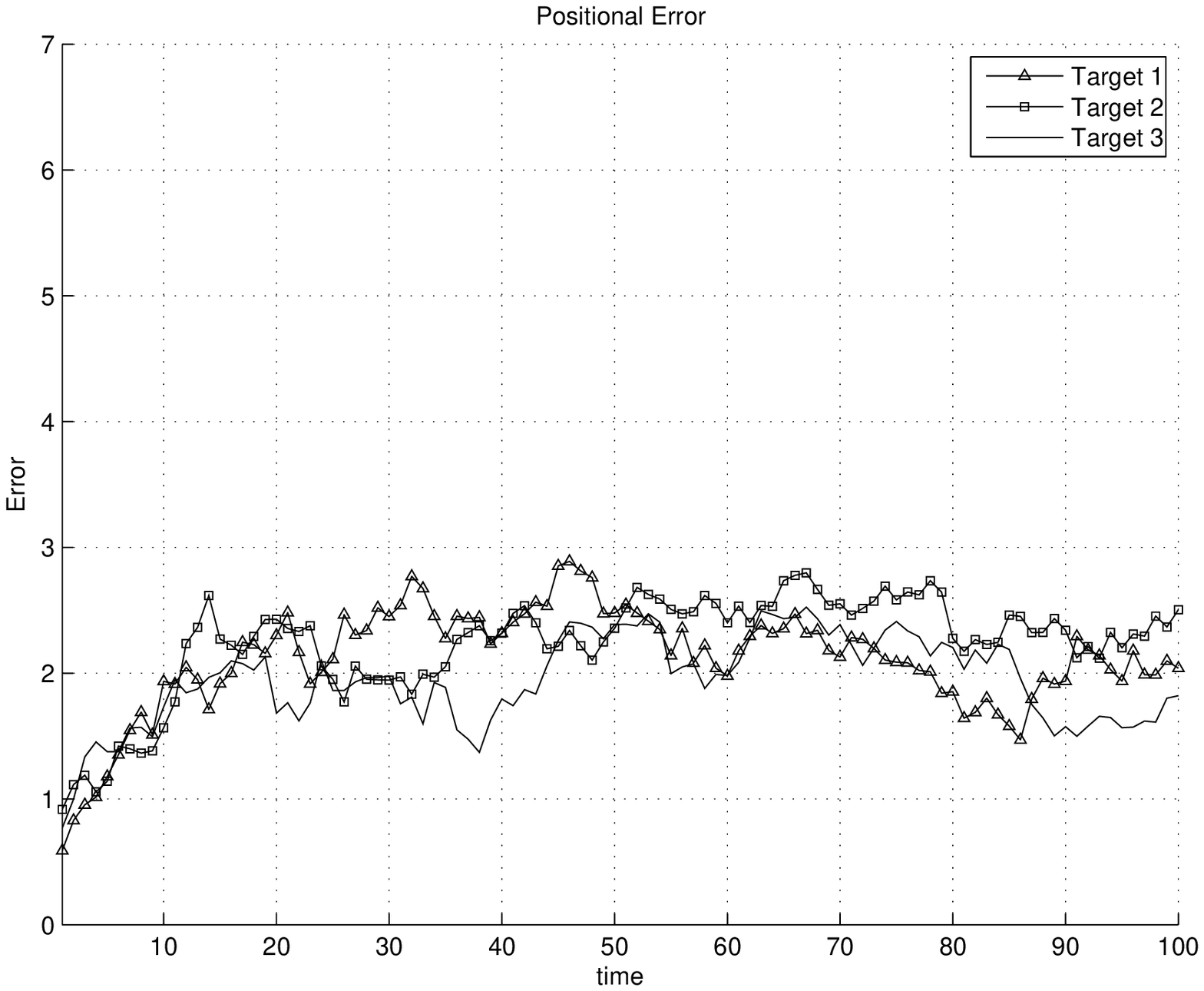}  
}%
\caption{MSE of the position estimates from $20$ Monte Carlo runs, obtained using MCJPDAF.}
\label{MCJPDAF_MSE_case_all}
\end{figure}

\begin{figure}[h]
\centering
\subfloat [Medium problem: Swap between target 2 and 3]
{\includegraphics[scale=0.4]{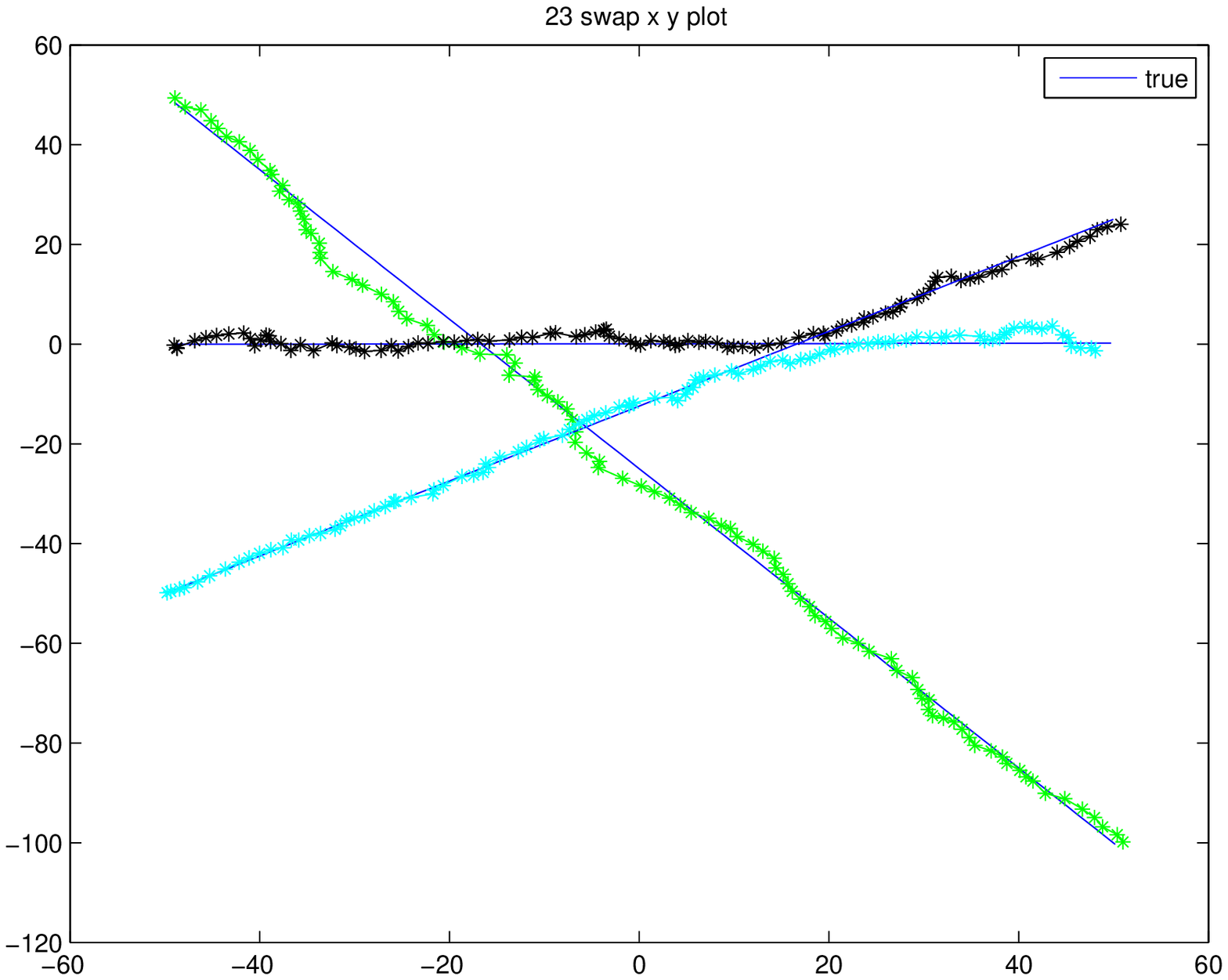}  
}\\
\subfloat [Difficult problem: Swap between target 2 and 3]
{\includegraphics[scale=0.4]{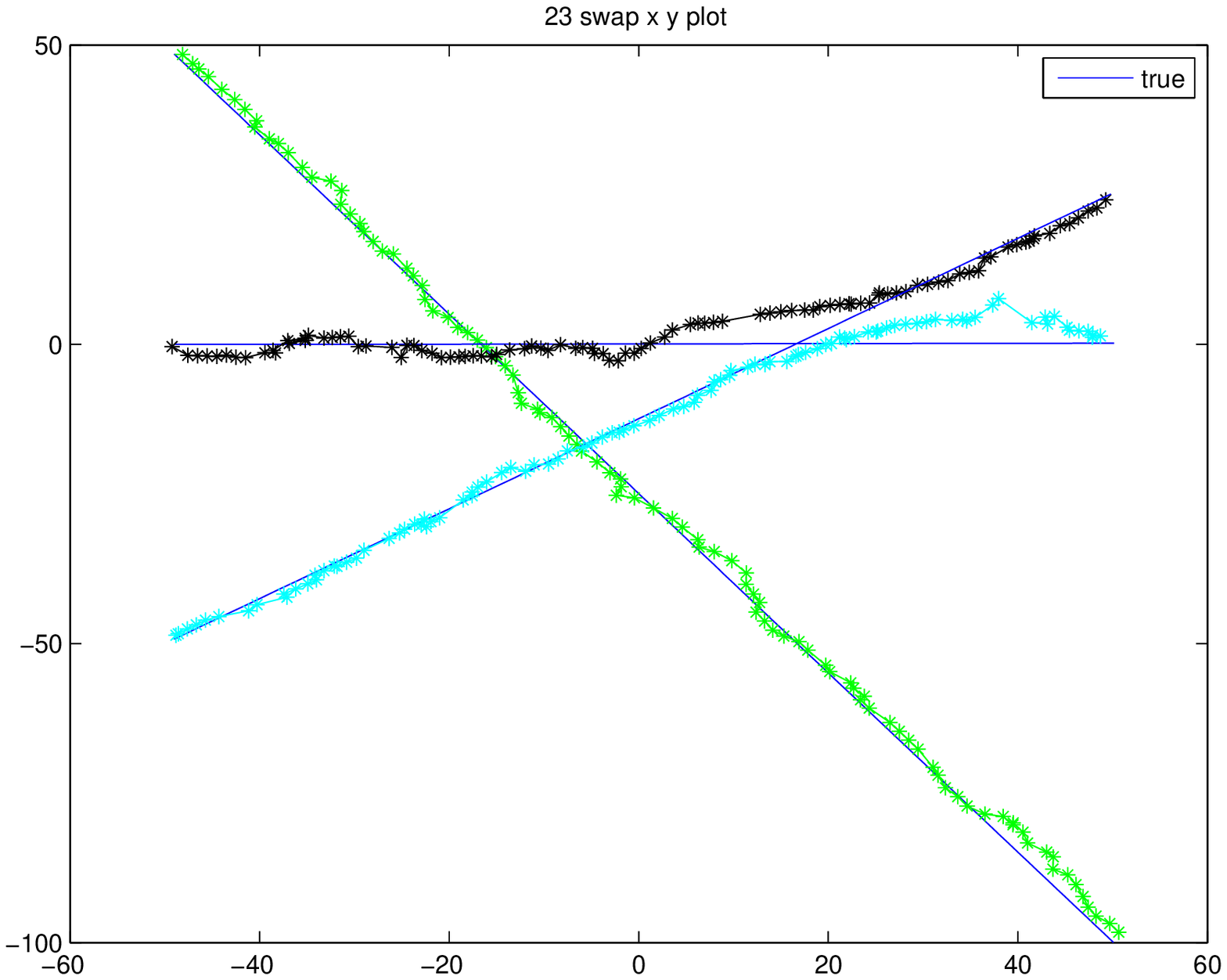}  
}%
\caption{Track of swapped target estimates.}
\label{MCJPDAF_swaps}
\end{figure}

 With the increase in clutter rate and decrease in target detection probability, the 2-$\sigma$ covariance region increased in area and the mean square error of position estimates also increased. This caused few swapped track estimates and are shown in Fig. \ref{MCJPDAF_swaps}. In 20 Monte Carlo runs, there was one swapped track estimate between target 2 and 3 in medium problem as well as difficult problem. The target position estimates with the 2-$\sigma$ covariance region is shown in Fig.\ref{MCJPDAF_cov_plot_case_all} and their MSE is shown in Fig.\ref{MCJPDAF_MSE_case_all}.

\section{Summary}
Data association problem arises due to the lack of information at the observer about the proper association between the targets and the received measurements. The problem becomes more involved when the targets move much closer and there are clutter and missed target detections at the observer. MC-JPDAF combines the JPDAF with particle filtering technique to accommodate non- linear and non-Gaussian models and to solve the data association problem. It incorporates clutter and missing measurements and also measurements from multiple observers. There are two types of representation for data association hypothesis, i.e. measurement to target association ($M{\rightarrow}T$) and target to measurement association ($T{\rightarrow}M$). Both carry same information and have one to one mapping between them. They can be converted from one type of representation to another and are used depending upon the situation. The infeasible hypotheses are neglected using a gating procedure to reduce computational complexity.  It calculates the posterior hypotheses probability of the remaining hypotheses. The filtered estimate of each hypothesis is calculated and is combined by weighting each with their corresponding posterior hypothesis probability. 

With the increase in clutter rate and decrease in target detection probability, mean square error and the number of diverged tracks gets increased slightly. MCJPDAF requires smaller number of particles for estimation.  It efficiently solves the data association problem, and there were almost no diverged tracks with moderate clutter rate and detection probability.

\chapter{Monte Carlo Multiple Model Joint Probabilistic Data Association Filter (MC-MMJPDAF)}
Monte Carlo Multiple Model Joint Probabilistic Data Association Filter (MC-MMJPDAF) is a technique proposed for tracking maneuvering multi-targets under data association uncertainty. It is an extension of MC-JPDAF for maneuvering targets. The original MC-JPDAF was proposed for slowly maneuvering targets and diverges for highly maneuvering targets. MC-MMJPDAF incorporates the technique used in Multiple Model Particle Filter discussed in chapter \ref{chap:MMPF}, which uses multiple models to account for different types of target dynamics similar to circular motion, accelerated motions etc. The resulting filter is capable of maneuvering, multi-target tracking.

The model description is almost the same as the MC-JPDAF except that each particle consists of a state vector $\mathbf{x}_{k,t}$ for target $k$, augmented by an index vector $A_{k,t}$ representing the model. Thus particles have continuous valued vector $\mathbf{x}_{k,t}$ of target kinematics variables, like position, velocity, acceleration, etc, and a discrete valued regime variable $A_{k,t}$ that represents the index of the model which generated $\mathbf{x}_{k,t}$ during the time period $(t-1^+,t]$. The regime variable can be one of the fixed set of $s$ models i.e., $A_{k,t}\in S=\{1,2, . . . .,s\}$. The posterior density of  $k^{th}$ target $p(x_{k,t} \mid y_t)$ is represented  using $N$ particles $\{b_t^n,w_t^n\}_{n=1}^N$, i.e., the augmented state vector and the weight. The posterior model probabilities $\{\pi_i(k,t)\}_{i=1}^s$ are approximately equal to the proportion of the samples from each model in the index set $\{A_{k,t}^n\}_{n=1}^N$. The combined augmented state for all targets is represented as $\{b_{t-1}^n, w_{t-1}^n\}_{n=1}^{N}$. The combined regime variable for all targets is represented as $A_{t}^n$

The algorithm for Monte Carlo Multiple Model Joint Probabilistic Data Association Filter (MC-MMJPDAF) is presented in Table.\ref{tab:MCMMJPDAF}.

\begin{table}[H] 
\caption{Monte Carlo Multiple Model Joint Probabilistic Data Association Filter (MC-MMJPDAF)} 
\centering          
\begin{tabular}{l}
  \hline
  \begin{minipage}{6in}
    \vskip 4pt
$[\{b_t^n, w_t^n\}_{n=1}^N]=$MC-MMJPDAF$[\{b_{t-1}^n, w_{t-1}^n\}_{n=1}^{N},\mathbf{y}_t]$
\begin{itemize}
	\item For $k=1..K$ perform Regime transition (Table.\ref{tab:RT}):\\
$[\{A_{k,t}^n\}_{n=1}^N]=$RT$[\{A_{k,t-1}^n\}_{n=1}^{N},\Pi]$	  
	\item For $k=1..K$ perform Regime Conditioned MC-JPDAF(Table.\ref{tab:RCMCJPDAF}):\\
$[\{\mathbf{x}_{k,t}^n, w_{k,t}^n\}_{n=1}^N]=$RC-SIS$[\{\mathbf{x}_{k,t-1}^n, A_{k,t}^n, w_{k,t-1}^n\}_{n=1}^{N},\mathbf{y}_t]$
        \item If required resample the particles and do roughening.
\end{itemize}
   \vskip 4pt
 \end{minipage}
 \\
  \hline
 \end{tabular}
\label{tab:MCMMJPDAF} 
\end{table}

\begin{table}[ht] 
\caption{Regime Conditioned MC-JPDAF} 
\centering          
\resizebox{!}{4in} {
\begin{tabular}{l}
  \hline
  \begin{minipage}{7in}
    \vskip 4pt
$[\{\mathbf{x}_{t}^n, w_{t}^n\}_{n=1}^N]=$RC-MCJPDAF$[\{\mathbf{x}_{t-1}^n, A_{t}^n, w_{t-1}^n\}_{n=1}^{N},y_t]$
\begin{itemize}
      \item Prediction step: FOR $k=1..K$, $n=1:N$, draw samples 
	      \begin{equation}
	      \mathbf{x}_{k,t}^{(n)}\sim q_k(\mathbf{x}_{k,t}\mid \mathbf{x}_{k,t-1}^{(n)},A_{k,t}^n,\mathbf{y}_t)
	      \end{equation}

		    \item Evaluate the predictive weights upto a normalizing constant
			    \begin{equation}
			   \alpha_{k,t}^{(n)}\propto w_{k,t-1}^{(n)}\frac{p_k(\mathbf{y}_t\mid \mathbf{x}_{k,t}^{(n)})p_k(\mathbf{x}_{k,t}^{(n)}\mid\mathbf{x}_{k,t-1}^{(n)},A_{k,t}^n)}{q_k(\mathbf{x}_{k,t}^{(n)}\mid\mathbf{x}_{k,t-1}^{(n)},A_{k,t}^n,\mathbf{y}_t)}  
			    \end{equation}
		    \item Normalize the predictive weights
			    \begin{equation}
				\displaystyle \sum_{n=1}^N \alpha_{k,t}^{(n)}=1
			    \end{equation}

		    \item FOR $k=1..K$, $i=1..N_o$, $j=1..M^i$, calculate the predictive likelihood
			    \begin{equation}
				  p_k(\mathbf{y}_{j,t}^i\mid \mathbf{y}_{1:t-1}) \approx \displaystyle \sum_{i=1}^N \alpha_{k,t}^{(n)} p_T^i(\mathbf{y}_{j,t}^i\mid \mathbf{x}_{k,t}^{(n)}) 
			    \end{equation}

		    \item FOR observer $i=1..N_o$, enumerate all valid target to measurement association hypotheses $\tilde{\lambda}^i_t$.

		    \item Perform gating on the valid target to measurement hypotheses by the following procedure:
			  \begin{itemize}
				\item For $k=1..K$, calculate the approximation for the predictive likelihood of target $k$ using \eqref{eq:appr_pr_lh}
				      \begin{eqnarray}
				      p_k(\mathbf{y}\mid \mathbf{y}_{1:t-1}) & \approx & \mathcal{N}(\mu_{\hat{\mathbf{y}}_k},\Sigma_{\hat{\mathbf{y}}_k}) \\ 
				      \mu_{\hat{\mathbf{y}}_k} & = & \displaystyle{\sum_{n=1}^N}\alpha_k^{(n)}\mathbf{g}(\mathbf{x}_k^{(n)},\mathbf{p_0}) \\
				      \Sigma_{\hat{\mathbf{y}}_k} & = & \Sigma_{\mathbf{y}}+\displaystyle{\sum_{n=1}^N}\alpha_k^{(n)}[\mathbf{g}(\mathbf{x}_k^{(n)},\mathbf{p_0})-\mu_{\hat{\mathbf{y}}_k}][\mathbf{g}(\mathbf{x}_k^{(n)},\mathbf{p_0})-\mu_{\hat{\mathbf{y}}_k}]^T 
				      \end{eqnarray}
				\item For $k=1..K$, $i=1..N_o$, $j=1..M^i$, calculate the squared distance $d_k^i(\mathbf{y}_j)$ between the predicted and observed measurements using measurement innovations.
				      \begin{equation}
				      d_k^2(\mathbf{y}_j)=(\mathbf{y}_j-\mu_{\hat{\mathbf{y}}_k})^T \Sigma_{\hat{\mathbf{y}}_k}^{-1}(\mathbf{y}_j-\mu_{\hat{\mathbf{y}}_k})
				      \end{equation}
				\item Perform chi-square hypothesis testing on the proposed target-measurement association hypotheses. Accept a hypothesis if its chi-square statistics $d_k^2$ satisfies the relation $d_k^2<\chi^2_\alpha$	
				to obtain the set of gated hypotheses $\tilde{\Lambda}_t^i$ at each observer $i$.

			  \end{itemize}
\end{itemize}
  \vskip 4pt
 \end{minipage}
 \\
  \hline
 \end{tabular}
}
\label{tab:RCMCJPDAF} 
\end{table}

\begin{table}[ht] 
\caption{Regime Conditioned MC-JPDAF Algorithm (contd..)} 
\centering          
\resizebox{!}{4in} {
\begin{tabular}{l}
  \hline
  \begin{minipage}{7in}
    \vskip 4pt
\begin{itemize}

		  \item Convert $T{\rightarrow}M$ hypotheses to $M{\rightarrow}T$ hypotheses and calculate the number of clutter measurements $M_C^i$ in each hypothesis. 

		  \item FOR observer $i=1..N_o$, calculate association prior of all hypotheses.
			  \begin{equation}
			  p(\tilde{\mathbf{r}}_k^{i}\mid\tilde{\mathbf{r}}_{k-1}^{i})\propto
			  \begin{cases}
			  1-P_D \;\qquad\text{if $j=0$} \\
			  0 \;\qquad\text{if $j>0$ and $j \in \{\tilde{r}_1^i\cdot\cdot\cdot\tilde{r}_{k-1}^i\}$} \\
			  \frac{P_D}{M_k^i} \;\qquad \text{otherwise}
			  \end{cases} 
			  \end{equation}
			  \begin{eqnarray}		      p(\tilde{\lambda}^i)=p(M_C^i)\displaystyle\prod_{k=1}^{K}p(\tilde{\mathbf{r}}_k^{i}\mid\tilde{\mathbf{r}}_{k-1}^{i}) 
			  \end{eqnarray}
		  \item FOR $i=1..N_o$, compute joint association posterior probability and normalize it at each observer $i$.
			  \begin{eqnarray}
			    p(\tilde{\lambda}_t^i\mid\mathbf{y}_{1:t}) & \propto p(\tilde{\lambda}_t^i) (V^i)^{-M_C^i}\displaystyle\prod_{j\in \mathcal{I}^i}p_{r_{j,t}^i}(\mathbf{y}_{j,t}^i\mid \mathbf{y}_{1:t-1}) 
			  \end{eqnarray}
			  \begin{eqnarray}
			    \displaystyle\sum p(\tilde{\lambda}_t^i\mid\mathbf{y}_{1:t}) & = 1 \qquad \text{at each observer $i$.}  
			  \end{eqnarray}
		  \item FOR $k=1..K$, $i=1..N_o$, $j=0..M^i$, calculate the marginal association posterior probability
		  \begin{equation}
		  \beta_{jk}^i = \displaystyle\sum_{\{\tilde{\lambda}_t^i\in\tilde{\Lambda}_t^i:\tilde{r}_{k,t}^i=j\}} p(\tilde{\lambda}_t^i\mid\mathbf{y}_{1:t}) 
		  \end{equation}
		  \item FOR $k=1..K$, compute target likelihood.
			  \begin{equation}		      p_k(\mathbf{y}_{t}\mid\mathbf{x}_{k,t}^{(n)})=\displaystyle\prod_{i=1}^{N_o}\left[\beta_{0k}^i+\displaystyle\sum_{j=1}^{M^i}\beta_{jk}^ip_T^i(\mathbf{y}_{j,t,}^i\mid \mathbf{x}_{k,t}^{(n)})\right]
			  \end{equation}

     \item Update step: FOR $k=1..K$, $n=1..N$, calculate and normalize particle weights.
	      \begin{equation}
	      w_{k,t}^{(n)}\varpropto w_{k,t-1}^{(n)}\dfrac{p_k(\mathbf{y}_t|\mathbf{x}_{k,t}^{(n)})p_k(\mathbf{x}_{k,t}^{(n)}|\mathbf{x}_{k,t-1}^{(n)},A_{k,t}^n)}{q(\mathbf{x}_{k,t}^{(n)}|\mathbf{x}_{k,t-1}^{(n)},A_{k,t}^n,\mathbf{y}_{t})}  ; \;\qquad\displaystyle \sum_{n=1}^N w_{k,t}^{(n)}=1
	      \end{equation}

     \item FOR $k=1..K$, if required, resample the particles $\{\mathbf{x}_{k,t}^{(n)}\}_{n=1}^N$ and do roughening.

\end{itemize}
  \vskip 4pt
 \end{minipage}
 \\
  \hline
 \end{tabular}
}
\label{tab:RCMCJPDAF_cntd} 
\end{table}

\section{Simulation Results}
\begin{figure}[h]
\centering
{\includegraphics[scale=0.5]{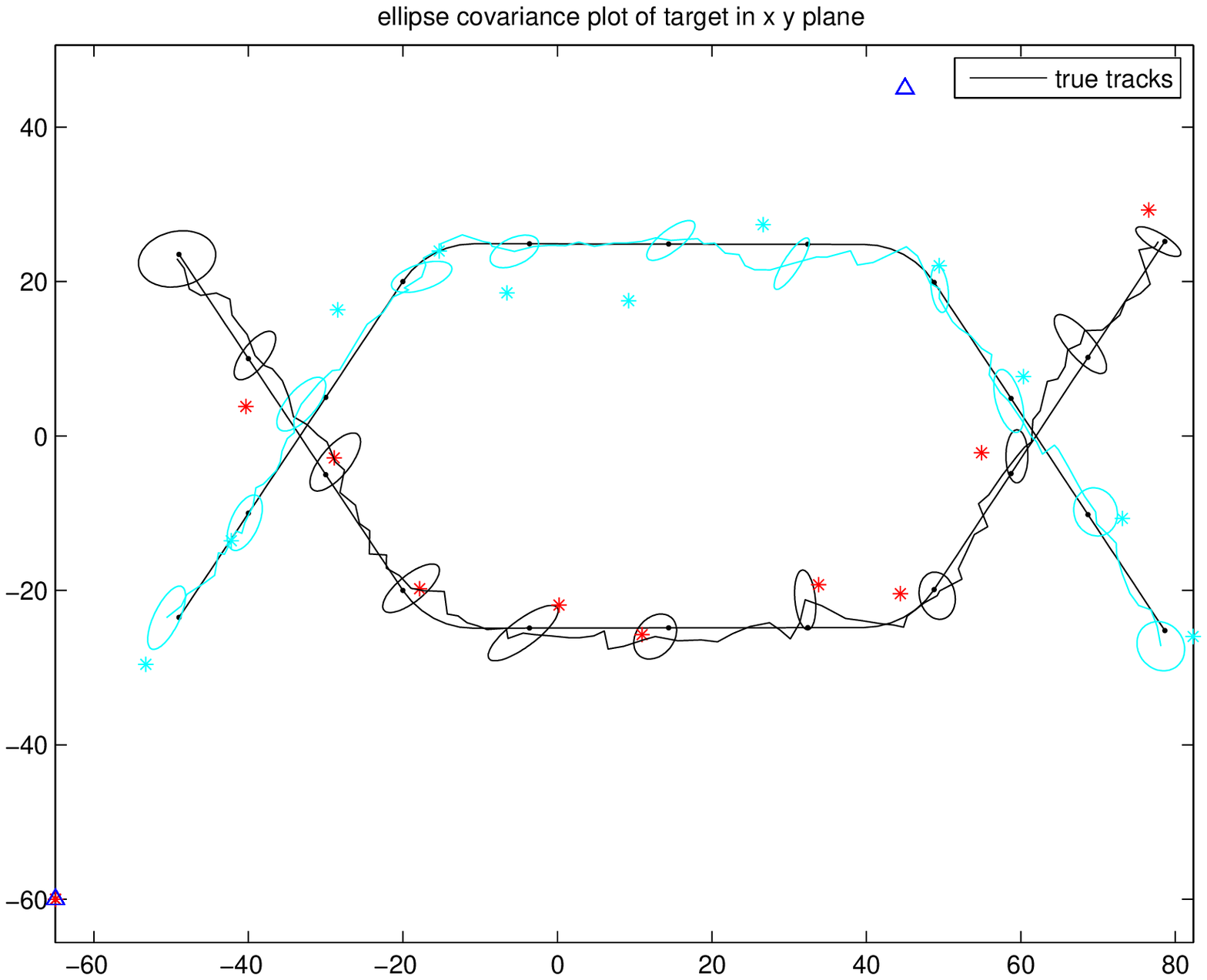}}    
\caption{Targets' true $xy$ track and their estimated track covariance obtained using MC-MMJPDAF for a single run: The location of the sensors are shown in blue triangles.}
\label{MCMMJPDAF_cov_plot}
\end{figure}

\begin{figure}[h]
\centering
{\label{MCMMJPDAF_T1_MSE}\includegraphics[scale=0.4]{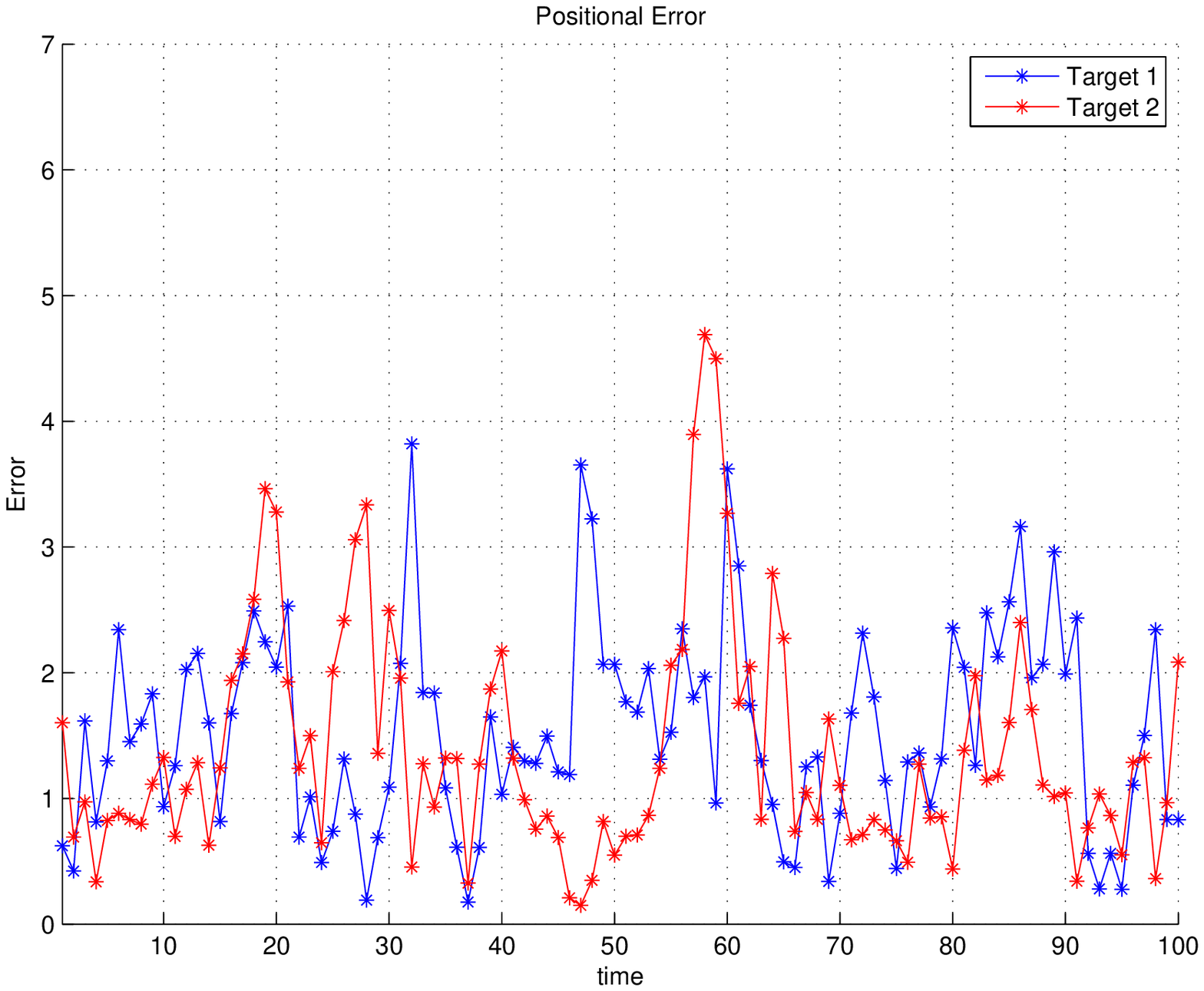}  
}
\caption{MSE of the position estimates from $100$ Monte Carlo runs, obtained using MC-MMJPDAF.}
\label{MCMMJPDAF_MSE}
\end{figure}

\begin{figure}[h] 
\centering
\subfloat [position $x$ ]
{\includegraphics[scale=0.4]{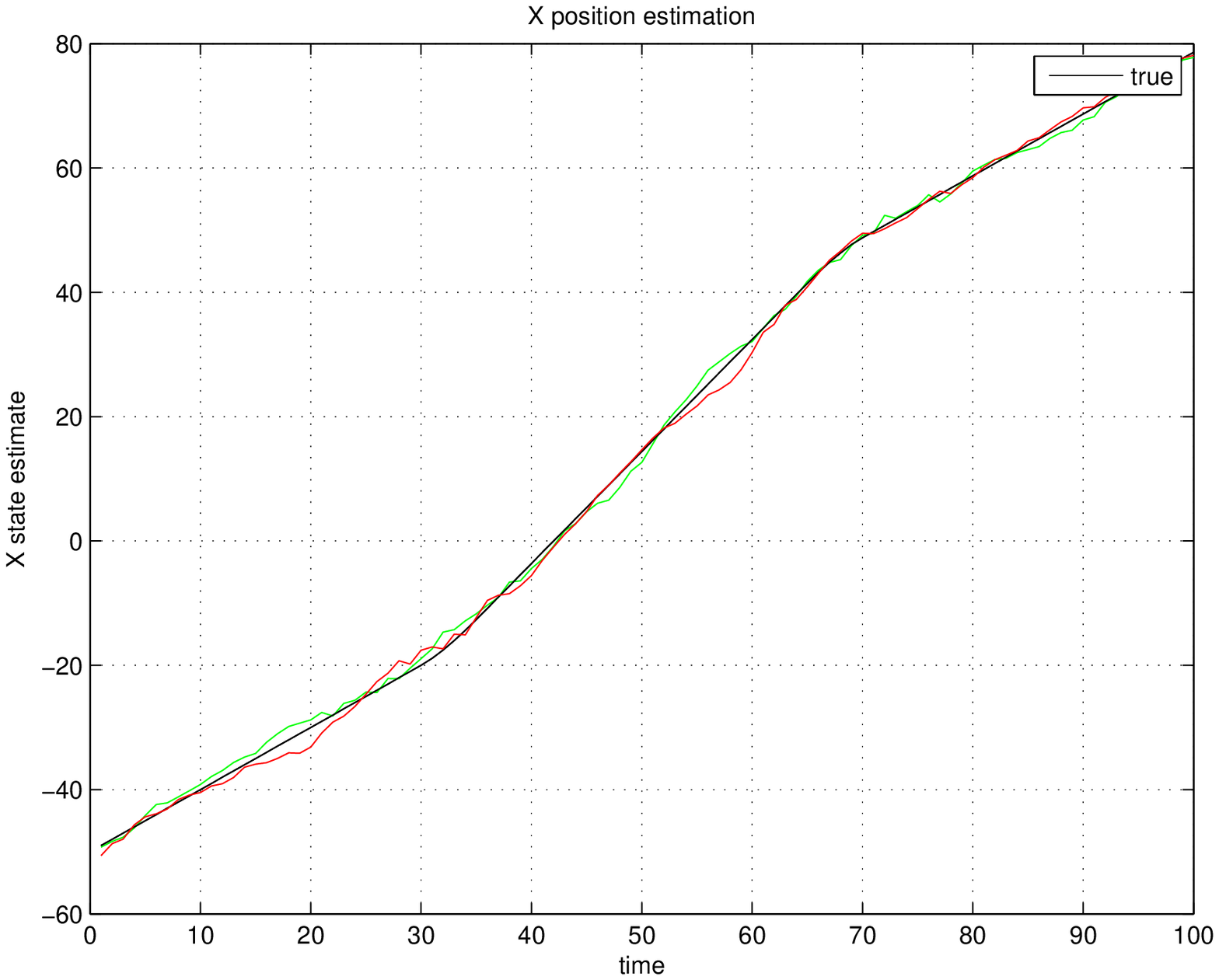}  
}\\
\subfloat [position $y$ ]
{\includegraphics[scale=0.4]{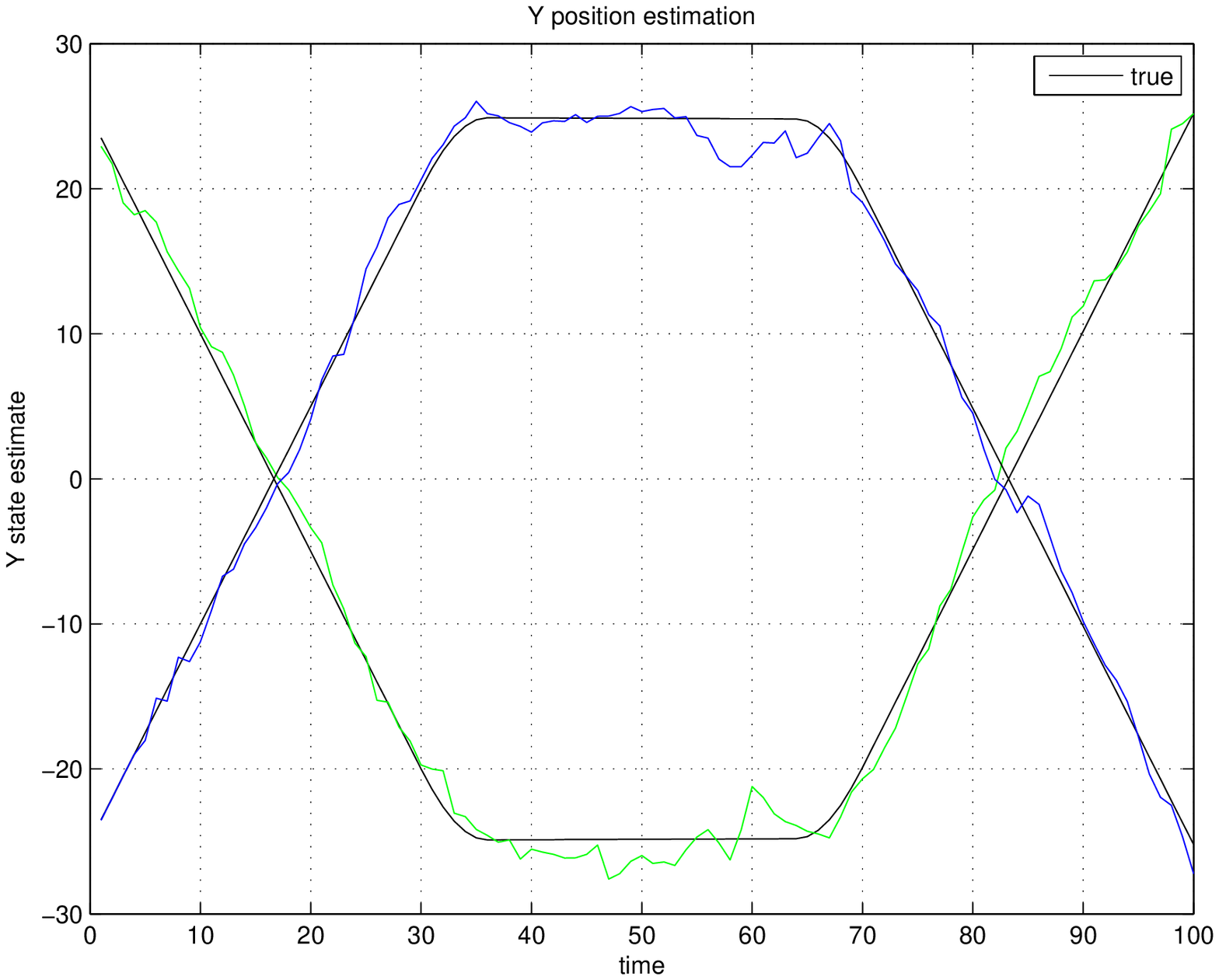}  
}\\
\subfloat [velocity $v_x$]
{\includegraphics[scale=0.4]{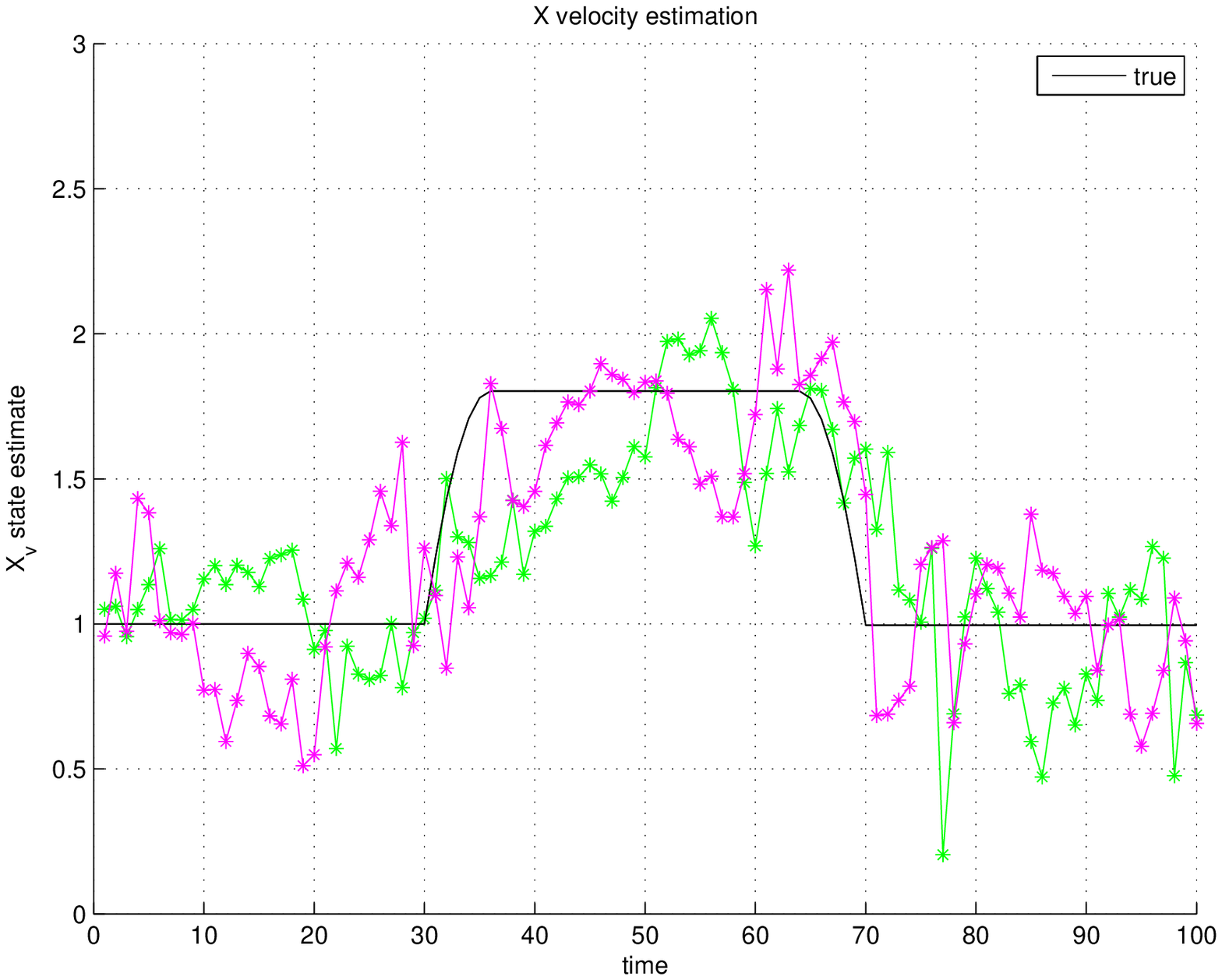}  
}%
\subfloat [velocity $v_y$]
{\includegraphics[scale=0.4]{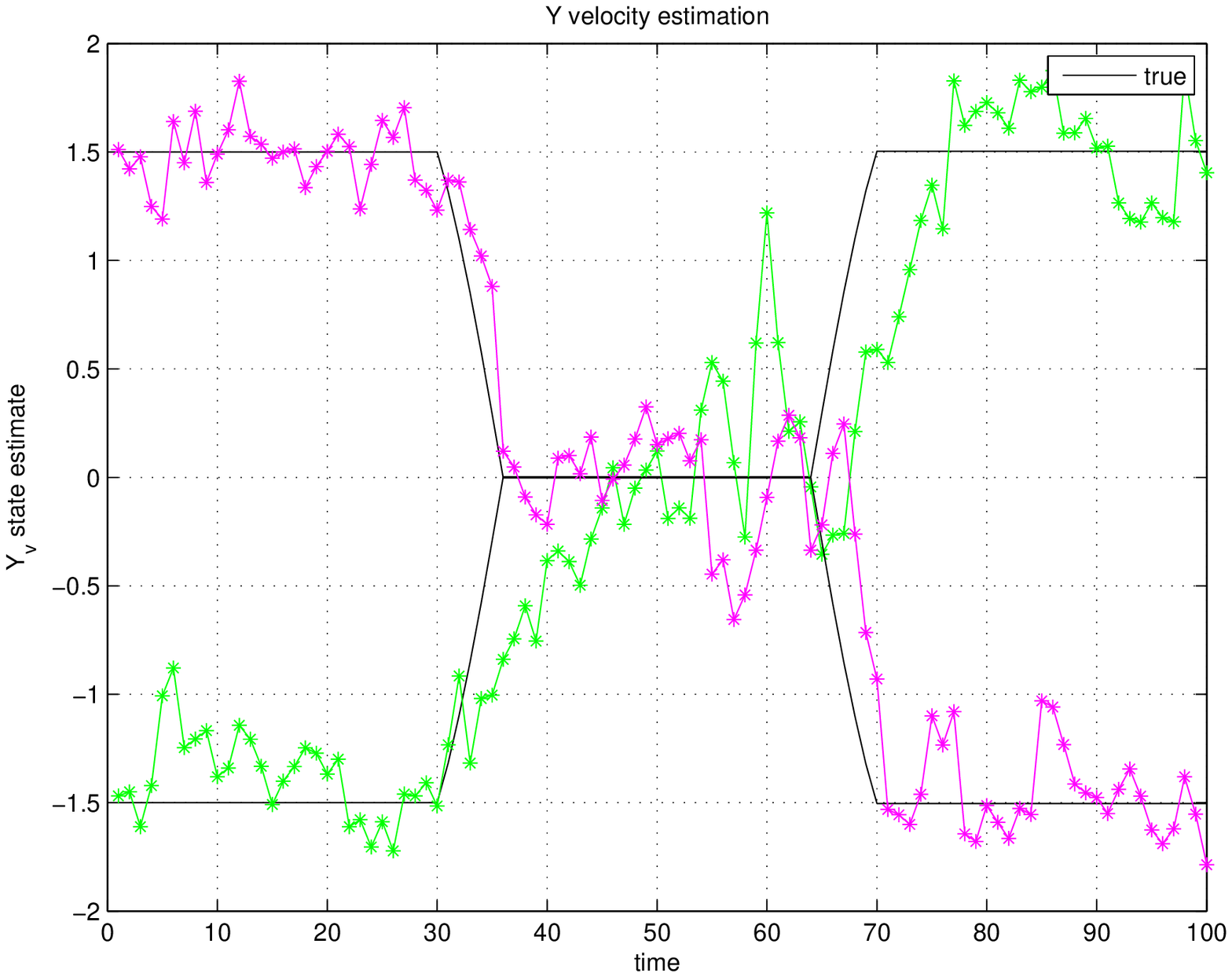}  
}\\
\caption{Targets' true states and their estimates from single run obtained using MC-MMJPDAF.}
\label{MCMMJPDAF_state}
\end{figure}

To verify the effectiveness of the algorithm, targets' motion scenario and their measurements are simulated according to the given models and the estimates using the algorithm is compared with the true trajectories. 
The augmented state vector consists of position $x,y$, velocities $v_x, v_y$ of the target, and the regime variable $A$, 
\begin{eqnarray}
\mathbf{x}_{k,t}=
\begin{bmatrix}
x_{k,t} & \dot{x}_{k,t} & y_{k,t}  & \dot{y}_{k,t} &A_{k,t}
 \end{bmatrix}^T 
\end{eqnarray}
The target has constant velocity motion as well as coordinated turn motions. The two targets move in constant velocity of $(1.0,1.5)$ and $(1.0,1.5)$. The first target undergoes a maneuver from $t=30$ to $t=36$ and $t=64$ to $t=70$ with a turn rate of $0.1641 rad/s$ anti-clockwise, and the second target undergoes a maneuver from $t=30$ to $t=36$ and $t=64$ to $t=70$ with a turn rate of $-0.1641 rad/s$ clockwise. The initial position of the targets are $(-50,25)$ and $(-50,-25)$ respectively. 
All the targets have state transition model $F$ such that:
\begin{eqnarray}
\mathbf{x}_{k,t} & = & F(\mathbf{x}_{k,t-1})+\mathbf{w}_{k,t-1} 
\end{eqnarray}
 The constant velocity model, anti-clockwise coordinated turn model and clockwise coordinated turn model are given by the transition matrix $F_1$, $F_2$ and $F_3$
\begin{equation}
 F_1=\begin{bmatrix}
1&T&0&0\\
0&1&0&0\\
0&0&1&T\\
0&0&0&1\\
\end{bmatrix}
\end{equation}
\begin{equation}
F_2=
\begin{bmatrix}
1    &\dfrac{\sin (\Omega T)}{\Omega}       &0    &-\dfrac{1 - \cos(\Omega  T)}{\Omega}\\
        0    &\cos(\Omega  T)          &0     &-\sin(\Omega  T)\\
        0    &\dfrac{1-\cos(\Omega  T)}{\Omega}   &1      &\dfrac{\sin(\Omega  T)}{\Omega}\\
        0    &\sin(\Omega  T)          &0      &\cos(\Omega  T) \\
\end{bmatrix}
\end{equation}
\begin{equation}
F_3=
\begin{bmatrix}
1    &\dfrac{\sin (-\Omega T)}{-\Omega}       &0    &-\dfrac{1 - \cos(-\Omega  T)}{-\Omega}\\
        0    &\cos(-\Omega  T)          &0     &-\sin(-\Omega  T)\\
        0    &\dfrac{1-\cos(-\Omega  T)}{-\Omega}   &1      &\dfrac{\sin(-\Omega  T)}{-\Omega}\\
        0    &\sin(-\Omega  T)          &0      &\cos(-\Omega  T) \\
\end{bmatrix}
\end{equation}
where $T$ is the sampling period of the target dynamics and $\Omega$ is the turn rate.
The measurement sensors are located at $(-65,-60)$, $(45,45)$ meters respectively. The $k$-th targets' range $r_{k}$ and bearing $\theta_{k}$ at time $t$ are available as the measurement $\mathbf{y}_{k,t}^i$ at time step of $T=1$ at each observer $i$. 
\begin{eqnarray}
\mathbf{y}_{k,t}^i=
\begin{bmatrix}
r_{k,t}\\
\theta_{k,t}\\
\end{bmatrix}
\end{eqnarray}
The errors in the range and bearing are such that $\sigma_R=5$ and $\sigma_\theta=0.02$. The maximum range detected by the sensor is $100 m$. The probability of detection of a target is $P_D=0.9$ and the clutter rate is $\lambda_C=0.5$. The exact association of the measurements to the targets is unknown at the observers.
The measurement model $h(\cdotp)$ for the target $k$ at the $i$-th observer is given by:
\begin{eqnarray}
\mathbf{y}_{k,t}^i=
h_k(\mathbf{x}_{k,t})^i+\mathbf{v}_{k,t}^i=
\begin{bmatrix}
\sqrt{(x_{k,t}-x_o^i)^2+(y_{k,t}-y_o^i)^2}\\
\tan^{-1}\left(\dfrac{y_{k,t}-y_o^i}{x_{k,t}-x_o^i}\right)
\end{bmatrix}
\end{eqnarray}
with $\mathbf{p}_0^i=(x_o^i,y_o^i)$. The maximum range of sensor is $R_{max}^i=100$ and the volume of measurement space is $V^i=2\pi R_{max}^i$.
The measurement error $\mathbf{v}_{k,t}^i$ is uncorrelated and has zero mean Gaussian distribution with covariance matrix $\Sigma_{\mathbf{y}_k}$. 
\begin{eqnarray}
\Sigma_{\mathbf{y}_k}=
\begin{bmatrix}
\sigma_R^{2} &0 \\
0 &\sigma_\theta^2 \\
\end{bmatrix} 
\end{eqnarray}
The measurement errors are assumed to be the same at all the observers. The initial state estimate is assumed to be a Gaussian vector with mean $ \hat{\mathbf{x}}_{k,0}=\mathbf{x}_{k,0}$ and error covariance $P_{k,0} = diag(5,0.1,5,0.1)$. Hence initial particles for each target $\{\mathbf{x}_{k,0}^{(n)}\}_{n=1}^N$ were generated based on the distribution 
\begin{equation}
 \mathbf{x}_{k,0} \sim \mathcal{N}(\mathbf{x}_{k,0},P_{k,0}) 
\end{equation}
In this implementation of the particle filter, the transitional prior which is a sub-optimal choice of importance density is  used to propose particles. Thus the importance density used is:\\
\begin{eqnarray*}
q(\mathbf{x}_{k,t}\mid \mathbf{x}_{k,t-1}^{(n)},A_{k,t}^n,y_{t})&=&p(\mathbf{x}_{k,t}\mid \mathbf{x}_{k,t-1}^{(n)},A_{k,t}^n)\\
&=&\left\{
 \begin{array}{rl}
  \mathcal{N}(F_{1}(\mathbf{x}_{k,t-1}),Q_{w}) & \text{if } A_{k,t}^n = 1\\
  \mathcal{N}(F_{2}(\mathbf{x}_{k,t-1}),Q_{w}) & \text{if } A_{k,t}^n = 2\\
  \mathcal{N}(F_{3}(\mathbf{x}_{k,t-1}),Q_{w}) & \text{if } A_{k,t}^n = 3\\
  \end{array} \right.
\end{eqnarray*}
The process noise used for estimation is such that $\sigma_x=\sigma_y=5\times10^{-2}$.
The mode transition probability matrix assumed by the filter for the target was
\begin{equation}
\pi_{ij}=\begin{bmatrix}
.8 &.1 &.1\\
.1 & .8 &.1 \\
.1 &.1 & .8
\end{bmatrix}
\end{equation}
The initial mode probability is assumed to be 
\begin{equation}
\pi_i(0)=\left\{
 \begin{array}{rl}
  1& \text{if } i=1\\
  0 & \text{if } i=2\\
  0 & \text{if } i=3\\
  \end{array} \right.
\end{equation}
The squared distance of the measurement with respect to the predicted measurement $d_k^2$, follows chi-square distribution with 2 degrees of freedom. The significance level used for the gating of hypotheses is $\alpha=0.01$. The chi-square critical value comes to be $\chi^2_\alpha=9.21$. A hypothesis is rejected if its chi-square statistics $d_k^2$ satisfies the relation $d_k^2>\chi^2_\alpha$. A total of $N=102$ particles were used. The simulation was carried for $100$ Monte Carlo runs and the estimates were obtained.
The true trajectories of the targets and their estimates of a single run are shown in Fig.\ref{MCMMJPDAF_cov_plot}. The ellipses indicate the 2-$\sigma$ region of the estimate covariances. The state estimates of the targets are shown in Fig.\ref{MCMMJPDAF_state}. The mean square error (MSE) of the position estimates are shown in Fig.\ref{MCMMJPDAF_MSE}. The results show that MC-MMJPDAF handles data association as well as target maneuver efficiently  . It had good track of the target states in all the Monte Carlo runs and there were no diverged track estimates. The missing measurements and clutters didn't have any significant effect in the estimates.

\section{Summary}
The proposed MC-MMJPDAF combines Multiple Model Particle Filter(MMPF) for highly maneuvering targets and Monte Carlo Joint Probabilistic Data Association Filter (MC-JPDAF) for multi-target tracking with data association uncertainty in presence of clutter and missed target measurements. The simulation results show that MC-MMJPDAF efficiently maintains good track of state estimates in maneuvering, multi-target tracking.

\begin{figure}[h]
\centering
\subfloat [Target 1]
{\includegraphics[scale=0.4]{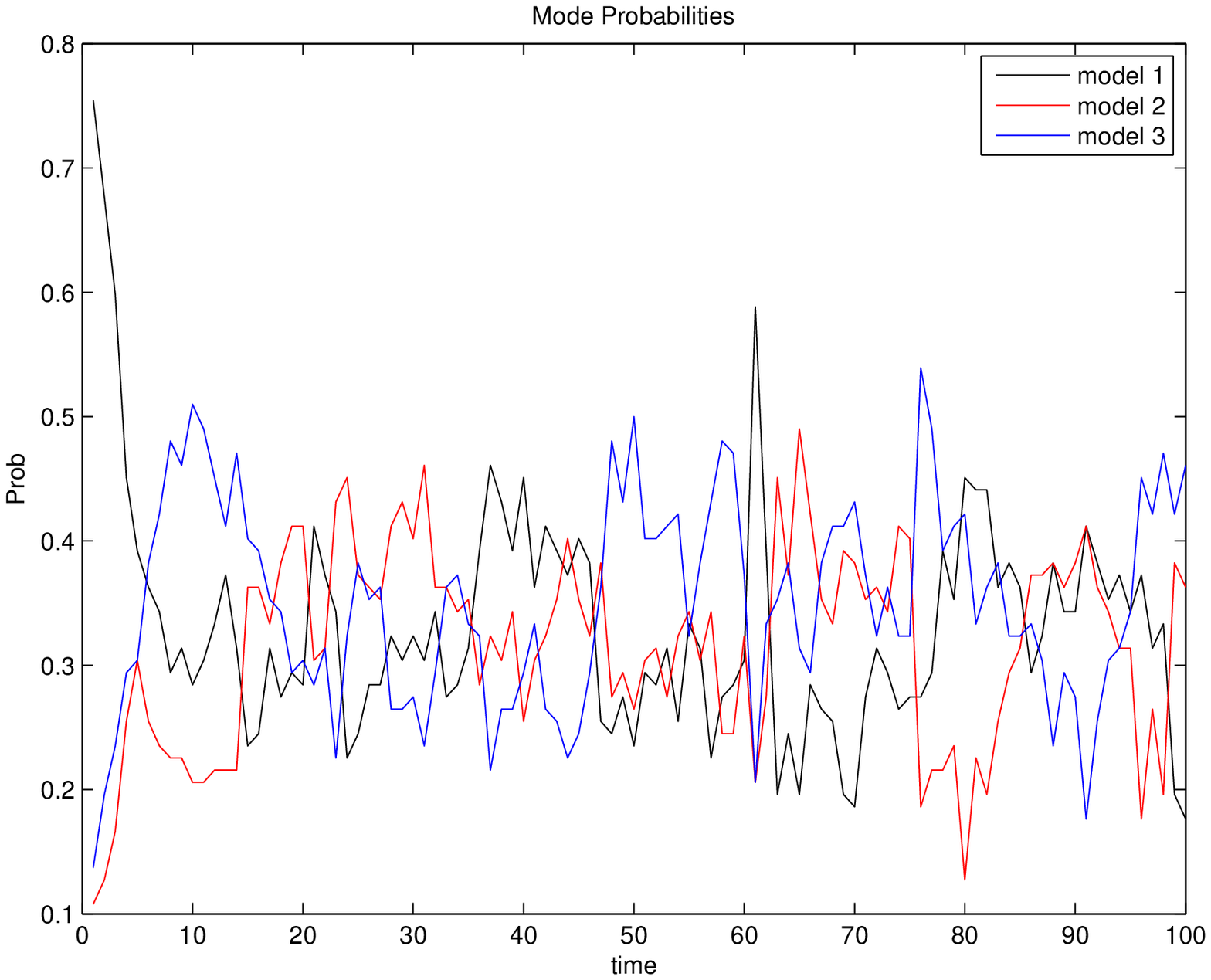}  
} 
\subfloat [Target 2]
{\includegraphics[scale=0.4]{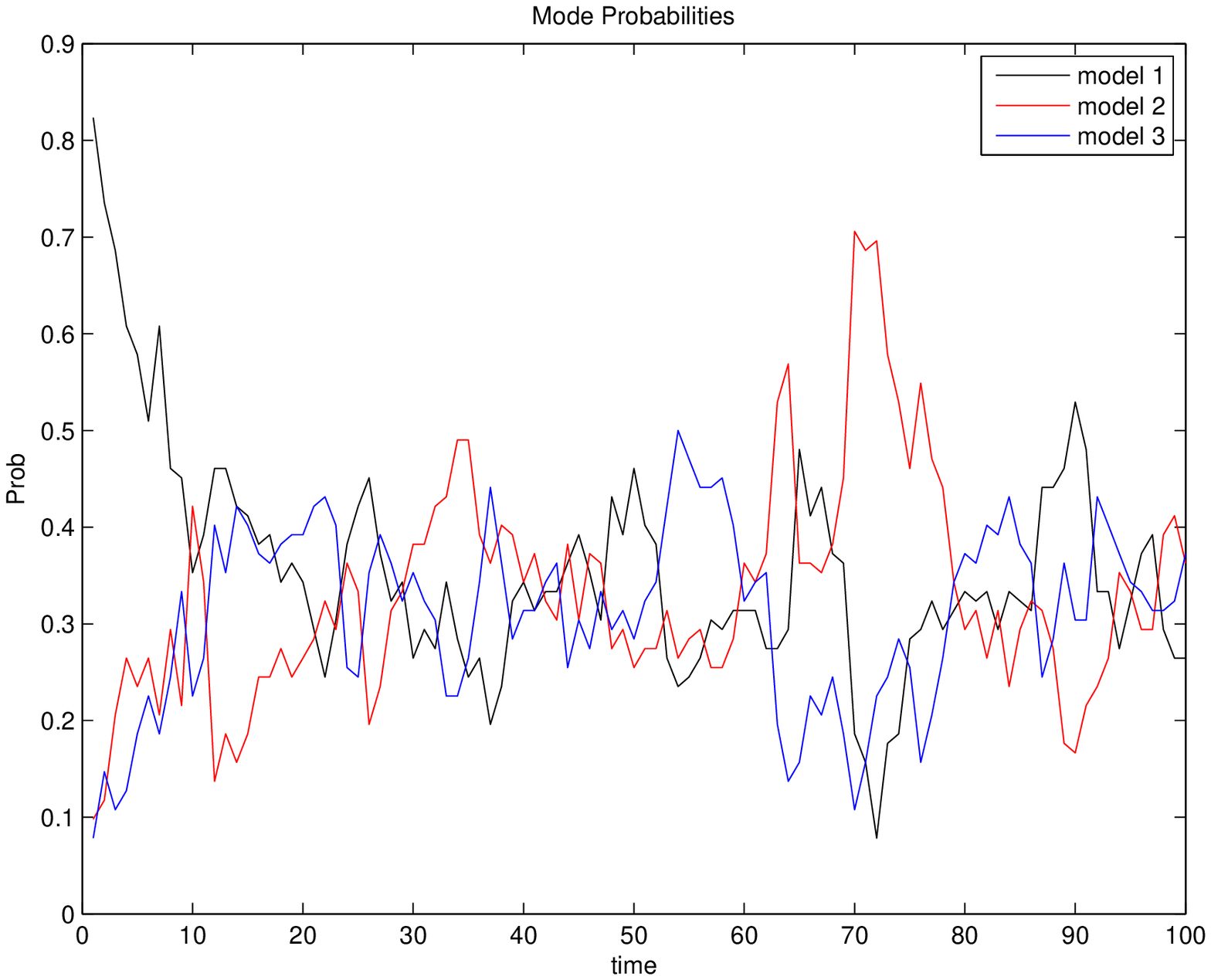}  
}\\   
\caption{Targets' mode estimate}
\end{figure}

\begin{figure}[h]
\centering
\subfloat [Sensor 1]
{\includegraphics[scale=0.4]{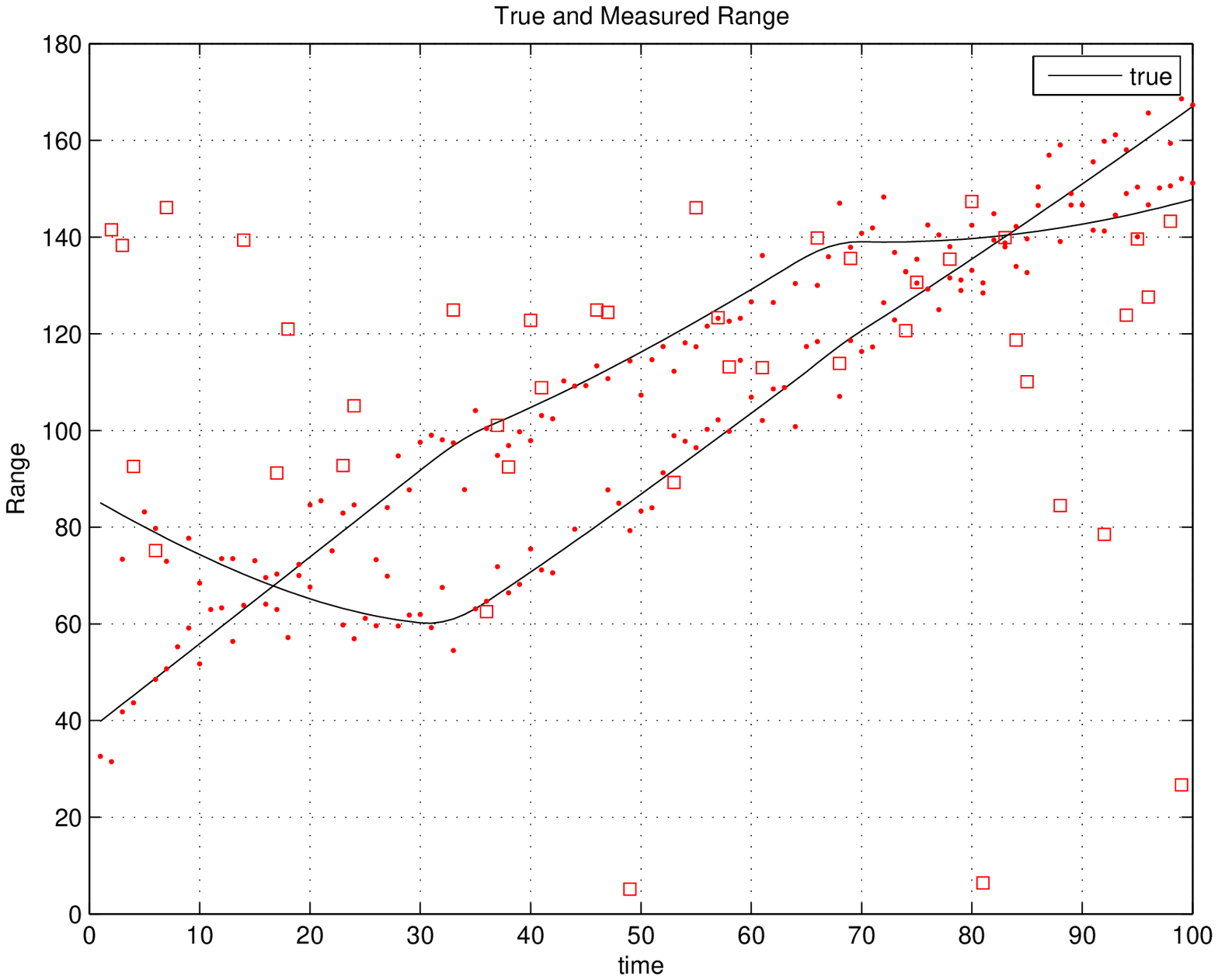}  
} 
\subfloat [Sensor 2]
{\includegraphics[scale=0.4]{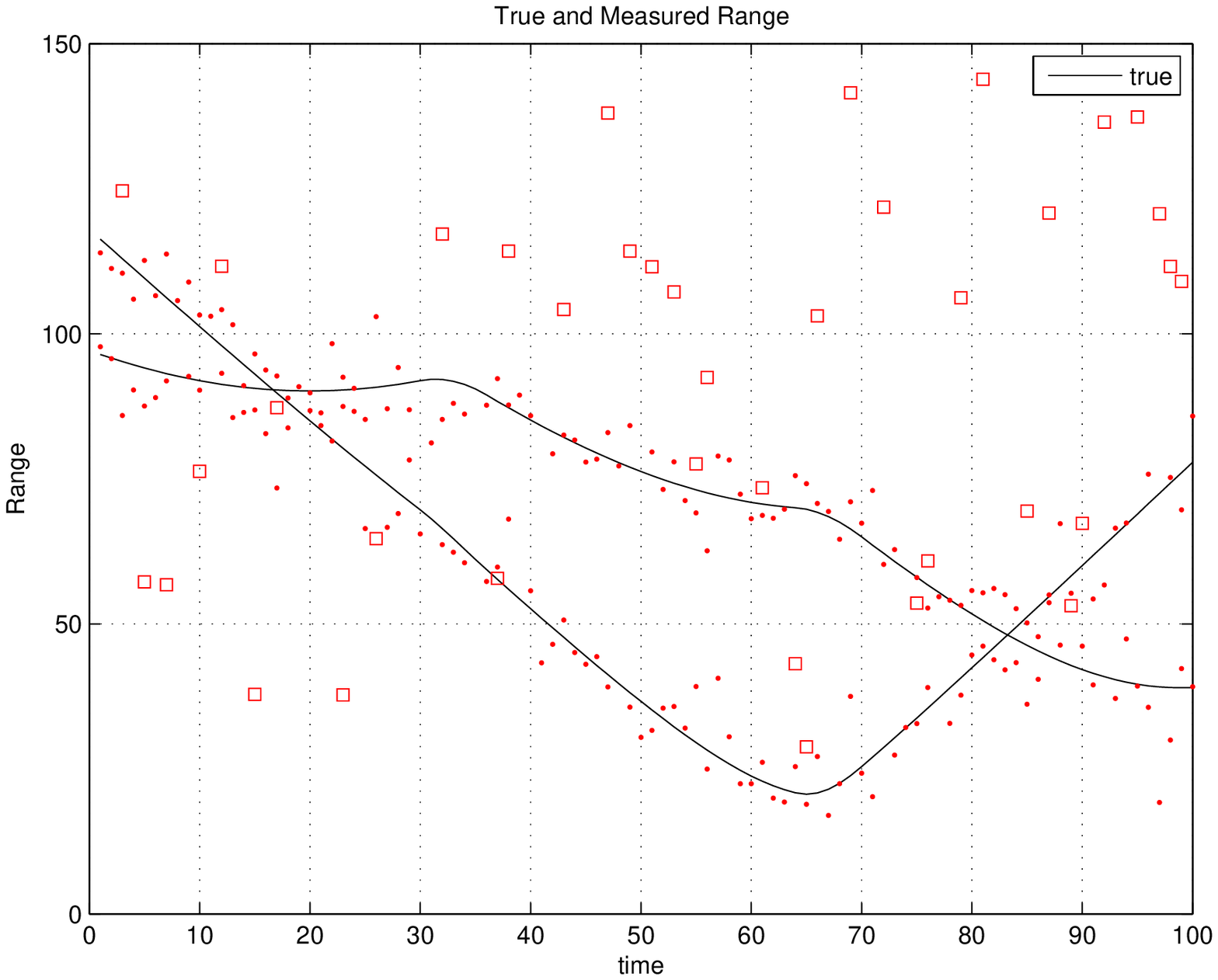}  
}\\   
\caption{Targets' true range and their measurements: The target measurements are shown in dots and the clutter measurements are shown in squares.}
\label{MCMMJPDAF_range_meas}
\end{figure}

\begin{figure}[h]
\centering
\subfloat [Sensor 1]
{\includegraphics[scale=0.4]{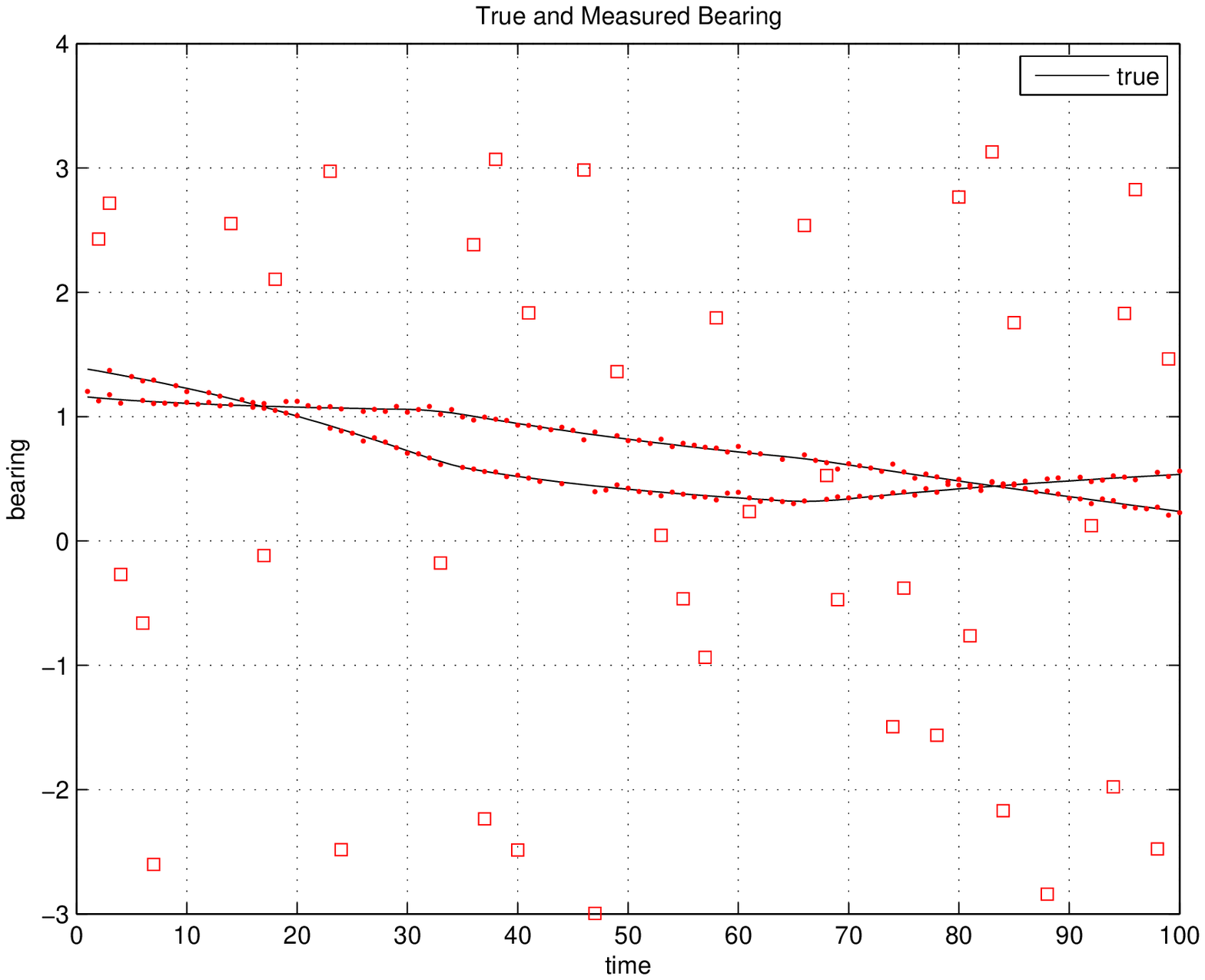}  
} 
\subfloat [Sensor 2]
{\includegraphics[scale=0.4]{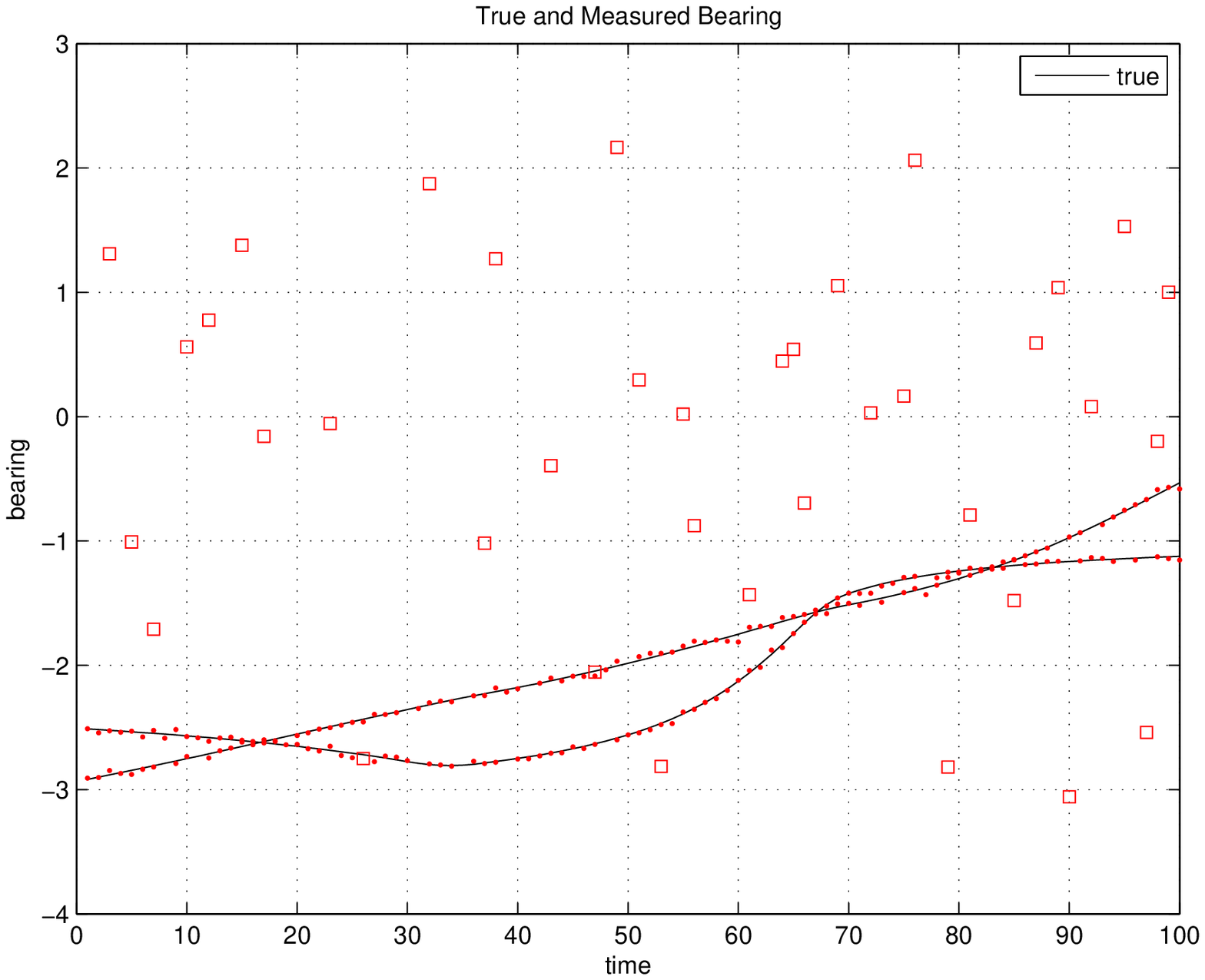}  
}\\   
\caption{Targets' true bearing and their measurements: The target measurements are shown in dots and the clutter measurements are shown in squares.}
\label{MCMMJPDAF_bearing_meas}
\end{figure}

 \begin{figure}[p]
 \centering 
   \includegraphics[scale=0.7]{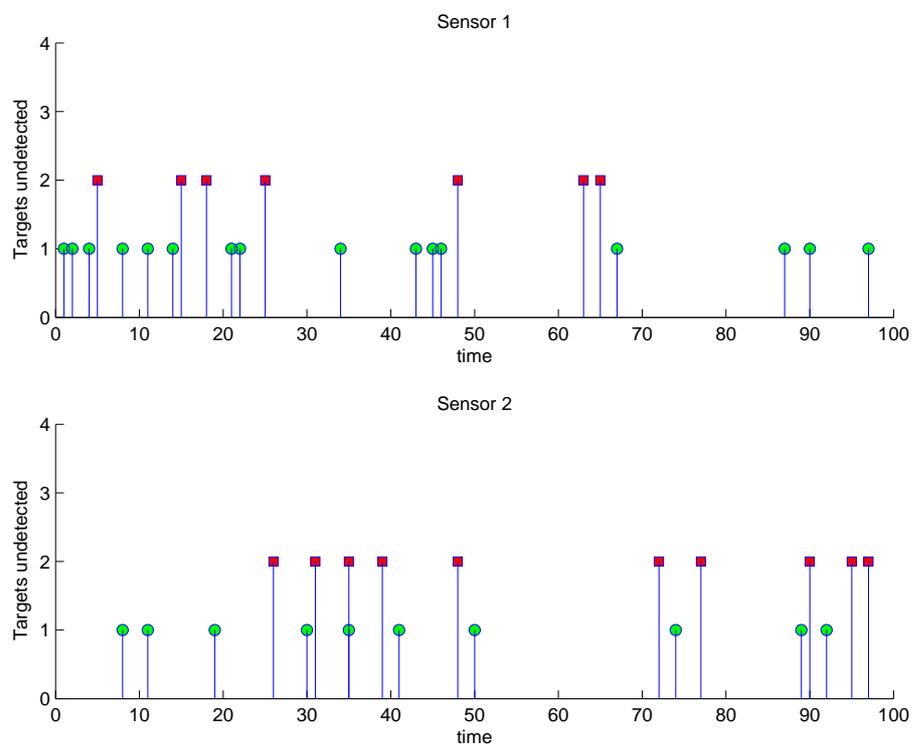} 
\caption{The index of the targets which were undetected at the sensors for a single run.}
\label{MCMMJPDAF_missed_targets}
 \end{figure}

\chapter{Conclusions}
Multi-target ttracking systems are non-linear and have Guassian distributions. Kalman filter (KF) based techniques rely on linear and Gaussian models for estimation. Their performance degrades as non-linearity becomes severe. Particle filter (PF) is an efficient numerical approximation method for the implementation of recursive Bayesian solution. It represents the probability distributions of target using particles and associated weights. It is capable of handling complex noise distributions and non-linearities in target's measurements and as well as target dynamics. The performance of particle filter depends on the number of particles and proposal distributions used. Simulations confirm that particle filter outperforms EKF at the expense of computational cost. The computations in particle filter are highly parellelizable and can be efficiently implemented in FPGA and GPU. 

In high dimensional sytems like multi-target tracking, the proportion of high likelihood particles are smaller and higher number of particles are required. Independent partition sampling and weighted resampling helps Indepedent Partition Particle Filter (IPPF) in better proposal of particles and hence track multiple targets with lesser number of particles.

For maneuvering multi-target tracking, Monte Carlo Multiple Model Joint Probabilistic Data Association Filter (MC-MMJPDAF) is proposed, which efficiently handles maneuvering multi-targets as well as data association uncertainity in presence of clutter measurements and missed target detections. It also incoorporates measurements from multiple observers. It combines Monte Carlo Joint Probabilistic Data Association Filter (MC-JPDAF) and Multiple Model Particle Filter (MMPF). Monte Carlo Joint Probabilistic Data Association Filter (MC-JPDAF) implements the standard Joint Probabilistic Data Association Filter (JPDA) for data association in particle filtering framework for non-linear and non-Gaussian systems. MC-JPDAF can track only slowly maneuvering targets and solves the data association efficiently. It requires smaller number of particles for estimation. MMPF is used for highly maneuvering targets. MMPF helps to track abrupt deviations in targets with less number of particles by incorporating multiple kinematic models. MMPF has better tracking capabilities than standard particle filter and Interacting Multiple Model Extended Kalman Filter (IMM-EKF). MC-MMJPDAF thus efficiently utilizes the multi-target data association capability of the MC-JPDA and the maneuvering target tracking capability of MMPF for tracking maneuvering multiple targets. The simulation results show that there were almost no diverged tracks with moderate clutter rate and target detection probability, and confirm the efficiency of the proposed technique. The particle filtering technique efficiently handles maneuvering, multi-target tracking, and has been verified with some field data.

\appendix
\begin{appendices} 
\chapter{Algorithm for sampling indices from a distribution}
\label{appendix:a}

Suppose there are $N$ particles with indices from $1$ to $N$, i.e.$\{\mathbf{x}^{(1)},\mathbf{x}^{(2)},\mathbf{x}^{(3)},....,\mathbf{x}^{(N)}\}$ and if it is required to sample $R$ particles from these given particles such that the distribution of the indices of these sampled particles follow a desired probability distribution $\rho(\cdotp)$, then any of the following two method can be used. The desired distribution $\rho(\cdotp)$ is specified using a set of indices from $1$ to $M$ and their corresponding weights $\rho^{(1)}, \rho^{(2)}, \rho^{(3)}, . . ., \rho^{(M)}$. The desired distribution function $\rho(\cdotp)$ is usually a function of the initial particles itself like their likelihood fuction or their cumulative distribution function. This technique is used in systematic resampling and weighted resampling of a set of particles.

\section{$O(NR)$ algorithm}
\begin{figure}[H]
 \centering 
   \includegraphics[scale=0.5]{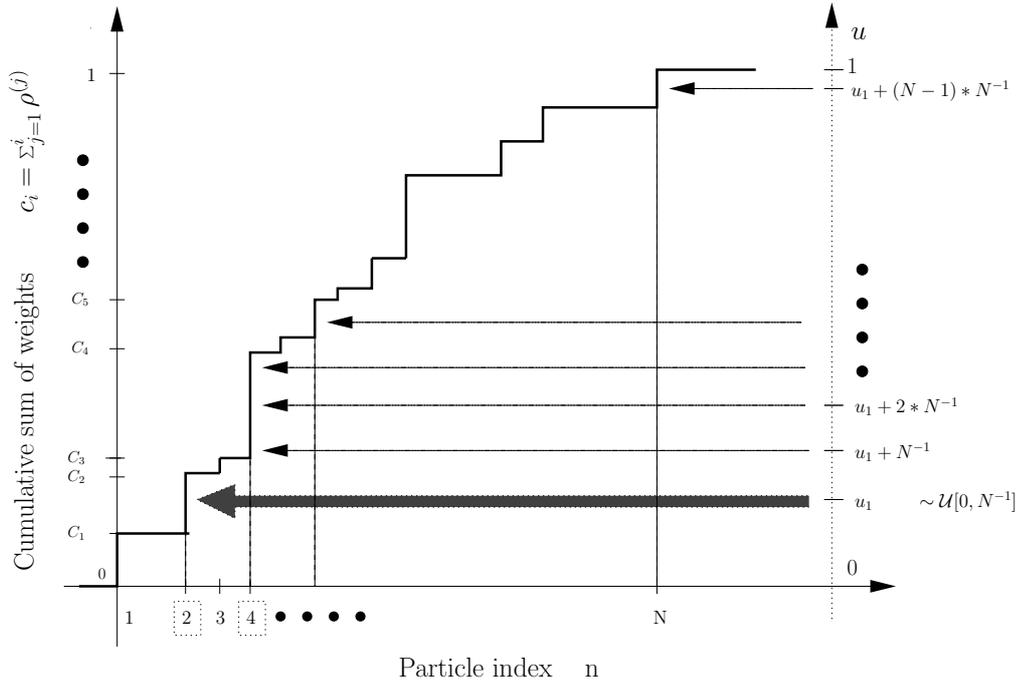}
\caption{Method 1: Generating indices from a given distribution}
\label{fig:Distribution_sampling_M}
 \end{figure}

Let $X$ have a probabilty distribution function $F_X(X)$. Given a uniform randon variable $Y$, the transformation $X\equiv F_X^{-1}(Y)$ will generate a random variable with probability distribution $F_X(X)$. This technique can be used to generate random variable with specified distributions from a uniform random variable. Hence this technique is used here to generate indices from $1$ to $M$ of distribution $\rho(\cdotp)$. This algorithm requires $R$ random number generations and $N$ comparisons in the worst case at every iteration. Hence the order of this algorithm is $O(NR)$. This algorithm has been derived from the algorithm described in \cite{4} for Regime Transition. The pseudo code of the algorithm is shown in Table.\ref{tab:Generate_Dist1}. The algorithm is illustrated in Fig. \ref{fig:Distribution_sampling_M}.
\begin{table}[H] 
\caption{Method 1: Generating indices from a given distribution} 
\centering          
\begin{tabular}{l}
  \hline
  \begin{minipage}{4in}
    \vskip 4pt
$[\{j(n)\}_{n=1}^R]=$Generate indices$[\{\rho^{(n)}\}_{n=1}^{N}, R]$
\begin{itemize}
	\item $c(0)=0$ 
	\item FOR $i=1:N$,
	\begin{itemize}
		\item $c(i)=c(i-1)+ \rho^{(i)}$ 		
	\end{itemize}
	\item END FOR
	\item FOR  $n=1:R$,
	\begin{itemize}
	      \item Draw $u_n\sim \mathcal{U} [0,1]$  		
	      \item m=1
	      \item WHILE $(c(m)<u_n)$
	      \begin{itemize}
		    \item$ m=m+1$
	      \end{itemize}
	      \item END WHILE
	      \item Set $j(n)=m$
	\end{itemize}
	\item END FOR
\end{itemize}
   \vskip 4pt
 \end{minipage}
 \\
  \hline
 \end{tabular}
\label{tab:Generate_Dist1} 
\end{table}

\section{$O(max(N,R))$ algorithm}
\begin{figure}[H]
 \centering 
   \includegraphics[scale=0.5]{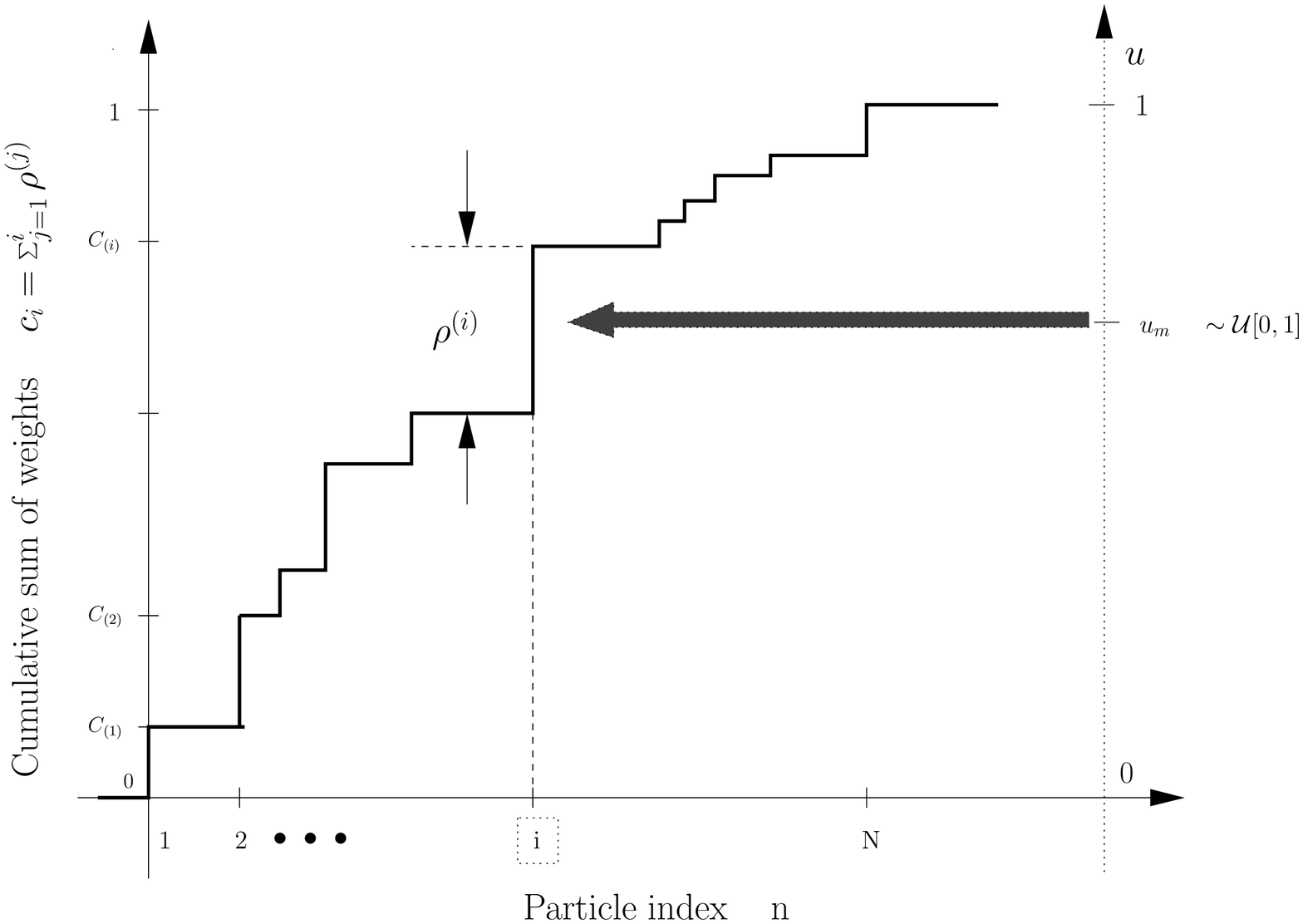}
\caption{Method 2: Generating indices from a given distribution}
\label{fig:Distribution_sampling_S}
 \end{figure}

This algorithm is based on the same principle of generating random variable with specified distributions from a uniform random variable as explained in the previous section. This method is simple to implement and relatively reduces the computational load. It requires $R+N$ comparisons and hence is of O(max(N,R)). This algorithm has been derived from the Systematic Resampling algorithm described in \cite{4}, for removing sample degeneracy in particle filters. The pseudo code of the algorithm is shown in Table.\ref{tab:Generate_Dist2}. The algorithm is illustrated in Fig. \ref{fig:Distribution_sampling_S}.
\begin{table}[H] 
\caption{Method 2: Generating indices from a given distribution} 
\centering          
\begin{tabular}{l}
  \hline
  \begin{minipage}{4in}
    \vskip 4pt
$[\{j(n)\}_{n=1}^R]=$Generate indices$[\{\rho^{(n)}\}_{n=1}^{N}, R]$
\begin{itemize}
	\item $c(0)=0$ 
	\item FOR $i=1:N$,
	\begin{itemize}
		\item $c(i)=c(i-1)+ \rho^{(i)}$ 		
	\end{itemize}
	\item END FOR
	\item Draw $u_1\sim \mathcal{U} [0, \frac{1}{R}]$  		
	\item m=1
	\item FOR  $n=1:R$,
	\begin{itemize}	      
	      \item $u_n=u_1+R^{-1}(n-1)$
	      \item WHILE $(c(m)<u_n)$
	      \begin{itemize}
		    \item$ m=m+1$
	      \end{itemize}
	      \item END WHILE
	      \item Set $j(n)=m$
	\end{itemize}
	\item END FOR
\end{itemize}
   \vskip 4pt
 \end{minipage}
 \\
  \hline
 \end{tabular}
\label{tab:Generate_Dist2} 
\end{table}

\end{appendices}

\end{document}